%% file: qcc_paper.tex
\documentstyle[aps]{revtex}
\begin{document}
\draft
\title{Characteristics of Quantum-Classical Correspondence for Two 
Interacting Spins}
\author{J. Emerson and L.E. Ballentine}
\address{Physics Department, Simon Fraser University, 
Burnaby, British Columbia, Canada V5A 1S6}
\date{\today}
\maketitle
\begin{abstract}
The conditions of quantum-classical correspondence for 
a system of two interacting spins are investigated. 
Differences between quantum expectation values and 
classical Liouville averages are examined for both   
regular and chaotic dynamics well beyond the 
short-time regime of narrow states. We find that quantum-classical 
differences initially grow exponentially with a characteristic 
exponent consistently larger than the largest Lyapunov exponent. 
We provide numerical evidence that the time of the break between 
the quantum and classical predictions scales as log(${\mathcal J}/ \hbar$), 
where ${\mathcal J}$ is a characteristic system action. However, this log 
break-time rule applies only while the quantum-classical deviations 
are smaller than ${\mathcal O}(\hbar)$.  
We find that the quantum observables remain well approximated 
by classical Liouville averages over long times 
even for the chaotic motions of a few degree-of-freedom system.  
To obtain this correspondence it is not necessary to  
introduce the decoherence effects of a many degree-of-freedom 
environment.  
\end{abstract}

\section{Introduction}
\label{sec1}

There is considerable interest in the interface between quantum and 
classical mechanics and the conditions that lead to the emergence 
of classical behaviour.  In order to characterize these conditions, 
it is important to differentiate 
two distinct regimes of 
quantum-classical correspondence \cite{Ball94}:   
\newline
\noindent
(i) Ehrenfest correspondence, in which 
the centroid of the wave packet
approximately follows a classical trajectory.  
\newline
\noindent
(ii) Liouville correspondence, in which the quantum probability
distributions are in approximate agreement with those of an 
appropriately constructed classical ensemble satisfying 
Liouville's equation. 

Regime (i) is relevant only when 
the width of the quantum state is small 
compared to the dimensions of the system; 
if the initial state is not narrow, this regime may be absent.
Regime (ii), which generally includes (i), applies to a much broader 
class of states, and this regime of correspondence may persist well 
after the Ehrenfest correspondence has broken down. 
The distinction between regimes (i) and (ii) has not always been 
made clear in the literature, though  
the conditions that delimit these two regimes, and in particular 
their scaling with system 
parameters, may be quite different. 

The theoretical study of quantum chaos 
has raised the question of whether the 
quantum-classical break occurs differently in chaotic states, in
states of regular motion, and in mixed phase-space systems.  
This is well understood only in the case of regime (i).  
There it is well-known \cite{BZ78,Haake87,Chi88}
that the time for a minimum-uncertainty wave packet to 
expand beyond the Ehrenfest regime scales as $\log({\mathcal J} / \hbar )$ 
for chaotic states, and as a power of ${\mathcal J} / \hbar $  
for regular states, where ${\mathcal J}$ denotes 
a characteristic system action.  

The breakdown of quantum-classical correspondence, 
in the case of regime (ii),  
is less well understood, though it has been argued that this regime 
may also be delimited by a $\log({\mathcal J}/ \hbar)$ break-time 
in classically chaotic states \cite{ZP94,Zurek98a}. 
Some numerical evidence in support of this conjecture has been 
reported in a study 
of the kicked rotor in the {\em anomolous diffusion} regime 
\cite{RBWG95}.
(On the other hand, in the regime of {\em quantum localization}, 
the break-time for the kicked rotor 
seems to scale as $({\mathcal J}/ \hbar)^2$ \cite{Haake91}.)
Since the $\log({\mathcal J}/ \hbar)$ time scale is rather short, 
it has been suggested 
that certain macroscopic objects would be predicted to exhibit 
non-classical behaviour 
on observable time scales \cite{ZP95a,Zurek98b}. 
These results highlight the importance of investigating  
the characteristics of quantum-classical correspondence in more detail. 

In this paper we study the classical and quantum dynamics of two
interacting spins.  This model is convenient because the Hilbert space of
the quantum system is finite-dimensional, and hence tractable for
computations.  
Spin models have been useful in the past for
exploring classical and quantum chaos \cite{Haake87,FP83,B91a,B91b,B93,RR98} 
and our model belongs to a class of spin models which show promise of 
experimental realization in the near future \cite{Mil99}. 
The classical limit is approached by taking the magnitude of
both spins to be very large relative to $\hbar$, while keeping their ratio 
fixed.  For our model 
a characteristic system action is given by ${\mathcal J} 
\simeq \hbar l$, where $l$ is a quantum number, 
and the classical limit is simply the limit of large 
quantum numbers, {\it i.e.} the limit $l \rightarrow \infty$. 

In the case of the chaotic dynamics for our model, we first show that 
the widths of both the quantum and classical states 
grow exponentially at a rate 
given approximately by the largest Lyapunov exponent (until saturation 
at the system dimension).  
We then show that the initially small quantum-classical differences 
also grow at an exponential rate, with an exponent $\lambda_{qc}$   
that is independent of the quantum numbers and at least twice 
as large as the largest Lyapunov exponent.  
We demonstrate how this exponential growth of differences 
leads to a log break-time rule,  
$t_b \simeq \lambda_{qc}^{-1} \ln( l p / \hbar )$, delimiting the regime 
of Liouville correspondence. The factor $p$, measured in units of $\hbar$,  
is some preset tolerance that defines a {\em break} between 
the quantum and classical expectation values.  
However, we also show that this logarithmic 
rule holds {\em only if} the tolerance $p$ for quantum-classical 
differences is chosen extremely small, 
in particular $p < {\mathcal O}(\hbar)$. 
For larger values of the tolerance, the break-time does not 
occur on this log time-scale and may not occur until the recurrence time.  
In this sense, log break-time rules describing Liouville correspondence 
are not robust.
These results 
demonstrate that, for chaotic states in the classical limit, 
quantum observables 
are described approximately 
by Liouville ensemble averages well beyond the 
Ehrenfest time-scale, after which both quantum and classical 
states have relaxed towards equilibrium distributions. 
This demonstration of correspondence  
is obtained for 
a few degree-of-freedom quantum system of coupled spins 
that is described by a 
pure state and subject only to unitary evolution.

This paper is organised as follows. In section II we describe 
the quantum and classical versions of our model. 
Since the model is novel we examine the behaviours of 
the classical dynamics in some detail.  
In section III we define the initial 
quantum states, which are SU(2) coherent states, 
and then define a corresponding classical density on the 
2-sphere which is a good analog for these  
states. We show in the Appendix that a perfect match is 
impossible: no distribution on ${\mathcal S}^2$ can reproduce the moments 
of the SU(2) coherent states exactly. In section IV we describe our numerical 
techniques. In section V we examine the quantum dynamics 
in regimes of classically chaotic and regular behaviour and 
demonstrate the close quantitative correspondence with the Liouville 
dynamics that persists well after the Ehrenfest break-time. 
In section VI we characterize the growth of 
quantum-classical differences in the time-domain. 
In section VII we characterize the scaling of the break-time 
for small quantum-classical differences 
and also examine the scaling of the maximum quantum-classical 
differences in the classical limit.

\section{The Model}
\label{sec2}

We consider the quantum and classical dynamics 
generated by a non-integrable model of two interacting spins, 
\begin{equation}
        H = a (S_z + L_z) + c S_x L_x \sum_{n = -\infty}^{\infty}\delta(t-n) 
\label{eqn:ham}
\end{equation}
where ${\bf S} = (S_x,S_y,S_z)$ and 
${\bf L} = (L_x,L_y,L_z)$.
The first two terms in (\ref{eqn:ham}) 
correspond to simple rotation of both spins 
about the $z$-axis. 
The sum over coupling terms describes an infinite 
sequence of $\delta$-function interactions 
at times $t=n$ for integer n. 
Each interaction term corresponds to an impulsive rotation 
of each spin about the $x$-axis by an angle proportional 
to the $x$-component of the other spin. 

\subsection{The Quantum Dynamics}


To obtain the quantum dynamics we interpret 
the Cartesian components of the spins 
as operators satisfying the usual angular momentum 
commutation relations,  
\begin{eqnarray*}
  & [ S_i,S_j ] = & i \epsilon_{ijk} S_k \\
  & [ L_i,L_j ] = & i \epsilon_{ijk} L_k \\ 
  & [ J_i,J_j ] = & i \epsilon_{ijk} J_k .
\end{eqnarray*}
In the above we have set $\hbar =1$ and introduced the total 
angular momentum vector ${\bf J} = {\bf S} + {\bf L}$.

The Hamiltonian (\ref{eqn:ham}) possesses kinematic 
constants of the motion, $ [ {\bf S}^2,H] = 0$ and $[ {\bf L}^2,H] =0 $,  
and the total state vector $|\psi\rangle$ 
can be represented in a finite Hilbert space of  
dimension $ (2s+1) \times (2l+1)$. 
This space  is spanned by the orthonormal 
vectors $|s,m_s \rangle \otimes |l,m_l \rangle$ 
where $m_s \in \{ s, s-1, \dots, -s \}$ 
and $m_l \in \{ l, l-1, \dots, -l \}$.  
These are the joint eigenvectors of the four spin operators 
\begin{eqnarray}
\label{eqn:basis}
	{\bf S}^2 |s, l, m_s,m_l \rangle & =& s(s+1) 
| s,l,m_s,m_l \rangle  \nonumber \\
	S_z | s, l,m_s,m_l \rangle & =& m_s 
| s,l,m_s,m_l \rangle  \\
	{\bf L}^2 | s,l,m_s,m_l \rangle &= & l(l+1) 
| s,l,m_s,m_l \rangle  \nonumber \\
	L_z | s,l,m_s,m_l \rangle &= & m_l 
| s,l,m_s,m_l \rangle  \nonumber .
\end{eqnarray}

The periodic sequence of interactions introduced by the 
$\delta$-function produces a quantum mapping. 
The time-evolution 
for a single iteration, from just before a kick to just before 
the next, is produced by the unitary transformation, 
\begin{equation}
\label{eqn:qmmap}
	| \psi(n+1) \rangle = F \; | \psi(n) \rangle,  
\end{equation}
where $F$ is the single-step Floquet operator, 
\begin{equation}
\label{eqn:floquet}
        F = \exp \left[ - i a (S_z + L_z) \right] 
                \exp \left[ - i c S_x L_x \right].  
\end{equation}
Since $a$ is a rotation its range is $2\pi$ radians. 
The quantum dynamics are thus specified by two 
parameters, $a$ and $c$, 
and two quantum numbers, $s$ and $l$. 

An explicit representation of the 
single-step Floquet operator can be obtained in the basis 
(\ref{eqn:basis}) by first re-expressing the 
interaction operator in (\ref{eqn:floquet})  
in terms of rotation operators, 
\begin{eqnarray}
 \exp \left[ - i c S_x \otimes L_x \right] & =  & 
[R^{(s)}(\theta,\phi) \otimes  R^{(l)}(\theta,\phi)] \;  
\exp \left[ - i c S_z \otimes L_z \right]	\nonumber \\
& &  \times [R^{(s)}(\theta,\phi) \otimes R^{(l)}(\theta,\phi)]^{-1}, 
\end{eqnarray}
using polar angle $\theta=\pi/2$ and azimuthal angle $\phi=0$. 
Then the only non-diagonal terms arise in the expressions  
for the rotation matrices, which take the form,   
\begin{equation}
	\langle j,m'| R^{(j)}(\theta,\phi) | j,m \rangle = \exp(-i m' \phi )  
		d^{(j)}_{m',m}(\theta).
\end{equation}
The matrix elements, 
\begin{equation}
d^{(j)}_{m',m}(\theta) = \langle j,m'| \exp( - i \theta J_y ) | j,m \rangle
\end{equation}
are given explicitly by Wigner's formula \cite{sakurai}.

We are interested in studying the different time-domain 
characteristics of quantum observables when the corresponding 
classical system exhibits either regular or chaotic dynamics. 
In order to compare quantum systems with different quantum numbers   
it is convenient to normalize subsystem observables 
by the subsystem magnitude 
 $\sqrt{\langle {\bf L}^2\rangle} = \sqrt{l(l+1)}$. 
We denote such normalized observables with a tilde, where
\begin{equation}  
	\langle \tilde{L}_z (n) \rangle  =  
{ \langle \psi(n)| L_z  |\psi(n) \rangle  \over \sqrt{l(l+1)} }
\end{equation}
and the normalized variance at time $n$ is defined as, 
\begin{equation}  
	\Delta {\tilde {\bf L}  }^2 (n)   
	= {\langle  {\bf L}^2 \rangle - 
\langle  {\bf L }(n) \rangle^2 \over l(l+1) } .  
\end{equation}

We are also interested in evaluating the properties 
of the quantum probability distributions. The 
probability distribution 
corresponding to the observable $L_z$ is given by the trace, 
\begin{equation}
P_z(m_l) = {\mathrm Tr} \left[ \rho^{(l)}(n) | l, m_l \rangle 
\langle l, m_l | \right] = \langle l,m_l |  \rho^{(l)}(n) |  l,m_l\rangle , 
\end{equation}
where $\rho^{(l)}(n)= {\mathrm Tr}^{(s)} \left[ \; | \psi(n) \rangle 
\langle \psi(n) | \; | s, m_s \rangle 
\langle s, m_s | \; \right]$ is the reduced state operator for 
the spin ${\bf L}$ at time $n$ and ${\mathrm Tr}^{(s)}$ denotes 
a trace over the factor space corresponding to the spin ${\bf S}$. 


\subsection{Classical Map} 

For the Hamiltonian (\ref{eqn:ham}) the corresponding 
classical equations of motion 
are obtained by interpreting the angular momentum components 
as dynamical variables satisfying,   
\begin{eqnarray*}
  & \{ S_i,S_j \} = & \epsilon_{ijk} S_k \\
  & \{ L_i,L_j \} = & \epsilon_{ijk} L_k \\ 
  & \{ J_i,J_j \} = & \epsilon_{ijk} J_k ,
\end{eqnarray*}
with $\{ \cdot,\cdot \}$ denoting the Poisson bracket. 
The periodic $\delta$-function in the coupling term 
can be used to define surfaces at $t=n$, for integer $n$, 
on which the time-evolution reduces to a stroboscopic mapping,  
\begin{eqnarray}
\label{eqn:map}
        \tilde{S}_x^{n+1} & = & \tilde{S}_x^n \cos( a) - 
        \left[ \tilde{S}_y^n \cos( \gamma r \tilde{L}_x^n ) - 
        \tilde{S}_z^n \sin( \gamma r 
        \tilde{L}_x^n)\right] \sin( a), \nonumber \\ 
        \tilde{S}_y^{n+1} & = & \left[ \tilde{S}_y^n \cos( 
        \gamma r \tilde{L}_x^n) - \tilde{S}_z^n \sin( \gamma r
        \tilde{L}_x^n ) \right] \cos( a) + 
        \tilde{S}_x^n \sin( a), \nonumber \\
        \tilde{S}_z^{n+1} & = &  \tilde{S}_z^n \cos( 
        \gamma r \tilde{L}_x^n) + \tilde{S}_y^n \sin( 
        \gamma r \tilde{L}_x^n), \\
        \tilde{L}_x^{n+1} & = & \tilde{L}_x^n \cos( a) - 
        \left[\tilde{L}_y^n\cos(\gamma \tilde{S}_x^n) - \tilde{L}_z^n 
        \sin(\gamma\tilde{S}_x^n )\right]\sin(a), \nonumber \\ 
        \tilde{L}_y^{n+1} & = & \left[ \tilde{L}_y^n \cos( \gamma 
        \tilde{S}_x^n) - \tilde{L}_z^n \sin( \gamma
        \tilde{S}_x^n) \right] \cos( a) + 
        \tilde{L}_x^n \sin( a), \nonumber \\
        \tilde{L}_z^{n+1} & = &  \tilde{L}_z^n \cos( 
        \gamma \tilde{S}_x^n )+ \tilde{L}_y^n \sin( \gamma 
        \tilde{S}_x^n), \nonumber 
\end{eqnarray}
where ${\tilde {\bf L}} = {\bf L} / |{\bf L}|$ , 
${\tilde {\bf S}} = {\bf S} / |{\bf S}|$   
and we have introduced the parameters 
$ \gamma = c |{\bf S}| $ and $ r = | {\bf L}| / | {\bf S}| $. 
The mapping equations (\ref{eqn:map}) describe 
the time-evolution of (\ref{eqn:ham}) 
from just before one kick to just before 
the next. 

Since the magnitudes of both spins are conserved, 
$ \{ {\bf S}^2,H \} = \{ {\bf L}^2,H \} =0$, 
the motion is actually confined to the four-dimensional 
manifold ${\mathcal P} ={\mathcal S}^2 \times {\mathcal S}^2$, 
which corresponds to the surfaces of two spheres. 
This is manifest when the mapping (\ref{eqn:map}) 
is expressed in terms of the four {\it canonical} coordinates 
${\bf x} = (S_z, \phi_s, L_z , \phi_l )$, where 
$\phi_s = \tan (S_y /S_x)$ and $\phi_l = \tan(L_y/L_x) $. 
We will refer to the mapping (\ref{eqn:map}) in canonical form 
using the shorthand notation ${\bf x}^{n+1} = {\bf F}({\bf x}^n)$. 
It is also useful to introduce a complete set of spherical coordinates 
$ \vec{\theta} = (\theta_s,\phi_s,\theta_l,\phi_l) $ where 
$\theta_s = \cos^{-1} (S_z / |{\bf S}|) $ and 
$\theta_l = \cos^{-1} (L_z / |{\bf L}|) $.

The classical flow (\ref{eqn:map}) on the reduced surface  
${\mathcal P}$ still has a rather large 
parameter space; the dynamics are  
determined from three independent dimensionless 
parameters: $a \in [0,2\pi)$, $\gamma \in (-\infty,\infty)$, 
and $r \ge 1$. 
The first of these, $a$, controls the angle of 
free-field rotation about the $z$-axis. 
The parameter $\gamma= c |{ \bf S }|$ 
is a dimensionless coupling strength 
and $r = |{\bf L}|/ |{ \bf S }| $ corresponds 
to the relative magnitude of the two spins.

We are particularly interested in the effect of increasing  
the coupling strength $\gamma$ for different fixed values of $r$. 
In Fig.\ \ref{regimes} 
we plot the dependence of the classical behaviour 
on these two parameters for the case $a=5$, which produces typical results. 
The data in this figure 
was generated by randomly sampling initial conditions on
${\mathcal P}$, using the canonical measure,
\begin{equation}
\label{eqn:measure}
d \mu( {\bf x} ) = d \tilde{S}_z d \phi_s d \tilde{L}_z d \phi_l ,
\end{equation}
and then calculating the largest Lyapunov exponent associated 
with each trajectory. Open circles 
correspond to regimes where at least $99\%$ of the 
initial conditions were found to exhibit regular behaviour 
and crosses correspond to regimes where at least $99\%$ of 
these randomly sampled initial conditions were found to exhibit 
chaotic behaviour.
Circles with crosses through them (the superposition of 
both symbols) correspond to regimes with a mixed phase space. 
For the case $a=5$ and with $r$ held constant, 
the scaled coupling strength $\gamma$ plays the role of a   
perturbation parameter: 
the classical behaviour varies from regular, to mixed, 
to predominantly chaotic as $|\gamma|$ is increased from zero.

The fixed points of the classical map (\ref{eqn:map}) provide 
useful information about the parameter dependence of the classical 
behaviour and, more importantly, in the case of mixed regimes, help   
locate the zones of regular behaviour in the 4-dimensional phase space. 
We find it sufficient to consider only 
the four trivial (parameter-independent) fixed points 
which lie at the poles along the $z$-axis: 
two of these points correspond to parallel spins,  
$(S_z,L_z) = \pm (|{\bf S}|,|{\bf L}|)$,  
and the remaining two points correspond to anti-parallel spins, 
$(S_z,L_z) = (\pm |{\bf S}|, \mp|{\bf L}|)$. 

The stability around these fixed points can be determined from the 
eigenvalues of the tangent map matrix,
${\bf M} = \partial {\bf F} / \partial {\bf x}$, 
where all derivatives are evaluated at the fixed point of interest. 
(It is easiest to derive $M$  
using the six {\it non-canonical} mapping equations (\ref{eqn:map})
since the tangent map for the {\it canonical} mapping equations 
exhibits a coordinate system singularity at these fixed points.) 
The eigenvalues corresponding to  
the four trivial fixed points are obtained from  
the characteristic equation, 
\begin{equation}
	[ \xi^2 - 2 \xi \cos a  + 1 ]^2 
	\pm \xi^2 \gamma^2 r \sin^2 a = 0, 
\end{equation}
with the minus (plus) sign corresponding to the parallel 
(anti-parallel) cases and we have suppressed the trivial 
factor $(1 -\xi)^2$ which arises since the six equations (\ref{eqn:map})  
are not independent.  
For the parallel fixed points we have the four eigenvalues, 
\begin{eqnarray}
\label{eqn:parallel} 
	\xi_{1,2}^P  & = & \cos a  \pm {1 \over 2} \sqrt{r \gamma^2 \sin^2 a} 
	+ {1 \over 2} 
	\sqrt{ \pm 4 \cos a  \sqrt{ \gamma^2 r \sin^2 a}  
	- \sin^2 a (4 - \gamma^2 r) }, \nonumber \\ 
	\xi_{3,4}^P  & = & \cos a  \pm {1 \over 2} \sqrt{r \gamma^2 \sin^2 a} 
	- {1 \over 2} 
	\sqrt{ \pm 4 \cos a  \sqrt{ \gamma^2 r \sin^2 a}  
	- \sin^2 a (4 - \gamma^2 r) },  
\end{eqnarray}
and the eigenvalues for the anti-parallel cases, $\xi^{AP}$, are obtained 
from (\ref{eqn:parallel}) 
through the substitution $r \rightarrow -r $. 
A fixed point becomes unstable if and only if $|\xi| > 1$ for 
at least one of the four eigenvalues. 

\subsubsection{Mixed Phase Space: $\gamma = 1.215$}

\begin{figure}
\input{fig1}
\vspace{0.2in}
\caption{Stroboscopic trajectories on the unit sphere launched from 
a regular zone of the mixed regime with   $\gamma=1.215$, 
$r=1.1$, $a=5$ and $\vec{\theta}(0) = (5^o,5^o,5^o,5^o)$.
}
\label{r1.1.ric2}
\end{figure}
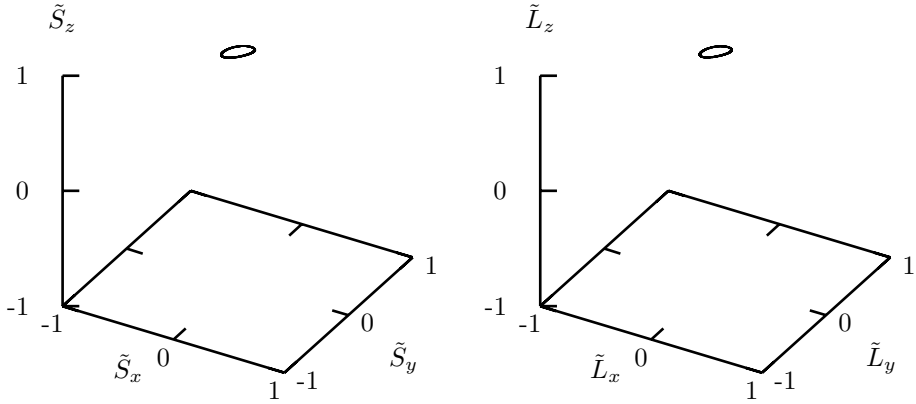

\begin{figure}
\input{fig2}
\vspace{0.2in}
\caption{
Same parameters as Fig.\ \ref{r1.1.ric2} 
but the trajectory is launched from a chaotic zone of the mixed 
regime with 
initial condition $\vec{\theta}(0) = (20^o,40^o,160^o,130^o)$.
}
\label{r1.1.cic2}
\end{figure}
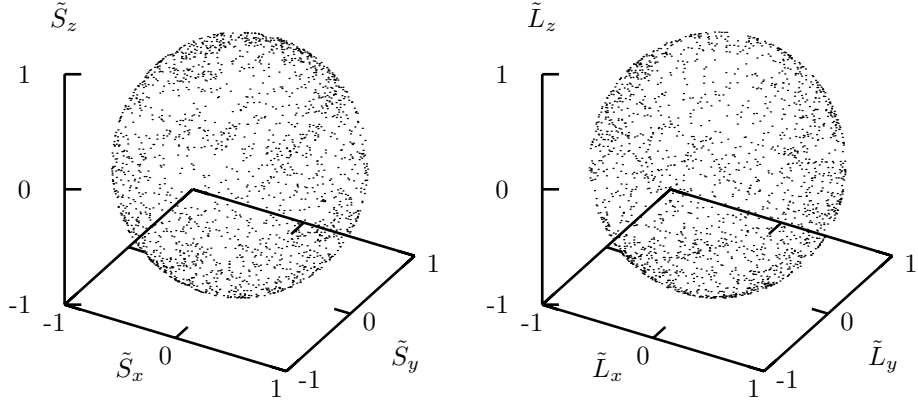

We are particularly interested in the behaviour of this model 
when the two spins are comparable in magnitude.
Choosing the value $r=1.1$ (with $a=5$ as before),  
we determined by 
numerical evaluation that the anti-parallel 
fixed points are unstable for $|\gamma| > 0$.  
In the case of the parallel fixed points, 
all four eigenvalues remain on the unit circle, $|\xi^{P}| =  1$, 
for $|\gamma| < 1.42$. This stability condition guarantees the presence 
of regular islands about the parallel fixed points \cite{LL}.  
In Fig.\ \ref{r1.1.ric2} we plot  
the trajectory corresponding to the parameters $a=5$, $r=1.1$, 
$\gamma=1.215$ and with initial 
condition $\vec{\theta}(0) =(5^o,5^o,5^o,5^o)$
which locates the trajectory  
near a stable fixed point 
of a mixed phase space (see Fig.\ \ref{regimes}.)
This trajectory clearly exhibits a periodic pattern 
which we have confirmed to be regular by computing 
the associated Lyapunov exponent ($\lambda_L=0$). In contrast, 
the trajectory plotted in Fig.\ \ref{r1.1.cic2} is launched 
with the same parameters but with initial condition 
$\vec{\theta}(0) = (20^o,40^o,160^o,130^o)$, which is close to one  
of the unstable anti-parallel fixed points.  This trajectory 
explores a much larger portion of the 
surface of the two spheres in a seemingly random manner. 
As expected, a computation of the largest 
associated Lyapunov exponent yields a positive number ($\lambda_L = 0.04$). 

\subsubsection{Global Chaos: $\gamma = 2.835$}

If we increase the coupling strength to the value $\gamma =2.835 $, 
with $a=5$ and $r=1.1$ as before, then 
all four trivial fixed points become unstable. 
By randomly sampling ${\mathcal P}$ with $3 \times 10^4$ 
initial conditions we find that less than $0.1$\% of 
the kinematically accessible surface ${\mathcal P}$ is covered with 
regular islands (see Fig.\ \ref{regimes}). 
This set of parameters produces 
a connected chaotic zone with 
largest Lyapunov exponent $\lambda_L=0.45$. 
We will refer to this type of regime as one of `global chaos' 
although the reader should note that our usage of this expression 
differs slightly from that in \cite{LL}. 

\subsubsection{The Limit $r \gg 1$}

Another interesting limit of our model arises when one of the spins 
is much larger than the other, $r \gg 1$. We expect that 
in this limit the larger spin (${\bf L}$) will act 
as a source of essentially external `driving' for the 
smaller spin (${\bf S}$). 
Referring to the coupling terms in the 
mapping (\ref{eqn:map}), 
the `driving' strength, or perturbation 
upon ${\bf S}$ from ${\bf L}$, is 
determined from the product $\gamma r = c |{\bf L}|$, 
which can be quite large, whereas 
the `back-reaction' strength, or perturbation upon ${\bf L}$ from ${\bf S}$, 
is governed only by the 
scaled coupling strength $\gamma = c|{\bf S}|$, which can be quite small. 
It is interesting to examine whether a dynamical regime exists 
where the larger system might approach regular behaviour while 
the smaller `driven' system is still subject to chaotic motion. 

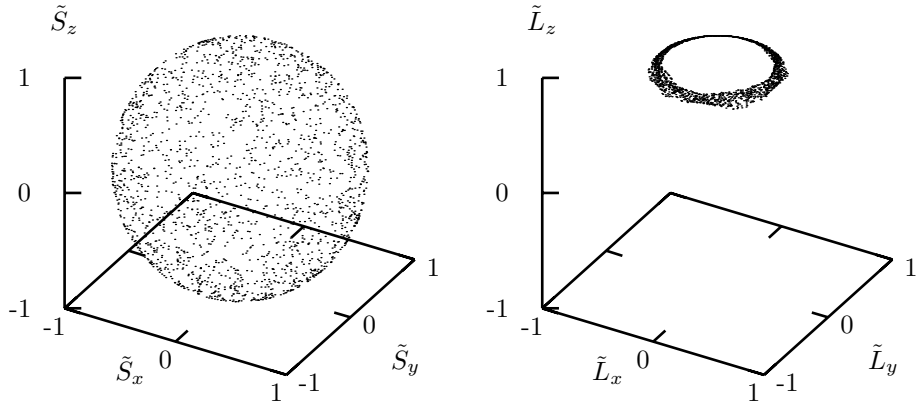
\begin{figure}
\input{fig3}
\vspace{0.2in}
\caption{A chaotic trajectory for mixed regime parameters  
 $\gamma=0.06$, $r=100$, and $a=5$ with  
$\vec{\theta}(0) = (27^o,27^o,27^o,27^o)$.
The motion of the larger spin appears to remain 
confined to a narrow band on the surface of the 
sphere. 
}
\label{r100.cic3}
\end{figure}

In Fig.\ \ref{r100.cic3} we plot a 
chaotic trajectory for $r=100$ with initial 
condition $\vec{\theta}(0) = (27^o,27^o,27^o,27^o)$
which is located in a chaotic zone ($\lambda_L= 0.026$) 
of a mixed phase space (with $a=5$ and $\gamma=0.06$). 
Although the small spin wanders 
chaotically over a large portion of its kinematically 
accessible shell ${\mathcal S}^2$, the motion of the large spin remains 
confined to a `narrow' band.  Although the 
band is narrow relative to the 
large spin's length, it is not small relative to the smaller spin's 
length. The trajectories are both plotted on 
the unit sphere, so the effective area explored by the 
large spin (relative to the effective area covered 
by the small spin) scales in proportion to $r^2$. 

\subsection{The Liouville Dynamics}

We are interested in comparing the 
quantum dynamics generated by (\ref{eqn:qmmap}) with 
the corresponding Liouville dynamics of a classical distribution. 
The time-evolution of a Liouville density is generated by the 
partial differential equation,    
\begin{equation}
\label{eqn:liouville}
{ \partial \rho_c({\bf x},t) \over \partial t } = - \{ \rho_c , H \},   
\end{equation}
where $H$ stands for the Hamiltonian (\ref{eqn:ham})  
and ${\bf x} = (S_z, \phi_s,L_z,\phi_l)$. 

The solution to (\ref{eqn:liouville}) can be expressed in 
the compact form, 
\begin{equation}
\label{eqn:soln}
	\rho_c({\bf x},t) =   
\int_{\mathcal P} {\mathrm d} \mu({\bf y}) \;
\delta({\bf x} - {\bf x}(t,{\bf y})) \; 
\rho_c({\bf y},0), 
\end{equation}
with measure $d \mu({\bf y})$ given by (\ref{eqn:measure}) and 
each time-dependent function ${\bf x}(t,{\bf y}) \in {\mathcal P}$ is  
solution of the equations of motion for (\ref{eqn:ham}) 
with initial condition$ {\bf y} \in {\mathcal P}$. 
This integral solution (\ref{eqn:soln}) simply expresses that 
Liouville's equation (\ref{eqn:liouville}) describes the dynamics 
of a classical density $\rho_c({\bf x},t)$ of points  
evolving in phase space under the Hamiltonian flow. 
We exploit this fact to numerically solve 
(\ref{eqn:liouville}) by 
randomly generating initial conditions  
consistent with an initial phase space 
distribution $\rho_c({\bf x},0)$ and 
then time-evolving each of these initial conditions 
using the equations of motion (\ref{eqn:map}). 
We then calculate the ensemble averages 
of dynamical variables,  
\begin{equation}
        \langle \tilde{L}_z(n) \rangle_c = 
\int_{\mathcal P} {\mathrm d} \mu({\bf x}) { L_z \over |{\bf L}| } \rho_c({\bf x},n).
\label{eqn:Lave}
\end{equation}
by summing over this distribution of trajectories at each time step.

\subsection{Correspondence Between Quantum and Classical Models}

For a quantum system specified by the four numbers $\{a,c,s,l\}$,  
the corresponding classical parameters $\{a,\gamma,r\}$ 
are determined if we associate the magnitudes of the 
classical angular momenta with the quantum spin magnitudes,  
\begin{eqnarray}
		|{\bf S}|_c & = & \sqrt{s(s+1)} \nonumber \\
		| {\bf L}|_c & = & \sqrt{l(l+1)}.
\end{eqnarray}
This prescription produces the classical parameters,
\begin{eqnarray}
		r & = & \sqrt{l(l+1) \over s(s+1)} \nonumber \\  
		\gamma & = & c \sqrt{s(s+1)}, 
\end{eqnarray}
with $a$ the same number for both models. 

We are interested in determining the behaviour of 
the quantum dynamics in the limit $s \rightarrow \infty$ 
and $l \rightarrow \infty$. This is accomplished by 
studying sequences of quantum models with $s$ and $l$ increasing though 
chosen such that the classical $r$ and $\gamma$ are held fixed. 
Since $s$ and $l$ 
are restricted to integer (or half-integer) values, the corresponding 
classical $r$ will actually vary slightly for each member of this sequence
(although $\gamma$ can be matched exactly by varying the quantum parameter 
$c$).  
In the limit $s \rightarrow \infty$ and $l \rightarrow \infty$ 
this variation becomes increasingly small since 
 $r= \sqrt{l(l+1)/s(s+1)} \rightarrow l/s$. 
For convenience, the classical $r$ corresponding to 
each member of the 
sequence of quantum models is identified by its value in this limit. 
We have examined the effect of the small variations in the value of $r$ 
on the classical behaviour and found the variation to be negligible.

\section{Initial States}

\subsection{Initial Quantum State}

We consider {\em initial} quantum states   
which are pure and separable, 
\begin{equation}
        | \psi(0) \rangle = | \psi_s(0) \rangle \otimes |\psi_l(0) \rangle. 
\end{equation}
For the initial state of each subsystem we use one of the directed 
angular momentum states,  
\begin{equation} 
\label{eqn:cs}
| \theta,\phi \rangle 
	= R^{(j)}(\theta,\phi) | j,j \rangle, 
\end{equation} 
which correspond to states of maximum polarization in the 
direction $(\theta,\phi)$. It has the properties:
\begin{eqnarray}
        \langle \theta, \phi | J_z | \theta, \phi \rangle & = & j \cos
        \theta \nonumber \\
        \langle \theta, 
\phi |J_x \pm i J_y | \theta, \phi \rangle & = &
        j e^{\pm i\phi} \sin \theta,
\end{eqnarray}
where $j$ in this section refers to either $l$ or $s$.

The  states (\ref{eqn:cs}) are the SU(2) coherent states, 
which, like their counterparts in the Euclidean 
phase space, are minimum uncertainty states \cite{cs}; 
the normalized variance of the quadratic operator,  
\begin{equation}    
\Delta {\bf \tilde J}^2  = { 
 \langle \theta, \phi | {\bf J}^2 | \theta, \phi \rangle 
- \langle \theta, \phi | {\bf J} | \theta, \phi \rangle^2 
\over  j(j+1) }  = {1 \over  (j+1)},
\end{equation}
is minimised for given $j$ and vanishes 
in the limit $j \rightarrow \infty$. 
The coherent states 
$| j,j\rangle $  and $| j,-j\rangle $ 
also saturate the inequality of the uncertainty relation,  
\begin{equation}
	\langle J_x^2 \rangle 
	\langle J_y^2 \rangle \ge { \langle J_z \rangle^2 \over 4 }, 
\end{equation}
although this inequality is not saturated 
for coherent states polarized along other axes.

\subsection{Initial Classical State 
and Correspondence in the Macroscopic Limit}

We compare the quantum dynamics 
with that of a classical Liouville density which is chosen to 
match the initial probability distributions of the 
quantum coherent state.  
For quantum systems with a Euclidean phase space it is always possible 
to construct a classical density with 
marginal probability distributions that match exactly 
the corresponding moments of the quantum coherent state.  
This follows from the fact that 
the marginal distributions for a coherent state are positive 
definite Gaussians, and therefore all of the moments can be matched 
{\it exactly} by choosing a Gaussian classical density. 
For the SU(2) coherent state, however, 
we show in the Appendix that no classical density has 
marginal distributions that can reproduce 
even the low order moments of the 
quantum probability distributions (except in the limit of infinite $j$). 
Thus from the outset it is clear that 
any choice of initial classical state will exhibit residual 
discrepancy in matching some of the initial quantum moments. 

We have examined the initial state and dynamical 
quantum-classical correspondence using several different 
classical distributions. These included the vector model 
distribution described in the Appendix and the Gaussian distribution 
used by Fox and Elston in correspondence studies of the 
kicked top \cite{Fox94b}. For a state polarized along the $z$-axis we 
chose the density,  
\begin{eqnarray}
\label{eqn:rho}
        \rho_c (\theta,\phi) \;  \sin \theta d \theta d \phi & = & 
\; C
\exp \left[ - { 2 \sin^2({\theta\over 2}) ) 
\over \sigma^2}\right] \sin \theta d \theta d\phi \\
& = & \; C \exp \left[ - {(1 -\tilde{J}_z)\over 
 \sigma^2 } \right] \; d \tilde{J}_z  d\phi, \nonumber 
\end{eqnarray}
with $ C = \left[ 2 \pi \sigma^2 \left( 1 - \exp( -2 \sigma^{-2})
 \right) \right]^{-1} $, 
instead of those previously considered,  
because it is periodic under $2\pi$ rotation. 
An initial state directed along  
$(\theta_o,\phi_o)$ is then produced by a rigid body 
rotation of (\ref{eqn:rho}) 
by an angle $\theta_o$ about the $y$-axis followed 
by rotation with angle $\phi_o$ about the $z$-axis. 

The variance $\sigma^2$ and the magnitude $|{\bf J}|_c$ 
are free parameters of the 
classical distribution that
should be chosen to fit the quantum probabilities as well as possible.  
It is shown in the Appendix that no classical density has marginal 
distributions which can match all of the quantum moments, 
so we concentrate only on matching the lowest order moments.   
Since the magnitude of the spin is a kinematic constant  
both classically and quantum mechanically, we choose 
the squared length of the classical spin to have the correct 
quantum value, 
\begin{equation}
\label{eqn:mag}
|{\bf J}|_c^2  =  \langle J_x^2 \rangle + 
\langle J_y^2 \rangle + \langle J_z^2 \rangle= j(j+1). 
\end{equation}

For a state polarized along the $z$-axis, we have 
$\langle J_x \rangle  = \langle J_y \rangle  = 0 $ 
and $\langle J_y^2 \rangle = \langle J_x^2 \rangle $ 
for both distributions as a consequence of the axial symmetry.  
Furthermore, as a consequence of (\ref{eqn:mag}), we will 
automatically satisfy the condition, 
\begin{equation}
	2 \langle J_x^2 \rangle_c + \langle J_z^2 \rangle_c =  j(j+1).
\end{equation}
Therefore we only need to consider the classical moments,  
\begin{eqnarray}
\label{eqn:cm}
	\langle J_z \rangle_c =  |{\bf J}| \; G(\sigma^2)  \\
	\langle J_x^2 \rangle_c 
=  |{\bf J}|^2 \sigma^2 \; G(\sigma^2), 
\end{eqnarray}
calculated from the density (\ref{eqn:rho}) 
in terms of the remaining free parameter, $\sigma^2$, where, 
\begin{equation}
	G(\sigma^2) = \left[ {1 + \exp(-2 \sigma^{-2}) 
	\over 1 - \exp (-2 \sigma^{-2}) }\right] - \sigma^2.
\end{equation}
We would like to match both of these classical moments with 
the corresponding quantum values,  
\begin{eqnarray}
\label{eqn:qmz}
	\langle J_z \rangle =  j, \\
	\langle J_x^2 \rangle =  j/2, 
\label{eqn:qmx2}
\end{eqnarray}
calculated for the coherent state (\ref{eqn:cs}). 
However, no choice of $\sigma^2$ will satisfy both 
constraints. 

If we choose $\sigma^2$ to satisfy (\ref{eqn:qmz}) exactly 
then we would obtain, 
\begin{equation}
\sigma^2 = \frac{1}{2j} - \frac{3}{8j^2}+{\mathcal O}(j^{-3}).
\end{equation}
If we choose $\sigma^2$ to satisfy (\ref{eqn:qmx2}) exactly 
then we would obtain, 
\begin{equation}
\sigma^2 = \frac{1}{2j} + \frac{1}{4j^2}+{\mathcal O}(j^{-3}).
\end{equation}
(These expansions are most easily derived from the approximation 
$G(\sigma^2) \simeq 1 - \sigma^2$, which has an exponentially 
small error for large $j$.)

We have chosen to compromise between these values by fixing 
$\sigma^2$ so that the ratio 
$\langle J_z \rangle_c / \langle J_x^2 \rangle_c $ 
has the correct quantum value. 
This leads to the choice, 
\begin{equation}
\label{eqn:sigma}
\sigma^2 = \frac{1}{2 \sqrt{j(j+1)}} =
\frac{1}{2j} - \frac{1}{4j^2}+{\mathcal O}(j^{-3}).
\end{equation}

These unavoidable initial differences between the classical 
and quantum moments will vanish in the ``classical'' 
limit. To see this explicitly it is convenient to 
introduce a measure of the quantum-classical differences, 
\begin{equation} 
	\delta J_z(n) = | \langle J_z(n) \rangle - \langle J_z(n) \rangle_c|, 
\end{equation}
defined at time $n$. 
For an initial state polarised in direction $(\theta,\phi)$, 
the choice (\ref{eqn:sigma}) produces the initial difference, 
\begin{equation} 
\label{eqn:initdiff}
	\delta J_z(0) 
= {  \cos (\theta) \over 8j } + {\mathcal O}(j^{-2}), 
\end{equation}
which vanishes as $j \rightarrow \infty$.

\section{Numerical Methods}
\label{sect:nummethods}


We have chosen to study the time-periodic spin 
Hamiltonian (\ref{eqn:ham}) because 
the time-dependence is then reduced to a simple mapping and  
the quantum state vector is confined to a finite dimensional 
Hilbert space. Consequently we can solve the exact time-evolution 
equations (\ref{eqn:qmmap}) numerically without introducing any 
artificial truncation of the Hilbert space. 
The principal source of numerical inaccuracy arises 
from the numerical evaluation of the matrix elements
of the rotation operator 
$ \langle j, m' | R(\theta,\phi) | j, m \rangle = \exp( - i \phi m' ) 
d_{m'm}^{(j)} (\theta)$.  The rotation operator is required 
both for 
calculation of the initial quatum coherent state,  
$ |\theta, \phi \rangle = R(\theta,\phi) | j,m=j \rangle $, 
and evaluation of the unitary Floquet operator. 
In order to maximise the precision of our results  
we calculated the matrix elements 
$d_{m'm}^{(j)}(\theta) = \langle j, m' | \exp(-i \theta J_y | j,m \rangle$  
using the recursion algorithm of Ref.\ \cite{Haake96} and 
then tested the accuracy of our results 
by introducing controlled numerical errors. 
For small quantum numbers ($j<50$) 
we are able to confirm the correctness of our coded algorithm 
by comparing these results with those 
obtained by direct evaluation of 
Wigner's formula for the matrix elements $d_{m'm}^{(j)}(\theta)$.

The time evolution of the Liouville density was simulated 
by numerically evaluating between $10^8$ and $10^9$ classical trajectories 
with randomly selected initial conditions weighted according to 
the initial distribution (\ref{eqn:rho}). 
Such a large number of trajectories was required in order 
to keep Monte Carlo errors small enough to  
resolve the initial normalized quantum-classical differences, which 
scale as $1/8j^2$, over the range of $j$ values we have examined.  

We identified initial conditions of the classical map as chaotic 
by numerically calculating the largest Lyapunov 
exponent, $\lambda_L$, using the formula, 
\begin{equation}
\label{eqn:liap}
\lambda_L =  { 1 \over N } \sum_{n=1}^N \ln d(n) 
\end{equation}
where $d(n) = \sum_i | \delta x_i(n) | $ , with $d(0) = 1$. 
The differential ${\bf \delta x}(n)$ is a difference 
vector between adjacent trajectories and thus evolves under the 
action of the tangent 
map $ {\bf \delta x}(n+1) = {\bf M} \cdot {\bf \delta x}(n)$, 
where ${\bf M}$ is evaluated along some fiducial trajectory \cite{LL}.

Since we are interested in studying quantum states, and 
corresponding classical distributions which have non-zero support 
on the sphere, it is also important to get an idea of the size of 
these regular and chaotic zones. By comparing the size of a given 
regular or chaotic zone to the variance of an  
initial state located within it, we can determine 
whether most of the state is contained within this zone. 
However, we can not perform 
this comparison by direct visual 
inspection since the relevant phase space 
is 4-dimensional.  
One strategy which we used to overcome this difficulty 
was to calculate the Lyapunov exponent for a large number of  
randomly sampled initial conditions and then 
project only those points which are regular (or chaotic) 
onto the plane spanned by $\tilde{S}_z = \cos \theta_s$ 
and $\tilde{L}_z = \cos \theta_l$. 
If the variance of the initial quantum state is located within,  
and several times smaller 
than, the dimensions of a zone devoid of any of these points, 
then the state in question 
can be safely identified as chaotic (or regular). 

\section{Characteristics of the Quantum and Liouville Dynamics}
\label{sec:qcc}



\subsection{Mixed Phase Space}

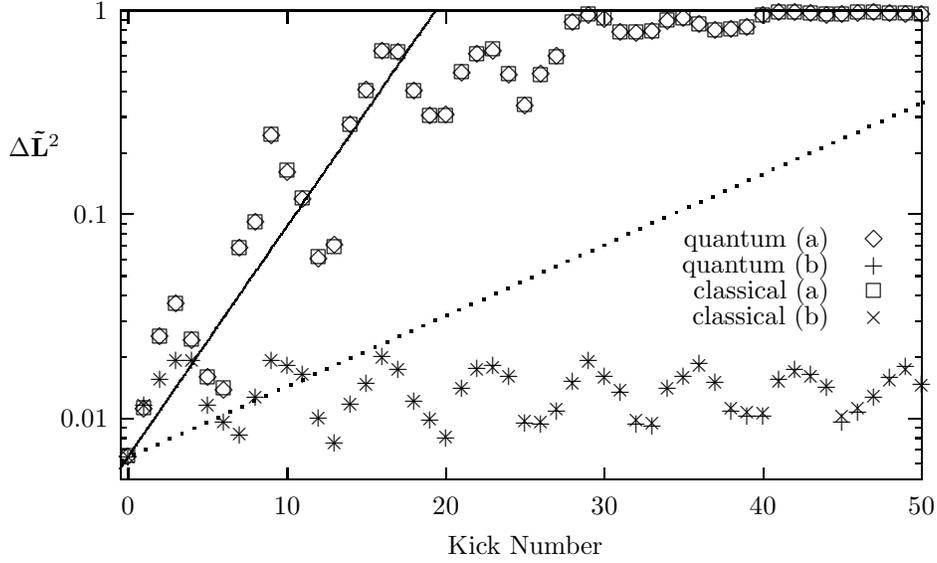
\begin{figure}
\begin{center}
\input{vargrowth.1.215.ric2}
\vspace{0.2in}
\caption{
Growth of normalized quantum and classical variances  
in a chaotic zone (a) 
and a regular zone (b) of the mixed phase space regime 
$\gamma = 1.215$ and $r\simeq 1.1$ with $l=154$.  
Quantum and  
and classical results are nearly indistinguishable on this scale.
In the chaotic case, the approximately exponential growth 
of both variances  
is governed by a much larger rate, $\lambda_{var}=0.13$ (solid line), 
than that predicted from 
the largest Lyapunov exponent, $\lambda_L=0.04$ (dotted line).
}
\label{vargrowth.1.215.ric2}
\end{center}
\end{figure}

We consider the 
time-development of initial quantum coherent states (\ref{eqn:cs})
evolved according to the mapping (\ref{eqn:qmmap})
using quantum numbers $s=140$ and $l=154$   
and associated classical  parameters 
$\gamma=1.215$, $r \simeq 1.1$, and $a=5$,  
which produce a mixed phase space (see Fig.\ \ref{regimes}). 
The classical results are generated by evolving the 
the initial ensemble  (\ref{eqn:rho}) 
using the mapping (\ref{eqn:map}).
In Fig.\ \ref{vargrowth.1.215.ric2} we compare  
the time-dependence of the normalized quantum variance, 
$\Delta {\tilde {\bf L}}^2  
= [\langle {\bf L}^2 \rangle - \langle {\bf L} \rangle^2 ] / l(l+1) $, 
with its classical counterpart,  
$\Delta {\tilde {\bf L}}^2_c 
= [\langle {\bf L}^2 \rangle_c - 
\langle {\bf L} \rangle_c^2 ] / |{\bf L}|^2 $. 
Squares (diamonds) correspond to the dynamics of 
an initial quantum (classical) 
state centered at $\vec{\theta}(0) = (20^o,40^o,160^o,130^o)$, 
which is located in the connected chaotic zone
near one of the unstable fixed points 
of the classical map. Crosses (plus signs) 
correspond to an initial quantum (classical) 
state centered on the initial condition 
$\vec{\theta}(0) =(5^o,5^o,5^o,5^o)$, which is located in 
the regular zone near one of the stable fixed points. 
For both initial conditions the quantum and classical 
results are nearly indistinguishable on the scale of the figure. 
In the case of the regular initial condition, the quantum 
variance remains narrow over long times and, 
like its classical counterpart, exhibits a regular oscillation. 
In the case of the chaotic initial condition 
the quantum variance also exhibits a periodic 
oscillation but this oscillation is superposed on a very rapid, approximately 
exponential, growth rate.  
This exponential growth persists until the variance approaches 
the system size, that is, when $\Delta {\tilde {\bf L}}^2  \simeq 1$ . 
The initial exponential growth of the quantum variance in classically chaotic 
regimes has been observed previously in several models and appears 
to be a generic feature of the quantum dynamics; this behaviour of 
the quantum variance is mimicked very accurately 
by the variance of an initially  well-matched 
classical distribution \cite{Ball98,Fox94b,Fox94a}. 

For well-localized states, in the classical case, the exponential growth 
of the distribution variance in chaotic zones is certainly related to 
the exponential divergence of the underlying trajectories, a property 
which characterizes classical chaos. To examine this connection  
we compare the observed exponential rate of growth of the widths of 
the classical (and quantum) state with the exponential rate predicted 
from the classical Lyapunov exponent. For the coherent states the 
initial variance can be calculated exactly, 
$\Delta {\tilde {\bf L}}^2(0) = 1/(l+1)$. 
Then, assuming exponential growth of this initial variance we get, 
\begin{equation}
\label{eqn:expvar} 
	\Delta {\tilde {\bf L}}^2(n)  
\simeq { 1 \over l  } \exp( 2 \lambda_w n)
\; \;\;\;\;\;\;\; {\mathrm for }  \;\;\;  n < t_{sat},
\end{equation}
where a factor of $2$ is included in the exponent since  
$\Delta {\tilde {\bf L}}^2 $ corresponds to a 
squared length. The dotted line in Fig. \ref{vargrowth.1.215.ric2} 
corresponds to the prediction (\ref{eqn:expvar}) 
with $\lambda_w = \lambda_L = 0.04$, the value of the 
largest classical Lyapunov exponent.  As can be seen from the figure, 
the actual growth rate of the classical (and quantum) variance   
of the chaotic initial state is significantly larger 
than that predicted using the largest Lyapunov exponent.   
For comparison purposes we also plot a  
solid line in Fig.\ \ref{vargrowth.1.215.ric2} corresponding 
to (\ref{eqn:expvar}) using $\lambda_w = 0.13 $, which provides 
a much closer approximation to the actual growth rate. 
We find, for a variety of initial conditions in the chaotic zone of 
this mixed regime, that the actual 
classical (and quantum) variance growth rate is consistently larger 
than the simple prediction (\ref{eqn:expvar}) using $\lambda_L$ for the 
growth rate.  This systematic bias requires some explanation. 

As pointed out in \cite{Fox94b}, the presence of 
some discrepancy can be expected from the fact that the 
Lyapunov exponent is defined  
as a geometric mean of the tangent map eigenvalues 
sampled over the entire connected chaotic zone (corresponding 
to the infinite time limit $n\rightarrow \infty$) whereas 
the {\it actual} growth rate of a given 
distribution over a small number of time-steps will be 
determined largely by a few eigenvalues 
of the local tangent map. In mixed regimes these local eigenvalues 
will vary considerably over the phase space manifold and the 
product of a few of these eigenvalues 
can be quite different from the geometric 
mean over the entire connected zone.  


However, we find that the actual growth rate is consistently {\it larger}  
than the Lyapunov exponent prediction. 
It is well known that in mixed regimes the remnant KAM tori 
can be `sticky'; these sticky regions can have a significant decreasing  
effect on a calculation of the Lyapunov exponent. 
In order to identify an initial condition as 
chaotic, we specifically choose initial states 
that are concentrated away from these KAM surfaces (regular islands).  
Such initial states will then be exposed mainly to the larger 
local expansion rates found away from these surfaces. 
This explanation is supported by our observations that, when  
we choose initial conditions closer to these remnant tori, we 
find that the growth rate of the variance is significantly reduced. 
These variance growth rates are still slightly larger than the 
Lyapunov rate, but this is not surprising since 
our initial distributions are concentrated over a 
significant fraction of the phase space and 
the growth of the distribution  
is probably more sensitive to contributions 
from those trajectories subject to large eigenvalues away from the 
KAM boundary than those stuck near the boundary. 
These explanations are further 
supported by the results of the following section, where  
we examine a phase space regime that is nearly devoid of regular islands.  
In these regimes we find that the Lyapunov exponent serves as a much 
better approximation to the variance growth rate. 

\subsection{Regime of Global Chaos}

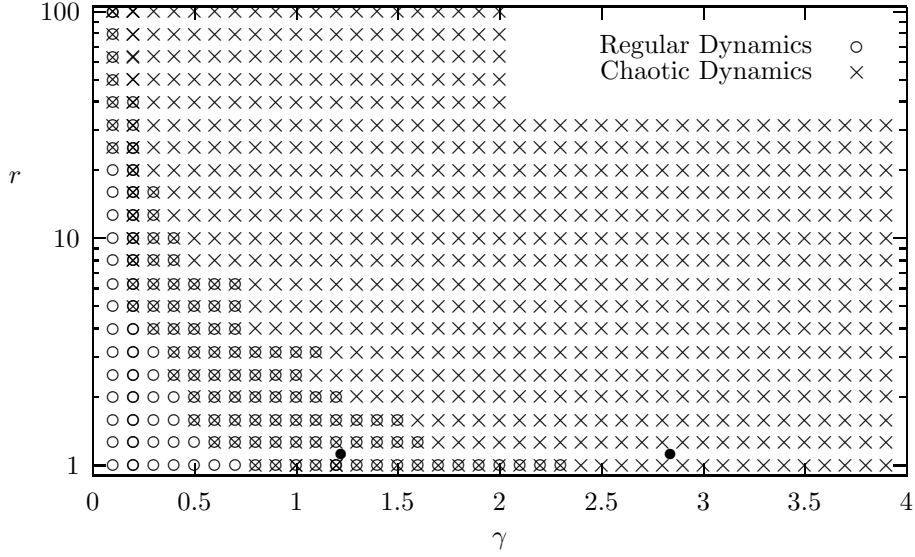
\begin{figure}
\begin{center}
\input{regimes}
\vspace{0.2in}
\caption{
Behaviour of the classical mapping for 
different values of $r=|{\bf L}|/ |{\bf S}|$ 
and $\gamma =c |{\bf S}|$ with $ a =5$.   
Circles correspond to parameter values for which at least 99\% 
of the surface area ${\mathcal P}$ produces regular dynamics and 
crosses correspond to parameter values for which 
the dynamics are at least 99\% chaotic.  
Superpositions of circles and crosses correspond to parameter values which 
produce a mixed phase space. We investigate quantum-classical correspondence 
for the parameter values $\gamma = 1.215$ (mixed regime) 
and $\gamma= 2.835$ (global chaos), with $r=1.1$, 
which are indicated by filled circles.  
}
\label{regimes}
\end{center}
\end{figure}

If we increase the dimensionless coupling strength to $\gamma=2.835$, 
with $a=5$ and $r \simeq 1.1$ as before, then the classical flow 
is predominantly chaotic on the surface ${\mathcal P}$
(see Fig.\ \ref{regimes}). 
Under these conditions 
we expect that generic initial classical distributions 
(with non-zero support) will spread to cover the full surface  
${\mathcal P}$ and then quickly relax close to microcanonical equilibrium. 
We find that the initially localised quantum states also exhibit  
these generic features when the quantum map is governed by 
parameters which produce these conditions classically.  

For the non-autonomous Hamiltonian system (\ref{eqn:map}) 
the total energy is not conserved, but 
the two invariants of motion ${\bf L}^2$ and ${\bf S}^2$ confine 
the dynamics 
to the 4-dimensional 
manifold ${\mathcal P} = {\mathcal S}^2 \times {\mathcal S}^2 $, which is 
the surface of two spheres.  
The corresponding microcanonical distribution
is a constant on this surface, 
with measure (\ref{eqn:measure}),  
and zero elsewhere.  
From this distribution we can calculate 
microcanonical equilibrium values 
for low order moments, where, for example,     
$\{ L_z \} = ( 4 \pi)^{-2} \int_{\mathcal P} L_z d \mu 
= 0$ and $\{ \Delta {\bf L}^2 \} = \{ {\bf L}^2 \}-\{ {\bf L} \}^2 = |{\bf L}|^2$. 
The symbols $\{ \cdot \}$ denote a microcanonical average. 

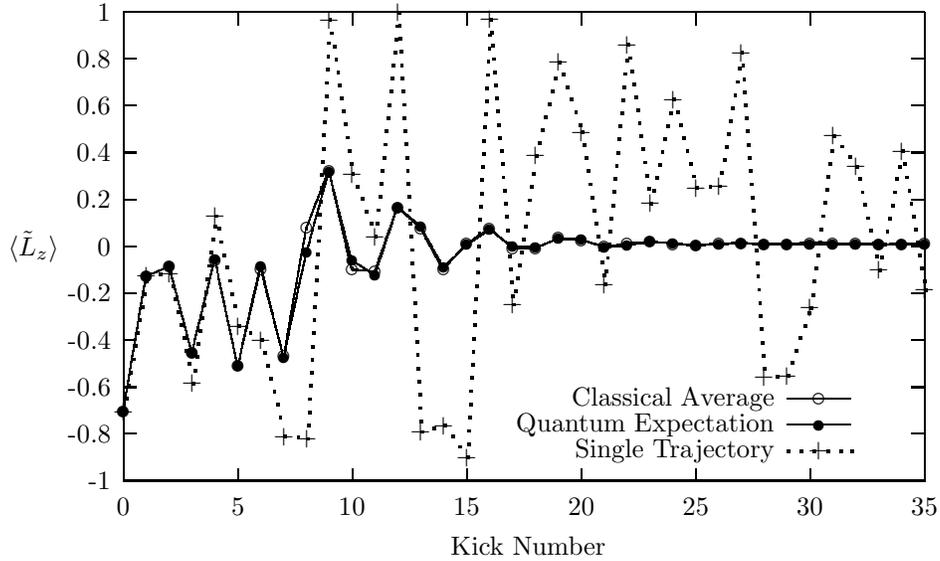
\begin{figure}
\begin{center}
\input{qmlmnm.2.835.154.cic1}
\vspace{0.2in}
\caption{Comparison of quantum expectation value and 
corresponding classical average $\langle L_z \rangle_c$
in the regime of global chaos $\gamma=2.835$ and $r\simeq 1.1$
with  $l=154$  
and initial condition $\vec{\theta}_o = (45^o,70^o,135^o,70^o) $. 
The points of the stroboscopic map 
are connected with lines to guide the eye.
The quantum expectation 
value and the Liouville average 
exhibit esentially the same rate of relaxation to 
microcanonical equilibrium, 
a behaviour which is qualitatively distinct from 
that of the single trajectory. 
}
\label{qmlmnm}
\end{center}
\end{figure}

To give a sense of the accuracy of the correspondence between 
the classical ensemble and the quantum dynamics 
in Fig.\ \ref{qmlmnm} we show a direct comparison 
of the dynamics of the quantum expectation value $
 \langle \tilde{L}_z \rangle $ 
with $l=154$ 
and the classical distribution average
$ \langle \tilde{L}_z \rangle_c $ 
for an initial coherent state and corresponding 
classical distribution centered at 
$\vec{\theta} = (45^o,70^o,135^o,70^o)$. 
To guide the eye in this figure we have drawn lines connecting 
the stroboscopic points of the mapping equations.   
The quantum expectation value exhibits essentially 
the same dynamics as the classical Liouville average, not only at 
early times, that is, in the 
initial Ehrenfest regime \cite{Ball94,HB94},   
but for times well into the equilibrium regime where the classical 
moment $ \langle L_z \rangle$ has relaxed close to the microcanonical 
equilibrium value $\{ L_z \} = 0 $. 
We have also provided results for a single trajectory launched 
from the same initial condition in order to emphasize the 
qualitatively distinct behaviour it exhibits.


In Fig.\ \ref{vargrowth.2.835.140} we show the 
exponential growth of the normalized quantum and 
classical variances on a semilog plot for 
the same set of parameters and quantum numbers.
Numerical data for (a) correspond to initial condition
$\vec{\theta}(0) = (20^o,40^o,160^o,130^o)$ 
and those for (b) correspond to $\vec{\theta}(0) = (45^o,70^o,135^o,70^o)$. 
As in the mixed regime case,  
the quantum-classical differences are nearly 
imperceptible on the scale of the figure,  
and the differences between the quantum and classical 
variance growth rates are many orders of magnitude 
smaller than the small differences in the growth rate arising 
from the different initial conditions. 

In contrast with the mixed regime case, 
in this regime of global chaos the prediction (\ref{eqn:expvar})  
with  $\lambda_w= \lambda_L=0.45$ 
now serves as a much better approximation 
of the exponential growth rate of the quantum variance, and associated 
relaxation rate of the quantum and classical states. 
In this regime the exponent $\lambda_w$ is also much larger than in 
the mixed regime case 
due to the stronger degree of classical chaos. As a result,   
the initially localised quantum and classical distributions 
saturate at system size much sooner. 

It is useful to apply (\ref{eqn:expvar}) 
to estimate the time-scale at which the quantum (and classical) 
distributions saturate at system size.   
From the condition 
$\Delta {\tilde {\bf L}}^2(t_{sat}) \simeq 1$ and 
using (\ref{eqn:expvar}) we obtain,  
\begin{equation}
\label{eqn:nsat}
	t_{sat} \simeq (2 \lambda_w)^{-1} \ln( l) 
\end{equation}
which serves as an estimate of this characteristic time-scale. 
In the  regimes for which the full surface ${\mathcal P}$ 
is predominately chaotic, 
we find that the actual 
exponential growth rate of the width of the quantum state, $\lambda_w$,   
is well approximated by the largest Lyapunov exponent $\lambda_L$.  
For $a=5$ and $r=1.1$, the approximation $\lambda_w \simeq \lambda_L$  
holds for coupling strengths $\gamma > 2$, for which 
more than 99\% of the surface ${\mathcal P}$ is covered by one connected 
chaotic zone (see Fig.\ \ref{regimes}).  

By comparing the quantum probability distribution to its classical 
counterpart, we can learn much more about the relaxation properties of the 
quantum dynamics. 
In order to compare each $m_l$ value of the quantum distribution, $P_z(m_l)$,  
with a corresponding piece of the continuous classical marginal 
probability distribution, 
\begin{equation}
P_c(L_z) = \int \! \! \int \! \! \int \! 
d \tilde{S}_z d \phi_s d \phi_l \; \rho_c(\theta_s,\phi_s, \theta_l, \phi_l), 
\end{equation}
we discretize the latter into $2j+1$ bins of width $\hbar=1$. 
This procedure produces a discrete classical probability distribution $P_z^c(m_l)$ 
which prescribes the probability 
of finding the spin component $L_z$ in the interval $[m_l+1/2,m_l-1/2]$ 
along the $z$-axis. 

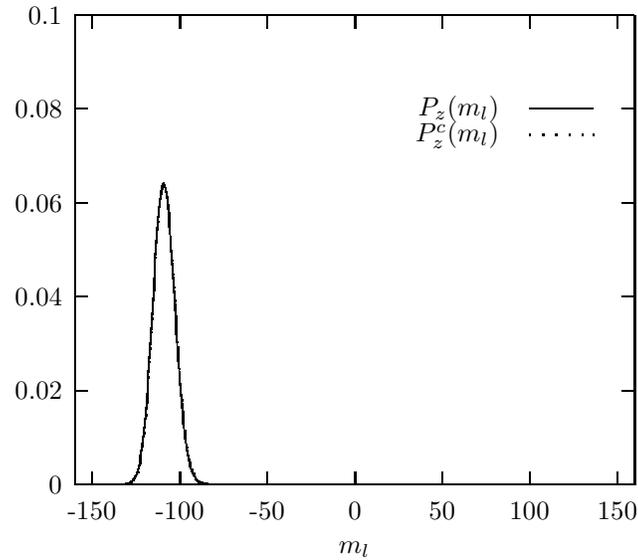
\begin{figure}
\begin{center}
\input{probdist00}
\vspace{0.2in}
\caption{
Initial probability distributions for $L_z$ 
for $\vec{\theta}(0) = (45^o,70^o,135^o,70^o) $ 
with $l=154$. 
The quantum and classical distributions are initially 
indistinguishable on the scale of the figure.
}
\label{probdist0}
\end{center}
\end{figure}

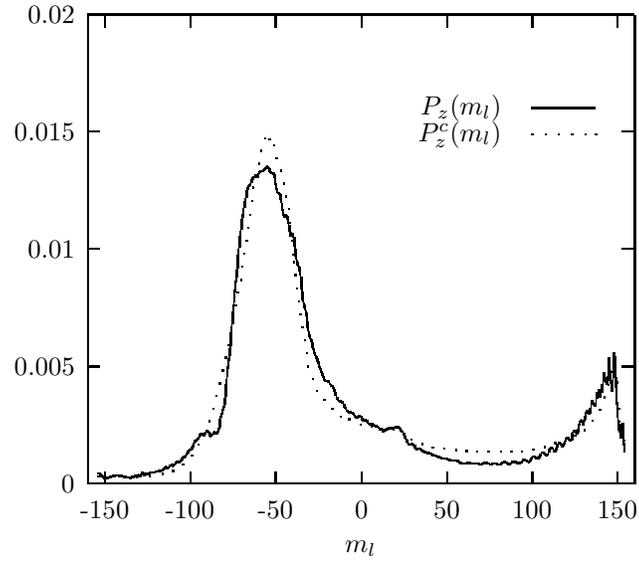
\begin{figure}
\begin{center}
\input{probdist06}
\vspace{0.2in}
\caption{
Same as Fig.\ \ref{probdist0} but the states have evolved to $n=6$ 
in the regime of global chaos $\gamma=2.835$ and $r\simeq 1.1$. 
Both the quantum and classical distribution have spread to system dimension 
and exhibit their largest differences on this saturation time-scale. 
}
\label{probdist6}
\end{center}
\end{figure}

To illustrate the time-development of these distributions 
we compare the quantum and classical probability distributions 
for three successive values of the kick number $n$,    
using the same quantum numbers and 
initial condition as in Fig.\ \ref{qmlmnm}. 
In Fig.\ \ref{probdist0} 
the initial quantum and classical states
are both well-localised and nearly 
indistinguishable on the scale of the figure. At time $n=6 \simeq t_{sat}$, 
shown in Fig.\ \ref{probdist6}, 
both distributions have grown to fill the accessible 
phase space. It is at this time 
that the most significant quantum-classical 
discrepancies appear. 

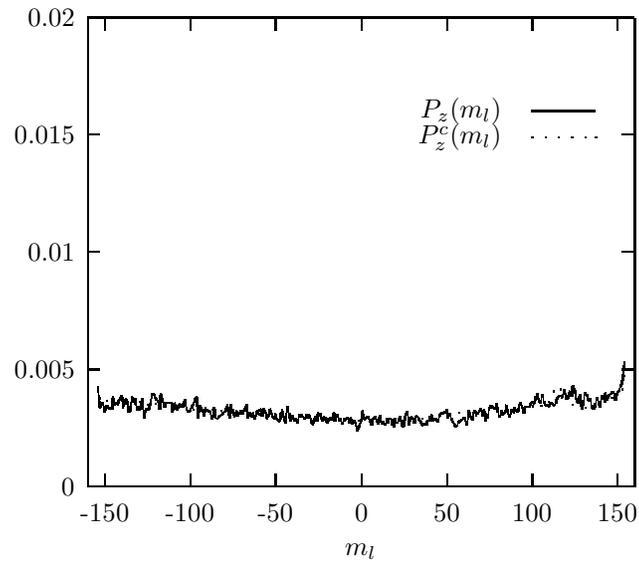
\begin{figure}
\begin{center}
\input{probdist15}
\vspace{0.2in}
\caption{
Same as Fig.\ \ref{probdist6}, but for $n=15$. Both quantum and classical 
distributions have relaxed close to the microcanonical equilibrium. 
}
\label{probdist15}
\end{center}
\end{figure}

For times greater than $t_{sat}$, however, these emergent 
quantum-classical discrepencies do not continue to grow,  
since both distributions begin relaxing towards equilibrium distributions. 
Since the dynamics are confined to a {\it compact} phase space,  
and in this parameter regime the remnant KAM tori 
fill a negligibly small fraction of the kinematicaly accessible phase space, 
we might expect the classical equilibrium distribution to be very 
close to the microcanonical distribution. 
Indeed such relaxation close to microcanonical equilibrium is 
apparent for both the quantum and the classical distribution 
at very early times, as demonstrated in Fig. \ref{probdist15}, 
corresponding to $n=15$.

Thus the signature of a classically hyperbolic flow,  
namely, the exponential relaxation of an 
arbitrary distribution (with non-zero measure) 
to microcanonical equilibrium \cite{Dorfman}, holds to good approximation  
in this model in a regime of global chaos.
More suprisingly, this classical signature  
is manifest also in the dynamics of the quantum distribution.
In the quantum case, however, as can be seen 
in Fig. \ref{probdist15}, the probability distribution 
is subject to small irreducible time-dependent fluctuations about the 
classical equilibrium.  We examine 
these quantum fluctuations in detail elsewhere \cite{EB00b}. 


\section{Time-Domain Characteristics of Quantum-Classical Differences}

\begin{figure}
\begin{center}
\input{delta.1.215.140.n200}
\vspace{0.2in}
\caption{
Time-dependence of quantum-classical differences in a regular 
zone (open circles) and a chaotic zone (filled circles) of mixed 
regime ($\gamma=1.215$ and $r \simeq 1.1$) 
with $l=154$. 
For the chaotic state 
$\delta L_z = | \langle L_z \rangle - \langle L_z \rangle_c| $ is 
compared with the Ehrenfest difference,
$ | \langle L_z \rangle - L_z| $, 
between the quantum expectation value and a single trajectory (plus signs), 
which grows until saturation at system dimension. 
The solid line corresponds to (\ref{eqn:expansatz}) 
using $\lambda_{qc} = 0.43$. 
The horizontal lines indicate two different values of the difference 
tolerance $p$ which may be used to determine the break-time;  
for $p=0.1$ (dotted line) $t_b$ occurs on a logarithmic time-scale,  
but for $p=15.4$ (sparse dotted line) 
$t_b$ is not defined over numerically accessible time-scales.  
}
\label{delta.1.215.140.n200}
\end{center}
\end{figure}
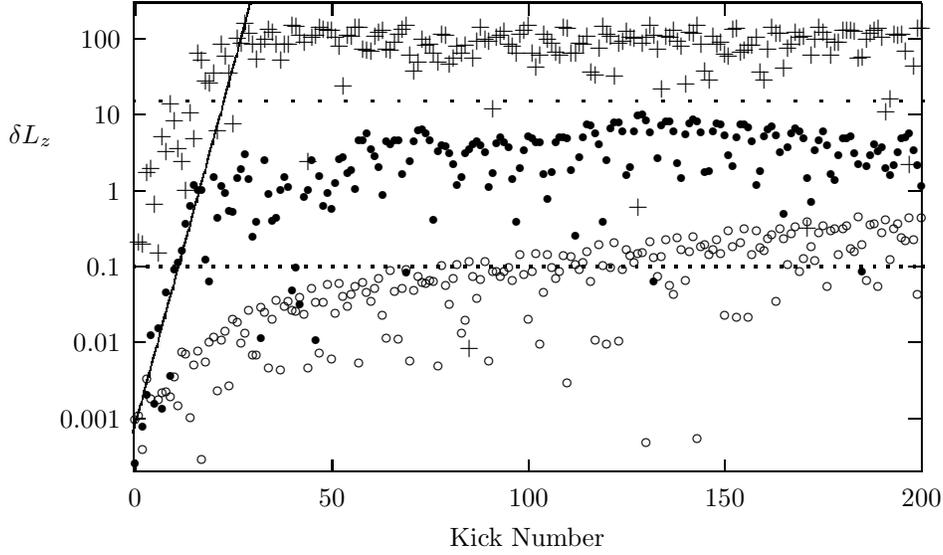

We consider the time dependence of quantum-classical differences 
defined along the $z$-axis of the spin ${\bf L}$,
\begin{equation}
\label{eqn:diff}
        \delta L_z(n) = 
| \langle L_z(n)  \rangle - \langle L_z(n) \rangle_c  |, 
\end{equation}
at the stroboscopic times $t=n$. 
In Fig.\ \ref{delta.1.215.140.n200}
we compare the time-dependence of 
$\delta L_z(n)$ on a semi-log plot 
for a chaotic state (filled circles), with 
$\vec{\theta}(0) = (20^o,40^o,160^o,130^o)$, 
and a regular state (open circles), 
 $\vec{\theta}(0) = (5^o,5^o,5^o,5^o)$, 
evolved using the same mixed regime parameters 
($\gamma=1.215$ and $r\simeq1.1$) and quantum numbers
($l=154$) as in Fig.\ \ref{vargrowth.1.215.ric2}. 

We are interested in the behaviour of the upper envelope of the data 
in Fig.\ \ref{delta.1.215.140.n200}. For the regular case, 
the upper envelope of the quantum-classical differences grows very 
slowly, as some polynomial function of time. 
For the chaotic case, on the other hand, at early times 
the difference measure (\ref{eqn:diff}) grows 
exponentially until saturation around 
$n=15$, which is 
well before reaching system dimension, $|{\bf L}| \simeq l = 154$.  
After this time, which we denote $t^*$, 
the quantum-classical differences exhibit no definite growth, 
and fluctuate about the equilibrium value $\delta L_z \sim 1 \ll |{\bf L}|$. 
In  Fig.\ \ref{delta.1.215.140.n200} we also include data for the 
time-dependence of the Ehrenfest difference 
$| \langle L_z \rangle - L_z| $, which is 
defined as the difference between 
the quantum expectation value and the dynamical 
variable of a single trajectory initially centered on the quantum state. 
In contrast to $\delta L_z$, the rapid growth of the Ehrenfest difference 
continues until saturation at the system dimension. 

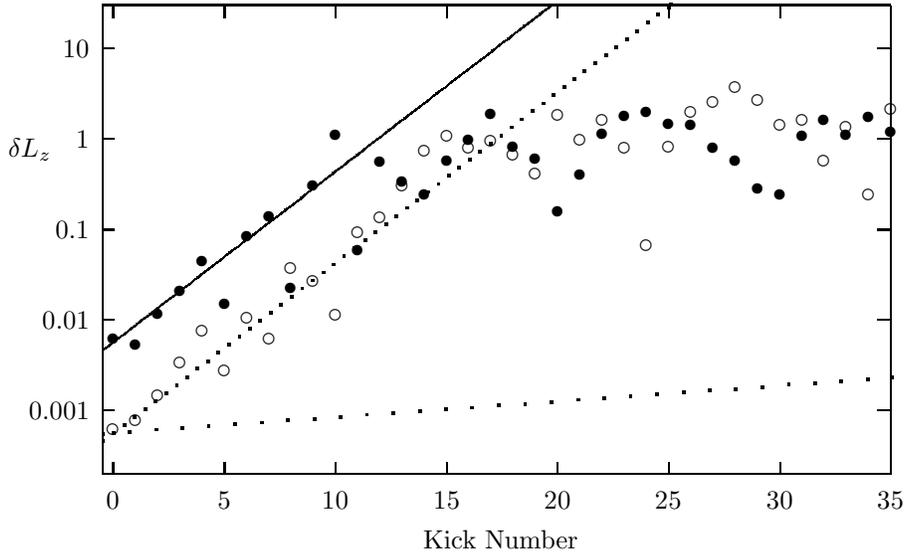
\begin{figure}
\begin{center}
\input{delta.1.215.dvsl}
\vspace{0.2in}
\caption{
Growth of the quantum-classical difference 
$\delta L_z$ 
in the chaotic zone 
of a mixed regime, $\gamma =1.215$ and $r\simeq 1.1$,    
with $l=22$ (filled circles) 
and $l=220$ (open circles). 
For $l=220$ the exponential growth rate 
(\ref{eqn:expansatz}) is plotted using the classical 
Lyapunov exponent, $\lambda_L =0.04$ (sparse dotted line),  
and for both $l$ values (\ref{eqn:expansatz}) is plotted using 
the exponent $\lambda_{qc}=0.43$ (solid line for $l=22$, 
dotted line for $l=220$), 
which is obtained from a fit of (\ref{eqn:tb}) to the 
corresponding break-time data in Fig.\ \ref{break_time_p0.1}.
}
\label{delta.1.215.dvsl}
\end{center}
\end{figure}

In Fig.\ \ref{delta.1.215.dvsl} we compare the time-dependence 
of the quantum-classical differences 
in the case of the chaotic 
initial condition $\vec{\theta}(0) = (20^o,40^o,160^o,130^o)$ 
for quantum numbers $l=22$ (filled circles) and $l=220$ (open circles), 
using the same parameters as in Fig.\ \ref{delta.1.215.140.n200}. 
This demonstrates the remarkable fact that the exponential growth 
terminates when the difference measure reaches an essentially 
fixed magnitude ($\delta L_z \sim 1$ as for the case $l=154$), 
although the system dimension differs by an order of magnitude 
in the two cases.  

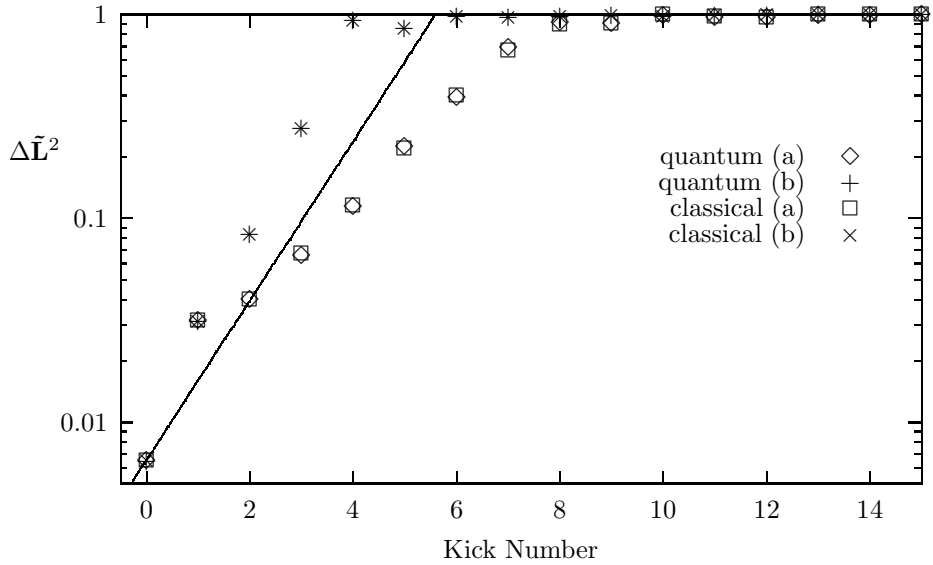
\begin{figure}
\begin{center}
\input{vargrowth.2.835.140}
\vspace{0.2in}
\caption{
Growth of normalized quantum and classical variances 
in the regime of global chaos, $\gamma =2.835$ and $r \simeq 1.1$ with l=154,
for the two initial conditions cited in the text.  
Quantum-classical differences are nearly imperceptible on this scale. 
In this regime the largest Lyapunov exponent $\lambda_L=0.45$ 
provides a much better estimate of the initial variance 
growth rate. 
}
\label{vargrowth.2.835.140}
\end{center}
\end{figure}

\begin{figure}
\begin{center}
\input{delta.2.835.140}
\vspace{0.2in}
\caption{
Growth of quantum-classical differences 
in the regime of global chaos $\gamma=2.835$ and $r\simeq 1.1$
with $l=154$ for the two initial conditions cited in text. 
The exponential growth rate 
(\ref{eqn:expansatz}) is plotted using the classical 
Lyapunov exponent, $\lambda_L =0.45$ (dotted line),  
and the exponent $\lambda_{qc}=1.1$ (solid line),   
which is obtained from a fit of (\ref{eqn:tb}) to the 
corresponding break-time data in Fig.\ \ref{break_time_p0.1}.
}
\label{delta.2.835.140}
\end{center}
\end{figure}
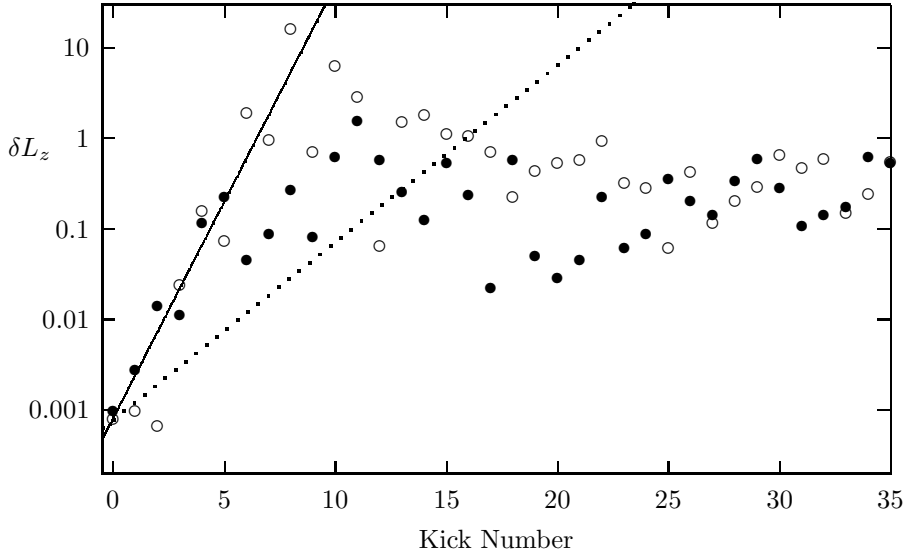

In Fig.\ \ref{delta.2.835.140} we consider the growth of the 
quantum-classical difference measure $\delta L_z(n)$ in 
a regime of global chaos, for $l=154$, 
and using the same set of parameters 
as those examined in Fig.\ \ref{vargrowth.2.835.140} 
($\gamma=2.835$ and $r\simeq1.1$).  
Again the upper envelope of the difference measure $\delta L_z(n)$ 
exhibits exponential growth at early times, though 
in this regime of global chaos the exponential growth 
persists only for a very 
short duration before saturation at $t^* \simeq 6$. 
The initial condition 
$\vec{\theta}(0) = (20^o,40^o,160^o,130^o)$ is a typical 
case (filled circles), 
where, as seen for the mixed regime parameters, 
the magnitude of the difference 
at the end of the exponential growth phase 
saturates at the value $\delta L_z(t^*) \simeq 1$, which does not scale 
with the system dimension (see Fig.\ \ref{deltamax.vs.l}).  
The initial condition 
$\vec{\theta}(0) = (45^o,70^o,135^o,70^o)$ (open circles) 
leads to an anomolously large deviation at the end  of the 
exponential growth phase, 
$\delta L_z(t^*) \simeq 10$, though still 
small relative to the system dimension $|{\bf L}| \simeq 154$. 
This deviation is transient however, and   
at later times the magnitude of quantum-classical differences fluctuates  
about the equilibrium value $\delta L_z \sim 1$. The quantum-classical 
differences are a factor 
of $1/l$ smaller than typical differences between the 
quantum expectation value and the single trajectory, 
which are of order system dimension 
(see Fig.\ \ref{qmlmnm}) as in the mixed regime case.

In all cases where the initial quantum and classical states are 
launched from a chaotic zone we find that the initial time-dependence 
of quantum-classical differences compares favorably with the 
exponential growth ansatz,
\begin{equation}
\label{eqn:expansatz}
	\delta L_z(n) 
\simeq { 1 \over 8 l } \exp ( \lambda_{qc} n )
 \; \;\;\;\;\;\;\; {\mathrm for }  \;\;\;  n < t^*,
\end{equation}
where the exponent $\lambda_{qc}$ is a new exponent subject 
to numerical measurement \cite{Ball98}.  
The prefactor $1 /  8 l  $
is obtained from (\ref{eqn:initdiff}) though   
we have dropped the $\cos \theta$ factor that specifies 
the {\em exact} initial difference for $L_z$.  
Since contributions from the initial differences in 
other mismatched moments  will generally mix under 
the dynamical flow, it is appropriate 
to consider an effective initial difference for 
the prefactor in (\ref{eqn:expansatz}).  
The prefactor $1/8l$ is obtained  
by accounting for the initial contributions from the 3 cartesian components,  
$ [ \delta^2 L_x(0) + \delta^2 L_y(0) + \delta^2 L_z(0 ]^{1/2} 
= 1/8l $. 

We are interested in whether the Lyapunov exponent 
$\lambda_L$ is a good approximation to $\lambda_{qc}$. 
In Fig.\ \ref{delta.1.215.dvsl} 
we plot (\ref{eqn:expansatz}) with $\lambda_{qc} = \lambda_L = 0.04$  
(dotted line) for $l=220$. Clearly 
the largest Lyapunov exponent severly underestimates  
the exponential growth rate of the quantum-classical differences, in this case 
by more than an order of magnitude.  
The growth rate of the state width, $\lambda_w = 0.13$ , is also 
several times smaller than the initial growth rate of the 
quantum-classical differences. 
In the case of Fig.\ \ref{delta.2.835.140}, corresponding to a 
regime of global chaos with a much larger Lyapunov exponent, 
we plot (\ref{eqn:expansatz}) with 
 $\lambda_{qc} = \lambda_L = 0.45$ (dotted line), 
demonstrating that, in this regime 
too the largest Lyapunov 
exponent underestimates the initial growth rate of the quantum-classical 
difference measure $\delta L_z(n)$. 


We also find, from inspection of our results, that the time $t^*$ 
at which the exponential growth (\ref{eqn:expansatz}) terminates
can be estimated from $t_{sat}$, the time-scale 
on which the distributions saturate at or near system size (\ref{eqn:nsat}). 
In the case of the chaotic initial condition of 
Fig.\ \ref{vargrowth.1.215.ric2}, for which $\gamma=1.215$, 
visual inspection of the figure suggests that 
$t_{sat} \simeq 18$. This should be compared with 
Fig.\ \ref{delta.1.215.140.n200}, where the exponential 
growth of $\delta L_z(n)$ ends rather abruptly at $t^* \simeq 15$. 
In Fig.\ \ref{vargrowth.2.835.140}, corresponding 
to a regime of global chaos ($\gamma=2.835$),  
the variance growth saturates much earlier, around   
$t_{sat} \simeq 6$ for both initial conditions. 
From Fig.\ \ref{delta.2.835.140} we can estimate that the 
initial exponential growth of the 
quantum-classical differences for these two initial conditions 
also ends around $n \simeq 6$. As we increase $\gamma$ further, 
we find that the exponential growth phase of quantum-classical 
differences $\delta L_z(n)$ is shortened, lasting only until 
the corresponding quantum and classical distributions 
saturate at system size. 
For $\gamma \simeq 12$, with $\lambda_L \simeq 1.65$, 
the chaos is sufficiently strong that the initial 
coherent state for $l=154$ spreads to cover ${\mathcal P}$ 
within a single time-step. Similarly the initial difference measure
$\delta L_z(0) \simeq 0.001$ grows to the magnitude $\delta L_z(1) \simeq 1$ 
within a single time-step and subsequently fluctuates about 
that equilibrium value. 
We have also inspected the variation of $t^*$ with the quantum numbers
and found it to be consistent with the logarithmic dependence 
of $t_{sat}$ in (\ref{eqn:nsat}).  



\section{Correspondence Scaling in the Classical Limit}
\label{sect:scaling}


We have assumed in (\ref{eqn:expansatz})  
that the exponent $\lambda_{qc}$ is independent of the quantum numbers. 
A convenient way of confirming this, and also estimating the numerical value 
of $\lambda_{qc}$, is by means of a break-time measure. The 
break-time is the time $t_b(l,p) $ at 
which quantum-classical differences exceed some fixed tolerance $p$,
with the classical parameters and initial condition held fixed.
Setting $\delta L_z ( t_b) = p $ in (\ref{eqn:expansatz}), we obtain  $t_b$ 
in terms of $p$, $l$ and $\lambda_{qc}$, 
\begin{equation}
\label{eqn:tb}
	t_b \simeq \lambda_{qc}^{-1} \ln ( 8 \; p \; l  )
\;\;\;\;\;\; {\mathrm provided }  \;\;\  p < {\mathcal O}(1).
\end{equation}
The restriction $p < {\mathcal O}(1)$, which 
plays a crucial role in limiting the robustness 
of the break-time measure (\ref{eqn:tb}), is explained and 
motivated further below. 

The explicit form we have obtained for the argument of the logarithm 
in (\ref{eqn:tb}) is a direct result of our estimate that the initial 
quantum-classical differences arising from the Cartesian components 
of the spin provide the dominant contribution to the prefactor  
of the exponential growth ansatz (\ref{eqn:expansatz}). 
Differences in the mismatched higher order moments, 
as well as intrinsic differences between 
the quantum dynamics and classical dynamics, 
may also contribute to this effective prefactor.  
We have checked that the initial value 
$\delta L_z(0) \simeq 1 /8l$ is an adequate 
estimate by comparing the intercept of 
the quantum-classical data on a semilog plot 
with the prefactor of (\ref{eqn:expansatz}) 
for a variety of $l$ values (see {\it e.g}.\ Fig.\ \ref{delta.1.215.dvsl}).  

\begin{figure}
\begin{center}
\input{break_time_p0.1}
\vspace{0.2in}
\caption{
Scaling of the break-time using tolerance $p=0.1$ as a function   
of increasing quantum number for the mixed regime parameters 
$\gamma = 1.215$ and $r \simeq 1.1$ with 
$\vec{\theta}(0) =(20^o,40^o,160^o,130^o)$  (filled circles) 
and for the global chaos parameters $\gamma =2.835$ and $r\simeq 1.1$ with 
$\vec{\theta}(0) =(45^o,70^o,135^o,70^o)$
(open circles). 
We also plot the results of fits  
to the log rule (\ref{eqn:tb}), 
which produced exponents $\lambda_{qc} = 0.43$ 
for $\gamma=1.215$ and $\lambda_{qc}=1.1$ for $\gamma = 2.835$.
}
\label{break_time_p0.1}
\end{center}
\end{figure}
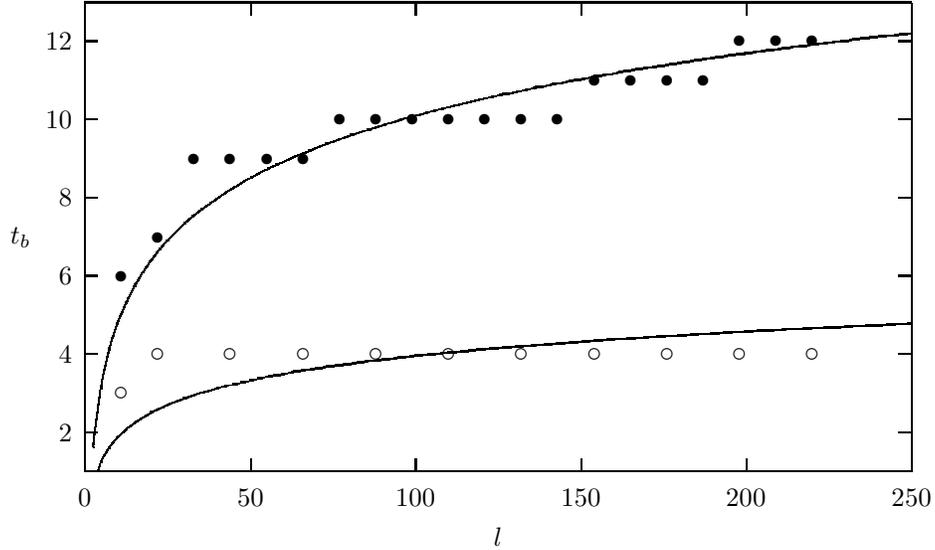

In Fig.\ \ref{break_time_p0.1} we examine the scaling of the break-time 
for $l$ values ranging from $11$ to $220$ and 
with fixed tolerance $p=0.1$. 
The break-time can assume only the integer values $t=n$ 
and thus the data exhibits a step-wise behaviour. 
For the mixed regime parameters, 
$\gamma=1.215$ and $r\simeq 1.1$ (filled circles), 
with initial condition $\vec{\theta}(0) = (20^o,40^o,160^o,130^o)$,  
a non-linear least squares fit to (\ref{eqn:tb}) gives 
$\lambda_{qc} = 0.43$. This fit result is 
plotted in the figure as a solid line. The close agreement between the data 
and the fit provides good evidence that 
the quantum-classical exponent $\lambda_{qc}$ is independent of 
the quantum numbers. 
To check this result against the time-dependent $\delta L_z(n)$ data, 
we have plotted the exponential curve (\ref{eqn:expansatz}) 
with $\lambda_{qc} = 0.43$ 
in Fig.\ \ref{delta.1.215.140.n200} using a solid line 
and  in Fig.\ \ref{delta.1.215.140.n200} using a solid line for $l=22$ and 
a dotted line for $l=220$. 
The exponent obtained from fitting (\ref{eqn:tb}) serves as an excellent 
approximation to the initial exponential growth (\ref{eqn:expansatz}) 
of the quantum-classical differences in each case. 

In Fig.\ \ref{break_time_p0.1} we also plot 
break-time results for the global chaos 
case $\gamma=2.835$ and $r\simeq 1.1$ (open circles) 
with initial condition $\vec{\theta}(0) = (45^o,70^o,135^o,70^o)$. 
In this regime the quantum-classical 
differences grow much more rapidly and, consequently, 
the break-time is very short and remains 
nearly constant over this range of computationally accessible quantum numbers. 
Due to this limited variation, in this regime we can not 
confirm (\ref{eqn:tb}), although the data is 
consistent with the predicted logarithmic dependence on $l$.
Moreover, the break-time 
results provide an effective method for estimating $\lambda_{qc}$ 
if we assume that (\ref{eqn:tb}) holds. 
The same fit procedure as detailed above yields the quantum-classical 
exponent $\lambda_{qc} = 1.1$. This fit result is 
plotted in  Fig.\ \ref{break_time_p0.1} as a solid line. More importantly, 
the exponential curve (\ref{eqn:expansatz}), plotted 
with fit result $\lambda_{qc} = 1.1$,  
can be seen to provide very good agreement with the initial growth rate 
of Fig.\ \ref{delta.2.835.140} for either initial condition, as expected. 

In the mixed regime ($\gamma=1.215$), 
the quantum-classical exponent $\lambda_{qc} = 0.43$ 
is an order of magnitude greater 
than the largest Lyapunov exponent $\lambda_L=0.04$ 
and about three times larger 
than the growth rate of the width $\lambda_w =0.13$. 
In the regime of global chaos ($\gamma=2.835$) 
the quantum-classical exponent $\lambda_{qc} = 1.1$ 
is a little more than twice as large as the 
largest Lyapunov exponent $\lambda_L=0.45$.

The condition $p< {\mathcal O}(1) $ is a very restrictive 
limitation on the domain of application of the log break-time 
(\ref{eqn:tb}) and it is worthwhile to explain the significance 
of this restriction. 
In the mixed regime case of Fig.\ \ref{delta.1.215.140.n200},  
with $l=154$, we have plotted the tolerance values 
$p=0.1$ (dotted line) and $p=15.4$ (sparse dotted line). 
The tolerance $p=0.1$ is exceeded at $t=11$, while 
the quantum-classical differences are still growing exponentially, 
leading to a log break-time for this tolerance value. 
For the tolerance $p=15.4 \ll |{\bf L}|$, on the other hand,  
the break-time does not occur on a measurable time-scale, whereas 
according to the logarithmic rule  (\ref{eqn:tb}), with $l = 154$ 
and $\lambda_{qc} = 0.43$, we should expect a rather 
short break-time $t_b \simeq 23$. Consequently the break-time  
(\ref{eqn:tb}), applied 
to delimiting the end of the Liouville regime, 
is not a robust measure of quantum-classical 
correspondence. 

Our definition of the break-time (\ref{eqn:tb})  
requires holding the tolerance $p$ fixed in absolute terms 
(and not as fraction of system dimension as in \cite{Haake87}) 
when comparing systems with different quantum numbers. 
Had we chosen to compare systems using a fixed relative tolerance, $f$,  
then the break-time would be of the form 
$t_b \simeq \lambda_{qc}^{-1} \ln ( 8 \; f \; l^2  )$ 
and subject to the restriction $f < {\mathcal O}(1/l)$. 
Since $f \rightarrow 0$ in the classical limit,  
this form emphasizes 
that the log break-time applies only to differences that are 
vanishing fraction of the system dimension in that limit.

Although we have provided numerical evidence 
(in Fig.\ \ref{delta.1.215.dvsl}) of one mixed regime case 
in which the largest quantum-classical differences 
occuring at the end of the exponential growth period 
remain essentially constant for varying quantum numbers, 
$\delta L_z (t^*) \sim {\mathcal O}(1)$, we find that 
this behaviour represents the typical case for all parameters and initial
conditions which produce chaos classically.  
To demonstrate this behaviour we consider the 
the scaling (with increasing quantum numbers) 
of the maximum values attained by $\delta L_z(n)$ 
over the first 200 kicks, $\delta L_z^{max}$. 
Since $t^* \ll 200$ over the range of $l$ values examined, 
the quantity $\delta L_z^{max}$ is a rigorous upper bound for 
$\delta L_z(t^*)$. 

\begin{figure}
\begin{center}
\input{deltamax.mod.l}
\vspace{0.2in}
\caption{
Maximum quantum-classical difference occuring over the first 200 kicks   
in the regime of global chaos ($\gamma=2.835$, $r\simeq 1.1$) 
plotted against increasing quantum number. These maximum values provide 
an upper bound on $\delta L_z(t^*)$ for each $l$. 
The data corresponding to the initial condition
$\vec{\theta}(0) =(20^o,40^o,160^o,130^o)$ (filled circles) 
represent a typical case in which 
the maximum quantum-classical differences 
do not vary significantly with $l$. 
The large deviations observed for the initial 
condition $\vec{\theta}(0) =(45^o,70^o,135^o,70^o)$ (open circles) 
are an exceptional case, with maximum differences growing rapidly for small 
quantum numbers but tending asymptotically toward independence of $l$. 
These curves provide an upper bound on the tolerance values 
$p$ for which the break-time measure scales logarithmicly with $l$. 
}
\label{deltamax.vs.l}
\end{center}
\end{figure}
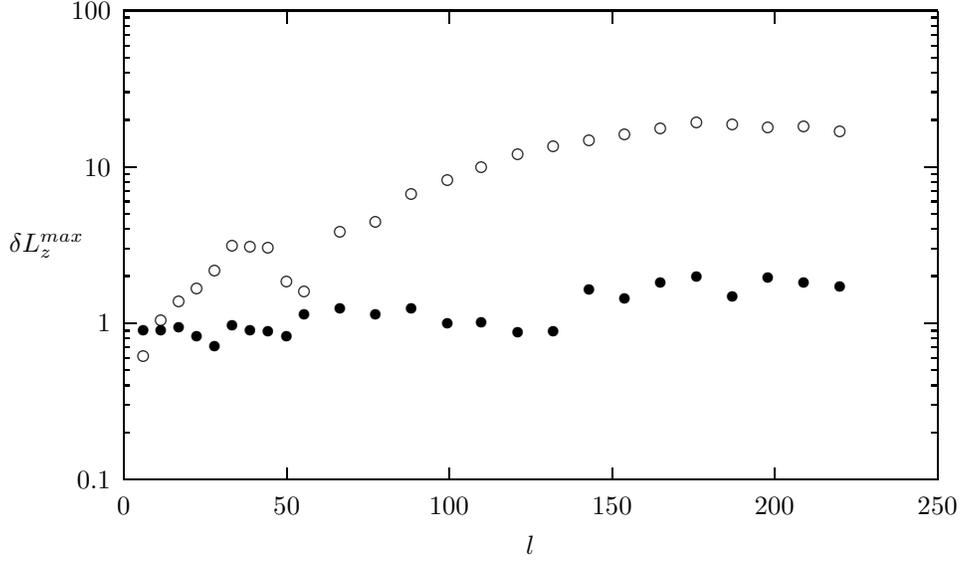

In Fig.\ \ref{deltamax.vs.l} we compare $\delta L_z^{max}$ 
for the two initial conditions of Fig.\ \ref{delta.2.835.140} 
and using the global chaos parameters ($\gamma =2.835$, $r \simeq 1.1$).  
The filled circles in Fig.\ \ref{deltamax.vs.l}  
correspond to the initial 
condition $\vec{\theta}(0) = (20^o,40^o,160^o,130^o)$.  
As in the mixed regime,  
the maximum deviations exhibit 
little or no scaling with increasing quantum number. This is the 
typical behaviour that we have observed  
for a variety of different initial conditions 
and parameter values. 
These results motivate the generic rule, 
\begin{equation}
\delta \tilde {L}_z(t^*) \le \delta \tilde{L}_z^{max} \sim {\mathcal O}(1/l). 
\end{equation}
Thus the magnitude of quantum-classical 
differences reached at the end of the exponential growth regime, 
expressed as a fraction of the system dimension, approaches  
zero in the classical limit. 

However, for a few combinations of parameters and initial conditions 
we do observe a `transient' discrepancy peak 
occuring at $ t \simeq t^* $ that exceeds ${\mathcal O}(1)$.  
This peak is quickly smoothed away 
by the subsequent relaxation of the quantum and classical distributions. 
This peak is apparent in 
Fig.\ \ref{delta.2.835.140} (open circles),  corresponding to 
the most conspicuous case that we have identified. This  
case is apparent as a small deviation in the normalized data of 
Fig.\ \ref{qmlmnm}. The scaling 
of the magnitude of this peak with increasing $l$ is 
plotted with open circles in Fig.\ \ref{deltamax.vs.l}. 
The magnitude of the peak initially increases rapidly but appears 
to become asymptotically independent of $l$. 
The other case that we have 
observed occurs for the classical parameters 
$\gamma =2.025$, with $r\simeq 1.1$ and $a=5$, and 
with initial condition $\vec{\theta}(0) = (20^o,40^o,160^o,130^o)$.  
We do not understand the 
mechanism leading to such transient peaks, although they are of 
considerable interest since they provide the most prominent 
examples of quantum-classical discrepancy that we have observed.

\section{Discussion}
\label{sec:discussion}

In this study of a non-integrable 
model of two interacting spins we have characterized the 
correspondence between quantum expectation values 
and classical ensemble averages for intially localised states.  
We have demonstrated that in chaotic states 
the quantum-classical differences initially grow 
exponentially with an exponent $\lambda_{qc}$ that is 
consistently larger than the largest Lyapunov exponent. 
In a study of the moments of the Henon-Heiles system, 
Ballentine and McRae \cite{Ball98,Ball00} have also shown 
that quantum-classical differences in chaotic states  
grow at an exponential rate with 
an exponent larger than the largest Lyapunov exponent. 
This exponential behaviour appears to be a generic feature of the 
short-time dynamics of quantum-classical differences in chaotic states. 


Since we have studied a spin system, we have been able to solve 
the quantum problem without truncation of the Hilbert space, 
subject only to numerical roundoff, and thus 
we are able to observe the dynamics of the quantum-classical differences 
well beyond the Ehrenfest regime. We have shown that the exponential growth 
phase of the quantum-classical differences 
terminates well before these differences
have reached system dimension. We find that the time-scale 
at which this occurs can be estimated  
from the time-scale at which the distribution widths 
approach the system dimension, 
$t_{sat} \simeq (2 \lambda_w)^{-1} \ln (l)$ for initial 
minimum uncertainty states. 
Due to the close correspondence in the growth rates of the 
quantum and classical distributions, this time-scale can 
be estimated from the classical physics alone. This is useful 
because the computational complexity of the problem does not grow  
with the system action in the classical case. Moreover, we find that 
the exponent $\lambda_w$ can be approximated by the largest 
Lyapunov exponent when the kinematic surface is predominantly chaotic. 

We have demonstrated that the exponent $\lambda_{qc}$ 
governing the initial growth rate of quantum-classical differences 
is independent of the quantum numbers, and that the effective prefactor to 
this exponential growth decreases as $1/l$. These results 
imply that a log break-time rule (\ref{eqn:tb})
delimits the dynamical regime of Liouville correspondence.  
However, the exponential growth of quantum-classical differences 
persists only for short times and small differences, and thus this 
log break-time rule applies only in a similarly restricted domain. 
In particular, we have found that the magnitude of the differences 
occuring at the end of the initial exponential growth phase  
does not scale with the system dimension. 
A typical magnitude for these differences, 
relative to the system dimension, is ${\mathcal O}(1/l)$. 
Therefore, $\log(l)$ break-time rules characterizing 
the end of the Liouville 
regime are not robust, since they 
apply to quantum-classical differences only in a restricted domain, 
{\it i.e}.\ to relative differences that are smaller 
than ${\mathcal O}(1/l)$. 

This restricted domain effect does not arise for the better known 
log break-time rules describing the end of the Ehrenfest 
regime \cite{Ball94,BZ78,Haake87}.  The Ehrenfest log 
break-time remains robust 
for arbitrarily large tolerances since the corresponding differences 
grow roughly exponentially until saturation at the system 
dimension \cite{Fox94b,Fox94a}. Consequently,  
a $\log(l)$ break-time indeed implies a {\em breakdown} of 
Ehrenfest correspondence.
However, the logarithmic break-time rule characterizing  
the end of the Liouville regime does not  
imply a breakdown of Liouville correspondence 
because it does not apply to the observation 
of quantum-classical discrepancies larger than ${\mathcal O}(1/l)$. 
The appearance of residual ${\mathcal O}(1/l)$ quantum-classical 
discrepancies in the description of a macroscopic body 
is, of course, consistent with quantum mechanics having a 
proper classical limit. 

We have found, however, that for certain exceptional 
combinations of parameters and initial conditions 
there are relative quantum-classical differences 
occuring at the end of the exponential growth phase that  
can be larger than ${\mathcal O}(1/l)$, though still 
much smaller than the system dimension. In absolute terms, these  
transient peaks seem to grow with the system dimension 
for small quantum numbers but become asymptotically 
independent of the system dimension for larger quantum numbers. 
Therefore, even in these least favorable 
cases, the {\em fractional} differences between quantum and 
classical dynamics approach zero in the limit $l \rightarrow \infty$. 
This vanishing of fractional differences is sufficient to 
ensure a classical limit for our model.  

Finally, contrary to the results found in the present model, 
it has been suggested that a log break-time delimiting 
the Liouville regime implies 
that certain isolated macroscopic bodies in chaotic 
motion should exhibit non-classical behaviour on observable time scales. 
However, since such non-classical behaviour is not observed in the chaotic 
motion of macroscopic bodies,   
it is argued that the observed classical behaviour emerges 
from quantum mechanics only when the quantum description is expanded to 
include interactions with the many degrees-of-freedom of the 
ubiquitous environment \cite{ZP95a,Zurek98b}. (This effect, 
called decoherence, rapidly evolves a pure system state 
into a mixture that is essentially devoid of non-classical properties.)    
However, 
in our model classical behaviour emerges in the macroscopic limit 
of a simple few degree-of-freedom quantum system that is 
described by a pure state and subject only to unitary evolution.  
Quantum-classical correspondence at both early and 
late times arises in spite of the log break-time because this break-time 
rule applies only when the quantum-classical difference threshold 
is chosen smaller than ${\mathcal O}(\hbar)$.  
In this sense we find that the decoherence 
effects of the environment are not necessary for correspondence 
in the macroscopic limit. Of course the effect of decoherence may be 
experimentally significant in the quantum and mesoscopic domains, but 
it is not required {\em as a matter of principle} to ensure a classical 
limit.

\section{Acknowledgements}

We wish to thank F.\ Haake and J.\ Weber for drawing our attention 
to the recursion algorithm for the rotation matrix elements 
published in \cite{Haake96}. J.\ E.\ would like to thank K. Kallio 
for stimulating discussions.

\section{Appendix}

Ideally we would like to construct an initial classical 
density that reproduces all of the moments of the 
initial quantum coherent states. This is possible 
in a Euclidean phase space, in which case all Weyl-ordered moments of 
the coherent state can be matched exactly by the moments of a 
Gaussian classical distribution. 
However, below we prove that no classical density $\rho_c(\theta,\phi)$
that describes an ensemble of spins of fixed length $|{\bf J}|$ 
can be constructed with marginal distributions that match   
those of the SU(2) coherent states (\ref{eqn:cs}). 
Specifically, we consider the set of  
distributions on ${\mathcal S}^2$ with continuous 
independent variables $\theta \in [ 0 , \pi ]$ 
and $ \phi \in [ 0,2\pi) $, 
measure  $ d \mu = \sin  \theta d \theta d \phi$, and subject to 
the usual normalization, 
\begin{equation}  
	\int_{{\mathcal S}^2} d \mu \; \rho_c(\theta,\phi) = 1.  
\end{equation}

For convenience we choose the coherent state 
to be polarized along the positive $z$-axis, $\rho = 
| j,j \rangle \langle j,j |$. This state is axially symmetric:  
rotations about the $z$-axis by an arbitrary 
angle $\phi$ leave the state operator invariant. 
Consequently we require axially symmetry of the  
corresponding classical distribution, 
\begin{equation}
\label{eqn:cazisymm}
\rho_c(\theta,\phi)= \rho_c(\theta).
\end{equation}

We use the expectation of the quadratic operator,  
  $ \langle {\bf J}^2 \rangle = j(j+1) $,
to fix the length of the classical spins,  
\begin{equation}
\label{eqn:length}
	|{\bf J}| = \sqrt{\langle J^2 \rangle_c} = \sqrt{j(j+1)}.  
\end{equation}
Furthermore, 
the coherent state $|j,j\rangle$ is an eigenstate of $J_z$ with 
moments along the $z$-axis given by
	$\langle J_z^n \rangle = j^n$
for integer $n$. 
Therefore we require that the classical distribution 
produces the moments, 
\begin{equation}
\label{eqn:cjz}
	\langle J_z^n \rangle_c = j^n. 
\end{equation}

These requirements are satisfied by 
the $\delta$-function distribution, 
\begin{equation}
\label{eqn:vm}
	 \rho_v(\theta) =  { \delta(\theta - 
	\theta_o) \over 2 \pi \sin \theta_o },   
\end{equation}
where $\cos \theta_o = j/|{\bf J}|$ defines $\theta_o$. 
This distribution is the 
familiar vector model of the old quantum theory 
corresponding to the intersection of a cone 
with the surface of the sphere. 

However, in order to derive an inconsistency between the quantum 
and classical moments we do not need to assume 
that the classical distribution is given explicitly 
by (\ref{eqn:vm}); we only need to make use of the  
the azimuthal invariance condition (\ref{eqn:cazisymm}), 
the length condition (\ref{eqn:length}), 
and the first two even moments of (\ref{eqn:cjz}).  

First we calculate some of the quantum coherent state moments 
along the $x$-axis (or any axis orthogonal to $z$),  
\begin{eqnarray}
\label{eqn:xmoments}
	\langle J_x^m \rangle & = & 0   \; \; \; 
{\mathrm for \; odd} \; m  \nonumber \\
	\langle J_x^2 \rangle & = & j/2 \nonumber \\ 
	\langle J_x^4 \rangle & = & 3j^2/4 -j/4 \nonumber.   
\end{eqnarray}
In the classical case, these moments are of the form,  
\begin{equation}
\label{eqn:jxm} 
\langle J_x^m \rangle_c  =  
\int d J_z \int d \phi \rho_c(\theta)  
|{\bf J}|^m \cos^m(\phi) \sin^m(\theta).  
\end{equation}
For $m$ odd the integral over $\phi$ vanishes, as required 
for correspondence with the odd quantum moments. For $m$ even 
we can evaluate (\ref{eqn:xmoments})
by expressing the r.h.s.\ as a linear combination 
of the $z$-axis moments (\ref{eqn:cjz}) of equal and lower order. 
For $m=2$ this requires substituting $\sin^2(\theta) = 1 - \cos^2(\theta) $ 
into (\ref{eqn:jxm}) and then integrating over $\phi$ to obtain 
\begin{eqnarray}
\langle J_x^2 \rangle_c  & = &   
\pi \int d J_z \rho_c(\theta) |{\bf J}|^2 
- \pi \int d J_z \rho_c(\theta) |{\bf J}|^2 \cos^2(\theta) \nonumber \\
& = & |{\bf J}| /2 - \langle J_z^2 \rangle /2. \nonumber 
\end{eqnarray}
Since $\langle J_z^2 \rangle$  is determined by 
(\ref{eqn:cjz}) and the length is fixed from (\ref{eqn:length})
we can deduce the classical value without knowing $\rho(\theta)$,  
\begin{equation}   
	\langle J_x^2 \rangle_c  =  j/2.  
\end{equation}
This agrees with the value of corresponding quantum moment. 
For $m=4$, however, by a similar procedure we deduce 
\begin{equation}   
	\langle J_x^4 \rangle_c =  3 j^2/8, 
\end{equation}
that differs from the quantum moment 
$ \langle J_x^4 \rangle $ 
by the factor, 
\begin{equation}
	\delta J_x^4 = 
| \langle J_x^4 \rangle - \langle J_x^4 \rangle_c | =  | 3 j^2/8 - j/4 |, 
\end{equation}
concluding our proof that 
no classical distribution on ${\mathcal S}^2$ can reproduce 
the quantum moments.

\newpage

\vspace{0.5in}

\vspace{0.5in} 

\vspace{0.5in} 

\vspace{0.5in} 

\vspace{0.5in}

\vspace{0.5in} 

\vspace{0.5in} 

\vspace{0.5in} 

\vspace{0.5in} 

\vspace{0.5in} 

\vspace{0.5in} 

\vspace{0.5in} 

\vspace{0.5in} 

\vspace{0.5in} 

\end{document}

%% file: regimes.tex
\setlength{\unitlength}{0.240900pt}
\ifx\plotpoint\undefined\newsavebox{\plotpoint}\fi
\sbox{\plotpoint}{\rule[-0.200pt]{0.400pt}{0.400pt}}%
\begin{picture}(1500,900)(0,0)
\font\gnuplot=cmr10 at 10pt
\gnuplot
\sbox{\plotpoint}{\rule[-0.200pt]{0.400pt}{0.400pt}}%
\put(161.0,123.0){\rule[-0.200pt]{2.409pt}{0.400pt}}
\put(1429.0,123.0){\rule[-0.200pt]{2.409pt}{0.400pt}}
\put(161.0,139.0){\rule[-0.200pt]{4.818pt}{0.400pt}}
\put(141,139){\makebox(0,0)[r]{1}}
\put(1419.0,139.0){\rule[-0.200pt]{4.818pt}{0.400pt}}
\put(161.0,247.0){\rule[-0.200pt]{2.409pt}{0.400pt}}
\put(1429.0,247.0){\rule[-0.200pt]{2.409pt}{0.400pt}}
\put(161.0,309.0){\rule[-0.200pt]{2.409pt}{0.400pt}}
\put(1429.0,309.0){\rule[-0.200pt]{2.409pt}{0.400pt}}
\put(161.0,354.0){\rule[-0.200pt]{2.409pt}{0.400pt}}
\put(1429.0,354.0){\rule[-0.200pt]{2.409pt}{0.400pt}}
\put(161.0,389.0){\rule[-0.200pt]{2.409pt}{0.400pt}}
\put(1429.0,389.0){\rule[-0.200pt]{2.409pt}{0.400pt}}
\put(161.0,417.0){\rule[-0.200pt]{2.409pt}{0.400pt}}
\put(1429.0,417.0){\rule[-0.200pt]{2.409pt}{0.400pt}}
\put(161.0,441.0){\rule[-0.200pt]{2.409pt}{0.400pt}}
\put(1429.0,441.0){\rule[-0.200pt]{2.409pt}{0.400pt}}
\put(161.0,461.0){\rule[-0.200pt]{2.409pt}{0.400pt}}
\put(1429.0,461.0){\rule[-0.200pt]{2.409pt}{0.400pt}}
\put(161.0,480.0){\rule[-0.200pt]{2.409pt}{0.400pt}}
\put(1429.0,480.0){\rule[-0.200pt]{2.409pt}{0.400pt}}
\put(161.0,496.0){\rule[-0.200pt]{4.818pt}{0.400pt}}
\put(141,496){\makebox(0,0)[r]{10}}
\put(1419.0,496.0){\rule[-0.200pt]{4.818pt}{0.400pt}}
\put(161.0,603.0){\rule[-0.200pt]{2.409pt}{0.400pt}}
\put(1429.0,603.0){\rule[-0.200pt]{2.409pt}{0.400pt}}
\put(161.0,666.0){\rule[-0.200pt]{2.409pt}{0.400pt}}
\put(1429.0,666.0){\rule[-0.200pt]{2.409pt}{0.400pt}}
\put(161.0,711.0){\rule[-0.200pt]{2.409pt}{0.400pt}}
\put(1429.0,711.0){\rule[-0.200pt]{2.409pt}{0.400pt}}
\put(161.0,745.0){\rule[-0.200pt]{2.409pt}{0.400pt}}
\put(1429.0,745.0){\rule[-0.200pt]{2.409pt}{0.400pt}}
\put(161.0,773.0){\rule[-0.200pt]{2.409pt}{0.400pt}}
\put(1429.0,773.0){\rule[-0.200pt]{2.409pt}{0.400pt}}
\put(161.0,797.0){\rule[-0.200pt]{2.409pt}{0.400pt}}
\put(1429.0,797.0){\rule[-0.200pt]{2.409pt}{0.400pt}}
\put(161.0,818.0){\rule[-0.200pt]{2.409pt}{0.400pt}}
\put(1429.0,818.0){\rule[-0.200pt]{2.409pt}{0.400pt}}
\put(161.0,836.0){\rule[-0.200pt]{2.409pt}{0.400pt}}
\put(1429.0,836.0){\rule[-0.200pt]{2.409pt}{0.400pt}}
\put(161.0,852.0){\rule[-0.200pt]{4.818pt}{0.400pt}}
\put(141,852){\makebox(0,0)[r]{100}}
\put(1419.0,852.0){\rule[-0.200pt]{4.818pt}{0.400pt}}
\put(161.0,123.0){\rule[-0.200pt]{0.400pt}{4.818pt}}
\put(161,82){\makebox(0,0){0}}
\put(161.0,840.0){\rule[-0.200pt]{0.400pt}{4.818pt}}
\put(321.0,123.0){\rule[-0.200pt]{0.400pt}{4.818pt}}
\put(321,82){\makebox(0,0){0.5}}
\put(321.0,840.0){\rule[-0.200pt]{0.400pt}{4.818pt}}
\put(481.0,123.0){\rule[-0.200pt]{0.400pt}{4.818pt}}
\put(481,82){\makebox(0,0){1}}
\put(481.0,840.0){\rule[-0.200pt]{0.400pt}{4.818pt}}
\put(640.0,123.0){\rule[-0.200pt]{0.400pt}{4.818pt}}
\put(640,82){\makebox(0,0){1.5}}
\put(640.0,840.0){\rule[-0.200pt]{0.400pt}{4.818pt}}
\put(800.0,123.0){\rule[-0.200pt]{0.400pt}{4.818pt}}
\put(800,82){\makebox(0,0){2}}
\put(800.0,840.0){\rule[-0.200pt]{0.400pt}{4.818pt}}
\put(960.0,123.0){\rule[-0.200pt]{0.400pt}{4.818pt}}
\put(960,82){\makebox(0,0){2.5}}
\put(960.0,840.0){\rule[-0.200pt]{0.400pt}{4.818pt}}
\put(1120.0,123.0){\rule[-0.200pt]{0.400pt}{4.818pt}}
\put(1120,82){\makebox(0,0){3}}
\put(1120.0,840.0){\rule[-0.200pt]{0.400pt}{4.818pt}}
\put(1279.0,123.0){\rule[-0.200pt]{0.400pt}{4.818pt}}
\put(1279,82){\makebox(0,0){3.5}}
\put(1279.0,840.0){\rule[-0.200pt]{0.400pt}{4.818pt}}
\put(1439.0,123.0){\rule[-0.200pt]{0.400pt}{4.818pt}}
\put(1439,82){\makebox(0,0){4}}
\put(1439.0,840.0){\rule[-0.200pt]{0.400pt}{4.818pt}}
\put(161.0,123.0){\rule[-0.200pt]{307.870pt}{0.400pt}}
\put(1439.0,123.0){\rule[-0.200pt]{0.400pt}{177.543pt}}
\put(161.0,860.0){\rule[-0.200pt]{307.870pt}{0.400pt}}
\put(40,591){\makebox(0,0){$ r $}}
\put(800,21){\makebox(0,0){$\gamma$}}
\put(161.0,123.0){\rule[-0.200pt]{0.400pt}{177.543pt}}
\put(1291,797){\makebox(0,0)[r]{Regular Dynamics}}
\put(193,139){\circle{18}}
\put(225,139){\circle{18}}
\put(193,175){\circle{18}}
\put(225,175){\circle{18}}
\put(193,210){\circle{18}}
\put(225,210){\circle{18}}
\put(193,496){\circle{18}}
\put(225,496){\circle{18}}
\put(193,852){\circle{18}}
\put(193,852){\circle{18}}
\put(193,532){\circle{18}}
\put(225,532){\circle{18}}
\put(193,568){\circle{18}}
\put(225,568){\circle{18}}
\put(193,247){\circle{18}}
\put(225,247){\circle{18}}
\put(193,281){\circle{18}}
\put(225,281){\circle{18}}
\put(193,603){\circle{18}}
\put(225,603){\circle{18}}
\put(193,638){\circle{18}}
\put(225,638){\circle{18}}
\put(193,317){\circle{18}}
\put(225,317){\circle{18}}
\put(193,674){\circle{18}}
\put(193,710){\circle{18}}
\put(193,354){\circle{18}}
\put(225,354){\circle{18}}
\put(193,389){\circle{18}}
\put(225,389){\circle{18}}
\put(193,745){\circle{18}}
\put(193,424){\circle{18}}
\put(225,424){\circle{18}}
\put(193,781){\circle{18}}
\put(193,817){\circle{18}}
\put(193,461){\circle{18}}
\put(225,461){\circle{18}}
\put(225,139){\circle{18}}
\put(257,139){\circle{18}}
\put(289,139){\circle{18}}
\put(321,139){\circle{18}}
\put(353,139){\circle{18}}
\put(385,139){\circle{18}}
\put(417,139){\circle{18}}
\put(449,139){\circle{18}}
\put(225,175){\circle{18}}
\put(257,175){\circle{18}}
\put(289,175){\circle{18}}
\put(321,175){\circle{18}}
\put(353,175){\circle{18}}
\put(385,175){\circle{18}}
\put(417,175){\circle{18}}
\put(449,175){\circle{18}}
\put(225,210){\circle{18}}
\put(257,210){\circle{18}}
\put(289,210){\circle{18}}
\put(321,210){\circle{18}}
\put(353,210){\circle{18}}
\put(385,210){\circle{18}}
\put(417,210){\circle{18}}
\put(449,210){\circle{18}}
\put(225,496){\circle{18}}
\put(257,496){\circle{18}}
\put(289,496){\circle{18}}
\put(225,532){\circle{18}}
\put(257,532){\circle{18}}
\put(225,568){\circle{18}}
\put(257,568){\circle{18}}
\put(225,247){\circle{18}}
\put(257,247){\circle{18}}
\put(289,247){\circle{18}}
\put(321,247){\circle{18}}
\put(353,247){\circle{18}}
\put(385,247){\circle{18}}
\put(417,247){\circle{18}}
\put(449,247){\circle{18}}
\put(225,281){\circle{18}}
\put(257,281){\circle{18}}
\put(289,281){\circle{18}}
\put(321,281){\circle{18}}
\put(353,281){\circle{18}}
\put(385,281){\circle{18}}
\put(417,281){\circle{18}}
\put(449,281){\circle{18}}
\put(225,603){\circle{18}}
\put(225,638){\circle{18}}
\put(225,317){\circle{18}}
\put(257,317){\circle{18}}
\put(289,317){\circle{18}}
\put(321,317){\circle{18}}
\put(353,317){\circle{18}}
\put(385,317){\circle{18}}
\put(417,317){\circle{18}}
\put(449,317){\circle{18}}
\put(225,674){\circle{18}}
\put(225,710){\circle{18}}
\put(225,354){\circle{18}}
\put(257,354){\circle{18}}
\put(289,354){\circle{18}}
\put(321,354){\circle{18}}
\put(353,354){\circle{18}}
\put(385,354){\circle{18}}
\put(225,389){\circle{18}}
\put(257,389){\circle{18}}
\put(289,389){\circle{18}}
\put(321,389){\circle{18}}
\put(353,389){\circle{18}}
\put(385,389){\circle{18}}
\put(225,424){\circle{18}}
\put(257,424){\circle{18}}
\put(289,424){\circle{18}}
\put(321,424){\circle{18}}
\put(353,424){\circle{18}}
\put(385,424){\circle{18}}
\put(225,461){\circle{18}}
\put(257,461){\circle{18}}
\put(289,461){\circle{18}}
\put(544,139){\circle{18}}
\put(481,139){\circle{18}}
\put(512,139){\circle{18}}
\put(544,139){\circle{18}}
\put(576,139){\circle{18}}
\put(608,139){\circle{18}}
\put(640,139){\circle{18}}
\put(672,139){\circle{18}}
\put(704,139){\circle{18}}
\put(736,139){\circle{18}}
\put(768,139){\circle{18}}
\put(481,175){\circle{18}}
\put(512,175){\circle{18}}
\put(544,175){\circle{18}}
\put(576,175){\circle{18}}
\put(608,175){\circle{18}}
\put(640,175){\circle{18}}
\put(672,175){\circle{18}}
\put(481,210){\circle{18}}
\put(512,210){\circle{18}}
\put(544,210){\circle{18}}
\put(576,210){\circle{18}}
\put(608,210){\circle{18}}
\put(640,210){\circle{18}}
\put(481,247){\circle{18}}
\put(512,247){\circle{18}}
\put(544,247){\circle{18}}
\put(481,281){\circle{18}}
\put(481,317){\circle{18}}
\put(512,317){\circle{18}}
\put(800,139){\circle{18}}
\put(832,139){\circle{18}}
\put(864,139){\circle{18}}
\put(896,139){\circle{18}}
\put(1361,797){\circle{18}}
\sbox{\plotpoint}{\rule[-0.500pt]{1.000pt}{1.000pt}}%
\put(1291,756){\makebox(0,0)[r]{Chaotic Dynamics}}
\put(225,496){\makebox(0,0){$\times$}}
\put(193,852){\makebox(0,0){$\times$}}
\put(225,852){\makebox(0,0){$\times$}}
\put(193,852){\makebox(0,0){$\times$}}
\put(225,852){\makebox(0,0){$\times$}}
\put(225,532){\makebox(0,0){$\times$}}
\put(225,568){\makebox(0,0){$\times$}}
\put(225,603){\makebox(0,0){$\times$}}
\put(193,638){\makebox(0,0){$\times$}}
\put(225,638){\makebox(0,0){$\times$}}
\put(193,674){\makebox(0,0){$\times$}}
\put(225,674){\makebox(0,0){$\times$}}
\put(193,710){\makebox(0,0){$\times$}}
\put(225,710){\makebox(0,0){$\times$}}
\put(193,745){\makebox(0,0){$\times$}}
\put(225,745){\makebox(0,0){$\times$}}
\put(193,781){\makebox(0,0){$\times$}}
\put(225,781){\makebox(0,0){$\times$}}
\put(193,817){\makebox(0,0){$\times$}}
\put(225,817){\makebox(0,0){$\times$}}
\put(225,461){\makebox(0,0){$\times$}}
\put(417,139){\makebox(0,0){$\times$}}
\put(449,139){\makebox(0,0){$\times$}}
\put(353,175){\makebox(0,0){$\times$}}
\put(385,175){\makebox(0,0){$\times$}}
\put(417,175){\makebox(0,0){$\times$}}
\put(449,175){\makebox(0,0){$\times$}}
\put(321,210){\makebox(0,0){$\times$}}
\put(353,210){\makebox(0,0){$\times$}}
\put(385,210){\makebox(0,0){$\times$}}
\put(417,210){\makebox(0,0){$\times$}}
\put(449,210){\makebox(0,0){$\times$}}
\put(225,496){\makebox(0,0){$\times$}}
\put(257,496){\makebox(0,0){$\times$}}
\put(289,496){\makebox(0,0){$\times$}}
\put(321,496){\makebox(0,0){$\times$}}
\put(353,496){\makebox(0,0){$\times$}}
\put(385,496){\makebox(0,0){$\times$}}
\put(417,496){\makebox(0,0){$\times$}}
\put(449,496){\makebox(0,0){$\times$}}
\put(225,852){\makebox(0,0){$\times$}}
\put(257,852){\makebox(0,0){$\times$}}
\put(289,852){\makebox(0,0){$\times$}}
\put(321,852){\makebox(0,0){$\times$}}
\put(353,852){\makebox(0,0){$\times$}}
\put(385,852){\makebox(0,0){$\times$}}
\put(417,852){\makebox(0,0){$\times$}}
\put(449,852){\makebox(0,0){$\times$}}
\put(225,532){\makebox(0,0){$\times$}}
\put(257,532){\makebox(0,0){$\times$}}
\put(289,532){\makebox(0,0){$\times$}}
\put(321,532){\makebox(0,0){$\times$}}
\put(353,532){\makebox(0,0){$\times$}}
\put(385,532){\makebox(0,0){$\times$}}
\put(417,532){\makebox(0,0){$\times$}}
\put(449,532){\makebox(0,0){$\times$}}
\put(225,568){\makebox(0,0){$\times$}}
\put(257,568){\makebox(0,0){$\times$}}
\put(289,568){\makebox(0,0){$\times$}}
\put(321,568){\makebox(0,0){$\times$}}
\put(353,568){\makebox(0,0){$\times$}}
\put(385,568){\makebox(0,0){$\times$}}
\put(417,568){\makebox(0,0){$\times$}}
\put(449,568){\makebox(0,0){$\times$}}
\put(321,247){\makebox(0,0){$\times$}}
\put(353,247){\makebox(0,0){$\times$}}
\put(385,247){\makebox(0,0){$\times$}}
\put(417,247){\makebox(0,0){$\times$}}
\put(449,247){\makebox(0,0){$\times$}}
\put(289,281){\makebox(0,0){$\times$}}
\put(321,281){\makebox(0,0){$\times$}}
\put(353,281){\makebox(0,0){$\times$}}
\put(385,281){\makebox(0,0){$\times$}}
\put(417,281){\makebox(0,0){$\times$}}
\put(449,281){\makebox(0,0){$\times$}}
\put(225,603){\makebox(0,0){$\times$}}
\put(257,603){\makebox(0,0){$\times$}}
\put(289,603){\makebox(0,0){$\times$}}
\put(321,603){\makebox(0,0){$\times$}}
\put(353,603){\makebox(0,0){$\times$}}
\put(385,603){\makebox(0,0){$\times$}}
\put(417,603){\makebox(0,0){$\times$}}
\put(449,603){\makebox(0,0){$\times$}}
\put(225,638){\makebox(0,0){$\times$}}
\put(257,638){\makebox(0,0){$\times$}}
\put(289,638){\makebox(0,0){$\times$}}
\put(321,638){\makebox(0,0){$\times$}}
\put(353,638){\makebox(0,0){$\times$}}
\put(385,638){\makebox(0,0){$\times$}}
\put(417,638){\makebox(0,0){$\times$}}
\put(449,638){\makebox(0,0){$\times$}}
\put(289,317){\makebox(0,0){$\times$}}
\put(321,317){\makebox(0,0){$\times$}}
\put(353,317){\makebox(0,0){$\times$}}
\put(385,317){\makebox(0,0){$\times$}}
\put(417,317){\makebox(0,0){$\times$}}
\put(449,317){\makebox(0,0){$\times$}}
\put(225,674){\makebox(0,0){$\times$}}
\put(257,674){\makebox(0,0){$\times$}}
\put(289,674){\makebox(0,0){$\times$}}
\put(321,674){\makebox(0,0){$\times$}}
\put(353,674){\makebox(0,0){$\times$}}
\put(385,674){\makebox(0,0){$\times$}}
\put(417,674){\makebox(0,0){$\times$}}
\put(449,674){\makebox(0,0){$\times$}}
\put(225,710){\makebox(0,0){$\times$}}
\put(257,710){\makebox(0,0){$\times$}}
\put(289,710){\makebox(0,0){$\times$}}
\put(321,710){\makebox(0,0){$\times$}}
\put(353,710){\makebox(0,0){$\times$}}
\put(385,710){\makebox(0,0){$\times$}}
\put(417,710){\makebox(0,0){$\times$}}
\put(449,710){\makebox(0,0){$\times$}}
\put(257,354){\makebox(0,0){$\times$}}
\put(289,354){\makebox(0,0){$\times$}}
\put(321,354){\makebox(0,0){$\times$}}
\put(353,354){\makebox(0,0){$\times$}}
\put(385,354){\makebox(0,0){$\times$}}
\put(417,354){\makebox(0,0){$\times$}}
\put(449,354){\makebox(0,0){$\times$}}
\put(225,389){\makebox(0,0){$\times$}}
\put(257,389){\makebox(0,0){$\times$}}
\put(289,389){\makebox(0,0){$\times$}}
\put(321,389){\makebox(0,0){$\times$}}
\put(353,389){\makebox(0,0){$\times$}}
\put(385,389){\makebox(0,0){$\times$}}
\put(417,389){\makebox(0,0){$\times$}}
\put(449,389){\makebox(0,0){$\times$}}
\put(225,745){\makebox(0,0){$\times$}}
\put(257,745){\makebox(0,0){$\times$}}
\put(289,745){\makebox(0,0){$\times$}}
\put(321,745){\makebox(0,0){$\times$}}
\put(353,745){\makebox(0,0){$\times$}}
\put(385,745){\makebox(0,0){$\times$}}
\put(417,745){\makebox(0,0){$\times$}}
\put(449,745){\makebox(0,0){$\times$}}
\put(225,424){\makebox(0,0){$\times$}}
\put(257,424){\makebox(0,0){$\times$}}
\put(289,424){\makebox(0,0){$\times$}}
\put(321,424){\makebox(0,0){$\times$}}
\put(353,424){\makebox(0,0){$\times$}}
\put(385,424){\makebox(0,0){$\times$}}
\put(417,424){\makebox(0,0){$\times$}}
\put(449,424){\makebox(0,0){$\times$}}
\put(225,781){\makebox(0,0){$\times$}}
\put(257,781){\makebox(0,0){$\times$}}
\put(289,781){\makebox(0,0){$\times$}}
\put(321,781){\makebox(0,0){$\times$}}
\put(353,781){\makebox(0,0){$\times$}}
\put(385,781){\makebox(0,0){$\times$}}
\put(417,781){\makebox(0,0){$\times$}}
\put(449,781){\makebox(0,0){$\times$}}
\put(225,817){\makebox(0,0){$\times$}}
\put(257,817){\makebox(0,0){$\times$}}
\put(289,817){\makebox(0,0){$\times$}}
\put(321,817){\makebox(0,0){$\times$}}
\put(353,817){\makebox(0,0){$\times$}}
\put(385,817){\makebox(0,0){$\times$}}
\put(417,817){\makebox(0,0){$\times$}}
\put(449,817){\makebox(0,0){$\times$}}
\put(225,461){\makebox(0,0){$\times$}}
\put(257,461){\makebox(0,0){$\times$}}
\put(289,461){\makebox(0,0){$\times$}}
\put(321,461){\makebox(0,0){$\times$}}
\put(353,461){\makebox(0,0){$\times$}}
\put(385,461){\makebox(0,0){$\times$}}
\put(417,461){\makebox(0,0){$\times$}}
\put(449,461){\makebox(0,0){$\times$}}
\put(481,139){\makebox(0,0){$\times$}}
\put(544,139){\makebox(0,0){$\times$}}
\put(481,139){\makebox(0,0){$\times$}}
\put(512,139){\makebox(0,0){$\times$}}
\put(544,139){\makebox(0,0){$\times$}}
\put(576,139){\makebox(0,0){$\times$}}
\put(608,139){\makebox(0,0){$\times$}}
\put(640,139){\makebox(0,0){$\times$}}
\put(672,139){\makebox(0,0){$\times$}}
\put(704,139){\makebox(0,0){$\times$}}
\put(736,139){\makebox(0,0){$\times$}}
\put(768,139){\makebox(0,0){$\times$}}
\put(481,175){\makebox(0,0){$\times$}}
\put(512,175){\makebox(0,0){$\times$}}
\put(544,175){\makebox(0,0){$\times$}}
\put(576,175){\makebox(0,0){$\times$}}
\put(608,175){\makebox(0,0){$\times$}}
\put(640,175){\makebox(0,0){$\times$}}
\put(672,175){\makebox(0,0){$\times$}}
\put(704,175){\makebox(0,0){$\times$}}
\put(736,175){\makebox(0,0){$\times$}}
\put(768,175){\makebox(0,0){$\times$}}
\put(481,210){\makebox(0,0){$\times$}}
\put(512,210){\makebox(0,0){$\times$}}
\put(544,210){\makebox(0,0){$\times$}}
\put(576,210){\makebox(0,0){$\times$}}
\put(608,210){\makebox(0,0){$\times$}}
\put(640,210){\makebox(0,0){$\times$}}
\put(672,210){\makebox(0,0){$\times$}}
\put(704,210){\makebox(0,0){$\times$}}
\put(736,210){\makebox(0,0){$\times$}}
\put(768,210){\makebox(0,0){$\times$}}
\put(481,496){\makebox(0,0){$\times$}}
\put(512,496){\makebox(0,0){$\times$}}
\put(544,496){\makebox(0,0){$\times$}}
\put(576,496){\makebox(0,0){$\times$}}
\put(608,496){\makebox(0,0){$\times$}}
\put(640,496){\makebox(0,0){$\times$}}
\put(672,496){\makebox(0,0){$\times$}}
\put(704,496){\makebox(0,0){$\times$}}
\put(736,496){\makebox(0,0){$\times$}}
\put(768,496){\makebox(0,0){$\times$}}
\put(481,852){\makebox(0,0){$\times$}}
\put(512,852){\makebox(0,0){$\times$}}
\put(544,852){\makebox(0,0){$\times$}}
\put(576,852){\makebox(0,0){$\times$}}
\put(608,852){\makebox(0,0){$\times$}}
\put(640,852){\makebox(0,0){$\times$}}
\put(672,852){\makebox(0,0){$\times$}}
\put(704,852){\makebox(0,0){$\times$}}
\put(736,852){\makebox(0,0){$\times$}}
\put(768,852){\makebox(0,0){$\times$}}
\put(481,532){\makebox(0,0){$\times$}}
\put(512,532){\makebox(0,0){$\times$}}
\put(544,532){\makebox(0,0){$\times$}}
\put(576,532){\makebox(0,0){$\times$}}
\put(608,532){\makebox(0,0){$\times$}}
\put(640,532){\makebox(0,0){$\times$}}
\put(672,532){\makebox(0,0){$\times$}}
\put(704,532){\makebox(0,0){$\times$}}
\put(736,532){\makebox(0,0){$\times$}}
\put(768,532){\makebox(0,0){$\times$}}
\put(481,568){\makebox(0,0){$\times$}}
\put(512,568){\makebox(0,0){$\times$}}
\put(544,568){\makebox(0,0){$\times$}}
\put(576,568){\makebox(0,0){$\times$}}
\put(608,568){\makebox(0,0){$\times$}}
\put(640,568){\makebox(0,0){$\times$}}
\put(672,568){\makebox(0,0){$\times$}}
\put(704,568){\makebox(0,0){$\times$}}
\put(736,568){\makebox(0,0){$\times$}}
\put(768,568){\makebox(0,0){$\times$}}
\put(481,247){\makebox(0,0){$\times$}}
\put(512,247){\makebox(0,0){$\times$}}
\put(544,247){\makebox(0,0){$\times$}}
\put(576,247){\makebox(0,0){$\times$}}
\put(608,247){\makebox(0,0){$\times$}}
\put(640,247){\makebox(0,0){$\times$}}
\put(672,247){\makebox(0,0){$\times$}}
\put(704,247){\makebox(0,0){$\times$}}
\put(736,247){\makebox(0,0){$\times$}}
\put(768,247){\makebox(0,0){$\times$}}
\put(481,281){\makebox(0,0){$\times$}}
\put(512,281){\makebox(0,0){$\times$}}
\put(544,281){\makebox(0,0){$\times$}}
\put(576,281){\makebox(0,0){$\times$}}
\put(608,281){\makebox(0,0){$\times$}}
\put(640,281){\makebox(0,0){$\times$}}
\put(672,281){\makebox(0,0){$\times$}}
\put(704,281){\makebox(0,0){$\times$}}
\put(736,281){\makebox(0,0){$\times$}}
\put(768,281){\makebox(0,0){$\times$}}
\put(481,603){\makebox(0,0){$\times$}}
\put(512,603){\makebox(0,0){$\times$}}
\put(544,603){\makebox(0,0){$\times$}}
\put(576,603){\makebox(0,0){$\times$}}
\put(608,603){\makebox(0,0){$\times$}}
\put(640,603){\makebox(0,0){$\times$}}
\put(672,603){\makebox(0,0){$\times$}}
\put(704,603){\makebox(0,0){$\times$}}
\put(736,603){\makebox(0,0){$\times$}}
\put(768,603){\makebox(0,0){$\times$}}
\put(481,638){\makebox(0,0){$\times$}}
\put(512,638){\makebox(0,0){$\times$}}
\put(544,638){\makebox(0,0){$\times$}}
\put(576,638){\makebox(0,0){$\times$}}
\put(608,638){\makebox(0,0){$\times$}}
\put(640,638){\makebox(0,0){$\times$}}
\put(672,638){\makebox(0,0){$\times$}}
\put(704,638){\makebox(0,0){$\times$}}
\put(736,638){\makebox(0,0){$\times$}}
\put(768,638){\makebox(0,0){$\times$}}
\put(481,317){\makebox(0,0){$\times$}}
\put(512,317){\makebox(0,0){$\times$}}
\put(544,317){\makebox(0,0){$\times$}}
\put(576,317){\makebox(0,0){$\times$}}
\put(608,317){\makebox(0,0){$\times$}}
\put(640,317){\makebox(0,0){$\times$}}
\put(672,317){\makebox(0,0){$\times$}}
\put(704,317){\makebox(0,0){$\times$}}
\put(736,317){\makebox(0,0){$\times$}}
\put(768,317){\makebox(0,0){$\times$}}
\put(481,674){\makebox(0,0){$\times$}}
\put(512,674){\makebox(0,0){$\times$}}
\put(544,674){\makebox(0,0){$\times$}}
\put(576,674){\makebox(0,0){$\times$}}
\put(608,674){\makebox(0,0){$\times$}}
\put(640,674){\makebox(0,0){$\times$}}
\put(672,674){\makebox(0,0){$\times$}}
\put(704,674){\makebox(0,0){$\times$}}
\put(736,674){\makebox(0,0){$\times$}}
\put(768,674){\makebox(0,0){$\times$}}
\put(481,710){\makebox(0,0){$\times$}}
\put(512,710){\makebox(0,0){$\times$}}
\put(544,710){\makebox(0,0){$\times$}}
\put(576,710){\makebox(0,0){$\times$}}
\put(608,710){\makebox(0,0){$\times$}}
\put(640,710){\makebox(0,0){$\times$}}
\put(672,710){\makebox(0,0){$\times$}}
\put(704,710){\makebox(0,0){$\times$}}
\put(736,710){\makebox(0,0){$\times$}}
\put(768,710){\makebox(0,0){$\times$}}
\put(481,354){\makebox(0,0){$\times$}}
\put(512,354){\makebox(0,0){$\times$}}
\put(544,354){\makebox(0,0){$\times$}}
\put(576,354){\makebox(0,0){$\times$}}
\put(608,354){\makebox(0,0){$\times$}}
\put(640,354){\makebox(0,0){$\times$}}
\put(672,354){\makebox(0,0){$\times$}}
\put(704,354){\makebox(0,0){$\times$}}
\put(736,354){\makebox(0,0){$\times$}}
\put(768,354){\makebox(0,0){$\times$}}
\put(481,389){\makebox(0,0){$\times$}}
\put(512,389){\makebox(0,0){$\times$}}
\put(544,389){\makebox(0,0){$\times$}}
\put(576,389){\makebox(0,0){$\times$}}
\put(608,389){\makebox(0,0){$\times$}}
\put(640,389){\makebox(0,0){$\times$}}
\put(672,389){\makebox(0,0){$\times$}}
\put(704,389){\makebox(0,0){$\times$}}
\put(736,389){\makebox(0,0){$\times$}}
\put(768,389){\makebox(0,0){$\times$}}
\put(481,745){\makebox(0,0){$\times$}}
\put(512,745){\makebox(0,0){$\times$}}
\put(544,745){\makebox(0,0){$\times$}}
\put(576,745){\makebox(0,0){$\times$}}
\put(608,745){\makebox(0,0){$\times$}}
\put(640,745){\makebox(0,0){$\times$}}
\put(672,745){\makebox(0,0){$\times$}}
\put(704,745){\makebox(0,0){$\times$}}
\put(736,745){\makebox(0,0){$\times$}}
\put(768,745){\makebox(0,0){$\times$}}
\put(481,424){\makebox(0,0){$\times$}}
\put(512,424){\makebox(0,0){$\times$}}
\put(544,424){\makebox(0,0){$\times$}}
\put(576,424){\makebox(0,0){$\times$}}
\put(608,424){\makebox(0,0){$\times$}}
\put(640,424){\makebox(0,0){$\times$}}
\put(672,424){\makebox(0,0){$\times$}}
\put(704,424){\makebox(0,0){$\times$}}
\put(736,424){\makebox(0,0){$\times$}}
\put(768,424){\makebox(0,0){$\times$}}
\put(481,781){\makebox(0,0){$\times$}}
\put(512,781){\makebox(0,0){$\times$}}
\put(544,781){\makebox(0,0){$\times$}}
\put(576,781){\makebox(0,0){$\times$}}
\put(608,781){\makebox(0,0){$\times$}}
\put(640,781){\makebox(0,0){$\times$}}
\put(672,781){\makebox(0,0){$\times$}}
\put(704,781){\makebox(0,0){$\times$}}
\put(736,781){\makebox(0,0){$\times$}}
\put(768,781){\makebox(0,0){$\times$}}
\put(481,817){\makebox(0,0){$\times$}}
\put(512,817){\makebox(0,0){$\times$}}
\put(544,817){\makebox(0,0){$\times$}}
\put(576,817){\makebox(0,0){$\times$}}
\put(608,817){\makebox(0,0){$\times$}}
\put(640,817){\makebox(0,0){$\times$}}
\put(672,817){\makebox(0,0){$\times$}}
\put(704,817){\makebox(0,0){$\times$}}
\put(736,817){\makebox(0,0){$\times$}}
\put(768,817){\makebox(0,0){$\times$}}
\put(481,461){\makebox(0,0){$\times$}}
\put(512,461){\makebox(0,0){$\times$}}
\put(544,461){\makebox(0,0){$\times$}}
\put(576,461){\makebox(0,0){$\times$}}
\put(608,461){\makebox(0,0){$\times$}}
\put(640,461){\makebox(0,0){$\times$}}
\put(672,461){\makebox(0,0){$\times$}}
\put(704,461){\makebox(0,0){$\times$}}
\put(736,461){\makebox(0,0){$\times$}}
\put(768,461){\makebox(0,0){$\times$}}
\put(800,139){\makebox(0,0){$\times$}}
\put(832,139){\makebox(0,0){$\times$}}
\put(864,139){\makebox(0,0){$\times$}}
\put(896,139){\makebox(0,0){$\times$}}
\put(928,139){\makebox(0,0){$\times$}}
\put(960,139){\makebox(0,0){$\times$}}
\put(992,139){\makebox(0,0){$\times$}}
\put(1024,139){\makebox(0,0){$\times$}}
\put(1056,139){\makebox(0,0){$\times$}}
\put(1088,139){\makebox(0,0){$\times$}}
\put(800,175){\makebox(0,0){$\times$}}
\put(832,175){\makebox(0,0){$\times$}}
\put(864,175){\makebox(0,0){$\times$}}
\put(896,175){\makebox(0,0){$\times$}}
\put(928,175){\makebox(0,0){$\times$}}
\put(960,175){\makebox(0,0){$\times$}}
\put(992,175){\makebox(0,0){$\times$}}
\put(1024,175){\makebox(0,0){$\times$}}
\put(1056,175){\makebox(0,0){$\times$}}
\put(1088,175){\makebox(0,0){$\times$}}
\put(800,210){\makebox(0,0){$\times$}}
\put(832,210){\makebox(0,0){$\times$}}
\put(864,210){\makebox(0,0){$\times$}}
\put(896,210){\makebox(0,0){$\times$}}
\put(928,210){\makebox(0,0){$\times$}}
\put(960,210){\makebox(0,0){$\times$}}
\put(992,210){\makebox(0,0){$\times$}}
\put(1024,210){\makebox(0,0){$\times$}}
\put(1056,210){\makebox(0,0){$\times$}}
\put(1088,210){\makebox(0,0){$\times$}}
\put(800,496){\makebox(0,0){$\times$}}
\put(832,496){\makebox(0,0){$\times$}}
\put(864,496){\makebox(0,0){$\times$}}
\put(896,496){\makebox(0,0){$\times$}}
\put(928,496){\makebox(0,0){$\times$}}
\put(960,496){\makebox(0,0){$\times$}}
\put(992,496){\makebox(0,0){$\times$}}
\put(1024,496){\makebox(0,0){$\times$}}
\put(1056,496){\makebox(0,0){$\times$}}
\put(1088,496){\makebox(0,0){$\times$}}
\put(800,852){\makebox(0,0){$\times$}}
\put(800,532){\makebox(0,0){$\times$}}
\put(832,532){\makebox(0,0){$\times$}}
\put(864,532){\makebox(0,0){$\times$}}
\put(896,532){\makebox(0,0){$\times$}}
\put(928,532){\makebox(0,0){$\times$}}
\put(960,532){\makebox(0,0){$\times$}}
\put(992,532){\makebox(0,0){$\times$}}
\put(1024,532){\makebox(0,0){$\times$}}
\put(1056,532){\makebox(0,0){$\times$}}
\put(1088,532){\makebox(0,0){$\times$}}
\put(800,568){\makebox(0,0){$\times$}}
\put(832,568){\makebox(0,0){$\times$}}
\put(864,568){\makebox(0,0){$\times$}}
\put(896,568){\makebox(0,0){$\times$}}
\put(928,568){\makebox(0,0){$\times$}}
\put(960,568){\makebox(0,0){$\times$}}
\put(992,568){\makebox(0,0){$\times$}}
\put(1024,568){\makebox(0,0){$\times$}}
\put(1056,568){\makebox(0,0){$\times$}}
\put(1088,568){\makebox(0,0){$\times$}}
\put(800,247){\makebox(0,0){$\times$}}
\put(832,247){\makebox(0,0){$\times$}}
\put(864,247){\makebox(0,0){$\times$}}
\put(896,247){\makebox(0,0){$\times$}}
\put(928,247){\makebox(0,0){$\times$}}
\put(960,247){\makebox(0,0){$\times$}}
\put(992,247){\makebox(0,0){$\times$}}
\put(1024,247){\makebox(0,0){$\times$}}
\put(1056,247){\makebox(0,0){$\times$}}
\put(1088,247){\makebox(0,0){$\times$}}
\put(800,281){\makebox(0,0){$\times$}}
\put(832,281){\makebox(0,0){$\times$}}
\put(864,281){\makebox(0,0){$\times$}}
\put(896,281){\makebox(0,0){$\times$}}
\put(928,281){\makebox(0,0){$\times$}}
\put(960,281){\makebox(0,0){$\times$}}
\put(992,281){\makebox(0,0){$\times$}}
\put(1024,281){\makebox(0,0){$\times$}}
\put(1056,281){\makebox(0,0){$\times$}}
\put(1088,281){\makebox(0,0){$\times$}}
\put(800,603){\makebox(0,0){$\times$}}
\put(832,603){\makebox(0,0){$\times$}}
\put(864,603){\makebox(0,0){$\times$}}
\put(896,603){\makebox(0,0){$\times$}}
\put(928,603){\makebox(0,0){$\times$}}
\put(960,603){\makebox(0,0){$\times$}}
\put(992,603){\makebox(0,0){$\times$}}
\put(1024,603){\makebox(0,0){$\times$}}
\put(1056,603){\makebox(0,0){$\times$}}
\put(1088,603){\makebox(0,0){$\times$}}
\put(800,638){\makebox(0,0){$\times$}}
\put(832,638){\makebox(0,0){$\times$}}
\put(864,638){\makebox(0,0){$\times$}}
\put(896,638){\makebox(0,0){$\times$}}
\put(928,638){\makebox(0,0){$\times$}}
\put(960,638){\makebox(0,0){$\times$}}
\put(992,638){\makebox(0,0){$\times$}}
\put(1024,638){\makebox(0,0){$\times$}}
\put(1056,638){\makebox(0,0){$\times$}}
\put(1088,638){\makebox(0,0){$\times$}}
\put(800,317){\makebox(0,0){$\times$}}
\put(832,317){\makebox(0,0){$\times$}}
\put(864,317){\makebox(0,0){$\times$}}
\put(896,317){\makebox(0,0){$\times$}}
\put(928,317){\makebox(0,0){$\times$}}
\put(960,317){\makebox(0,0){$\times$}}
\put(992,317){\makebox(0,0){$\times$}}
\put(1024,317){\makebox(0,0){$\times$}}
\put(1056,317){\makebox(0,0){$\times$}}
\put(1088,317){\makebox(0,0){$\times$}}
\put(800,674){\makebox(0,0){$\times$}}
\put(832,674){\makebox(0,0){$\times$}}
\put(864,674){\makebox(0,0){$\times$}}
\put(896,674){\makebox(0,0){$\times$}}
\put(928,674){\makebox(0,0){$\times$}}
\put(960,674){\makebox(0,0){$\times$}}
\put(992,674){\makebox(0,0){$\times$}}
\put(1024,674){\makebox(0,0){$\times$}}
\put(1056,674){\makebox(0,0){$\times$}}
\put(1088,674){\makebox(0,0){$\times$}}
\put(800,710){\makebox(0,0){$\times$}}
\put(800,354){\makebox(0,0){$\times$}}
\put(832,354){\makebox(0,0){$\times$}}
\put(864,354){\makebox(0,0){$\times$}}
\put(896,354){\makebox(0,0){$\times$}}
\put(928,354){\makebox(0,0){$\times$}}
\put(960,354){\makebox(0,0){$\times$}}
\put(992,354){\makebox(0,0){$\times$}}
\put(1024,354){\makebox(0,0){$\times$}}
\put(1056,354){\makebox(0,0){$\times$}}
\put(1088,354){\makebox(0,0){$\times$}}
\put(800,389){\makebox(0,0){$\times$}}
\put(832,389){\makebox(0,0){$\times$}}
\put(864,389){\makebox(0,0){$\times$}}
\put(896,389){\makebox(0,0){$\times$}}
\put(928,389){\makebox(0,0){$\times$}}
\put(960,389){\makebox(0,0){$\times$}}
\put(992,389){\makebox(0,0){$\times$}}
\put(1024,389){\makebox(0,0){$\times$}}
\put(1056,389){\makebox(0,0){$\times$}}
\put(1088,389){\makebox(0,0){$\times$}}
\put(800,745){\makebox(0,0){$\times$}}
\put(800,424){\makebox(0,0){$\times$}}
\put(832,424){\makebox(0,0){$\times$}}
\put(864,424){\makebox(0,0){$\times$}}
\put(896,424){\makebox(0,0){$\times$}}
\put(928,424){\makebox(0,0){$\times$}}
\put(960,424){\makebox(0,0){$\times$}}
\put(992,424){\makebox(0,0){$\times$}}
\put(1024,424){\makebox(0,0){$\times$}}
\put(1056,424){\makebox(0,0){$\times$}}
\put(1088,424){\makebox(0,0){$\times$}}
\put(800,781){\makebox(0,0){$\times$}}
\put(800,817){\makebox(0,0){$\times$}}
\put(800,461){\makebox(0,0){$\times$}}
\put(832,461){\makebox(0,0){$\times$}}
\put(864,461){\makebox(0,0){$\times$}}
\put(896,461){\makebox(0,0){$\times$}}
\put(928,461){\makebox(0,0){$\times$}}
\put(960,461){\makebox(0,0){$\times$}}
\put(992,461){\makebox(0,0){$\times$}}
\put(1024,461){\makebox(0,0){$\times$}}
\put(1056,461){\makebox(0,0){$\times$}}
\put(1088,461){\makebox(0,0){$\times$}}
\put(1120,139){\makebox(0,0){$\times$}}
\put(1151,139){\makebox(0,0){$\times$}}
\put(1183,139){\makebox(0,0){$\times$}}
\put(1215,139){\makebox(0,0){$\times$}}
\put(1247,139){\makebox(0,0){$\times$}}
\put(1279,139){\makebox(0,0){$\times$}}
\put(1311,139){\makebox(0,0){$\times$}}
\put(1343,139){\makebox(0,0){$\times$}}
\put(1375,139){\makebox(0,0){$\times$}}
\put(1407,139){\makebox(0,0){$\times$}}
\put(1120,175){\makebox(0,0){$\times$}}
\put(1151,175){\makebox(0,0){$\times$}}
\put(1183,175){\makebox(0,0){$\times$}}
\put(1215,175){\makebox(0,0){$\times$}}
\put(1247,175){\makebox(0,0){$\times$}}
\put(1279,175){\makebox(0,0){$\times$}}
\put(1311,175){\makebox(0,0){$\times$}}
\put(1343,175){\makebox(0,0){$\times$}}
\put(1375,175){\makebox(0,0){$\times$}}
\put(1407,175){\makebox(0,0){$\times$}}
\put(1120,210){\makebox(0,0){$\times$}}
\put(1151,210){\makebox(0,0){$\times$}}
\put(1183,210){\makebox(0,0){$\times$}}
\put(1215,210){\makebox(0,0){$\times$}}
\put(1247,210){\makebox(0,0){$\times$}}
\put(1279,210){\makebox(0,0){$\times$}}
\put(1311,210){\makebox(0,0){$\times$}}
\put(1343,210){\makebox(0,0){$\times$}}
\put(1375,210){\makebox(0,0){$\times$}}
\put(1407,210){\makebox(0,0){$\times$}}
\put(1120,496){\makebox(0,0){$\times$}}
\put(1151,496){\makebox(0,0){$\times$}}
\put(1183,496){\makebox(0,0){$\times$}}
\put(1215,496){\makebox(0,0){$\times$}}
\put(1247,496){\makebox(0,0){$\times$}}
\put(1279,496){\makebox(0,0){$\times$}}
\put(1311,496){\makebox(0,0){$\times$}}
\put(1343,496){\makebox(0,0){$\times$}}
\put(1375,496){\makebox(0,0){$\times$}}
\put(1407,496){\makebox(0,0){$\times$}}
\put(1120,532){\makebox(0,0){$\times$}}
\put(1151,532){\makebox(0,0){$\times$}}
\put(1183,532){\makebox(0,0){$\times$}}
\put(1215,532){\makebox(0,0){$\times$}}
\put(1247,532){\makebox(0,0){$\times$}}
\put(1279,532){\makebox(0,0){$\times$}}
\put(1311,532){\makebox(0,0){$\times$}}
\put(1343,532){\makebox(0,0){$\times$}}
\put(1375,532){\makebox(0,0){$\times$}}
\put(1407,532){\makebox(0,0){$\times$}}
\put(1120,568){\makebox(0,0){$\times$}}
\put(1151,568){\makebox(0,0){$\times$}}
\put(1183,568){\makebox(0,0){$\times$}}
\put(1215,568){\makebox(0,0){$\times$}}
\put(1247,568){\makebox(0,0){$\times$}}
\put(1279,568){\makebox(0,0){$\times$}}
\put(1311,568){\makebox(0,0){$\times$}}
\put(1343,568){\makebox(0,0){$\times$}}
\put(1375,568){\makebox(0,0){$\times$}}
\put(1407,568){\makebox(0,0){$\times$}}
\put(1120,247){\makebox(0,0){$\times$}}
\put(1151,247){\makebox(0,0){$\times$}}
\put(1183,247){\makebox(0,0){$\times$}}
\put(1215,247){\makebox(0,0){$\times$}}
\put(1247,247){\makebox(0,0){$\times$}}
\put(1279,247){\makebox(0,0){$\times$}}
\put(1311,247){\makebox(0,0){$\times$}}
\put(1343,247){\makebox(0,0){$\times$}}
\put(1375,247){\makebox(0,0){$\times$}}
\put(1407,247){\makebox(0,0){$\times$}}
\put(1120,281){\makebox(0,0){$\times$}}
\put(1151,281){\makebox(0,0){$\times$}}
\put(1183,281){\makebox(0,0){$\times$}}
\put(1215,281){\makebox(0,0){$\times$}}
\put(1247,281){\makebox(0,0){$\times$}}
\put(1279,281){\makebox(0,0){$\times$}}
\put(1311,281){\makebox(0,0){$\times$}}
\put(1343,281){\makebox(0,0){$\times$}}
\put(1375,281){\makebox(0,0){$\times$}}
\put(1407,281){\makebox(0,0){$\times$}}
\put(1120,603){\makebox(0,0){$\times$}}
\put(1151,603){\makebox(0,0){$\times$}}
\put(1183,603){\makebox(0,0){$\times$}}
\put(1215,603){\makebox(0,0){$\times$}}
\put(1247,603){\makebox(0,0){$\times$}}
\put(1279,603){\makebox(0,0){$\times$}}
\put(1311,603){\makebox(0,0){$\times$}}
\put(1343,603){\makebox(0,0){$\times$}}
\put(1375,603){\makebox(0,0){$\times$}}
\put(1407,603){\makebox(0,0){$\times$}}
\put(1120,638){\makebox(0,0){$\times$}}
\put(1151,638){\makebox(0,0){$\times$}}
\put(1183,638){\makebox(0,0){$\times$}}
\put(1215,638){\makebox(0,0){$\times$}}
\put(1247,638){\makebox(0,0){$\times$}}
\put(1279,638){\makebox(0,0){$\times$}}
\put(1311,638){\makebox(0,0){$\times$}}
\put(1343,638){\makebox(0,0){$\times$}}
\put(1375,638){\makebox(0,0){$\times$}}
\put(1407,638){\makebox(0,0){$\times$}}
\put(1120,317){\makebox(0,0){$\times$}}
\put(1151,317){\makebox(0,0){$\times$}}
\put(1183,317){\makebox(0,0){$\times$}}
\put(1215,317){\makebox(0,0){$\times$}}
\put(1247,317){\makebox(0,0){$\times$}}
\put(1279,317){\makebox(0,0){$\times$}}
\put(1311,317){\makebox(0,0){$\times$}}
\put(1343,317){\makebox(0,0){$\times$}}
\put(1375,317){\makebox(0,0){$\times$}}
\put(1407,317){\makebox(0,0){$\times$}}
\put(1120,674){\makebox(0,0){$\times$}}
\put(1151,674){\makebox(0,0){$\times$}}
\put(1183,674){\makebox(0,0){$\times$}}
\put(1215,674){\makebox(0,0){$\times$}}
\put(1247,674){\makebox(0,0){$\times$}}
\put(1279,674){\makebox(0,0){$\times$}}
\put(1311,674){\makebox(0,0){$\times$}}
\put(1343,674){\makebox(0,0){$\times$}}
\put(1375,674){\makebox(0,0){$\times$}}
\put(1407,674){\makebox(0,0){$\times$}}
\put(1120,354){\makebox(0,0){$\times$}}
\put(1151,354){\makebox(0,0){$\times$}}
\put(1183,354){\makebox(0,0){$\times$}}
\put(1215,354){\makebox(0,0){$\times$}}
\put(1247,354){\makebox(0,0){$\times$}}
\put(1279,354){\makebox(0,0){$\times$}}
\put(1311,354){\makebox(0,0){$\times$}}
\put(1343,354){\makebox(0,0){$\times$}}
\put(1375,354){\makebox(0,0){$\times$}}
\put(1407,354){\makebox(0,0){$\times$}}
\put(1120,389){\makebox(0,0){$\times$}}
\put(1151,389){\makebox(0,0){$\times$}}
\put(1183,389){\makebox(0,0){$\times$}}
\put(1215,389){\makebox(0,0){$\times$}}
\put(1247,389){\makebox(0,0){$\times$}}
\put(1279,389){\makebox(0,0){$\times$}}
\put(1311,389){\makebox(0,0){$\times$}}
\put(1343,389){\makebox(0,0){$\times$}}
\put(1375,389){\makebox(0,0){$\times$}}
\put(1407,389){\makebox(0,0){$\times$}}
\put(1120,424){\makebox(0,0){$\times$}}
\put(1151,424){\makebox(0,0){$\times$}}
\put(1183,424){\makebox(0,0){$\times$}}
\put(1215,424){\makebox(0,0){$\times$}}
\put(1247,424){\makebox(0,0){$\times$}}
\put(1279,424){\makebox(0,0){$\times$}}
\put(1311,424){\makebox(0,0){$\times$}}
\put(1343,424){\makebox(0,0){$\times$}}
\put(1375,424){\makebox(0,0){$\times$}}
\put(1407,424){\makebox(0,0){$\times$}}
\put(1120,461){\makebox(0,0){$\times$}}
\put(1151,461){\makebox(0,0){$\times$}}
\put(1183,461){\makebox(0,0){$\times$}}
\put(1215,461){\makebox(0,0){$\times$}}
\put(1247,461){\makebox(0,0){$\times$}}
\put(1279,461){\makebox(0,0){$\times$}}
\put(1311,461){\makebox(0,0){$\times$}}
\put(1343,461){\makebox(0,0){$\times$}}
\put(1375,461){\makebox(0,0){$\times$}}
\put(1407,461){\makebox(0,0){$\times$}}
\put(1361,756){\makebox(0,0){$\times$}}
\sbox{\plotpoint}{\rule[-0.600pt]{1.200pt}{1.200pt}}%
\put(551,157){\circle*{18}}
\put(1068,157){\circle*{18}}
\end{picture}

%% file: fig1.tex
\begingroup%
  \makeatletter%
  \newcommand{\GNUPLOTspecial}{%
    \@sanitize\catcode`\%=14\relax\special}%
  \setlength{\unitlength}{0.1bp}%
\begin{picture}(3600,2160)(0,0)%
\special{psfile=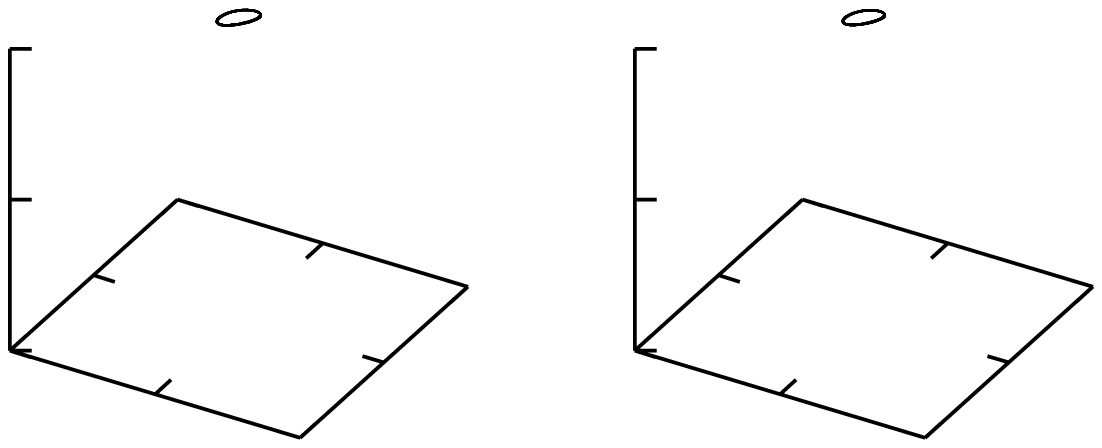 llx=0 lly=0 urx=720 ury=504 rwi=7200}
\put(2508,1906){\makebox(0,0){$\tilde{L}_z$}}%
\put(2382,1688){\makebox(0,0)[r]{1}}%
\put(2382,1254){\makebox(0,0)[r]{0}}%
\put(2382,819){\makebox(0,0)[r]{-1}}%
\put(3794,623){\makebox(0,0){$\tilde{L}_y$}}%
\put(3874,974){\makebox(0,0)[l]{1}}%
\put(3632,756){\makebox(0,0)[l]{0}}%
\put(3391,539){\makebox(0,0)[l]{-1}}%
\put(2756,585){\makebox(0,0){$\tilde{L}_x$}}%
\put(3306,501){\makebox(0,0){1}}%
\put(2888,626){\makebox(0,0){0}}%
\put(2470,752){\makebox(0,0){-1}}%
\put(708,1906){\makebox(0,0){$\tilde{S}_z$}}%
\put(582,1688){\makebox(0,0)[r]{1}}%
\put(582,1254){\makebox(0,0)[r]{0}}%
\put(582,819){\makebox(0,0)[r]{-1}}%
\put(1994,623){\makebox(0,0){$\tilde{S}_y$}}%
\put(2074,974){\makebox(0,0)[l]{1}}%
\put(1832,756){\makebox(0,0)[l]{0}}%
\put(1591,539){\makebox(0,0)[l]{-1}}%
\put(956,585){\makebox(0,0){$\tilde{S}_x$}}%
\put(1506,501){\makebox(0,0){1}}%
\put(1088,626){\makebox(0,0){0}}%
\put(670,752){\makebox(0,0){-1}}%
\end{picture}%
\endgroup
 

%% file: fig2.tex
\begingroup%
  \makeatletter%
  \newcommand{\GNUPLOTspecial}{%
    \@sanitize\catcode`\%=14\relax\special}%
  \setlength{\unitlength}{0.1bp}%
\begin{picture}(3600,2160)(0,0)%
\special{psfile=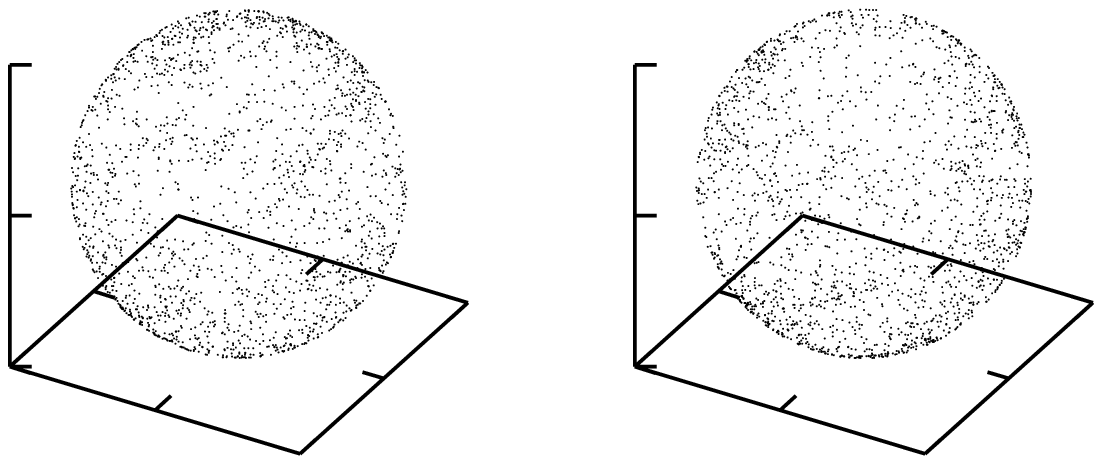 llx=0 lly=0 urx=720 ury=504 rwi=7200}
\put(2508,1906){\makebox(0,0){$\tilde{L}_z$}}%
\put(2382,1688){\makebox(0,0)[r]{1}}%
\put(2382,1254){\makebox(0,0)[r]{0}}%
\put(2382,819){\makebox(0,0)[r]{-1}}%
\put(3794,623){\makebox(0,0){$\tilde{L}_y$}}%
\put(3874,974){\makebox(0,0)[l]{1}}%
\put(3632,756){\makebox(0,0)[l]{0}}%
\put(3391,539){\makebox(0,0)[l]{-1}}%
\put(2756,585){\makebox(0,0){$\tilde{L}_x$}}%
\put(3306,501){\makebox(0,0){1}}%
\put(2888,626){\makebox(0,0){0}}%
\put(2470,752){\makebox(0,0){-1}}%
\put(708,1906){\makebox(0,0){$\tilde{S}_z$}}%
\put(582,1688){\makebox(0,0)[r]{1}}%
\put(582,1254){\makebox(0,0)[r]{0}}%
\put(582,819){\makebox(0,0)[r]{-1}}%
\put(1994,623){\makebox(0,0){$\tilde{S}_y$}}%
\put(2074,974){\makebox(0,0)[l]{1}}%
\put(1832,756){\makebox(0,0)[l]{0}}%
\put(1591,539){\makebox(0,0)[l]{-1}}%
\put(956,585){\makebox(0,0){$\tilde{S}_x$}}%
\put(1506,501){\makebox(0,0){1}}%
\put(1088,626){\makebox(0,0){0}}%
\put(670,752){\makebox(0,0){-1}}%
\end{picture}%
\endgroup
 

%% file: fig3.tex
\begingroup%
  \makeatletter%
  \newcommand{\GNUPLOTspecial}{%
    \@sanitize\catcode`\%=14\relax\special}%
  \setlength{\unitlength}{0.1bp}%
\begin{picture}(3600,2160)(0,0)%
\special{psfile=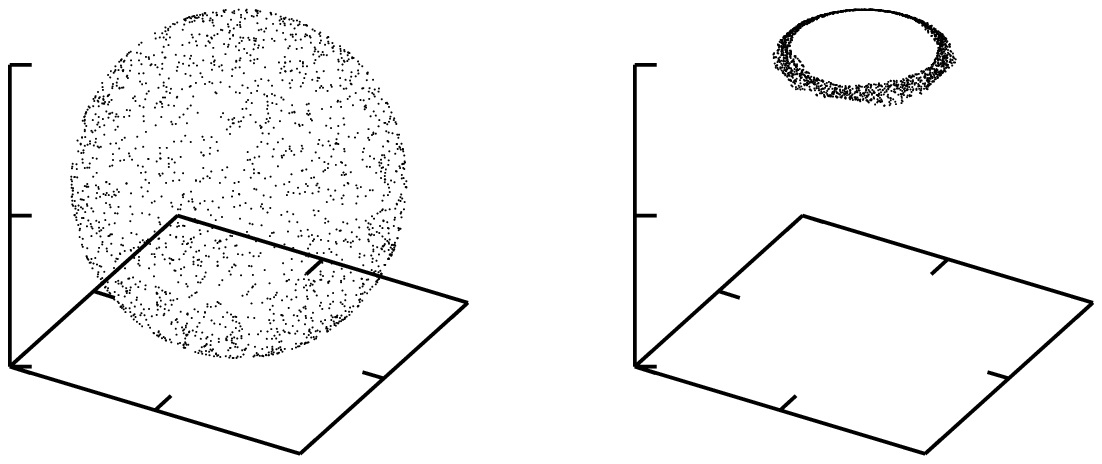 llx=0 lly=0 urx=720 ury=504 rwi=7200}
\put(2508,1906){\makebox(0,0){$\tilde{L}_z$}}%
\put(2382,1688){\makebox(0,0)[r]{1}}%
\put(2382,1254){\makebox(0,0)[r]{0}}%
\put(2382,819){\makebox(0,0)[r]{-1}}%
\put(3794,623){\makebox(0,0){$\tilde{L}_y$}}%
\put(3874,974){\makebox(0,0)[l]{1}}%
\put(3632,756){\makebox(0,0)[l]{0}}%
\put(3391,539){\makebox(0,0)[l]{-1}}%
\put(2756,585){\makebox(0,0){$\tilde{L}_x$}}%
\put(3306,501){\makebox(0,0){1}}%
\put(2888,626){\makebox(0,0){0}}%
\put(2470,752){\makebox(0,0){-1}}%
\put(708,1906){\makebox(0,0){$\tilde{S}_z$}}%
\put(582,1688){\makebox(0,0)[r]{1}}%
\put(582,1254){\makebox(0,0)[r]{0}}%
\put(582,819){\makebox(0,0)[r]{-1}}%
\put(1994,623){\makebox(0,0){$\tilde{S}_y$}}%
\put(2074,974){\makebox(0,0)[l]{1}}%
\put(1832,756){\makebox(0,0)[l]{0}}%
\put(1591,539){\makebox(0,0)[l]{-1}}%
\put(956,585){\makebox(0,0){$\tilde{S}_x$}}%
\put(1506,501){\makebox(0,0){1}}%
\put(1088,626){\makebox(0,0){0}}%
\put(670,752){\makebox(0,0){-1}}%
\end{picture}%
\endgroup
 

%% file: vargrowth.1.215.ric2.tex
\setlength{\unitlength}{0.240900pt}
\ifx\plotpoint\undefined\newsavebox{\plotpoint}\fi
\sbox{\plotpoint}{\rule[-0.200pt]{0.400pt}{0.400pt}}%
\begin{picture}(1500,900)(0,0)
\font\gnuplot=cmr10 at 10pt
\gnuplot
\sbox{\plotpoint}{\rule[-0.200pt]{0.400pt}{0.400pt}}%
\put(181.0,123.0){\rule[-0.200pt]{2.409pt}{0.400pt}}
\put(1429.0,123.0){\rule[-0.200pt]{2.409pt}{0.400pt}}
\put(181.0,148.0){\rule[-0.200pt]{2.409pt}{0.400pt}}
\put(1429.0,148.0){\rule[-0.200pt]{2.409pt}{0.400pt}}
\put(181.0,170.0){\rule[-0.200pt]{2.409pt}{0.400pt}}
\put(1429.0,170.0){\rule[-0.200pt]{2.409pt}{0.400pt}}
\put(181.0,188.0){\rule[-0.200pt]{2.409pt}{0.400pt}}
\put(1429.0,188.0){\rule[-0.200pt]{2.409pt}{0.400pt}}
\put(181.0,205.0){\rule[-0.200pt]{2.409pt}{0.400pt}}
\put(1429.0,205.0){\rule[-0.200pt]{2.409pt}{0.400pt}}
\put(181.0,219.0){\rule[-0.200pt]{4.818pt}{0.400pt}}
\put(161,219){\makebox(0,0)[r]{0.01}}
\put(1419.0,219.0){\rule[-0.200pt]{4.818pt}{0.400pt}}
\put(181.0,316.0){\rule[-0.200pt]{2.409pt}{0.400pt}}
\put(1429.0,316.0){\rule[-0.200pt]{2.409pt}{0.400pt}}
\put(181.0,372.0){\rule[-0.200pt]{2.409pt}{0.400pt}}
\put(1429.0,372.0){\rule[-0.200pt]{2.409pt}{0.400pt}}
\put(181.0,412.0){\rule[-0.200pt]{2.409pt}{0.400pt}}
\put(1429.0,412.0){\rule[-0.200pt]{2.409pt}{0.400pt}}
\put(181.0,443.0){\rule[-0.200pt]{2.409pt}{0.400pt}}
\put(1429.0,443.0){\rule[-0.200pt]{2.409pt}{0.400pt}}
\put(181.0,469.0){\rule[-0.200pt]{2.409pt}{0.400pt}}
\put(1429.0,469.0){\rule[-0.200pt]{2.409pt}{0.400pt}}
\put(181.0,490.0){\rule[-0.200pt]{2.409pt}{0.400pt}}
\put(1429.0,490.0){\rule[-0.200pt]{2.409pt}{0.400pt}}
\put(181.0,509.0){\rule[-0.200pt]{2.409pt}{0.400pt}}
\put(1429.0,509.0){\rule[-0.200pt]{2.409pt}{0.400pt}}
\put(181.0,525.0){\rule[-0.200pt]{2.409pt}{0.400pt}}
\put(1429.0,525.0){\rule[-0.200pt]{2.409pt}{0.400pt}}
\put(181.0,540.0){\rule[-0.200pt]{4.818pt}{0.400pt}}
\put(161,540){\makebox(0,0)[r]{0.1}}
\put(1419.0,540.0){\rule[-0.200pt]{4.818pt}{0.400pt}}
\put(181.0,636.0){\rule[-0.200pt]{2.409pt}{0.400pt}}
\put(1429.0,636.0){\rule[-0.200pt]{2.409pt}{0.400pt}}
\put(181.0,693.0){\rule[-0.200pt]{2.409pt}{0.400pt}}
\put(1429.0,693.0){\rule[-0.200pt]{2.409pt}{0.400pt}}
\put(181.0,733.0){\rule[-0.200pt]{2.409pt}{0.400pt}}
\put(1429.0,733.0){\rule[-0.200pt]{2.409pt}{0.400pt}}
\put(181.0,764.0){\rule[-0.200pt]{2.409pt}{0.400pt}}
\put(1429.0,764.0){\rule[-0.200pt]{2.409pt}{0.400pt}}
\put(181.0,789.0){\rule[-0.200pt]{2.409pt}{0.400pt}}
\put(1429.0,789.0){\rule[-0.200pt]{2.409pt}{0.400pt}}
\put(181.0,810.0){\rule[-0.200pt]{2.409pt}{0.400pt}}
\put(1429.0,810.0){\rule[-0.200pt]{2.409pt}{0.400pt}}
\put(181.0,829.0){\rule[-0.200pt]{2.409pt}{0.400pt}}
\put(1429.0,829.0){\rule[-0.200pt]{2.409pt}{0.400pt}}
\put(181.0,845.0){\rule[-0.200pt]{2.409pt}{0.400pt}}
\put(1429.0,845.0){\rule[-0.200pt]{2.409pt}{0.400pt}}
\put(181.0,860.0){\rule[-0.200pt]{4.818pt}{0.400pt}}
\put(161,860){\makebox(0,0)[r]{1}}
\put(1419.0,860.0){\rule[-0.200pt]{4.818pt}{0.400pt}}
\put(193.0,123.0){\rule[-0.200pt]{0.400pt}{4.818pt}}
\put(193,82){\makebox(0,0){0}}
\put(193.0,840.0){\rule[-0.200pt]{0.400pt}{4.818pt}}
\put(443.0,123.0){\rule[-0.200pt]{0.400pt}{4.818pt}}
\put(443,82){\makebox(0,0){10}}
\put(443.0,840.0){\rule[-0.200pt]{0.400pt}{4.818pt}}
\put(692.0,123.0){\rule[-0.200pt]{0.400pt}{4.818pt}}
\put(692,82){\makebox(0,0){20}}
\put(692.0,840.0){\rule[-0.200pt]{0.400pt}{4.818pt}}
\put(941.0,123.0){\rule[-0.200pt]{0.400pt}{4.818pt}}
\put(941,82){\makebox(0,0){30}}
\put(941.0,840.0){\rule[-0.200pt]{0.400pt}{4.818pt}}
\put(1190.0,123.0){\rule[-0.200pt]{0.400pt}{4.818pt}}
\put(1190,82){\makebox(0,0){40}}
\put(1190.0,840.0){\rule[-0.200pt]{0.400pt}{4.818pt}}
\put(1439.0,123.0){\rule[-0.200pt]{0.400pt}{4.818pt}}
\put(1439,82){\makebox(0,0){50}}
\put(1439.0,840.0){\rule[-0.200pt]{0.400pt}{4.818pt}}
\put(181.0,123.0){\rule[-0.200pt]{303.052pt}{0.400pt}}
\put(1439.0,123.0){\rule[-0.200pt]{0.400pt}{177.543pt}}
\put(181.0,860.0){\rule[-0.200pt]{303.052pt}{0.400pt}}
\put(40,651){\makebox(0,0){ $ \Delta {\bf {\tilde L}}^2  $}}
\put(810,21){\makebox(0,0){ { Kick Number }}}
\put(181.0,123.0){\rule[-0.200pt]{0.400pt}{177.543pt}}
\put(181,140){\usebox{\plotpoint}}
\multiput(181.58,140.00)(0.493,0.734){23}{\rule{0.119pt}{0.685pt}}
\multiput(180.17,140.00)(13.000,17.579){2}{\rule{0.400pt}{0.342pt}}
\multiput(194.58,159.00)(0.492,0.755){21}{\rule{0.119pt}{0.700pt}}
\multiput(193.17,159.00)(12.000,16.547){2}{\rule{0.400pt}{0.350pt}}
\multiput(206.58,177.00)(0.493,0.734){23}{\rule{0.119pt}{0.685pt}}
\multiput(205.17,177.00)(13.000,17.579){2}{\rule{0.400pt}{0.342pt}}
\multiput(219.58,196.00)(0.493,0.695){23}{\rule{0.119pt}{0.654pt}}
\multiput(218.17,196.00)(13.000,16.643){2}{\rule{0.400pt}{0.327pt}}
\multiput(232.58,214.00)(0.493,0.734){23}{\rule{0.119pt}{0.685pt}}
\multiput(231.17,214.00)(13.000,17.579){2}{\rule{0.400pt}{0.342pt}}
\multiput(245.58,233.00)(0.492,0.755){21}{\rule{0.119pt}{0.700pt}}
\multiput(244.17,233.00)(12.000,16.547){2}{\rule{0.400pt}{0.350pt}}
\multiput(257.58,251.00)(0.493,0.734){23}{\rule{0.119pt}{0.685pt}}
\multiput(256.17,251.00)(13.000,17.579){2}{\rule{0.400pt}{0.342pt}}
\multiput(270.58,270.00)(0.493,0.695){23}{\rule{0.119pt}{0.654pt}}
\multiput(269.17,270.00)(13.000,16.643){2}{\rule{0.400pt}{0.327pt}}
\multiput(283.58,288.00)(0.492,0.755){21}{\rule{0.119pt}{0.700pt}}
\multiput(282.17,288.00)(12.000,16.547){2}{\rule{0.400pt}{0.350pt}}
\multiput(295.58,306.00)(0.493,0.734){23}{\rule{0.119pt}{0.685pt}}
\multiput(294.17,306.00)(13.000,17.579){2}{\rule{0.400pt}{0.342pt}}
\multiput(308.58,325.00)(0.493,0.695){23}{\rule{0.119pt}{0.654pt}}
\multiput(307.17,325.00)(13.000,16.643){2}{\rule{0.400pt}{0.327pt}}
\multiput(321.58,343.00)(0.492,0.798){21}{\rule{0.119pt}{0.733pt}}
\multiput(320.17,343.00)(12.000,17.478){2}{\rule{0.400pt}{0.367pt}}
\multiput(333.58,362.00)(0.493,0.695){23}{\rule{0.119pt}{0.654pt}}
\multiput(332.17,362.00)(13.000,16.643){2}{\rule{0.400pt}{0.327pt}}
\multiput(346.58,380.00)(0.493,0.734){23}{\rule{0.119pt}{0.685pt}}
\multiput(345.17,380.00)(13.000,17.579){2}{\rule{0.400pt}{0.342pt}}
\multiput(359.58,399.00)(0.493,0.695){23}{\rule{0.119pt}{0.654pt}}
\multiput(358.17,399.00)(13.000,16.643){2}{\rule{0.400pt}{0.327pt}}
\multiput(372.58,417.00)(0.492,0.798){21}{\rule{0.119pt}{0.733pt}}
\multiput(371.17,417.00)(12.000,17.478){2}{\rule{0.400pt}{0.367pt}}
\multiput(384.58,436.00)(0.493,0.695){23}{\rule{0.119pt}{0.654pt}}
\multiput(383.17,436.00)(13.000,16.643){2}{\rule{0.400pt}{0.327pt}}
\multiput(397.58,454.00)(0.493,0.695){23}{\rule{0.119pt}{0.654pt}}
\multiput(396.17,454.00)(13.000,16.643){2}{\rule{0.400pt}{0.327pt}}
\multiput(410.58,472.00)(0.492,0.798){21}{\rule{0.119pt}{0.733pt}}
\multiput(409.17,472.00)(12.000,17.478){2}{\rule{0.400pt}{0.367pt}}
\multiput(422.58,491.00)(0.493,0.695){23}{\rule{0.119pt}{0.654pt}}
\multiput(421.17,491.00)(13.000,16.643){2}{\rule{0.400pt}{0.327pt}}
\multiput(435.58,509.00)(0.493,0.734){23}{\rule{0.119pt}{0.685pt}}
\multiput(434.17,509.00)(13.000,17.579){2}{\rule{0.400pt}{0.342pt}}
\multiput(448.58,528.00)(0.493,0.695){23}{\rule{0.119pt}{0.654pt}}
\multiput(447.17,528.00)(13.000,16.643){2}{\rule{0.400pt}{0.327pt}}
\multiput(461.58,546.00)(0.492,0.798){21}{\rule{0.119pt}{0.733pt}}
\multiput(460.17,546.00)(12.000,17.478){2}{\rule{0.400pt}{0.367pt}}
\multiput(473.58,565.00)(0.493,0.695){23}{\rule{0.119pt}{0.654pt}}
\multiput(472.17,565.00)(13.000,16.643){2}{\rule{0.400pt}{0.327pt}}
\multiput(486.58,583.00)(0.493,0.734){23}{\rule{0.119pt}{0.685pt}}
\multiput(485.17,583.00)(13.000,17.579){2}{\rule{0.400pt}{0.342pt}}
\multiput(499.58,602.00)(0.492,0.755){21}{\rule{0.119pt}{0.700pt}}
\multiput(498.17,602.00)(12.000,16.547){2}{\rule{0.400pt}{0.350pt}}
\multiput(511.58,620.00)(0.493,0.695){23}{\rule{0.119pt}{0.654pt}}
\multiput(510.17,620.00)(13.000,16.643){2}{\rule{0.400pt}{0.327pt}}
\multiput(524.58,638.00)(0.493,0.734){23}{\rule{0.119pt}{0.685pt}}
\multiput(523.17,638.00)(13.000,17.579){2}{\rule{0.400pt}{0.342pt}}
\multiput(537.58,657.00)(0.493,0.695){23}{\rule{0.119pt}{0.654pt}}
\multiput(536.17,657.00)(13.000,16.643){2}{\rule{0.400pt}{0.327pt}}
\multiput(550.58,675.00)(0.492,0.798){21}{\rule{0.119pt}{0.733pt}}
\multiput(549.17,675.00)(12.000,17.478){2}{\rule{0.400pt}{0.367pt}}
\multiput(562.58,694.00)(0.493,0.695){23}{\rule{0.119pt}{0.654pt}}
\multiput(561.17,694.00)(13.000,16.643){2}{\rule{0.400pt}{0.327pt}}
\multiput(575.58,712.00)(0.493,0.734){23}{\rule{0.119pt}{0.685pt}}
\multiput(574.17,712.00)(13.000,17.579){2}{\rule{0.400pt}{0.342pt}}
\multiput(588.58,731.00)(0.492,0.755){21}{\rule{0.119pt}{0.700pt}}
\multiput(587.17,731.00)(12.000,16.547){2}{\rule{0.400pt}{0.350pt}}
\multiput(600.58,749.00)(0.493,0.734){23}{\rule{0.119pt}{0.685pt}}
\multiput(599.17,749.00)(13.000,17.579){2}{\rule{0.400pt}{0.342pt}}
\multiput(613.58,768.00)(0.493,0.695){23}{\rule{0.119pt}{0.654pt}}
\multiput(612.17,768.00)(13.000,16.643){2}{\rule{0.400pt}{0.327pt}}
\multiput(626.58,786.00)(0.492,0.798){21}{\rule{0.119pt}{0.733pt}}
\multiput(625.17,786.00)(12.000,17.478){2}{\rule{0.400pt}{0.367pt}}
\multiput(638.58,805.00)(0.493,0.695){23}{\rule{0.119pt}{0.654pt}}
\multiput(637.17,805.00)(13.000,16.643){2}{\rule{0.400pt}{0.327pt}}
\multiput(651.58,823.00)(0.493,0.695){23}{\rule{0.119pt}{0.654pt}}
\multiput(650.17,823.00)(13.000,16.643){2}{\rule{0.400pt}{0.327pt}}
\multiput(664.58,841.00)(0.493,0.734){23}{\rule{0.119pt}{0.685pt}}
\multiput(663.17,841.00)(13.000,17.579){2}{\rule{0.400pt}{0.342pt}}
\put(677,860){\usebox{\plotpoint}}
\sbox{\plotpoint}{\rule[-0.500pt]{1.000pt}{1.000pt}}%
\put(181,153){\usebox{\plotpoint}}
\put(181.00,153.00){\usebox{\plotpoint}}
\put(199.94,161.48){\usebox{\plotpoint}}
\put(218.89,169.95){\usebox{\plotpoint}}
\put(237.89,178.27){\usebox{\plotpoint}}
\put(256.75,186.88){\usebox{\plotpoint}}
\put(275.75,195.21){\usebox{\plotpoint}}
\put(294.62,203.81){\usebox{\plotpoint}}
\put(313.61,212.16){\usebox{\plotpoint}}
\put(332.48,220.74){\usebox{\plotpoint}}
\put(351.47,229.10){\usebox{\plotpoint}}
\put(370.52,237.32){\usebox{\plotpoint}}
\put(389.33,246.05){\usebox{\plotpoint}}
\put(408.38,254.25){\usebox{\plotpoint}}
\put(427.18,262.99){\usebox{\plotpoint}}
\put(446.24,271.19){\usebox{\plotpoint}}
\put(465.16,279.73){\usebox{\plotpoint}}
\put(484.13,288.14){\usebox{\plotpoint}}
\put(503.04,296.68){\usebox{\plotpoint}}
\put(522.02,305.08){\usebox{\plotpoint}}
\put(540.86,313.78){\usebox{\plotpoint}}
\put(559.87,322.11){\usebox{\plotpoint}}
\put(578.75,330.73){\usebox{\plotpoint}}
\put(597.75,339.06){\usebox{\plotpoint}}
\put(616.63,347.68){\usebox{\plotpoint}}
\put(635.64,356.02){\usebox{\plotpoint}}
\put(654.52,364.63){\usebox{\plotpoint}}
\put(673.63,372.70){\usebox{\plotpoint}}
\put(692.38,381.56){\usebox{\plotpoint}}
\put(711.49,389.65){\usebox{\plotpoint}}
\put(730.25,398.50){\usebox{\plotpoint}}
\put(749.35,406.59){\usebox{\plotpoint}}
\put(768.26,415.13){\usebox{\plotpoint}}
\put(787.20,423.54){\usebox{\plotpoint}}
\put(806.12,432.06){\usebox{\plotpoint}}
\put(825.06,440.49){\usebox{\plotpoint}}
\put(843.98,448.99){\usebox{\plotpoint}}
\put(862.92,457.43){\usebox{\plotpoint}}
\put(881.88,465.87){\usebox{\plotpoint}}
\put(900.85,474.27){\usebox{\plotpoint}}
\put(919.76,482.81){\usebox{\plotpoint}}
\put(938.74,491.22){\usebox{\plotpoint}}
\put(957.65,499.76){\usebox{\plotpoint}}
\put(976.50,508.46){\usebox{\plotpoint}}
\put(995.54,516.71){\usebox{\plotpoint}}
\put(1014.38,525.41){\usebox{\plotpoint}}
\put(1033.42,533.66){\usebox{\plotpoint}}
\put(1052.27,542.36){\usebox{\plotpoint}}
\put(1071.31,550.61){\usebox{\plotpoint}}
\put(1090.16,559.30){\usebox{\plotpoint}}
\put(1109.35,567.17){\usebox{\plotpoint}}
\put(1128.02,576.24){\usebox{\plotpoint}}
\put(1147.21,584.11){\usebox{\plotpoint}}
\put(1165.88,593.18){\usebox{\plotpoint}}
\put(1185.08,601.04){\usebox{\plotpoint}}
\put(1203.84,609.92){\usebox{\plotpoint}}
\put(1222.94,617.98){\usebox{\plotpoint}}
\put(1241.70,626.85){\usebox{\plotpoint}}
\put(1260.80,634.92){\usebox{\plotpoint}}
\put(1279.65,643.61){\usebox{\plotpoint}}
\put(1298.69,651.87){\usebox{\plotpoint}}
\put(1317.54,660.56){\usebox{\plotpoint}}
\put(1336.57,668.82){\usebox{\plotpoint}}
\put(1355.42,677.50){\usebox{\plotpoint}}
\put(1374.46,685.77){\usebox{\plotpoint}}
\put(1393.31,694.45){\usebox{\plotpoint}}
\put(1412.16,703.15){\usebox{\plotpoint}}
\put(1431.20,711.40){\usebox{\plotpoint}}
\put(1439,715){\usebox{\plotpoint}}
\sbox{\plotpoint}{\rule[-0.400pt]{0.800pt}{0.800pt}}%
\put(1294,500){\makebox(0,0)[r]{quantum (a)}}
\put(193,158){\raisebox{-.8pt}{\makebox(0,0){$\Diamond$}}}
\put(218,233){\raisebox{-.8pt}{\makebox(0,0){$\Diamond$}}}
\put(243,346){\raisebox{-.8pt}{\makebox(0,0){$\Diamond$}}}
\put(268,398){\raisebox{-.8pt}{\makebox(0,0){$\Diamond$}}}
\put(293,341){\raisebox{-.8pt}{\makebox(0,0){$\Diamond$}}}
\put(318,283){\raisebox{-.8pt}{\makebox(0,0){$\Diamond$}}}
\put(343,263){\raisebox{-.8pt}{\makebox(0,0){$\Diamond$}}}
\put(368,485){\raisebox{-.8pt}{\makebox(0,0){$\Diamond$}}}
\put(393,526){\raisebox{-.8pt}{\makebox(0,0){$\Diamond$}}}
\put(418,663){\raisebox{-.8pt}{\makebox(0,0){$\Diamond$}}}
\put(443,604){\raisebox{-.8pt}{\makebox(0,0){$\Diamond$}}}
\put(467,563){\raisebox{-.8pt}{\makebox(0,0){$\Diamond$}}}
\put(492,469){\raisebox{-.8pt}{\makebox(0,0){$\Diamond$}}}
\put(517,490){\raisebox{-.8pt}{\makebox(0,0){$\Diamond$}}}
\put(542,679){\raisebox{-.8pt}{\makebox(0,0){$\Diamond$}}}
\put(567,734){\raisebox{-.8pt}{\makebox(0,0){$\Diamond$}}}
\put(592,795){\raisebox{-.8pt}{\makebox(0,0){$\Diamond$}}}
\put(617,793){\raisebox{-.8pt}{\makebox(0,0){$\Diamond$}}}
\put(642,732){\raisebox{-.8pt}{\makebox(0,0){$\Diamond$}}}
\put(667,694){\raisebox{-.8pt}{\makebox(0,0){$\Diamond$}}}
\put(692,695){\raisebox{-.8pt}{\makebox(0,0){$\Diamond$}}}
\put(717,761){\raisebox{-.8pt}{\makebox(0,0){$\Diamond$}}}
\put(741,790){\raisebox{-.8pt}{\makebox(0,0){$\Diamond$}}}
\put(766,795){\raisebox{-.8pt}{\makebox(0,0){$\Diamond$}}}
\put(791,757){\raisebox{-.8pt}{\makebox(0,0){$\Diamond$}}}
\put(816,709){\raisebox{-.8pt}{\makebox(0,0){$\Diamond$}}}
\put(841,759){\raisebox{-.8pt}{\makebox(0,0){$\Diamond$}}}
\put(866,787){\raisebox{-.8pt}{\makebox(0,0){$\Diamond$}}}
\put(891,841){\raisebox{-.8pt}{\makebox(0,0){$\Diamond$}}}
\put(916,852){\raisebox{-.8pt}{\makebox(0,0){$\Diamond$}}}
\put(941,845){\raisebox{-.8pt}{\makebox(0,0){$\Diamond$}}}
\put(966,825){\raisebox{-.8pt}{\makebox(0,0){$\Diamond$}}}
\put(991,824){\raisebox{-.8pt}{\makebox(0,0){$\Diamond$}}}
\put(1016,827){\raisebox{-.8pt}{\makebox(0,0){$\Diamond$}}}
\put(1040,842){\raisebox{-.8pt}{\makebox(0,0){$\Diamond$}}}
\put(1065,847){\raisebox{-.8pt}{\makebox(0,0){$\Diamond$}}}
\put(1090,837){\raisebox{-.8pt}{\makebox(0,0){$\Diamond$}}}
\put(1115,828){\raisebox{-.8pt}{\makebox(0,0){$\Diamond$}}}
\put(1140,830){\raisebox{-.8pt}{\makebox(0,0){$\Diamond$}}}
\put(1165,833){\raisebox{-.8pt}{\makebox(0,0){$\Diamond$}}}
\put(1190,851){\raisebox{-.8pt}{\makebox(0,0){$\Diamond$}}}
\put(1215,856){\raisebox{-.8pt}{\makebox(0,0){$\Diamond$}}}
\put(1240,856){\raisebox{-.8pt}{\makebox(0,0){$\Diamond$}}}
\put(1265,855){\raisebox{-.8pt}{\makebox(0,0){$\Diamond$}}}
\put(1290,852){\raisebox{-.8pt}{\makebox(0,0){$\Diamond$}}}
\put(1314,853){\raisebox{-.8pt}{\makebox(0,0){$\Diamond$}}}
\put(1339,855){\raisebox{-.8pt}{\makebox(0,0){$\Diamond$}}}
\put(1364,856){\raisebox{-.8pt}{\makebox(0,0){$\Diamond$}}}
\put(1389,855){\raisebox{-.8pt}{\makebox(0,0){$\Diamond$}}}
\put(1414,853){\raisebox{-.8pt}{\makebox(0,0){$\Diamond$}}}
\put(1439,853){\raisebox{-.8pt}{\makebox(0,0){$\Diamond$}}}
\put(1364,500){\raisebox{-.8pt}{\makebox(0,0){$\Diamond$}}}
\sbox{\plotpoint}{\rule[-0.500pt]{1.000pt}{1.000pt}}%
\put(1294,459){\makebox(0,0)[r]{quantum (b)}}
\put(193,159){\makebox(0,0){$+$}}
\put(218,240){\makebox(0,0){$+$}}
\put(243,281){\makebox(0,0){$+$}}
\put(268,310){\makebox(0,0){$+$}}
\put(293,311){\makebox(0,0){$+$}}
\put(318,240){\makebox(0,0){$+$}}
\put(343,213){\makebox(0,0){$+$}}
\put(368,193){\makebox(0,0){$+$}}
\put(393,253){\makebox(0,0){$+$}}
\put(418,311){\makebox(0,0){$+$}}
\put(443,302){\makebox(0,0){$+$}}
\put(467,288){\makebox(0,0){$+$}}
\put(492,219){\makebox(0,0){$+$}}
\put(517,180){\makebox(0,0){$+$}}
\put(542,242){\makebox(0,0){$+$}}
\put(567,274){\makebox(0,0){$+$}}
\put(592,316){\makebox(0,0){$+$}}
\put(617,296){\makebox(0,0){$+$}}
\put(642,246){\makebox(0,0){$+$}}
\put(667,216){\makebox(0,0){$+$}}
\put(692,188){\makebox(0,0){$+$}}
\put(717,267){\makebox(0,0){$+$}}
\put(741,298){\makebox(0,0){$+$}}
\put(766,302){\makebox(0,0){$+$}}
\put(791,286){\makebox(0,0){$+$}}
\put(816,212){\makebox(0,0){$+$}}
\put(841,211){\makebox(0,0){$+$}}
\put(866,231){\makebox(0,0){$+$}}
\put(891,277){\makebox(0,0){$+$}}
\put(916,310){\makebox(0,0){$+$}}
\put(941,285){\makebox(0,0){$+$}}
\put(966,261){\makebox(0,0){$+$}}
\put(991,210){\makebox(0,0){$+$}}
\put(1016,207){\makebox(0,0){$+$}}
\put(1040,266){\makebox(0,0){$+$}}
\put(1065,285){\makebox(0,0){$+$}}
\put(1090,305){\makebox(0,0){$+$}}
\put(1115,276){\makebox(0,0){$+$}}
\put(1140,230){\makebox(0,0){$+$}}
\put(1165,223){\makebox(0,0){$+$}}
\put(1190,223){\makebox(0,0){$+$}}
\put(1215,281){\makebox(0,0){$+$}}
\put(1240,296){\makebox(0,0){$+$}}
\put(1265,288){\makebox(0,0){$+$}}
\put(1290,268){\makebox(0,0){$+$}}
\put(1314,213){\makebox(0,0){$+$}}
\put(1339,229){\makebox(0,0){$+$}}
\put(1364,252){\makebox(0,0){$+$}}
\put(1389,283){\makebox(0,0){$+$}}
\put(1414,301){\makebox(0,0){$+$}}
\put(1439,273){\makebox(0,0){$+$}}
\put(1364,459){\makebox(0,0){$+$}}
\sbox{\plotpoint}{\rule[-0.600pt]{1.200pt}{1.200pt}}%
\put(1294,418){\makebox(0,0)[r]{classical (a)}}
\put(193,159){\raisebox{-.8pt}{\makebox(0,0){$\Box$}}}
\put(218,234){\raisebox{-.8pt}{\makebox(0,0){$\Box$}}}
\put(243,346){\raisebox{-.8pt}{\makebox(0,0){$\Box$}}}
\put(268,399){\raisebox{-.8pt}{\makebox(0,0){$\Box$}}}
\put(293,342){\raisebox{-.8pt}{\makebox(0,0){$\Box$}}}
\put(318,282){\raisebox{-.8pt}{\makebox(0,0){$\Box$}}}
\put(343,265){\raisebox{-.8pt}{\makebox(0,0){$\Box$}}}
\put(368,485){\raisebox{-.8pt}{\makebox(0,0){$\Box$}}}
\put(393,526){\raisebox{-.8pt}{\makebox(0,0){$\Box$}}}
\put(418,664){\raisebox{-.8pt}{\makebox(0,0){$\Box$}}}
\put(443,607){\raisebox{-.8pt}{\makebox(0,0){$\Box$}}}
\put(467,564){\raisebox{-.8pt}{\makebox(0,0){$\Box$}}}
\put(492,471){\raisebox{-.8pt}{\makebox(0,0){$\Box$}}}
\put(517,487){\raisebox{-.8pt}{\makebox(0,0){$\Box$}}}
\put(542,679){\raisebox{-.8pt}{\makebox(0,0){$\Box$}}}
\put(567,733){\raisebox{-.8pt}{\makebox(0,0){$\Box$}}}
\put(592,795){\raisebox{-.8pt}{\makebox(0,0){$\Box$}}}
\put(617,794){\raisebox{-.8pt}{\makebox(0,0){$\Box$}}}
\put(642,732){\raisebox{-.8pt}{\makebox(0,0){$\Box$}}}
\put(667,694){\raisebox{-.8pt}{\makebox(0,0){$\Box$}}}
\put(692,693){\raisebox{-.8pt}{\makebox(0,0){$\Box$}}}
\put(717,762){\raisebox{-.8pt}{\makebox(0,0){$\Box$}}}
\put(741,791){\raisebox{-.8pt}{\makebox(0,0){$\Box$}}}
\put(766,798){\raisebox{-.8pt}{\makebox(0,0){$\Box$}}}
\put(791,759){\raisebox{-.8pt}{\makebox(0,0){$\Box$}}}
\put(816,710){\raisebox{-.8pt}{\makebox(0,0){$\Box$}}}
\put(841,758){\raisebox{-.8pt}{\makebox(0,0){$\Box$}}}
\put(866,786){\raisebox{-.8pt}{\makebox(0,0){$\Box$}}}
\put(891,841){\raisebox{-.8pt}{\makebox(0,0){$\Box$}}}
\put(916,853){\raisebox{-.8pt}{\makebox(0,0){$\Box$}}}
\put(941,846){\raisebox{-.8pt}{\makebox(0,0){$\Box$}}}
\put(966,825){\raisebox{-.8pt}{\makebox(0,0){$\Box$}}}
\put(991,825){\raisebox{-.8pt}{\makebox(0,0){$\Box$}}}
\put(1016,827){\raisebox{-.8pt}{\makebox(0,0){$\Box$}}}
\put(1040,843){\raisebox{-.8pt}{\makebox(0,0){$\Box$}}}
\put(1065,847){\raisebox{-.8pt}{\makebox(0,0){$\Box$}}}
\put(1090,838){\raisebox{-.8pt}{\makebox(0,0){$\Box$}}}
\put(1115,828){\raisebox{-.8pt}{\makebox(0,0){$\Box$}}}
\put(1140,829){\raisebox{-.8pt}{\makebox(0,0){$\Box$}}}
\put(1165,832){\raisebox{-.8pt}{\makebox(0,0){$\Box$}}}
\put(1190,851){\raisebox{-.8pt}{\makebox(0,0){$\Box$}}}
\put(1215,856){\raisebox{-.8pt}{\makebox(0,0){$\Box$}}}
\put(1240,856){\raisebox{-.8pt}{\makebox(0,0){$\Box$}}}
\put(1265,855){\raisebox{-.8pt}{\makebox(0,0){$\Box$}}}
\put(1290,853){\raisebox{-.8pt}{\makebox(0,0){$\Box$}}}
\put(1314,853){\raisebox{-.8pt}{\makebox(0,0){$\Box$}}}
\put(1339,856){\raisebox{-.8pt}{\makebox(0,0){$\Box$}}}
\put(1364,856){\raisebox{-.8pt}{\makebox(0,0){$\Box$}}}
\put(1389,855){\raisebox{-.8pt}{\makebox(0,0){$\Box$}}}
\put(1414,854){\raisebox{-.8pt}{\makebox(0,0){$\Box$}}}
\put(1439,853){\raisebox{-.8pt}{\makebox(0,0){$\Box$}}}
\put(1364,418){\raisebox{-.8pt}{\makebox(0,0){$\Box$}}}
\sbox{\plotpoint}{\rule[-0.500pt]{1.000pt}{1.000pt}}%
\put(1294,377){\makebox(0,0)[r]{classical (b)}}
\put(193,159){\makebox(0,0){$\times$}}
\put(218,240){\makebox(0,0){$\times$}}
\put(243,281){\makebox(0,0){$\times$}}
\put(268,310){\makebox(0,0){$\times$}}
\put(293,311){\makebox(0,0){$\times$}}
\put(318,240){\makebox(0,0){$\times$}}
\put(343,213){\makebox(0,0){$\times$}}
\put(368,194){\makebox(0,0){$\times$}}
\put(393,253){\makebox(0,0){$\times$}}
\put(418,311){\makebox(0,0){$\times$}}
\put(443,301){\makebox(0,0){$\times$}}
\put(467,288){\makebox(0,0){$\times$}}
\put(492,220){\makebox(0,0){$\times$}}
\put(517,181){\makebox(0,0){$\times$}}
\put(542,242){\makebox(0,0){$\times$}}
\put(567,273){\makebox(0,0){$\times$}}
\put(592,315){\makebox(0,0){$\times$}}
\put(617,296){\makebox(0,0){$\times$}}
\put(642,247){\makebox(0,0){$\times$}}
\put(667,218){\makebox(0,0){$\times$}}
\put(692,190){\makebox(0,0){$\times$}}
\put(717,267){\makebox(0,0){$\times$}}
\put(741,297){\makebox(0,0){$\times$}}
\put(766,300){\makebox(0,0){$\times$}}
\put(791,286){\makebox(0,0){$\times$}}
\put(816,215){\makebox(0,0){$\times$}}
\put(841,214){\makebox(0,0){$\times$}}
\put(866,233){\makebox(0,0){$\times$}}
\put(891,275){\makebox(0,0){$\times$}}
\put(916,308){\makebox(0,0){$\times$}}
\put(941,284){\makebox(0,0){$\times$}}
\put(966,263){\makebox(0,0){$\times$}}
\put(991,216){\makebox(0,0){$\times$}}
\put(1016,212){\makebox(0,0){$\times$}}
\put(1040,266){\makebox(0,0){$\times$}}
\put(1065,283){\makebox(0,0){$\times$}}
\put(1090,302){\makebox(0,0){$\times$}}
\put(1115,276){\makebox(0,0){$\times$}}
\put(1140,235){\makebox(0,0){$\times$}}
\put(1165,229){\makebox(0,0){$\times$}}
\put(1190,228){\makebox(0,0){$\times$}}
\put(1215,278){\makebox(0,0){$\times$}}
\put(1240,293){\makebox(0,0){$\times$}}
\put(1265,285){\makebox(0,0){$\times$}}
\put(1290,269){\makebox(0,0){$\times$}}
\put(1314,223){\makebox(0,0){$\times$}}
\put(1339,236){\makebox(0,0){$\times$}}
\put(1364,254){\makebox(0,0){$\times$}}
\put(1389,279){\makebox(0,0){$\times$}}
\put(1414,296){\makebox(0,0){$\times$}}
\put(1439,272){\makebox(0,0){$\times$}}
\put(1364,377){\makebox(0,0){$\times$}}
\end{picture}

%% file: qmlmnm.2.835.154.cic1.tex
\setlength{\unitlength}{0.240900pt}
\ifx\plotpoint\undefined\newsavebox{\plotpoint}\fi
\sbox{\plotpoint}{\rule[-0.200pt]{0.400pt}{0.400pt}}%
\begin{picture}(1500,900)(0,0)
\font\gnuplot=cmr10 at 10pt
\gnuplot
\sbox{\plotpoint}{\rule[-0.200pt]{0.400pt}{0.400pt}}%
\put(181.0,123.0){\rule[-0.200pt]{4.818pt}{0.400pt}}
\put(161,123){\makebox(0,0)[r]{-1}}
\put(1419.0,123.0){\rule[-0.200pt]{4.818pt}{0.400pt}}
\put(181.0,197.0){\rule[-0.200pt]{4.818pt}{0.400pt}}
\put(161,197){\makebox(0,0)[r]{-0.8}}
\put(1419.0,197.0){\rule[-0.200pt]{4.818pt}{0.400pt}}
\put(181.0,270.0){\rule[-0.200pt]{4.818pt}{0.400pt}}
\put(161,270){\makebox(0,0)[r]{-0.6}}
\put(1419.0,270.0){\rule[-0.200pt]{4.818pt}{0.400pt}}
\put(181.0,344.0){\rule[-0.200pt]{4.818pt}{0.400pt}}
\put(161,344){\makebox(0,0)[r]{-0.4}}
\put(1419.0,344.0){\rule[-0.200pt]{4.818pt}{0.400pt}}
\put(181.0,418.0){\rule[-0.200pt]{4.818pt}{0.400pt}}
\put(161,418){\makebox(0,0)[r]{-0.2}}
\put(1419.0,418.0){\rule[-0.200pt]{4.818pt}{0.400pt}}
\put(181.0,491.0){\rule[-0.200pt]{4.818pt}{0.400pt}}
\put(161,491){\makebox(0,0)[r]{0}}
\put(1419.0,491.0){\rule[-0.200pt]{4.818pt}{0.400pt}}
\put(181.0,565.0){\rule[-0.200pt]{4.818pt}{0.400pt}}
\put(161,565){\makebox(0,0)[r]{0.2}}
\put(1419.0,565.0){\rule[-0.200pt]{4.818pt}{0.400pt}}
\put(181.0,639.0){\rule[-0.200pt]{4.818pt}{0.400pt}}
\put(161,639){\makebox(0,0)[r]{0.4}}
\put(1419.0,639.0){\rule[-0.200pt]{4.818pt}{0.400pt}}
\put(181.0,713.0){\rule[-0.200pt]{4.818pt}{0.400pt}}
\put(161,713){\makebox(0,0)[r]{0.6}}
\put(1419.0,713.0){\rule[-0.200pt]{4.818pt}{0.400pt}}
\put(181.0,786.0){\rule[-0.200pt]{4.818pt}{0.400pt}}
\put(161,786){\makebox(0,0)[r]{0.8}}
\put(1419.0,786.0){\rule[-0.200pt]{4.818pt}{0.400pt}}
\put(181.0,860.0){\rule[-0.200pt]{4.818pt}{0.400pt}}
\put(161,860){\makebox(0,0)[r]{1}}
\put(1419.0,860.0){\rule[-0.200pt]{4.818pt}{0.400pt}}
\put(181.0,123.0){\rule[-0.200pt]{0.400pt}{4.818pt}}
\put(181,82){\makebox(0,0){0}}
\put(181.0,840.0){\rule[-0.200pt]{0.400pt}{4.818pt}}
\put(361.0,123.0){\rule[-0.200pt]{0.400pt}{4.818pt}}
\put(361,82){\makebox(0,0){5}}
\put(361.0,840.0){\rule[-0.200pt]{0.400pt}{4.818pt}}
\put(540.0,123.0){\rule[-0.200pt]{0.400pt}{4.818pt}}
\put(540,82){\makebox(0,0){10}}
\put(540.0,840.0){\rule[-0.200pt]{0.400pt}{4.818pt}}
\put(720.0,123.0){\rule[-0.200pt]{0.400pt}{4.818pt}}
\put(720,82){\makebox(0,0){15}}
\put(720.0,840.0){\rule[-0.200pt]{0.400pt}{4.818pt}}
\put(900.0,123.0){\rule[-0.200pt]{0.400pt}{4.818pt}}
\put(900,82){\makebox(0,0){20}}
\put(900.0,840.0){\rule[-0.200pt]{0.400pt}{4.818pt}}
\put(1080.0,123.0){\rule[-0.200pt]{0.400pt}{4.818pt}}
\put(1080,82){\makebox(0,0){25}}
\put(1080.0,840.0){\rule[-0.200pt]{0.400pt}{4.818pt}}
\put(1259.0,123.0){\rule[-0.200pt]{0.400pt}{4.818pt}}
\put(1259,82){\makebox(0,0){30}}
\put(1259.0,840.0){\rule[-0.200pt]{0.400pt}{4.818pt}}
\put(1439.0,123.0){\rule[-0.200pt]{0.400pt}{4.818pt}}
\put(1439,82){\makebox(0,0){35}}
\put(1439.0,840.0){\rule[-0.200pt]{0.400pt}{4.818pt}}
\put(181.0,123.0){\rule[-0.200pt]{303.052pt}{0.400pt}}
\put(1439.0,123.0){\rule[-0.200pt]{0.400pt}{177.543pt}}
\put(181.0,860.0){\rule[-0.200pt]{303.052pt}{0.400pt}}
\put(40,491){\makebox(0,0){ {$ \langle {\tilde L}_z \rangle $ }}}
\put(810,21){\makebox(0,0){ { Kick Number }}}
\put(181.0,123.0){\rule[-0.200pt]{0.400pt}{177.543pt}}
\put(1203,252){\makebox(0,0)[r]{Classical Average}}
\put(1223.0,252.0){\rule[-0.200pt]{24.090pt}{0.400pt}}
\put(181,232){\usebox{\plotpoint}}
\multiput(181.58,232.00)(0.498,2.982){69}{\rule{0.120pt}{2.467pt}}
\multiput(180.17,232.00)(36.000,207.880){2}{\rule{0.400pt}{1.233pt}}
\multiput(217.00,445.58)(1.137,0.494){29}{\rule{1.000pt}{0.119pt}}
\multiput(217.00,444.17)(33.924,16.000){2}{\rule{0.500pt}{0.400pt}}
\multiput(253.58,454.31)(0.498,-1.902){69}{\rule{0.120pt}{1.611pt}}
\multiput(252.17,457.66)(36.000,-132.656){2}{\rule{0.400pt}{0.806pt}}
\multiput(289.58,325.00)(0.498,2.028){69}{\rule{0.120pt}{1.711pt}}
\multiput(288.17,325.00)(36.000,141.449){2}{\rule{0.400pt}{0.856pt}}
\multiput(325.58,461.93)(0.498,-2.322){69}{\rule{0.120pt}{1.944pt}}
\multiput(324.17,465.96)(36.000,-161.964){2}{\rule{0.400pt}{0.972pt}}
\multiput(361.58,304.00)(0.498,2.126){69}{\rule{0.120pt}{1.789pt}}
\multiput(360.17,304.00)(36.000,148.287){2}{\rule{0.400pt}{0.894pt}}
\multiput(397.58,449.27)(0.498,-1.916){69}{\rule{0.120pt}{1.622pt}}
\multiput(396.17,452.63)(36.000,-133.633){2}{\rule{0.400pt}{0.811pt}}
\multiput(433.58,319.00)(0.498,2.841){69}{\rule{0.120pt}{2.356pt}}
\multiput(432.17,319.00)(36.000,198.111){2}{\rule{0.400pt}{1.178pt}}
\multiput(469.58,522.00)(0.498,1.264){67}{\rule{0.120pt}{1.106pt}}
\multiput(468.17,522.00)(35.000,85.705){2}{\rule{0.400pt}{0.553pt}}
\multiput(504.58,602.44)(0.498,-2.168){69}{\rule{0.120pt}{1.822pt}}
\multiput(503.17,606.22)(36.000,-151.218){2}{\rule{0.400pt}{0.911pt}}
\multiput(540.00,453.95)(7.830,-0.447){3}{\rule{4.900pt}{0.108pt}}
\multiput(540.00,454.17)(25.830,-3.000){2}{\rule{2.450pt}{0.400pt}}
\multiput(576.58,452.00)(0.498,1.411){69}{\rule{0.120pt}{1.222pt}}
\multiput(575.17,452.00)(36.000,98.463){2}{\rule{0.400pt}{0.611pt}}
\multiput(612.00,551.92)(0.545,-0.497){63}{\rule{0.536pt}{0.120pt}}
\multiput(612.00,552.17)(34.887,-33.000){2}{\rule{0.268pt}{0.400pt}}
\multiput(648.58,516.59)(0.498,-0.906){69}{\rule{0.120pt}{0.822pt}}
\multiput(647.17,518.29)(36.000,-63.293){2}{\rule{0.400pt}{0.411pt}}
\multiput(684.58,455.00)(0.498,0.569){69}{\rule{0.120pt}{0.556pt}}
\multiput(683.17,455.00)(36.000,39.847){2}{\rule{0.400pt}{0.278pt}}
\multiput(720.00,496.58)(0.752,0.496){45}{\rule{0.700pt}{0.120pt}}
\multiput(720.00,495.17)(34.547,24.000){2}{\rule{0.350pt}{0.400pt}}
\multiput(756.00,518.92)(0.580,-0.497){59}{\rule{0.565pt}{0.120pt}}
\multiput(756.00,519.17)(34.828,-31.000){2}{\rule{0.282pt}{0.400pt}}
\multiput(828.00,489.58)(1.137,0.494){29}{\rule{1.000pt}{0.119pt}}
\multiput(828.00,488.17)(33.924,16.000){2}{\rule{0.500pt}{0.400pt}}
\multiput(864.00,503.94)(5.160,-0.468){5}{\rule{3.700pt}{0.113pt}}
\multiput(864.00,504.17)(28.320,-4.000){2}{\rule{1.850pt}{0.400pt}}
\multiput(900.00,499.93)(2.067,-0.489){15}{\rule{1.700pt}{0.118pt}}
\multiput(900.00,500.17)(32.472,-9.000){2}{\rule{0.850pt}{0.400pt}}
\multiput(936.00,492.60)(5.160,0.468){5}{\rule{3.700pt}{0.113pt}}
\multiput(936.00,491.17)(28.320,4.000){2}{\rule{1.850pt}{0.400pt}}
\multiput(972.00,496.60)(5.160,0.468){5}{\rule{3.700pt}{0.113pt}}
\multiput(972.00,495.17)(28.320,4.000){2}{\rule{1.850pt}{0.400pt}}
\multiput(1008.00,498.93)(3.938,-0.477){7}{\rule{2.980pt}{0.115pt}}
\multiput(1008.00,499.17)(29.815,-5.000){2}{\rule{1.490pt}{0.400pt}}
\put(1044,493.17){\rule{7.300pt}{0.400pt}}
\multiput(1044.00,494.17)(20.848,-2.000){2}{\rule{3.650pt}{0.400pt}}
\multiput(1080.00,493.61)(7.830,0.447){3}{\rule{4.900pt}{0.108pt}}
\multiput(1080.00,492.17)(25.830,3.000){2}{\rule{2.450pt}{0.400pt}}
\put(792.0,489.0){\rule[-0.200pt]{8.672pt}{0.400pt}}
\put(1151,494.67){\rule{8.672pt}{0.400pt}}
\multiput(1151.00,495.17)(18.000,-1.000){2}{\rule{4.336pt}{0.400pt}}
\put(1116.0,496.0){\rule[-0.200pt]{8.431pt}{0.400pt}}
\put(1223,495.17){\rule{7.300pt}{0.400pt}}
\multiput(1223.00,494.17)(20.848,2.000){2}{\rule{3.650pt}{0.400pt}}
\put(1259,495.67){\rule{8.672pt}{0.400pt}}
\multiput(1259.00,496.17)(18.000,-1.000){2}{\rule{4.336pt}{0.400pt}}
\put(1187.0,495.0){\rule[-0.200pt]{8.672pt}{0.400pt}}
\put(1331,494.17){\rule{7.300pt}{0.400pt}}
\multiput(1331.00,495.17)(20.848,-2.000){2}{\rule{3.650pt}{0.400pt}}
\put(1295.0,496.0){\rule[-0.200pt]{8.672pt}{0.400pt}}
\put(1403,494.17){\rule{7.300pt}{0.400pt}}
\multiput(1403.00,493.17)(20.848,2.000){2}{\rule{3.650pt}{0.400pt}}
\put(1367.0,494.0){\rule[-0.200pt]{8.672pt}{0.400pt}}
\put(1439,496){\usebox{\plotpoint}}
\put(181,232){\circle{18}}
\put(217,445){\circle{18}}
\put(253,461){\circle{18}}
\put(289,325){\circle{18}}
\put(325,470){\circle{18}}
\put(361,304){\circle{18}}
\put(397,456){\circle{18}}
\put(433,319){\circle{18}}
\put(469,522){\circle{18}}
\put(504,610){\circle{18}}
\put(540,455){\circle{18}}
\put(576,452){\circle{18}}
\put(612,553){\circle{18}}
\put(648,520){\circle{18}}
\put(684,455){\circle{18}}
\put(720,496){\circle{18}}
\put(756,520){\circle{18}}
\put(792,489){\circle{18}}
\put(828,489){\circle{18}}
\put(864,505){\circle{18}}
\put(900,501){\circle{18}}
\put(936,492){\circle{18}}
\put(972,496){\circle{18}}
\put(1008,500){\circle{18}}
\put(1044,495){\circle{18}}
\put(1080,493){\circle{18}}
\put(1116,496){\circle{18}}
\put(1151,496){\circle{18}}
\put(1187,495){\circle{18}}
\put(1223,495){\circle{18}}
\put(1259,497){\circle{18}}
\put(1295,496){\circle{18}}
\put(1331,496){\circle{18}}
\put(1367,494){\circle{18}}
\put(1403,494){\circle{18}}
\put(1439,496){\circle{18}}
\put(1273,252){\circle{18}}
\put(1203,211){\makebox(0,0)[r]{Quantum Expectation}}
\put(1223.0,211.0){\rule[-0.200pt]{24.090pt}{0.400pt}}
\put(181,232){\usebox{\plotpoint}}
\multiput(181.58,232.00)(0.498,2.982){69}{\rule{0.120pt}{2.467pt}}
\multiput(180.17,232.00)(36.000,207.880){2}{\rule{0.400pt}{1.233pt}}
\multiput(217.00,445.58)(1.137,0.494){29}{\rule{1.000pt}{0.119pt}}
\multiput(217.00,444.17)(33.924,16.000){2}{\rule{0.500pt}{0.400pt}}
\multiput(253.58,454.31)(0.498,-1.902){69}{\rule{0.120pt}{1.611pt}}
\multiput(252.17,457.66)(36.000,-132.656){2}{\rule{0.400pt}{0.806pt}}
\multiput(289.58,325.00)(0.498,2.042){69}{\rule{0.120pt}{1.722pt}}
\multiput(288.17,325.00)(36.000,142.425){2}{\rule{0.400pt}{0.861pt}}
\multiput(325.58,462.88)(0.498,-2.337){69}{\rule{0.120pt}{1.956pt}}
\multiput(324.17,466.94)(36.000,-162.941){2}{\rule{0.400pt}{0.978pt}}
\multiput(361.58,304.00)(0.498,2.196){69}{\rule{0.120pt}{1.844pt}}
\multiput(360.17,304.00)(36.000,153.172){2}{\rule{0.400pt}{0.922pt}}
\multiput(397.58,453.94)(0.498,-2.014){69}{\rule{0.120pt}{1.700pt}}
\multiput(396.17,457.47)(36.000,-140.472){2}{\rule{0.400pt}{0.850pt}}
\multiput(433.58,317.00)(0.498,2.322){69}{\rule{0.120pt}{1.944pt}}
\multiput(432.17,317.00)(36.000,161.964){2}{\rule{0.400pt}{0.972pt}}
\multiput(469.58,483.00)(0.498,1.812){67}{\rule{0.120pt}{1.540pt}}
\multiput(468.17,483.00)(35.000,122.804){2}{\rule{0.400pt}{0.770pt}}
\multiput(504.58,602.17)(0.498,-1.944){69}{\rule{0.120pt}{1.644pt}}
\multiput(503.17,605.59)(36.000,-135.587){2}{\rule{0.400pt}{0.822pt}}
\multiput(540.00,468.92)(0.752,-0.496){45}{\rule{0.700pt}{0.120pt}}
\multiput(540.00,469.17)(34.547,-24.000){2}{\rule{0.350pt}{0.400pt}}
\multiput(576.58,446.00)(0.498,1.495){69}{\rule{0.120pt}{1.289pt}}
\multiput(575.17,446.00)(36.000,104.325){2}{\rule{0.400pt}{0.644pt}}
\multiput(612.00,551.92)(0.600,-0.497){57}{\rule{0.580pt}{0.120pt}}
\multiput(612.00,552.17)(34.796,-30.000){2}{\rule{0.290pt}{0.400pt}}
\multiput(648.58,519.63)(0.498,-0.892){69}{\rule{0.120pt}{0.811pt}}
\multiput(647.17,521.32)(36.000,-62.316){2}{\rule{0.400pt}{0.406pt}}
\multiput(684.00,459.58)(0.513,0.498){67}{\rule{0.511pt}{0.120pt}}
\multiput(684.00,458.17)(34.939,35.000){2}{\rule{0.256pt}{0.400pt}}
\multiput(720.00,494.58)(0.752,0.496){45}{\rule{0.700pt}{0.120pt}}
\multiput(720.00,493.17)(34.547,24.000){2}{\rule{0.350pt}{0.400pt}}
\multiput(756.00,516.92)(0.667,-0.497){51}{\rule{0.633pt}{0.120pt}}
\multiput(756.00,517.17)(34.685,-27.000){2}{\rule{0.317pt}{0.400pt}}
\put(792,489.67){\rule{8.672pt}{0.400pt}}
\multiput(792.00,490.17)(18.000,-1.000){2}{\rule{4.336pt}{0.400pt}}
\multiput(828.00,490.58)(1.305,0.494){25}{\rule{1.129pt}{0.119pt}}
\multiput(828.00,489.17)(33.658,14.000){2}{\rule{0.564pt}{0.400pt}}
\put(864,502.17){\rule{7.300pt}{0.400pt}}
\multiput(864.00,503.17)(20.848,-2.000){2}{\rule{3.650pt}{0.400pt}}
\multiput(900.00,500.92)(1.530,-0.492){21}{\rule{1.300pt}{0.119pt}}
\multiput(900.00,501.17)(33.302,-12.000){2}{\rule{0.650pt}{0.400pt}}
\multiput(936.00,490.61)(7.830,0.447){3}{\rule{4.900pt}{0.108pt}}
\multiput(936.00,489.17)(25.830,3.000){2}{\rule{2.450pt}{0.400pt}}
\multiput(972.00,493.59)(3.203,0.482){9}{\rule{2.500pt}{0.116pt}}
\multiput(972.00,492.17)(30.811,6.000){2}{\rule{1.250pt}{0.400pt}}
\multiput(1008.00,497.95)(7.830,-0.447){3}{\rule{4.900pt}{0.108pt}}
\multiput(1008.00,498.17)(25.830,-3.000){2}{\rule{2.450pt}{0.400pt}}
\multiput(1044.00,494.95)(7.830,-0.447){3}{\rule{4.900pt}{0.108pt}}
\multiput(1044.00,495.17)(25.830,-3.000){2}{\rule{2.450pt}{0.400pt}}
\put(1080,493.17){\rule{7.300pt}{0.400pt}}
\multiput(1080.00,492.17)(20.848,2.000){2}{\rule{3.650pt}{0.400pt}}
\put(1116,495.17){\rule{7.100pt}{0.400pt}}
\multiput(1116.00,494.17)(20.264,2.000){2}{\rule{3.550pt}{0.400pt}}
\multiput(1151.00,495.95)(7.830,-0.447){3}{\rule{4.900pt}{0.108pt}}
\multiput(1151.00,496.17)(25.830,-3.000){2}{\rule{2.450pt}{0.400pt}}
\put(1223,493.67){\rule{8.672pt}{0.400pt}}
\multiput(1223.00,493.17)(18.000,1.000){2}{\rule{4.336pt}{0.400pt}}
\put(1187.0,494.0){\rule[-0.200pt]{8.672pt}{0.400pt}}
\put(1295,493.67){\rule{8.672pt}{0.400pt}}
\multiput(1295.00,494.17)(18.000,-1.000){2}{\rule{4.336pt}{0.400pt}}
\put(1259.0,495.0){\rule[-0.200pt]{8.672pt}{0.400pt}}
\put(1367,493.67){\rule{8.672pt}{0.400pt}}
\multiput(1367.00,493.17)(18.000,1.000){2}{\rule{4.336pt}{0.400pt}}
\put(1403,493.67){\rule{8.672pt}{0.400pt}}
\multiput(1403.00,494.17)(18.000,-1.000){2}{\rule{4.336pt}{0.400pt}}
\put(1331.0,494.0){\rule[-0.200pt]{8.672pt}{0.400pt}}
\put(1439,494){\usebox{\plotpoint}}
\put(181,232){\circle*{18}}
\put(217,445){\circle*{18}}
\put(253,461){\circle*{18}}
\put(289,325){\circle*{18}}
\put(325,471){\circle*{18}}
\put(361,304){\circle*{18}}
\put(397,461){\circle*{18}}
\put(433,317){\circle*{18}}
\put(469,483){\circle*{18}}
\put(504,609){\circle*{18}}
\put(540,470){\circle*{18}}
\put(576,446){\circle*{18}}
\put(612,553){\circle*{18}}
\put(648,523){\circle*{18}}
\put(684,459){\circle*{18}}
\put(720,494){\circle*{18}}
\put(756,518){\circle*{18}}
\put(792,491){\circle*{18}}
\put(828,490){\circle*{18}}
\put(864,504){\circle*{18}}
\put(900,502){\circle*{18}}
\put(936,490){\circle*{18}}
\put(972,493){\circle*{18}}
\put(1008,499){\circle*{18}}
\put(1044,496){\circle*{18}}
\put(1080,493){\circle*{18}}
\put(1116,495){\circle*{18}}
\put(1151,497){\circle*{18}}
\put(1187,494){\circle*{18}}
\put(1223,494){\circle*{18}}
\put(1259,495){\circle*{18}}
\put(1295,495){\circle*{18}}
\put(1331,494){\circle*{18}}
\put(1367,494){\circle*{18}}
\put(1403,495){\circle*{18}}
\put(1439,494){\circle*{18}}
\put(1273,211){\circle*{18}}
\sbox{\plotpoint}{\rule[-0.500pt]{1.000pt}{1.000pt}}%
\put(1203,170){\makebox(0,0)[r]{Single Trajectory}}
\multiput(1223,170)(20.756,0.000){5}{\usebox{\plotpoint}}
\put(1323,170){\usebox{\plotpoint}}
\put(181,232){\usebox{\plotpoint}}
\multiput(181,232)(3.459,20.465){11}{\usebox{\plotpoint}}
\multiput(217,445)(20.629,2.292){2}{\usebox{\plotpoint}}
\multiput(253,449)(4.228,-20.320){8}{\usebox{\plotpoint}}
\multiput(289,276)(2.815,20.564){13}{\usebox{\plotpoint}}
\multiput(325,539)(4.228,-20.320){8}{\usebox{\plotpoint}}
\multiput(361,366)(17.710,-10.823){3}{\usebox{\plotpoint}}
\multiput(397,344)(4.813,-20.190){7}{\usebox{\plotpoint}}
\multiput(433,193)(20.629,-2.292){2}{\usebox{\plotpoint}}
\multiput(469,189)(1.102,20.726){31}{\usebox{\plotpoint}}
\multiput(504,847)(3.054,-20.530){12}{\usebox{\plotpoint}}
\multiput(540,605)(7.157,-19.483){5}{\usebox{\plotpoint}}
\multiput(576,507)(2.112,20.648){17}{\usebox{\plotpoint}}
\multiput(612,859)(1.132,-20.725){32}{\usebox{\plotpoint}}
\multiput(648,200)(19.998,5.555){2}{\usebox{\plotpoint}}
\multiput(684,210)(12.128,-16.844){3}{\usebox{\plotpoint}}
\multiput(720,160)(1.085,20.727){33}{\usebox{\plotpoint}}
\multiput(756,848)(1.662,-20.689){22}{\usebox{\plotpoint}}
\multiput(792,400)(3.156,20.514){11}{\usebox{\plotpoint}}
\multiput(828,634)(4.937,20.160){7}{\usebox{\plotpoint}}
\multiput(864,781)(6.456,-19.726){6}{\usebox{\plotpoint}}
\multiput(900,671)(3.091,-20.524){12}{\usebox{\plotpoint}}
\multiput(936,432)(1.978,20.661){18}{\usebox{\plotpoint}}
\multiput(972,808)(2.970,-20.542){12}{\usebox{\plotpoint}}
\multiput(1008,559)(4.476,20.267){8}{\usebox{\plotpoint}}
\multiput(1044,722)(5.204,-20.093){7}{\usebox{\plotpoint}}
\multiput(1080,583)(20.684,1.724){2}{\usebox{\plotpoint}}
\multiput(1116,586)(3.412,20.473){10}{\usebox{\plotpoint}}
\multiput(1151,796)(1.461,-20.704){25}{\usebox{\plotpoint}}
\put(1205.96,287.05){\usebox{\plotpoint}}
\multiput(1223,288)(6.619,19.672){6}{\usebox{\plotpoint}}
\multiput(1259,395)(2.743,20.573){13}{\usebox{\plotpoint}}
\multiput(1295,665)(12.453,-16.604){3}{\usebox{\plotpoint}}
\multiput(1331,617)(4.476,-20.267){8}{\usebox{\plotpoint}}
\multiput(1367,454)(3.924,20.381){9}{\usebox{\plotpoint}}
\multiput(1403,641)(3.382,-20.478){11}{\usebox{\plotpoint}}
\put(1439,423){\usebox{\plotpoint}}
\put(181,232){\makebox(0,0){$+$}}
\put(217,445){\makebox(0,0){$+$}}
\put(253,449){\makebox(0,0){$+$}}
\put(289,276){\makebox(0,0){$+$}}
\put(325,539){\makebox(0,0){$+$}}
\put(361,366){\makebox(0,0){$+$}}
\put(397,344){\makebox(0,0){$+$}}
\put(433,193){\makebox(0,0){$+$}}
\put(469,189){\makebox(0,0){$+$}}
\put(504,847){\makebox(0,0){$+$}}
\put(540,605){\makebox(0,0){$+$}}
\put(576,507){\makebox(0,0){$+$}}
\put(612,859){\makebox(0,0){$+$}}
\put(648,200){\makebox(0,0){$+$}}
\put(684,210){\makebox(0,0){$+$}}
\put(720,160){\makebox(0,0){$+$}}
\put(756,848){\makebox(0,0){$+$}}
\put(792,400){\makebox(0,0){$+$}}
\put(828,634){\makebox(0,0){$+$}}
\put(864,781){\makebox(0,0){$+$}}
\put(900,671){\makebox(0,0){$+$}}
\put(936,432){\makebox(0,0){$+$}}
\put(972,808){\makebox(0,0){$+$}}
\put(1008,559){\makebox(0,0){$+$}}
\put(1044,722){\makebox(0,0){$+$}}
\put(1080,583){\makebox(0,0){$+$}}
\put(1116,586){\makebox(0,0){$+$}}
\put(1151,796){\makebox(0,0){$+$}}
\put(1187,286){\makebox(0,0){$+$}}
\put(1223,288){\makebox(0,0){$+$}}
\put(1259,395){\makebox(0,0){$+$}}
\put(1295,665){\makebox(0,0){$+$}}
\put(1331,617){\makebox(0,0){$+$}}
\put(1367,454){\makebox(0,0){$+$}}
\put(1403,641){\makebox(0,0){$+$}}
\put(1439,423){\makebox(0,0){$+$}}
\put(1273,170){\makebox(0,0){$+$}}
\end{picture}

%% file: vargrowth.2.835.140.tex
\setlength{\unitlength}{0.240900pt}
\ifx\plotpoint\undefined\newsavebox{\plotpoint}\fi
\sbox{\plotpoint}{\rule[-0.200pt]{0.400pt}{0.400pt}}%
\begin{picture}(1500,900)(0,0)
\font\gnuplot=cmr10 at 10pt
\gnuplot
\sbox{\plotpoint}{\rule[-0.200pt]{0.400pt}{0.400pt}}%
\put(181.0,123.0){\rule[-0.200pt]{2.409pt}{0.400pt}}
\put(1429.0,123.0){\rule[-0.200pt]{2.409pt}{0.400pt}}
\put(181.0,148.0){\rule[-0.200pt]{2.409pt}{0.400pt}}
\put(1429.0,148.0){\rule[-0.200pt]{2.409pt}{0.400pt}}
\put(181.0,170.0){\rule[-0.200pt]{2.409pt}{0.400pt}}
\put(1429.0,170.0){\rule[-0.200pt]{2.409pt}{0.400pt}}
\put(181.0,188.0){\rule[-0.200pt]{2.409pt}{0.400pt}}
\put(1429.0,188.0){\rule[-0.200pt]{2.409pt}{0.400pt}}
\put(181.0,205.0){\rule[-0.200pt]{2.409pt}{0.400pt}}
\put(1429.0,205.0){\rule[-0.200pt]{2.409pt}{0.400pt}}
\put(181.0,219.0){\rule[-0.200pt]{4.818pt}{0.400pt}}
\put(161,219){\makebox(0,0)[r]{0.01}}
\put(1419.0,219.0){\rule[-0.200pt]{4.818pt}{0.400pt}}
\put(181.0,316.0){\rule[-0.200pt]{2.409pt}{0.400pt}}
\put(1429.0,316.0){\rule[-0.200pt]{2.409pt}{0.400pt}}
\put(181.0,372.0){\rule[-0.200pt]{2.409pt}{0.400pt}}
\put(1429.0,372.0){\rule[-0.200pt]{2.409pt}{0.400pt}}
\put(181.0,412.0){\rule[-0.200pt]{2.409pt}{0.400pt}}
\put(1429.0,412.0){\rule[-0.200pt]{2.409pt}{0.400pt}}
\put(181.0,443.0){\rule[-0.200pt]{2.409pt}{0.400pt}}
\put(1429.0,443.0){\rule[-0.200pt]{2.409pt}{0.400pt}}
\put(181.0,469.0){\rule[-0.200pt]{2.409pt}{0.400pt}}
\put(1429.0,469.0){\rule[-0.200pt]{2.409pt}{0.400pt}}
\put(181.0,490.0){\rule[-0.200pt]{2.409pt}{0.400pt}}
\put(1429.0,490.0){\rule[-0.200pt]{2.409pt}{0.400pt}}
\put(181.0,509.0){\rule[-0.200pt]{2.409pt}{0.400pt}}
\put(1429.0,509.0){\rule[-0.200pt]{2.409pt}{0.400pt}}
\put(181.0,525.0){\rule[-0.200pt]{2.409pt}{0.400pt}}
\put(1429.0,525.0){\rule[-0.200pt]{2.409pt}{0.400pt}}
\put(181.0,540.0){\rule[-0.200pt]{4.818pt}{0.400pt}}
\put(161,540){\makebox(0,0)[r]{0.1}}
\put(1419.0,540.0){\rule[-0.200pt]{4.818pt}{0.400pt}}
\put(181.0,636.0){\rule[-0.200pt]{2.409pt}{0.400pt}}
\put(1429.0,636.0){\rule[-0.200pt]{2.409pt}{0.400pt}}
\put(181.0,693.0){\rule[-0.200pt]{2.409pt}{0.400pt}}
\put(1429.0,693.0){\rule[-0.200pt]{2.409pt}{0.400pt}}
\put(181.0,733.0){\rule[-0.200pt]{2.409pt}{0.400pt}}
\put(1429.0,733.0){\rule[-0.200pt]{2.409pt}{0.400pt}}
\put(181.0,764.0){\rule[-0.200pt]{2.409pt}{0.400pt}}
\put(1429.0,764.0){\rule[-0.200pt]{2.409pt}{0.400pt}}
\put(181.0,789.0){\rule[-0.200pt]{2.409pt}{0.400pt}}
\put(1429.0,789.0){\rule[-0.200pt]{2.409pt}{0.400pt}}
\put(181.0,810.0){\rule[-0.200pt]{2.409pt}{0.400pt}}
\put(1429.0,810.0){\rule[-0.200pt]{2.409pt}{0.400pt}}
\put(181.0,829.0){\rule[-0.200pt]{2.409pt}{0.400pt}}
\put(1429.0,829.0){\rule[-0.200pt]{2.409pt}{0.400pt}}
\put(181.0,845.0){\rule[-0.200pt]{2.409pt}{0.400pt}}
\put(1429.0,845.0){\rule[-0.200pt]{2.409pt}{0.400pt}}
\put(181.0,860.0){\rule[-0.200pt]{4.818pt}{0.400pt}}
\put(161,860){\makebox(0,0)[r]{1}}
\put(1419.0,860.0){\rule[-0.200pt]{4.818pt}{0.400pt}}
\put(222.0,123.0){\rule[-0.200pt]{0.400pt}{4.818pt}}
\put(222,82){\makebox(0,0){0}}
\put(222.0,840.0){\rule[-0.200pt]{0.400pt}{4.818pt}}
\put(384.0,123.0){\rule[-0.200pt]{0.400pt}{4.818pt}}
\put(384,82){\makebox(0,0){2}}
\put(384.0,840.0){\rule[-0.200pt]{0.400pt}{4.818pt}}
\put(546.0,123.0){\rule[-0.200pt]{0.400pt}{4.818pt}}
\put(546,82){\makebox(0,0){4}}
\put(546.0,840.0){\rule[-0.200pt]{0.400pt}{4.818pt}}
\put(709.0,123.0){\rule[-0.200pt]{0.400pt}{4.818pt}}
\put(709,82){\makebox(0,0){6}}
\put(709.0,840.0){\rule[-0.200pt]{0.400pt}{4.818pt}}
\put(871.0,123.0){\rule[-0.200pt]{0.400pt}{4.818pt}}
\put(871,82){\makebox(0,0){8}}
\put(871.0,840.0){\rule[-0.200pt]{0.400pt}{4.818pt}}
\put(1033.0,123.0){\rule[-0.200pt]{0.400pt}{4.818pt}}
\put(1033,82){\makebox(0,0){10}}
\put(1033.0,840.0){\rule[-0.200pt]{0.400pt}{4.818pt}}
\put(1196.0,123.0){\rule[-0.200pt]{0.400pt}{4.818pt}}
\put(1196,82){\makebox(0,0){12}}
\put(1196.0,840.0){\rule[-0.200pt]{0.400pt}{4.818pt}}
\put(1358.0,123.0){\rule[-0.200pt]{0.400pt}{4.818pt}}
\put(1358,82){\makebox(0,0){14}}
\put(1358.0,840.0){\rule[-0.200pt]{0.400pt}{4.818pt}}
\put(181.0,123.0){\rule[-0.200pt]{303.052pt}{0.400pt}}
\put(1439.0,123.0){\rule[-0.200pt]{0.400pt}{177.543pt}}
\put(181.0,860.0){\rule[-0.200pt]{303.052pt}{0.400pt}}
\put(40,651){\makebox(0,0){ $ \Delta {\bf {\tilde L}}^2 $}}
\put(810,21){\makebox(0,0){ {Kick Number }}}
\put(181.0,123.0){\rule[-0.200pt]{0.400pt}{177.543pt}}
\multiput(199.59,123.00)(0.485,0.874){11}{\rule{0.117pt}{0.786pt}}
\multiput(198.17,123.00)(7.000,10.369){2}{\rule{0.400pt}{0.393pt}}
\multiput(206.58,135.00)(0.493,0.774){23}{\rule{0.119pt}{0.715pt}}
\multiput(205.17,135.00)(13.000,18.515){2}{\rule{0.400pt}{0.358pt}}
\multiput(219.58,155.00)(0.493,0.734){23}{\rule{0.119pt}{0.685pt}}
\multiput(218.17,155.00)(13.000,17.579){2}{\rule{0.400pt}{0.342pt}}
\multiput(232.58,174.00)(0.493,0.774){23}{\rule{0.119pt}{0.715pt}}
\multiput(231.17,174.00)(13.000,18.515){2}{\rule{0.400pt}{0.358pt}}
\multiput(245.58,194.00)(0.492,0.798){21}{\rule{0.119pt}{0.733pt}}
\multiput(244.17,194.00)(12.000,17.478){2}{\rule{0.400pt}{0.367pt}}
\multiput(257.58,213.00)(0.493,0.774){23}{\rule{0.119pt}{0.715pt}}
\multiput(256.17,213.00)(13.000,18.515){2}{\rule{0.400pt}{0.358pt}}
\multiput(270.58,233.00)(0.493,0.774){23}{\rule{0.119pt}{0.715pt}}
\multiput(269.17,233.00)(13.000,18.515){2}{\rule{0.400pt}{0.358pt}}
\multiput(283.58,253.00)(0.492,0.798){21}{\rule{0.119pt}{0.733pt}}
\multiput(282.17,253.00)(12.000,17.478){2}{\rule{0.400pt}{0.367pt}}
\multiput(295.58,272.00)(0.493,0.774){23}{\rule{0.119pt}{0.715pt}}
\multiput(294.17,272.00)(13.000,18.515){2}{\rule{0.400pt}{0.358pt}}
\multiput(308.58,292.00)(0.493,0.734){23}{\rule{0.119pt}{0.685pt}}
\multiput(307.17,292.00)(13.000,17.579){2}{\rule{0.400pt}{0.342pt}}
\multiput(321.58,311.00)(0.492,0.841){21}{\rule{0.119pt}{0.767pt}}
\multiput(320.17,311.00)(12.000,18.409){2}{\rule{0.400pt}{0.383pt}}
\multiput(333.58,331.00)(0.493,0.774){23}{\rule{0.119pt}{0.715pt}}
\multiput(332.17,331.00)(13.000,18.515){2}{\rule{0.400pt}{0.358pt}}
\multiput(346.58,351.00)(0.493,0.734){23}{\rule{0.119pt}{0.685pt}}
\multiput(345.17,351.00)(13.000,17.579){2}{\rule{0.400pt}{0.342pt}}
\multiput(359.58,370.00)(0.493,0.774){23}{\rule{0.119pt}{0.715pt}}
\multiput(358.17,370.00)(13.000,18.515){2}{\rule{0.400pt}{0.358pt}}
\multiput(372.58,390.00)(0.492,0.798){21}{\rule{0.119pt}{0.733pt}}
\multiput(371.17,390.00)(12.000,17.478){2}{\rule{0.400pt}{0.367pt}}
\multiput(384.58,409.00)(0.493,0.774){23}{\rule{0.119pt}{0.715pt}}
\multiput(383.17,409.00)(13.000,18.515){2}{\rule{0.400pt}{0.358pt}}
\multiput(397.58,429.00)(0.493,0.774){23}{\rule{0.119pt}{0.715pt}}
\multiput(396.17,429.00)(13.000,18.515){2}{\rule{0.400pt}{0.358pt}}
\multiput(410.58,449.00)(0.492,0.798){21}{\rule{0.119pt}{0.733pt}}
\multiput(409.17,449.00)(12.000,17.478){2}{\rule{0.400pt}{0.367pt}}
\multiput(422.58,468.00)(0.493,0.774){23}{\rule{0.119pt}{0.715pt}}
\multiput(421.17,468.00)(13.000,18.515){2}{\rule{0.400pt}{0.358pt}}
\multiput(435.58,488.00)(0.493,0.734){23}{\rule{0.119pt}{0.685pt}}
\multiput(434.17,488.00)(13.000,17.579){2}{\rule{0.400pt}{0.342pt}}
\multiput(448.58,507.00)(0.493,0.774){23}{\rule{0.119pt}{0.715pt}}
\multiput(447.17,507.00)(13.000,18.515){2}{\rule{0.400pt}{0.358pt}}
\multiput(461.58,527.00)(0.492,0.841){21}{\rule{0.119pt}{0.767pt}}
\multiput(460.17,527.00)(12.000,18.409){2}{\rule{0.400pt}{0.383pt}}
\multiput(473.58,547.00)(0.493,0.734){23}{\rule{0.119pt}{0.685pt}}
\multiput(472.17,547.00)(13.000,17.579){2}{\rule{0.400pt}{0.342pt}}
\multiput(486.58,566.00)(0.493,0.774){23}{\rule{0.119pt}{0.715pt}}
\multiput(485.17,566.00)(13.000,18.515){2}{\rule{0.400pt}{0.358pt}}
\multiput(499.58,586.00)(0.492,0.798){21}{\rule{0.119pt}{0.733pt}}
\multiput(498.17,586.00)(12.000,17.478){2}{\rule{0.400pt}{0.367pt}}
\multiput(511.58,605.00)(0.493,0.774){23}{\rule{0.119pt}{0.715pt}}
\multiput(510.17,605.00)(13.000,18.515){2}{\rule{0.400pt}{0.358pt}}
\multiput(524.58,625.00)(0.493,0.774){23}{\rule{0.119pt}{0.715pt}}
\multiput(523.17,625.00)(13.000,18.515){2}{\rule{0.400pt}{0.358pt}}
\multiput(537.58,645.00)(0.493,0.734){23}{\rule{0.119pt}{0.685pt}}
\multiput(536.17,645.00)(13.000,17.579){2}{\rule{0.400pt}{0.342pt}}
\multiput(550.58,664.00)(0.492,0.841){21}{\rule{0.119pt}{0.767pt}}
\multiput(549.17,664.00)(12.000,18.409){2}{\rule{0.400pt}{0.383pt}}
\multiput(562.58,684.00)(0.493,0.734){23}{\rule{0.119pt}{0.685pt}}
\multiput(561.17,684.00)(13.000,17.579){2}{\rule{0.400pt}{0.342pt}}
\multiput(575.58,703.00)(0.493,0.774){23}{\rule{0.119pt}{0.715pt}}
\multiput(574.17,703.00)(13.000,18.515){2}{\rule{0.400pt}{0.358pt}}
\multiput(588.58,723.00)(0.492,0.841){21}{\rule{0.119pt}{0.767pt}}
\multiput(587.17,723.00)(12.000,18.409){2}{\rule{0.400pt}{0.383pt}}
\multiput(600.58,743.00)(0.493,0.734){23}{\rule{0.119pt}{0.685pt}}
\multiput(599.17,743.00)(13.000,17.579){2}{\rule{0.400pt}{0.342pt}}
\multiput(613.58,762.00)(0.493,0.774){23}{\rule{0.119pt}{0.715pt}}
\multiput(612.17,762.00)(13.000,18.515){2}{\rule{0.400pt}{0.358pt}}
\multiput(626.58,782.00)(0.492,0.798){21}{\rule{0.119pt}{0.733pt}}
\multiput(625.17,782.00)(12.000,17.478){2}{\rule{0.400pt}{0.367pt}}
\multiput(638.58,801.00)(0.493,0.774){23}{\rule{0.119pt}{0.715pt}}
\multiput(637.17,801.00)(13.000,18.515){2}{\rule{0.400pt}{0.358pt}}
\multiput(651.58,821.00)(0.493,0.774){23}{\rule{0.119pt}{0.715pt}}
\multiput(650.17,821.00)(13.000,18.515){2}{\rule{0.400pt}{0.358pt}}
\multiput(664.58,841.00)(0.492,0.798){21}{\rule{0.119pt}{0.733pt}}
\multiput(663.17,841.00)(12.000,17.478){2}{\rule{0.400pt}{0.367pt}}
\put(1257,636){\makebox(0,0)[r]{quantum (a)}}
\put(222,158){\raisebox{-.8pt}{\makebox(0,0){$\Diamond$}}}
\put(303,378){\raisebox{-.8pt}{\makebox(0,0){$\Diamond$}}}
\put(384,412){\raisebox{-.8pt}{\makebox(0,0){$\Diamond$}}}
\put(465,481){\raisebox{-.8pt}{\makebox(0,0){$\Diamond$}}}
\put(546,557){\raisebox{-.8pt}{\makebox(0,0){$\Diamond$}}}
\put(627,652){\raisebox{-.8pt}{\makebox(0,0){$\Diamond$}}}
\put(709,729){\raisebox{-.8pt}{\makebox(0,0){$\Diamond$}}}
\put(790,808){\raisebox{-.8pt}{\makebox(0,0){$\Diamond$}}}
\put(871,847){\raisebox{-.8pt}{\makebox(0,0){$\Diamond$}}}
\put(952,845){\raisebox{-.8pt}{\makebox(0,0){$\Diamond$}}}
\put(1033,857){\raisebox{-.8pt}{\makebox(0,0){$\Diamond$}}}
\put(1114,855){\raisebox{-.8pt}{\makebox(0,0){$\Diamond$}}}
\put(1196,855){\raisebox{-.8pt}{\makebox(0,0){$\Diamond$}}}
\put(1277,858){\raisebox{-.8pt}{\makebox(0,0){$\Diamond$}}}
\put(1358,858){\raisebox{-.8pt}{\makebox(0,0){$\Diamond$}}}
\put(1439,860){\raisebox{-.8pt}{\makebox(0,0){$\Diamond$}}}
\put(1327,636){\raisebox{-.8pt}{\makebox(0,0){$\Diamond$}}}
\sbox{\plotpoint}{\rule[-0.400pt]{0.800pt}{0.800pt}}%
\put(1257,595){\makebox(0,0)[r]{quantum (b)}}
\put(222,158){\makebox(0,0){$+$}}
\put(303,378){\makebox(0,0){$+$}}
\put(384,515){\makebox(0,0){$+$}}
\put(465,681){\makebox(0,0){$+$}}
\put(546,851){\makebox(0,0){$+$}}
\put(627,838){\makebox(0,0){$+$}}
\put(709,858){\makebox(0,0){$+$}}
\put(790,856){\makebox(0,0){$+$}}
\put(871,857){\makebox(0,0){$+$}}
\put(952,859){\makebox(0,0){$+$}}
\put(1033,858){\makebox(0,0){$+$}}
\put(1114,859){\makebox(0,0){$+$}}
\put(1196,859){\makebox(0,0){$+$}}
\put(1277,860){\makebox(0,0){$+$}}
\put(1358,859){\makebox(0,0){$+$}}
\put(1439,860){\makebox(0,0){$+$}}
\put(1327,595){\makebox(0,0){$+$}}
\sbox{\plotpoint}{\rule[-0.500pt]{1.000pt}{1.000pt}}%
\put(1257,554){\makebox(0,0)[r]{classical (a)}}
\put(222,159){\raisebox{-.8pt}{\makebox(0,0){$\Box$}}}
\put(303,378){\raisebox{-.8pt}{\makebox(0,0){$\Box$}}}
\put(384,412){\raisebox{-.8pt}{\makebox(0,0){$\Box$}}}
\put(465,483){\raisebox{-.8pt}{\makebox(0,0){$\Box$}}}
\put(546,559){\raisebox{-.8pt}{\makebox(0,0){$\Box$}}}
\put(627,649){\raisebox{-.8pt}{\makebox(0,0){$\Box$}}}
\put(709,732){\raisebox{-.8pt}{\makebox(0,0){$\Box$}}}
\put(790,802){\raisebox{-.8pt}{\makebox(0,0){$\Box$}}}
\put(871,843){\raisebox{-.8pt}{\makebox(0,0){$\Box$}}}
\put(952,845){\raisebox{-.8pt}{\makebox(0,0){$\Box$}}}
\put(1033,859){\raisebox{-.8pt}{\makebox(0,0){$\Box$}}}
\put(1114,856){\raisebox{-.8pt}{\makebox(0,0){$\Box$}}}
\put(1196,855){\raisebox{-.8pt}{\makebox(0,0){$\Box$}}}
\put(1277,859){\raisebox{-.8pt}{\makebox(0,0){$\Box$}}}
\put(1358,859){\raisebox{-.8pt}{\makebox(0,0){$\Box$}}}
\put(1439,860){\raisebox{-.8pt}{\makebox(0,0){$\Box$}}}
\put(1327,554){\raisebox{-.8pt}{\makebox(0,0){$\Box$}}}
\sbox{\plotpoint}{\rule[-0.600pt]{1.200pt}{1.200pt}}%
\put(1257,513){\makebox(0,0)[r]{classical (b)}}
\put(222,159){\makebox(0,0){$\times$}}
\put(303,378){\makebox(0,0){$\times$}}
\put(384,515){\makebox(0,0){$\times$}}
\put(465,681){\makebox(0,0){$\times$}}
\put(546,851){\makebox(0,0){$\times$}}
\put(627,838){\makebox(0,0){$\times$}}
\put(709,858){\makebox(0,0){$\times$}}
\put(790,856){\makebox(0,0){$\times$}}
\put(871,857){\makebox(0,0){$\times$}}
\put(952,859){\makebox(0,0){$\times$}}
\put(1033,858){\makebox(0,0){$\times$}}
\put(1114,859){\makebox(0,0){$\times$}}
\put(1196,860){\makebox(0,0){$\times$}}
\put(1277,860){\makebox(0,0){$\times$}}
\put(1358,860){\makebox(0,0){$\times$}}
\put(1439,860){\makebox(0,0){$\times$}}
\put(1327,513){\makebox(0,0){$\times$}}
\end{picture}

%% file: probdist00.tex
\setlength{\unitlength}{0.240900pt}
\ifx\plotpoint\undefined\newsavebox{\plotpoint}\fi
\begin{picture}(1081,900)(0,0)
\font\gnuplot=cmr10 at 10pt
\gnuplot
\sbox{\plotpoint}{\rule[-0.200pt]{0.400pt}{0.400pt}}%
\put(181.0,123.0){\rule[-0.200pt]{4.818pt}{0.400pt}}
\put(161,123){\makebox(0,0)[r]{0}}
\put(1040.0,123.0){\rule[-0.200pt]{4.818pt}{0.400pt}}
\put(181.0,270.0){\rule[-0.200pt]{4.818pt}{0.400pt}}
\put(161,270){\makebox(0,0)[r]{0.02}}
\put(1040.0,270.0){\rule[-0.200pt]{4.818pt}{0.400pt}}
\put(181.0,418.0){\rule[-0.200pt]{4.818pt}{0.400pt}}
\put(161,418){\makebox(0,0)[r]{0.04}}
\put(1040.0,418.0){\rule[-0.200pt]{4.818pt}{0.400pt}}
\put(181.0,565.0){\rule[-0.200pt]{4.818pt}{0.400pt}}
\put(161,565){\makebox(0,0)[r]{0.06}}
\put(1040.0,565.0){\rule[-0.200pt]{4.818pt}{0.400pt}}
\put(181.0,713.0){\rule[-0.200pt]{4.818pt}{0.400pt}}
\put(161,713){\makebox(0,0)[r]{0.08}}
\put(1040.0,713.0){\rule[-0.200pt]{4.818pt}{0.400pt}}
\put(181.0,860.0){\rule[-0.200pt]{4.818pt}{0.400pt}}
\put(161,860){\makebox(0,0)[r]{0.1}}
\put(1040.0,860.0){\rule[-0.200pt]{4.818pt}{0.400pt}}
\put(208.0,123.0){\rule[-0.200pt]{0.400pt}{4.818pt}}
\put(208,82){\makebox(0,0){-150}}
\put(208.0,840.0){\rule[-0.200pt]{0.400pt}{4.818pt}}
\put(346.0,123.0){\rule[-0.200pt]{0.400pt}{4.818pt}}
\put(346,82){\makebox(0,0){-100}}
\put(346.0,840.0){\rule[-0.200pt]{0.400pt}{4.818pt}}
\put(483.0,123.0){\rule[-0.200pt]{0.400pt}{4.818pt}}
\put(483,82){\makebox(0,0){-50}}
\put(483.0,840.0){\rule[-0.200pt]{0.400pt}{4.818pt}}
\put(621.0,123.0){\rule[-0.200pt]{0.400pt}{4.818pt}}
\put(621,82){\makebox(0,0){0}}
\put(621.0,840.0){\rule[-0.200pt]{0.400pt}{4.818pt}}
\put(758.0,123.0){\rule[-0.200pt]{0.400pt}{4.818pt}}
\put(758,82){\makebox(0,0){50}}
\put(758.0,840.0){\rule[-0.200pt]{0.400pt}{4.818pt}}
\put(895.0,123.0){\rule[-0.200pt]{0.400pt}{4.818pt}}
\put(895,82){\makebox(0,0){100}}
\put(895.0,840.0){\rule[-0.200pt]{0.400pt}{4.818pt}}
\put(1033.0,123.0){\rule[-0.200pt]{0.400pt}{4.818pt}}
\put(1033,82){\makebox(0,0){150}}
\put(1033.0,840.0){\rule[-0.200pt]{0.400pt}{4.818pt}}
\put(181.0,123.0){\rule[-0.200pt]{211.751pt}{0.400pt}}
\put(1060.0,123.0){\rule[-0.200pt]{0.400pt}{177.543pt}}
\put(181.0,860.0){\rule[-0.200pt]{211.751pt}{0.400pt}}
\put(620,21){\makebox(0,0){$m_l$}}
\put(181.0,123.0){\rule[-0.200pt]{0.400pt}{177.543pt}}
\put(875,713){\makebox(0,0)[r]{ { $P_z(m_l)$ } }}
\put(895.0,713.0){\rule[-0.200pt]{24.090pt}{0.400pt}}
\put(1044,123){\usebox{\plotpoint}}
\put(387,122.67){\rule{0.723pt}{0.400pt}}
\multiput(388.50,122.17)(-1.500,1.000){2}{\rule{0.361pt}{0.400pt}}
\put(390.0,123.0){\rule[-0.200pt]{157.549pt}{0.400pt}}
\put(382,123.67){\rule{0.482pt}{0.400pt}}
\multiput(383.00,123.17)(-1.000,1.000){2}{\rule{0.241pt}{0.400pt}}
\put(379,124.67){\rule{0.723pt}{0.400pt}}
\multiput(380.50,124.17)(-1.500,1.000){2}{\rule{0.361pt}{0.400pt}}
\put(376,125.67){\rule{0.723pt}{0.400pt}}
\multiput(377.50,125.17)(-1.500,1.000){2}{\rule{0.361pt}{0.400pt}}
\put(373,127.17){\rule{0.700pt}{0.400pt}}
\multiput(374.55,126.17)(-1.547,2.000){2}{\rule{0.350pt}{0.400pt}}
\put(371.17,129){\rule{0.400pt}{0.700pt}}
\multiput(372.17,129.00)(-2.000,1.547){2}{\rule{0.400pt}{0.350pt}}
\multiput(369.95,132.00)(-0.447,0.909){3}{\rule{0.108pt}{0.767pt}}
\multiput(370.17,132.00)(-3.000,3.409){2}{\rule{0.400pt}{0.383pt}}
\multiput(366.95,137.00)(-0.447,1.355){3}{\rule{0.108pt}{1.033pt}}
\multiput(367.17,137.00)(-3.000,4.855){2}{\rule{0.400pt}{0.517pt}}
\multiput(363.95,144.00)(-0.447,1.802){3}{\rule{0.108pt}{1.300pt}}
\multiput(364.17,144.00)(-3.000,6.302){2}{\rule{0.400pt}{0.650pt}}
\put(360.17,153){\rule{0.400pt}{2.500pt}}
\multiput(361.17,153.00)(-2.000,6.811){2}{\rule{0.400pt}{1.250pt}}
\multiput(358.95,165.00)(-0.447,3.141){3}{\rule{0.108pt}{2.100pt}}
\multiput(359.17,165.00)(-3.000,10.641){2}{\rule{0.400pt}{1.050pt}}
\multiput(355.95,180.00)(-0.447,4.258){3}{\rule{0.108pt}{2.767pt}}
\multiput(356.17,180.00)(-3.000,14.258){2}{\rule{0.400pt}{1.383pt}}
\multiput(352.95,200.00)(-0.447,5.374){3}{\rule{0.108pt}{3.433pt}}
\multiput(353.17,200.00)(-3.000,17.874){2}{\rule{0.400pt}{1.717pt}}
\put(349.17,225){\rule{0.400pt}{5.900pt}}
\multiput(350.17,225.00)(-2.000,16.754){2}{\rule{0.400pt}{2.950pt}}
\multiput(347.95,254.00)(-0.447,7.607){3}{\rule{0.108pt}{4.767pt}}
\multiput(348.17,254.00)(-3.000,25.107){2}{\rule{0.400pt}{2.383pt}}
\multiput(344.95,289.00)(-0.447,8.276){3}{\rule{0.108pt}{5.167pt}}
\multiput(345.17,289.00)(-3.000,27.276){2}{\rule{0.400pt}{2.583pt}}
\multiput(341.95,327.00)(-0.447,9.393){3}{\rule{0.108pt}{5.833pt}}
\multiput(342.17,327.00)(-3.000,30.893){2}{\rule{0.400pt}{2.917pt}}
\put(338.17,370){\rule{0.400pt}{8.900pt}}
\multiput(339.17,370.00)(-2.000,25.528){2}{\rule{0.400pt}{4.450pt}}
\multiput(336.95,414.00)(-0.447,9.616){3}{\rule{0.108pt}{5.967pt}}
\multiput(337.17,414.00)(-3.000,31.616){2}{\rule{0.400pt}{2.983pt}}
\multiput(333.95,458.00)(-0.447,9.169){3}{\rule{0.108pt}{5.700pt}}
\multiput(334.17,458.00)(-3.000,30.169){2}{\rule{0.400pt}{2.850pt}}
\multiput(330.95,500.00)(-0.447,8.053){3}{\rule{0.108pt}{5.033pt}}
\multiput(331.17,500.00)(-3.000,26.553){2}{\rule{0.400pt}{2.517pt}}
\put(327.17,537){\rule{0.400pt}{6.100pt}}
\multiput(328.17,537.00)(-2.000,17.339){2}{\rule{0.400pt}{3.050pt}}
\multiput(325.95,567.00)(-0.447,4.258){3}{\rule{0.108pt}{2.767pt}}
\multiput(326.17,567.00)(-3.000,14.258){2}{\rule{0.400pt}{1.383pt}}
\multiput(322.95,587.00)(-0.447,1.802){3}{\rule{0.108pt}{1.300pt}}
\multiput(323.17,587.00)(-3.000,6.302){2}{\rule{0.400pt}{0.650pt}}
\multiput(318.92,594.95)(-0.462,-0.447){3}{\rule{0.500pt}{0.108pt}}
\multiput(319.96,595.17)(-1.962,-3.000){2}{\rule{0.250pt}{0.400pt}}
\put(316.17,578){\rule{0.400pt}{3.100pt}}
\multiput(317.17,586.57)(-2.000,-8.566){2}{\rule{0.400pt}{1.550pt}}
\multiput(314.95,563.19)(-0.447,-5.597){3}{\rule{0.108pt}{3.567pt}}
\multiput(315.17,570.60)(-3.000,-18.597){2}{\rule{0.400pt}{1.783pt}}
\multiput(311.95,531.66)(-0.447,-7.830){3}{\rule{0.108pt}{4.900pt}}
\multiput(312.17,541.83)(-3.000,-25.830){2}{\rule{0.400pt}{2.450pt}}
\multiput(308.95,492.34)(-0.447,-9.169){3}{\rule{0.108pt}{5.700pt}}
\multiput(309.17,504.17)(-3.000,-30.169){2}{\rule{0.400pt}{2.850pt}}
\put(305.17,427){\rule{0.400pt}{9.500pt}}
\multiput(306.17,454.28)(-2.000,-27.282){2}{\rule{0.400pt}{4.750pt}}
\multiput(303.95,400.02)(-0.447,-10.509){3}{\rule{0.108pt}{6.500pt}}
\multiput(304.17,413.51)(-3.000,-34.509){2}{\rule{0.400pt}{3.250pt}}
\multiput(300.95,352.57)(-0.447,-10.286){3}{\rule{0.108pt}{6.367pt}}
\multiput(301.17,365.79)(-3.000,-33.786){2}{\rule{0.400pt}{3.183pt}}
\multiput(297.95,307.79)(-0.447,-9.393){3}{\rule{0.108pt}{5.833pt}}
\multiput(298.17,319.89)(-3.000,-30.893){2}{\rule{0.400pt}{2.917pt}}
\put(294.17,250){\rule{0.400pt}{7.900pt}}
\multiput(295.17,272.60)(-2.000,-22.603){2}{\rule{0.400pt}{3.950pt}}
\multiput(292.95,231.87)(-0.447,-6.937){3}{\rule{0.108pt}{4.367pt}}
\multiput(293.17,240.94)(-3.000,-22.937){2}{\rule{0.400pt}{2.183pt}}
\multiput(289.95,202.64)(-0.447,-5.820){3}{\rule{0.108pt}{3.700pt}}
\multiput(290.17,210.32)(-3.000,-19.320){2}{\rule{0.400pt}{1.850pt}}
\multiput(286.95,179.52)(-0.447,-4.258){3}{\rule{0.108pt}{2.767pt}}
\multiput(287.17,185.26)(-3.000,-14.258){2}{\rule{0.400pt}{1.383pt}}
\put(283.17,155){\rule{0.400pt}{3.300pt}}
\multiput(284.17,164.15)(-2.000,-9.151){2}{\rule{0.400pt}{1.650pt}}
\multiput(281.95,148.50)(-0.447,-2.248){3}{\rule{0.108pt}{1.567pt}}
\multiput(282.17,151.75)(-3.000,-7.748){2}{\rule{0.400pt}{0.783pt}}
\multiput(278.95,139.16)(-0.447,-1.579){3}{\rule{0.108pt}{1.167pt}}
\multiput(279.17,141.58)(-3.000,-5.579){2}{\rule{0.400pt}{0.583pt}}
\multiput(275.95,132.82)(-0.447,-0.909){3}{\rule{0.108pt}{0.767pt}}
\multiput(276.17,134.41)(-3.000,-3.409){2}{\rule{0.400pt}{0.383pt}}
\put(272.17,128){\rule{0.400pt}{0.700pt}}
\multiput(273.17,129.55)(-2.000,-1.547){2}{\rule{0.400pt}{0.350pt}}
\put(269,126.17){\rule{0.700pt}{0.400pt}}
\multiput(270.55,127.17)(-1.547,-2.000){2}{\rule{0.350pt}{0.400pt}}
\put(266,124.17){\rule{0.700pt}{0.400pt}}
\multiput(267.55,125.17)(-1.547,-2.000){2}{\rule{0.350pt}{0.400pt}}
\put(384.0,124.0){\rule[-0.200pt]{0.723pt}{0.400pt}}
\put(261,122.67){\rule{0.482pt}{0.400pt}}
\multiput(262.00,123.17)(-1.000,-1.000){2}{\rule{0.241pt}{0.400pt}}
\put(263.0,124.0){\rule[-0.200pt]{0.723pt}{0.400pt}}
\put(197.0,123.0){\rule[-0.200pt]{15.418pt}{0.400pt}}
\put(875,672){\makebox(0,0)[r]{ { $P_z^c(m_l)$ } }}
\multiput(895,672)(20.756,0.000){5}{\usebox{\plotpoint}}
\put(995,672){\usebox{\plotpoint}}
\put(1044,123){\usebox{\plotpoint}}
\put(1044.00,123.00){\usebox{\plotpoint}}
\put(1023.24,123.00){\usebox{\plotpoint}}
\put(1002.49,123.00){\usebox{\plotpoint}}
\put(981.73,123.00){\usebox{\plotpoint}}
\put(960.98,123.00){\usebox{\plotpoint}}
\put(940.22,123.00){\usebox{\plotpoint}}
\put(919.47,123.00){\usebox{\plotpoint}}
\put(898.71,123.00){\usebox{\plotpoint}}
\put(877.96,123.00){\usebox{\plotpoint}}
\put(857.20,123.00){\usebox{\plotpoint}}
\put(836.45,123.00){\usebox{\plotpoint}}
\put(815.69,123.00){\usebox{\plotpoint}}
\put(794.93,123.00){\usebox{\plotpoint}}
\put(774.18,123.00){\usebox{\plotpoint}}
\put(753.42,123.00){\usebox{\plotpoint}}
\put(732.67,123.00){\usebox{\plotpoint}}
\put(711.91,123.00){\usebox{\plotpoint}}
\put(691.16,123.00){\usebox{\plotpoint}}
\put(670.40,123.00){\usebox{\plotpoint}}
\put(649.65,123.00){\usebox{\plotpoint}}
\put(628.89,123.00){\usebox{\plotpoint}}
\put(608.13,123.00){\usebox{\plotpoint}}
\put(587.38,123.00){\usebox{\plotpoint}}
\put(566.62,123.00){\usebox{\plotpoint}}
\put(545.87,123.00){\usebox{\plotpoint}}
\put(525.11,123.00){\usebox{\plotpoint}}
\put(504.36,123.00){\usebox{\plotpoint}}
\put(483.60,123.00){\usebox{\plotpoint}}
\put(462.85,123.00){\usebox{\plotpoint}}
\put(442.09,123.00){\usebox{\plotpoint}}
\put(421.34,123.00){\usebox{\plotpoint}}
\put(400.58,123.00){\usebox{\plotpoint}}
\put(380.31,125.56){\usebox{\plotpoint}}
\put(367.35,140.53){\usebox{\plotpoint}}
\put(360.97,160.17){\usebox{\plotpoint}}
\put(357.26,180.59){\usebox{\plotpoint}}
\multiput(354,201)(-2.473,20.608){2}{\usebox{\plotpoint}}
\put(349.87,242.38){\usebox{\plotpoint}}
\multiput(349,255)(-1.999,20.659){2}{\usebox{\plotpoint}}
\put(344.58,304.42){\usebox{\plotpoint}}
\multiput(343,325)(-1.479,20.703){3}{\usebox{\plotpoint}}
\multiput(340,367)(-0.942,20.734){2}{\usebox{\plotpoint}}
\multiput(338,411)(-1.479,20.703){2}{\usebox{\plotpoint}}
\multiput(335,453)(-1.479,20.703){2}{\usebox{\plotpoint}}
\put(330.63,511.49){\usebox{\plotpoint}}
\multiput(329,531)(-1.428,20.706){2}{\usebox{\plotpoint}}
\put(325.38,573.52){\usebox{\plotpoint}}
\put(320.38,593.21){\usebox{\plotpoint}}
\multiput(316,582)(-2.292,-20.629){2}{\usebox{\plotpoint}}
\put(311.09,534.66){\usebox{\plotpoint}}
\multiput(310,523)(-1.412,-20.707){2}{\usebox{\plotpoint}}
\multiput(307,479)(-0.882,-20.737){2}{\usebox{\plotpoint}}
\multiput(305,432)(-1.268,-20.717){3}{\usebox{\plotpoint}}
\multiput(302,383)(-1.322,-20.713){2}{\usebox{\plotpoint}}
\multiput(299,336)(-1.381,-20.710){2}{\usebox{\plotpoint}}
\multiput(296,291)(-1.063,-20.728){2}{\usebox{\plotpoint}}
\multiput(294,252)(-1.824,-20.675){2}{\usebox{\plotpoint}}
\put(289.37,203.33){\usebox{\plotpoint}}
\put(286.92,182.72){\usebox{\plotpoint}}
\put(284.16,162.15){\usebox{\plotpoint}}
\put(279.95,141.86){\usebox{\plotpoint}}
\put(268.86,124.95){\usebox{\plotpoint}}
\put(248.42,123.00){\usebox{\plotpoint}}
\put(227.66,123.00){\usebox{\plotpoint}}
\put(206.91,123.00){\usebox{\plotpoint}}
\put(197,123){\usebox{\plotpoint}}
\end{picture}

%% file: probdist06.tex
\setlength{\unitlength}{0.240900pt}
\ifx\plotpoint\undefined\newsavebox{\plotpoint}\fi
\begin{picture}(1081,900)(0,0)
\font\gnuplot=cmr10 at 10pt
\gnuplot
\sbox{\plotpoint}{\rule[-0.200pt]{0.400pt}{0.400pt}}%
\put(201.0,123.0){\rule[-0.200pt]{4.818pt}{0.400pt}}
\put(181,123){\makebox(0,0)[r]{0}}
\put(1040.0,123.0){\rule[-0.200pt]{4.818pt}{0.400pt}}
\put(201.0,307.0){\rule[-0.200pt]{4.818pt}{0.400pt}}
\put(181,307){\makebox(0,0)[r]{0.005}}
\put(1040.0,307.0){\rule[-0.200pt]{4.818pt}{0.400pt}}
\put(201.0,492.0){\rule[-0.200pt]{4.818pt}{0.400pt}}
\put(181,492){\makebox(0,0)[r]{0.01}}
\put(1040.0,492.0){\rule[-0.200pt]{4.818pt}{0.400pt}}
\put(201.0,676.0){\rule[-0.200pt]{4.818pt}{0.400pt}}
\put(181,676){\makebox(0,0)[r]{0.015}}
\put(1040.0,676.0){\rule[-0.200pt]{4.818pt}{0.400pt}}
\put(201.0,860.0){\rule[-0.200pt]{4.818pt}{0.400pt}}
\put(181,860){\makebox(0,0)[r]{0.02}}
\put(1040.0,860.0){\rule[-0.200pt]{4.818pt}{0.400pt}}
\put(228.0,123.0){\rule[-0.200pt]{0.400pt}{4.818pt}}
\put(228,82){\makebox(0,0){-150}}
\put(228.0,840.0){\rule[-0.200pt]{0.400pt}{4.818pt}}
\put(362.0,123.0){\rule[-0.200pt]{0.400pt}{4.818pt}}
\put(362,82){\makebox(0,0){-100}}
\put(362.0,840.0){\rule[-0.200pt]{0.400pt}{4.818pt}}
\put(496.0,123.0){\rule[-0.200pt]{0.400pt}{4.818pt}}
\put(496,82){\makebox(0,0){-50}}
\put(496.0,840.0){\rule[-0.200pt]{0.400pt}{4.818pt}}
\put(631.0,123.0){\rule[-0.200pt]{0.400pt}{4.818pt}}
\put(631,82){\makebox(0,0){0}}
\put(631.0,840.0){\rule[-0.200pt]{0.400pt}{4.818pt}}
\put(765.0,123.0){\rule[-0.200pt]{0.400pt}{4.818pt}}
\put(765,82){\makebox(0,0){50}}
\put(765.0,840.0){\rule[-0.200pt]{0.400pt}{4.818pt}}
\put(899.0,123.0){\rule[-0.200pt]{0.400pt}{4.818pt}}
\put(899,82){\makebox(0,0){100}}
\put(899.0,840.0){\rule[-0.200pt]{0.400pt}{4.818pt}}
\put(1033.0,123.0){\rule[-0.200pt]{0.400pt}{4.818pt}}
\put(1033,82){\makebox(0,0){150}}
\put(1033.0,840.0){\rule[-0.200pt]{0.400pt}{4.818pt}}
\put(201.0,123.0){\rule[-0.200pt]{206.933pt}{0.400pt}}
\put(1060.0,123.0){\rule[-0.200pt]{0.400pt}{177.543pt}}
\put(201.0,860.0){\rule[-0.200pt]{206.933pt}{0.400pt}}
\put(630,21){\makebox(0,0){$m_l$}}
\put(201.0,123.0){\rule[-0.200pt]{0.400pt}{177.543pt}}
\put(879,713){\makebox(0,0)[r]{ { $P_z(m_l)$ } }}
\put(899.0,713.0){\rule[-0.200pt]{24.090pt}{0.400pt}}
\put(1044,173){\usebox{\plotpoint}}
\multiput(1042.95,173.00)(-0.447,9.839){3}{\rule{0.108pt}{6.100pt}}
\multiput(1043.17,173.00)(-3.000,32.339){2}{\rule{0.400pt}{3.050pt}}
\put(1039.17,194){\rule{0.400pt}{4.900pt}}
\multiput(1040.17,207.83)(-2.000,-13.830){2}{\rule{0.400pt}{2.450pt}}
\multiput(1037.95,194.00)(-0.447,2.695){3}{\rule{0.108pt}{1.833pt}}
\multiput(1038.17,194.00)(-3.000,9.195){2}{\rule{0.400pt}{0.917pt}}
\multiput(1034.95,207.00)(-0.447,6.044){3}{\rule{0.108pt}{3.833pt}}
\multiput(1035.17,207.00)(-3.000,20.044){2}{\rule{0.400pt}{1.917pt}}
\multiput(1031.95,235.00)(-0.447,12.072){3}{\rule{0.108pt}{7.433pt}}
\multiput(1032.17,235.00)(-3.000,39.572){2}{\rule{0.400pt}{3.717pt}}
\put(1028.17,290){\rule{0.400pt}{7.700pt}}
\multiput(1029.17,290.00)(-2.000,22.018){2}{\rule{0.400pt}{3.850pt}}
\multiput(1026.95,294.38)(-0.447,-13.188){3}{\rule{0.108pt}{8.100pt}}
\multiput(1027.17,311.19)(-3.000,-43.188){2}{\rule{0.400pt}{4.050pt}}
\multiput(1023.95,268.00)(-0.447,3.588){3}{\rule{0.108pt}{2.367pt}}
\multiput(1024.17,268.00)(-3.000,12.088){2}{\rule{0.400pt}{1.183pt}}
\put(1020.17,285){\rule{0.400pt}{7.300pt}}
\multiput(1021.17,285.00)(-2.000,20.848){2}{\rule{0.400pt}{3.650pt}}
\multiput(1018.95,297.34)(-0.447,-9.169){3}{\rule{0.108pt}{5.700pt}}
\multiput(1019.17,309.17)(-3.000,-30.169){2}{\rule{0.400pt}{2.850pt}}
\multiput(1015.95,279.00)(-0.447,0.909){3}{\rule{0.108pt}{0.767pt}}
\multiput(1016.17,279.00)(-3.000,3.409){2}{\rule{0.400pt}{0.383pt}}
\put(1012.17,284){\rule{0.400pt}{2.700pt}}
\multiput(1013.17,284.00)(-2.000,7.396){2}{\rule{0.400pt}{1.350pt}}
\multiput(1010.95,281.64)(-0.447,-5.820){3}{\rule{0.108pt}{3.700pt}}
\multiput(1011.17,289.32)(-3.000,-19.320){2}{\rule{0.400pt}{1.850pt}}
\multiput(1007.95,263.50)(-0.447,-2.248){3}{\rule{0.108pt}{1.567pt}}
\multiput(1008.17,266.75)(-3.000,-7.748){2}{\rule{0.400pt}{0.783pt}}
\put(1004.17,259){\rule{0.400pt}{2.300pt}}
\multiput(1005.17,259.00)(-2.000,6.226){2}{\rule{0.400pt}{1.150pt}}
\multiput(1002.95,265.71)(-0.447,-1.355){3}{\rule{0.108pt}{1.033pt}}
\multiput(1003.17,267.86)(-3.000,-4.855){2}{\rule{0.400pt}{0.517pt}}
\multiput(999.95,252.07)(-0.447,-4.034){3}{\rule{0.108pt}{2.633pt}}
\multiput(1000.17,257.53)(-3.000,-13.534){2}{\rule{0.400pt}{1.317pt}}
\put(996.17,244){\rule{0.400pt}{0.900pt}}
\multiput(997.17,244.00)(-2.000,2.132){2}{\rule{0.400pt}{0.450pt}}
\multiput(994.95,248.00)(-0.447,0.909){3}{\rule{0.108pt}{0.767pt}}
\multiput(995.17,248.00)(-3.000,3.409){2}{\rule{0.400pt}{0.383pt}}
\multiput(991.95,243.18)(-0.447,-3.588){3}{\rule{0.108pt}{2.367pt}}
\multiput(992.17,248.09)(-3.000,-12.088){2}{\rule{0.400pt}{1.183pt}}
\put(988.17,232){\rule{0.400pt}{0.900pt}}
\multiput(989.17,234.13)(-2.000,-2.132){2}{\rule{0.400pt}{0.450pt}}
\multiput(986.95,232.00)(-0.447,1.132){3}{\rule{0.108pt}{0.900pt}}
\multiput(987.17,232.00)(-3.000,4.132){2}{\rule{0.400pt}{0.450pt}}
\multiput(983.95,229.84)(-0.447,-2.918){3}{\rule{0.108pt}{1.967pt}}
\multiput(984.17,233.92)(-3.000,-9.918){2}{\rule{0.400pt}{0.983pt}}
\multiput(980.95,218.05)(-0.447,-2.025){3}{\rule{0.108pt}{1.433pt}}
\multiput(981.17,221.03)(-3.000,-7.025){2}{\rule{0.400pt}{0.717pt}}
\put(977.17,214){\rule{0.400pt}{1.900pt}}
\multiput(978.17,214.00)(-2.000,5.056){2}{\rule{0.400pt}{0.950pt}}
\multiput(975.95,220.37)(-0.447,-0.685){3}{\rule{0.108pt}{0.633pt}}
\multiput(976.17,221.69)(-3.000,-2.685){2}{\rule{0.400pt}{0.317pt}}
\multiput(972.95,209.73)(-0.447,-3.365){3}{\rule{0.108pt}{2.233pt}}
\multiput(973.17,214.36)(-3.000,-11.365){2}{\rule{0.400pt}{1.117pt}}
\put(969,201.17){\rule{0.482pt}{0.400pt}}
\multiput(970.00,202.17)(-1.000,-2.000){2}{\rule{0.241pt}{0.400pt}}
\multiput(967.95,201.00)(-0.447,1.132){3}{\rule{0.108pt}{0.900pt}}
\multiput(968.17,201.00)(-3.000,4.132){2}{\rule{0.400pt}{0.450pt}}
\multiput(964.95,204.37)(-0.447,-0.685){3}{\rule{0.108pt}{0.633pt}}
\multiput(965.17,205.69)(-3.000,-2.685){2}{\rule{0.400pt}{0.317pt}}
\put(961.17,192){\rule{0.400pt}{2.300pt}}
\multiput(962.17,198.23)(-2.000,-6.226){2}{\rule{0.400pt}{1.150pt}}
\put(958,190.17){\rule{0.700pt}{0.400pt}}
\multiput(959.55,191.17)(-1.547,-2.000){2}{\rule{0.350pt}{0.400pt}}
\multiput(956.95,190.00)(-0.447,0.909){3}{\rule{0.108pt}{0.767pt}}
\multiput(957.17,190.00)(-3.000,3.409){2}{\rule{0.400pt}{0.383pt}}
\multiput(951.95,190.16)(-0.447,-1.579){3}{\rule{0.108pt}{1.167pt}}
\multiput(952.17,192.58)(-3.000,-5.579){2}{\rule{0.400pt}{0.583pt}}
\multiput(948.95,182.71)(-0.447,-1.355){3}{\rule{0.108pt}{1.033pt}}
\multiput(949.17,184.86)(-3.000,-4.855){2}{\rule{0.400pt}{0.517pt}}
\put(945.17,180){\rule{0.400pt}{0.700pt}}
\multiput(946.17,180.00)(-2.000,1.547){2}{\rule{0.400pt}{0.350pt}}
\multiput(943.95,183.00)(-0.447,0.909){3}{\rule{0.108pt}{0.767pt}}
\multiput(944.17,183.00)(-3.000,3.409){2}{\rule{0.400pt}{0.383pt}}
\multiput(940.95,185.37)(-0.447,-0.685){3}{\rule{0.108pt}{0.633pt}}
\multiput(941.17,186.69)(-3.000,-2.685){2}{\rule{0.400pt}{0.317pt}}
\put(937.17,175){\rule{0.400pt}{1.900pt}}
\multiput(938.17,180.06)(-2.000,-5.056){2}{\rule{0.400pt}{0.950pt}}
\put(934,173.17){\rule{0.700pt}{0.400pt}}
\multiput(935.55,174.17)(-1.547,-2.000){2}{\rule{0.350pt}{0.400pt}}
\multiput(932.95,173.00)(-0.447,1.132){3}{\rule{0.108pt}{0.900pt}}
\multiput(933.17,173.00)(-3.000,4.132){2}{\rule{0.400pt}{0.450pt}}
\put(928,178.67){\rule{0.723pt}{0.400pt}}
\multiput(929.50,178.17)(-1.500,1.000){2}{\rule{0.361pt}{0.400pt}}
\put(926.17,174){\rule{0.400pt}{1.300pt}}
\multiput(927.17,177.30)(-2.000,-3.302){2}{\rule{0.400pt}{0.650pt}}
\multiput(924.95,170.26)(-0.447,-1.132){3}{\rule{0.108pt}{0.900pt}}
\multiput(925.17,172.13)(-3.000,-4.132){2}{\rule{0.400pt}{0.450pt}}
\put(953.0,195.0){\rule[-0.200pt]{0.482pt}{0.400pt}}
\put(918.17,168){\rule{0.400pt}{0.700pt}}
\multiput(919.17,168.00)(-2.000,1.547){2}{\rule{0.400pt}{0.350pt}}
\put(915,170.67){\rule{0.723pt}{0.400pt}}
\multiput(916.50,170.17)(-1.500,1.000){2}{\rule{0.361pt}{0.400pt}}
\multiput(913.95,169.37)(-0.447,-0.685){3}{\rule{0.108pt}{0.633pt}}
\multiput(914.17,170.69)(-3.000,-2.685){2}{\rule{0.400pt}{0.317pt}}
\put(910.17,163){\rule{0.400pt}{1.100pt}}
\multiput(911.17,165.72)(-2.000,-2.717){2}{\rule{0.400pt}{0.550pt}}
\put(907,161.67){\rule{0.723pt}{0.400pt}}
\multiput(908.50,162.17)(-1.500,-1.000){2}{\rule{0.361pt}{0.400pt}}
\put(904,162.17){\rule{0.700pt}{0.400pt}}
\multiput(905.55,161.17)(-1.547,2.000){2}{\rule{0.350pt}{0.400pt}}
\put(902,164.17){\rule{0.482pt}{0.400pt}}
\multiput(903.00,163.17)(-1.000,2.000){2}{\rule{0.241pt}{0.400pt}}
\multiput(899.92,164.95)(-0.462,-0.447){3}{\rule{0.500pt}{0.108pt}}
\multiput(900.96,165.17)(-1.962,-3.000){2}{\rule{0.250pt}{0.400pt}}
\multiput(897.95,159.82)(-0.447,-0.909){3}{\rule{0.108pt}{0.767pt}}
\multiput(898.17,161.41)(-3.000,-3.409){2}{\rule{0.400pt}{0.383pt}}
\put(894,156.67){\rule{0.482pt}{0.400pt}}
\multiput(895.00,157.17)(-1.000,-1.000){2}{\rule{0.241pt}{0.400pt}}
\put(891,157.17){\rule{0.700pt}{0.400pt}}
\multiput(892.55,156.17)(-1.547,2.000){2}{\rule{0.350pt}{0.400pt}}
\put(888,158.67){\rule{0.723pt}{0.400pt}}
\multiput(889.50,158.17)(-1.500,1.000){2}{\rule{0.361pt}{0.400pt}}
\put(920.0,168.0){\rule[-0.200pt]{0.723pt}{0.400pt}}
\multiput(883.92,158.95)(-0.462,-0.447){3}{\rule{0.500pt}{0.108pt}}
\multiput(884.96,159.17)(-1.962,-3.000){2}{\rule{0.250pt}{0.400pt}}
\put(880,155.17){\rule{0.700pt}{0.400pt}}
\multiput(881.55,156.17)(-1.547,-2.000){2}{\rule{0.350pt}{0.400pt}}
\put(886.0,160.0){\rule[-0.200pt]{0.482pt}{0.400pt}}
\put(875,154.67){\rule{0.482pt}{0.400pt}}
\multiput(876.00,154.17)(-1.000,1.000){2}{\rule{0.241pt}{0.400pt}}
\put(872,156.17){\rule{0.700pt}{0.400pt}}
\multiput(873.55,155.17)(-1.547,2.000){2}{\rule{0.350pt}{0.400pt}}
\put(877.0,155.0){\rule[-0.200pt]{0.723pt}{0.400pt}}
\put(867.17,155){\rule{0.400pt}{0.700pt}}
\multiput(868.17,156.55)(-2.000,-1.547){2}{\rule{0.400pt}{0.350pt}}
\put(864,153.17){\rule{0.700pt}{0.400pt}}
\multiput(865.55,154.17)(-1.547,-2.000){2}{\rule{0.350pt}{0.400pt}}
\put(861,152.67){\rule{0.723pt}{0.400pt}}
\multiput(862.50,152.17)(-1.500,1.000){2}{\rule{0.361pt}{0.400pt}}
\put(859,154.17){\rule{0.482pt}{0.400pt}}
\multiput(860.00,153.17)(-1.000,2.000){2}{\rule{0.241pt}{0.400pt}}
\put(856,155.67){\rule{0.723pt}{0.400pt}}
\multiput(857.50,155.17)(-1.500,1.000){2}{\rule{0.361pt}{0.400pt}}
\put(853,155.67){\rule{0.723pt}{0.400pt}}
\multiput(854.50,156.17)(-1.500,-1.000){2}{\rule{0.361pt}{0.400pt}}
\put(851,154.17){\rule{0.482pt}{0.400pt}}
\multiput(852.00,155.17)(-1.000,-2.000){2}{\rule{0.241pt}{0.400pt}}
\put(848,152.67){\rule{0.723pt}{0.400pt}}
\multiput(849.50,153.17)(-1.500,-1.000){2}{\rule{0.361pt}{0.400pt}}
\put(869.0,158.0){\rule[-0.200pt]{0.723pt}{0.400pt}}
\put(843,153.17){\rule{0.482pt}{0.400pt}}
\multiput(844.00,152.17)(-1.000,2.000){2}{\rule{0.241pt}{0.400pt}}
\put(845.0,153.0){\rule[-0.200pt]{0.723pt}{0.400pt}}
\put(837,153.67){\rule{0.723pt}{0.400pt}}
\multiput(838.50,154.17)(-1.500,-1.000){2}{\rule{0.361pt}{0.400pt}}
\put(835,152.67){\rule{0.482pt}{0.400pt}}
\multiput(836.00,153.17)(-1.000,-1.000){2}{\rule{0.241pt}{0.400pt}}
\put(832,152.67){\rule{0.723pt}{0.400pt}}
\multiput(833.50,152.17)(-1.500,1.000){2}{\rule{0.361pt}{0.400pt}}
\put(829,153.67){\rule{0.723pt}{0.400pt}}
\multiput(830.50,153.17)(-1.500,1.000){2}{\rule{0.361pt}{0.400pt}}
\put(826,154.67){\rule{0.723pt}{0.400pt}}
\multiput(827.50,154.17)(-1.500,1.000){2}{\rule{0.361pt}{0.400pt}}
\put(824,154.67){\rule{0.482pt}{0.400pt}}
\multiput(825.00,155.17)(-1.000,-1.000){2}{\rule{0.241pt}{0.400pt}}
\put(821,153.17){\rule{0.700pt}{0.400pt}}
\multiput(822.55,154.17)(-1.547,-2.000){2}{\rule{0.350pt}{0.400pt}}
\put(840.0,155.0){\rule[-0.200pt]{0.723pt}{0.400pt}}
\put(816,152.67){\rule{0.482pt}{0.400pt}}
\multiput(817.00,152.17)(-1.000,1.000){2}{\rule{0.241pt}{0.400pt}}
\put(818.0,153.0){\rule[-0.200pt]{0.723pt}{0.400pt}}
\put(805,153.67){\rule{0.723pt}{0.400pt}}
\multiput(806.50,153.17)(-1.500,1.000){2}{\rule{0.361pt}{0.400pt}}
\put(808.0,154.0){\rule[-0.200pt]{1.927pt}{0.400pt}}
\put(800,154.67){\rule{0.482pt}{0.400pt}}
\multiput(801.00,154.17)(-1.000,1.000){2}{\rule{0.241pt}{0.400pt}}
\put(797,154.67){\rule{0.723pt}{0.400pt}}
\multiput(798.50,155.17)(-1.500,-1.000){2}{\rule{0.361pt}{0.400pt}}
\put(802.0,155.0){\rule[-0.200pt]{0.723pt}{0.400pt}}
\put(786,154.67){\rule{0.723pt}{0.400pt}}
\multiput(787.50,154.17)(-1.500,1.000){2}{\rule{0.361pt}{0.400pt}}
\put(784,156.17){\rule{0.482pt}{0.400pt}}
\multiput(785.00,155.17)(-1.000,2.000){2}{\rule{0.241pt}{0.400pt}}
\put(781,157.67){\rule{0.723pt}{0.400pt}}
\multiput(782.50,157.17)(-1.500,1.000){2}{\rule{0.361pt}{0.400pt}}
\put(789.0,155.0){\rule[-0.200pt]{1.927pt}{0.400pt}}
\put(770,158.67){\rule{0.723pt}{0.400pt}}
\multiput(771.50,158.17)(-1.500,1.000){2}{\rule{0.361pt}{0.400pt}}
\put(767,159.67){\rule{0.723pt}{0.400pt}}
\multiput(768.50,159.17)(-1.500,1.000){2}{\rule{0.361pt}{0.400pt}}
\put(765,160.67){\rule{0.482pt}{0.400pt}}
\multiput(766.00,160.17)(-1.000,1.000){2}{\rule{0.241pt}{0.400pt}}
\put(773.0,159.0){\rule[-0.200pt]{1.927pt}{0.400pt}}
\put(757,161.67){\rule{0.482pt}{0.400pt}}
\multiput(758.00,161.17)(-1.000,1.000){2}{\rule{0.241pt}{0.400pt}}
\put(754,162.67){\rule{0.723pt}{0.400pt}}
\multiput(755.50,162.17)(-1.500,1.000){2}{\rule{0.361pt}{0.400pt}}
\put(751,163.67){\rule{0.723pt}{0.400pt}}
\multiput(752.50,163.17)(-1.500,1.000){2}{\rule{0.361pt}{0.400pt}}
\put(749,165.17){\rule{0.482pt}{0.400pt}}
\multiput(750.00,164.17)(-1.000,2.000){2}{\rule{0.241pt}{0.400pt}}
\put(759.0,162.0){\rule[-0.200pt]{1.445pt}{0.400pt}}
\put(743,166.67){\rule{0.723pt}{0.400pt}}
\multiput(744.50,166.17)(-1.500,1.000){2}{\rule{0.361pt}{0.400pt}}
\put(741,168.17){\rule{0.482pt}{0.400pt}}
\multiput(742.00,167.17)(-1.000,2.000){2}{\rule{0.241pt}{0.400pt}}
\put(738,169.67){\rule{0.723pt}{0.400pt}}
\multiput(739.50,169.17)(-1.500,1.000){2}{\rule{0.361pt}{0.400pt}}
\put(746.0,167.0){\rule[-0.200pt]{0.723pt}{0.400pt}}
\put(733,170.67){\rule{0.482pt}{0.400pt}}
\multiput(734.00,170.17)(-1.000,1.000){2}{\rule{0.241pt}{0.400pt}}
\put(730,171.67){\rule{0.723pt}{0.400pt}}
\multiput(731.50,171.17)(-1.500,1.000){2}{\rule{0.361pt}{0.400pt}}
\put(727,173.17){\rule{0.700pt}{0.400pt}}
\multiput(728.55,172.17)(-1.547,2.000){2}{\rule{0.350pt}{0.400pt}}
\put(724,175.17){\rule{0.700pt}{0.400pt}}
\multiput(725.55,174.17)(-1.547,2.000){2}{\rule{0.350pt}{0.400pt}}
\put(722,176.67){\rule{0.482pt}{0.400pt}}
\multiput(723.00,176.17)(-1.000,1.000){2}{\rule{0.241pt}{0.400pt}}
\put(719,176.67){\rule{0.723pt}{0.400pt}}
\multiput(720.50,177.17)(-1.500,-1.000){2}{\rule{0.361pt}{0.400pt}}
\put(716,177.17){\rule{0.700pt}{0.400pt}}
\multiput(717.55,176.17)(-1.547,2.000){2}{\rule{0.350pt}{0.400pt}}
\put(714.17,179){\rule{0.400pt}{0.900pt}}
\multiput(715.17,179.00)(-2.000,2.132){2}{\rule{0.400pt}{0.450pt}}
\put(711,183.17){\rule{0.700pt}{0.400pt}}
\multiput(712.55,182.17)(-1.547,2.000){2}{\rule{0.350pt}{0.400pt}}
\put(708,183.67){\rule{0.723pt}{0.400pt}}
\multiput(709.50,184.17)(-1.500,-1.000){2}{\rule{0.361pt}{0.400pt}}
\put(735.0,171.0){\rule[-0.200pt]{0.723pt}{0.400pt}}
\multiput(704.95,184.00)(-0.447,0.685){3}{\rule{0.108pt}{0.633pt}}
\multiput(705.17,184.00)(-3.000,2.685){2}{\rule{0.400pt}{0.317pt}}
\multiput(701.95,188.00)(-0.447,1.355){3}{\rule{0.108pt}{1.033pt}}
\multiput(702.17,188.00)(-3.000,4.855){2}{\rule{0.400pt}{0.517pt}}
\put(698.17,195){\rule{0.400pt}{1.300pt}}
\multiput(699.17,195.00)(-2.000,3.302){2}{\rule{0.400pt}{0.650pt}}
\multiput(696.95,201.00)(-0.447,0.685){3}{\rule{0.108pt}{0.633pt}}
\multiput(697.17,201.00)(-3.000,2.685){2}{\rule{0.400pt}{0.317pt}}
\multiput(692.92,205.61)(-0.462,0.447){3}{\rule{0.500pt}{0.108pt}}
\multiput(693.96,204.17)(-1.962,3.000){2}{\rule{0.250pt}{0.400pt}}
\put(690.17,208){\rule{0.400pt}{0.700pt}}
\multiput(691.17,208.00)(-2.000,1.547){2}{\rule{0.400pt}{0.350pt}}
\put(687,210.67){\rule{0.723pt}{0.400pt}}
\multiput(688.50,210.17)(-1.500,1.000){2}{\rule{0.361pt}{0.400pt}}
\put(684,210.67){\rule{0.723pt}{0.400pt}}
\multiput(685.50,211.17)(-1.500,-1.000){2}{\rule{0.361pt}{0.400pt}}
\put(682,209.67){\rule{0.482pt}{0.400pt}}
\multiput(683.00,210.17)(-1.000,-1.000){2}{\rule{0.241pt}{0.400pt}}
\put(706.0,184.0){\rule[-0.200pt]{0.482pt}{0.400pt}}
\put(676,209.67){\rule{0.723pt}{0.400pt}}
\multiput(677.50,209.17)(-1.500,1.000){2}{\rule{0.361pt}{0.400pt}}
\put(673,209.17){\rule{0.700pt}{0.400pt}}
\multiput(674.55,210.17)(-1.547,-2.000){2}{\rule{0.350pt}{0.400pt}}
\put(671,207.67){\rule{0.482pt}{0.400pt}}
\multiput(672.00,208.17)(-1.000,-1.000){2}{\rule{0.241pt}{0.400pt}}
\put(668,206.67){\rule{0.723pt}{0.400pt}}
\multiput(669.50,207.17)(-1.500,-1.000){2}{\rule{0.361pt}{0.400pt}}
\put(679.0,210.0){\rule[-0.200pt]{0.723pt}{0.400pt}}
\put(663,206.67){\rule{0.482pt}{0.400pt}}
\multiput(664.00,206.17)(-1.000,1.000){2}{\rule{0.241pt}{0.400pt}}
\put(660,207.67){\rule{0.723pt}{0.400pt}}
\multiput(661.50,207.17)(-1.500,1.000){2}{\rule{0.361pt}{0.400pt}}
\multiput(657.92,209.61)(-0.462,0.447){3}{\rule{0.500pt}{0.108pt}}
\multiput(658.96,208.17)(-1.962,3.000){2}{\rule{0.250pt}{0.400pt}}
\put(655,211.67){\rule{0.482pt}{0.400pt}}
\multiput(656.00,211.17)(-1.000,1.000){2}{\rule{0.241pt}{0.400pt}}
\put(652,212.67){\rule{0.723pt}{0.400pt}}
\multiput(653.50,212.17)(-1.500,1.000){2}{\rule{0.361pt}{0.400pt}}
\put(649,214.17){\rule{0.700pt}{0.400pt}}
\multiput(650.55,213.17)(-1.547,2.000){2}{\rule{0.350pt}{0.400pt}}
\put(647,215.67){\rule{0.482pt}{0.400pt}}
\multiput(648.00,215.17)(-1.000,1.000){2}{\rule{0.241pt}{0.400pt}}
\put(644,215.67){\rule{0.723pt}{0.400pt}}
\multiput(645.50,216.17)(-1.500,-1.000){2}{\rule{0.361pt}{0.400pt}}
\multiput(641.92,216.61)(-0.462,0.447){3}{\rule{0.500pt}{0.108pt}}
\multiput(642.96,215.17)(-1.962,3.000){2}{\rule{0.250pt}{0.400pt}}
\put(639.17,219){\rule{0.400pt}{0.900pt}}
\multiput(640.17,219.00)(-2.000,2.132){2}{\rule{0.400pt}{0.450pt}}
\put(636,221.67){\rule{0.723pt}{0.400pt}}
\multiput(637.50,222.17)(-1.500,-1.000){2}{\rule{0.361pt}{0.400pt}}
\put(633,221.67){\rule{0.723pt}{0.400pt}}
\multiput(634.50,221.17)(-1.500,1.000){2}{\rule{0.361pt}{0.400pt}}
\put(631.17,223){\rule{0.400pt}{1.100pt}}
\multiput(632.17,223.00)(-2.000,2.717){2}{\rule{0.400pt}{0.550pt}}
\put(628,227.67){\rule{0.723pt}{0.400pt}}
\multiput(629.50,227.17)(-1.500,1.000){2}{\rule{0.361pt}{0.400pt}}
\put(625,227.17){\rule{0.700pt}{0.400pt}}
\multiput(626.55,228.17)(-1.547,-2.000){2}{\rule{0.350pt}{0.400pt}}
\put(665.0,207.0){\rule[-0.200pt]{0.723pt}{0.400pt}}
\put(620.17,227){\rule{0.400pt}{0.900pt}}
\multiput(621.17,227.00)(-2.000,2.132){2}{\rule{0.400pt}{0.450pt}}
\multiput(617.92,231.61)(-0.462,0.447){3}{\rule{0.500pt}{0.108pt}}
\multiput(618.96,230.17)(-1.962,3.000){2}{\rule{0.250pt}{0.400pt}}
\put(614,233.67){\rule{0.723pt}{0.400pt}}
\multiput(615.50,233.17)(-1.500,1.000){2}{\rule{0.361pt}{0.400pt}}
\put(622.0,227.0){\rule[-0.200pt]{0.723pt}{0.400pt}}
\put(609,235.17){\rule{0.700pt}{0.400pt}}
\multiput(610.55,234.17)(-1.547,2.000){2}{\rule{0.350pt}{0.400pt}}
\multiput(607.95,237.00)(-0.447,0.685){3}{\rule{0.108pt}{0.633pt}}
\multiput(608.17,237.00)(-3.000,2.685){2}{\rule{0.400pt}{0.317pt}}
\put(604.17,241){\rule{0.400pt}{1.100pt}}
\multiput(605.17,241.00)(-2.000,2.717){2}{\rule{0.400pt}{0.550pt}}
\multiput(602.95,246.00)(-0.447,0.909){3}{\rule{0.108pt}{0.767pt}}
\multiput(603.17,246.00)(-3.000,3.409){2}{\rule{0.400pt}{0.383pt}}
\put(598,250.67){\rule{0.723pt}{0.400pt}}
\multiput(599.50,250.17)(-1.500,1.000){2}{\rule{0.361pt}{0.400pt}}
\put(596,251.67){\rule{0.482pt}{0.400pt}}
\multiput(597.00,251.17)(-1.000,1.000){2}{\rule{0.241pt}{0.400pt}}
\multiput(593.92,253.61)(-0.462,0.447){3}{\rule{0.500pt}{0.108pt}}
\multiput(594.96,252.17)(-1.962,3.000){2}{\rule{0.250pt}{0.400pt}}
\multiput(591.95,256.00)(-0.447,1.355){3}{\rule{0.108pt}{1.033pt}}
\multiput(592.17,256.00)(-3.000,4.855){2}{\rule{0.400pt}{0.517pt}}
\put(588.17,263){\rule{0.400pt}{1.900pt}}
\multiput(589.17,263.00)(-2.000,5.056){2}{\rule{0.400pt}{0.950pt}}
\multiput(586.95,272.00)(-0.447,0.909){3}{\rule{0.108pt}{0.767pt}}
\multiput(587.17,272.00)(-3.000,3.409){2}{\rule{0.400pt}{0.383pt}}
\put(582,277.17){\rule{0.700pt}{0.400pt}}
\multiput(583.55,276.17)(-1.547,2.000){2}{\rule{0.350pt}{0.400pt}}
\multiput(580.95,279.00)(-0.447,0.909){3}{\rule{0.108pt}{0.767pt}}
\multiput(581.17,279.00)(-3.000,3.409){2}{\rule{0.400pt}{0.383pt}}
\put(612.0,235.0){\rule[-0.200pt]{0.482pt}{0.400pt}}
\multiput(572.95,284.00)(-0.447,1.355){3}{\rule{0.108pt}{1.033pt}}
\multiput(573.17,284.00)(-3.000,4.855){2}{\rule{0.400pt}{0.517pt}}
\put(569.17,291){\rule{0.400pt}{1.100pt}}
\multiput(570.17,291.00)(-2.000,2.717){2}{\rule{0.400pt}{0.550pt}}
\multiput(567.95,296.00)(-0.447,1.579){3}{\rule{0.108pt}{1.167pt}}
\multiput(568.17,296.00)(-3.000,5.579){2}{\rule{0.400pt}{0.583pt}}
\multiput(564.95,304.00)(-0.447,2.025){3}{\rule{0.108pt}{1.433pt}}
\multiput(565.17,304.00)(-3.000,7.025){2}{\rule{0.400pt}{0.717pt}}
\put(561.17,314){\rule{0.400pt}{1.300pt}}
\multiput(562.17,314.00)(-2.000,3.302){2}{\rule{0.400pt}{0.650pt}}
\multiput(559.95,320.00)(-0.447,0.909){3}{\rule{0.108pt}{0.767pt}}
\multiput(560.17,320.00)(-3.000,3.409){2}{\rule{0.400pt}{0.383pt}}
\multiput(556.95,325.00)(-0.447,2.472){3}{\rule{0.108pt}{1.700pt}}
\multiput(557.17,325.00)(-3.000,8.472){2}{\rule{0.400pt}{0.850pt}}
\put(553.17,337){\rule{0.400pt}{1.900pt}}
\multiput(554.17,337.00)(-2.000,5.056){2}{\rule{0.400pt}{0.950pt}}
\multiput(551.95,346.00)(-0.447,1.355){3}{\rule{0.108pt}{1.033pt}}
\multiput(552.17,346.00)(-3.000,4.855){2}{\rule{0.400pt}{0.517pt}}
\multiput(548.95,353.00)(-0.447,3.141){3}{\rule{0.108pt}{2.100pt}}
\multiput(549.17,353.00)(-3.000,10.641){2}{\rule{0.400pt}{1.050pt}}
\put(545.17,368){\rule{0.400pt}{3.900pt}}
\multiput(546.17,368.00)(-2.000,10.905){2}{\rule{0.400pt}{1.950pt}}
\multiput(543.95,387.00)(-0.447,2.472){3}{\rule{0.108pt}{1.700pt}}
\multiput(544.17,387.00)(-3.000,8.472){2}{\rule{0.400pt}{0.850pt}}
\multiput(540.95,399.00)(-0.447,2.695){3}{\rule{0.108pt}{1.833pt}}
\multiput(541.17,399.00)(-3.000,9.195){2}{\rule{0.400pt}{0.917pt}}
\put(537.17,412){\rule{0.400pt}{5.900pt}}
\multiput(538.17,412.00)(-2.000,16.754){2}{\rule{0.400pt}{2.950pt}}
\multiput(535.95,441.00)(-0.447,4.704){3}{\rule{0.108pt}{3.033pt}}
\multiput(536.17,441.00)(-3.000,15.704){2}{\rule{0.400pt}{1.517pt}}
\multiput(531.92,463.61)(-0.462,0.447){3}{\rule{0.500pt}{0.108pt}}
\multiput(532.96,462.17)(-1.962,3.000){2}{\rule{0.250pt}{0.400pt}}
\multiput(529.95,466.00)(-0.447,2.918){3}{\rule{0.108pt}{1.967pt}}
\multiput(530.17,466.00)(-3.000,9.918){2}{\rule{0.400pt}{0.983pt}}
\put(526.17,480){\rule{0.400pt}{5.500pt}}
\multiput(527.17,480.00)(-2.000,15.584){2}{\rule{0.400pt}{2.750pt}}
\multiput(524.95,507.00)(-0.447,1.802){3}{\rule{0.108pt}{1.300pt}}
\multiput(525.17,507.00)(-3.000,6.302){2}{\rule{0.400pt}{0.650pt}}
\multiput(521.95,513.37)(-0.447,-0.685){3}{\rule{0.108pt}{0.633pt}}
\multiput(522.17,514.69)(-3.000,-2.685){2}{\rule{0.400pt}{0.317pt}}
\put(518.17,512){\rule{0.400pt}{2.500pt}}
\multiput(519.17,512.00)(-2.000,6.811){2}{\rule{0.400pt}{1.250pt}}
\multiput(516.95,524.00)(-0.447,4.034){3}{\rule{0.108pt}{2.633pt}}
\multiput(517.17,524.00)(-3.000,13.534){2}{\rule{0.400pt}{1.317pt}}
\put(512,543.17){\rule{0.700pt}{0.400pt}}
\multiput(513.55,542.17)(-1.547,2.000){2}{\rule{0.350pt}{0.400pt}}
\put(510.17,542){\rule{0.400pt}{0.700pt}}
\multiput(511.17,543.55)(-2.000,-1.547){2}{\rule{0.400pt}{0.350pt}}
\multiput(508.95,542.00)(-0.447,3.141){3}{\rule{0.108pt}{2.100pt}}
\multiput(509.17,542.00)(-3.000,10.641){2}{\rule{0.400pt}{1.050pt}}
\multiput(505.95,557.00)(-0.447,3.811){3}{\rule{0.108pt}{2.500pt}}
\multiput(506.17,557.00)(-3.000,12.811){2}{\rule{0.400pt}{1.250pt}}
\put(502.17,575){\rule{0.400pt}{0.900pt}}
\multiput(503.17,575.00)(-2.000,2.132){2}{\rule{0.400pt}{0.450pt}}
\multiput(499.92,579.61)(-0.462,0.447){3}{\rule{0.500pt}{0.108pt}}
\multiput(500.96,578.17)(-1.962,3.000){2}{\rule{0.250pt}{0.400pt}}
\multiput(497.95,582.00)(-0.447,3.141){3}{\rule{0.108pt}{2.100pt}}
\multiput(498.17,582.00)(-3.000,10.641){2}{\rule{0.400pt}{1.050pt}}
\put(494.17,597){\rule{0.400pt}{2.700pt}}
\multiput(495.17,597.00)(-2.000,7.396){2}{\rule{0.400pt}{1.350pt}}
\put(491,609.67){\rule{0.723pt}{0.400pt}}
\multiput(492.50,609.17)(-1.500,1.000){2}{\rule{0.361pt}{0.400pt}}
\put(488,611.17){\rule{0.700pt}{0.400pt}}
\multiput(489.55,610.17)(-1.547,2.000){2}{\rule{0.350pt}{0.400pt}}
\put(486.17,613){\rule{0.400pt}{1.100pt}}
\multiput(487.17,613.00)(-2.000,2.717){2}{\rule{0.400pt}{0.550pt}}
\multiput(484.95,618.00)(-0.447,0.685){3}{\rule{0.108pt}{0.633pt}}
\multiput(485.17,618.00)(-3.000,2.685){2}{\rule{0.400pt}{0.317pt}}
\multiput(481.95,619.37)(-0.447,-0.685){3}{\rule{0.108pt}{0.633pt}}
\multiput(482.17,620.69)(-3.000,-2.685){2}{\rule{0.400pt}{0.317pt}}
\multiput(478.95,615.37)(-0.447,-0.685){3}{\rule{0.108pt}{0.633pt}}
\multiput(479.17,616.69)(-3.000,-2.685){2}{\rule{0.400pt}{0.317pt}}
\put(574.0,284.0){\rule[-0.200pt]{1.204pt}{0.400pt}}
\put(472,612.67){\rule{0.723pt}{0.400pt}}
\multiput(473.50,613.17)(-1.500,-1.000){2}{\rule{0.361pt}{0.400pt}}
\multiput(470.95,609.82)(-0.447,-0.909){3}{\rule{0.108pt}{0.767pt}}
\multiput(471.17,611.41)(-3.000,-3.409){2}{\rule{0.400pt}{0.383pt}}
\put(467.17,604){\rule{0.400pt}{0.900pt}}
\multiput(468.17,606.13)(-2.000,-2.132){2}{\rule{0.400pt}{0.450pt}}
\put(464,603.67){\rule{0.723pt}{0.400pt}}
\multiput(465.50,603.17)(-1.500,1.000){2}{\rule{0.361pt}{0.400pt}}
\put(461,603.67){\rule{0.723pt}{0.400pt}}
\multiput(462.50,604.17)(-1.500,-1.000){2}{\rule{0.361pt}{0.400pt}}
\put(459.17,597){\rule{0.400pt}{1.500pt}}
\multiput(460.17,600.89)(-2.000,-3.887){2}{\rule{0.400pt}{0.750pt}}
\multiput(457.95,593.82)(-0.447,-0.909){3}{\rule{0.108pt}{0.767pt}}
\multiput(458.17,595.41)(-3.000,-3.409){2}{\rule{0.400pt}{0.383pt}}
\put(453,590.17){\rule{0.700pt}{0.400pt}}
\multiput(454.55,591.17)(-1.547,-2.000){2}{\rule{0.350pt}{0.400pt}}
\put(451.17,577){\rule{0.400pt}{2.700pt}}
\multiput(452.17,584.40)(-2.000,-7.396){2}{\rule{0.400pt}{1.350pt}}
\multiput(449.95,567.73)(-0.447,-3.365){3}{\rule{0.108pt}{2.233pt}}
\multiput(450.17,572.36)(-3.000,-11.365){2}{\rule{0.400pt}{1.117pt}}
\multiput(446.95,553.39)(-0.447,-2.695){3}{\rule{0.108pt}{1.833pt}}
\multiput(447.17,557.19)(-3.000,-9.195){2}{\rule{0.400pt}{0.917pt}}
\put(443.17,531){\rule{0.400pt}{3.500pt}}
\multiput(444.17,540.74)(-2.000,-9.736){2}{\rule{0.400pt}{1.750pt}}
\multiput(441.95,515.09)(-0.447,-6.044){3}{\rule{0.108pt}{3.833pt}}
\multiput(442.17,523.04)(-3.000,-20.044){2}{\rule{0.400pt}{1.917pt}}
\multiput(438.95,485.98)(-0.447,-6.490){3}{\rule{0.108pt}{4.100pt}}
\multiput(439.17,494.49)(-3.000,-21.490){2}{\rule{0.400pt}{2.050pt}}
\put(435.17,447){\rule{0.400pt}{5.300pt}}
\multiput(436.17,462.00)(-2.000,-15.000){2}{\rule{0.400pt}{2.650pt}}
\multiput(433.95,430.53)(-0.447,-6.267){3}{\rule{0.108pt}{3.967pt}}
\multiput(434.17,438.77)(-3.000,-20.767){2}{\rule{0.400pt}{1.983pt}}
\multiput(430.95,398.21)(-0.447,-7.607){3}{\rule{0.108pt}{4.767pt}}
\multiput(431.17,408.11)(-3.000,-25.107){2}{\rule{0.400pt}{2.383pt}}
\multiput(427.95,364.32)(-0.447,-7.160){3}{\rule{0.108pt}{4.500pt}}
\multiput(428.17,373.66)(-3.000,-23.660){2}{\rule{0.400pt}{2.250pt}}
\put(424.17,324){\rule{0.400pt}{5.300pt}}
\multiput(425.17,339.00)(-2.000,-15.000){2}{\rule{0.400pt}{2.650pt}}
\multiput(422.95,309.75)(-0.447,-5.374){3}{\rule{0.108pt}{3.433pt}}
\multiput(423.17,316.87)(-3.000,-17.874){2}{\rule{0.400pt}{1.717pt}}
\multiput(419.95,282.53)(-0.447,-6.267){3}{\rule{0.108pt}{3.967pt}}
\multiput(420.17,290.77)(-3.000,-20.767){2}{\rule{0.400pt}{1.983pt}}
\put(416.17,247){\rule{0.400pt}{4.700pt}}
\multiput(417.17,260.24)(-2.000,-13.245){2}{\rule{0.400pt}{2.350pt}}
\multiput(414.95,239.94)(-0.447,-2.472){3}{\rule{0.108pt}{1.700pt}}
\multiput(415.17,243.47)(-3.000,-8.472){2}{\rule{0.400pt}{0.850pt}}
\multiput(411.95,229.05)(-0.447,-2.025){3}{\rule{0.108pt}{1.433pt}}
\multiput(412.17,232.03)(-3.000,-7.025){2}{\rule{0.400pt}{0.717pt}}
\put(408.17,212){\rule{0.400pt}{2.700pt}}
\multiput(409.17,219.40)(-2.000,-7.396){2}{\rule{0.400pt}{1.350pt}}
\multiput(406.95,206.05)(-0.447,-2.025){3}{\rule{0.108pt}{1.433pt}}
\multiput(407.17,209.03)(-3.000,-7.025){2}{\rule{0.400pt}{0.717pt}}
\put(475.0,614.0){\rule[-0.200pt]{0.482pt}{0.400pt}}
\put(400,202.17){\rule{0.482pt}{0.400pt}}
\multiput(401.00,201.17)(-1.000,2.000){2}{\rule{0.241pt}{0.400pt}}
\multiput(398.95,200.82)(-0.447,-0.909){3}{\rule{0.108pt}{0.767pt}}
\multiput(399.17,202.41)(-3.000,-3.409){2}{\rule{0.400pt}{0.383pt}}
\put(394,197.17){\rule{0.700pt}{0.400pt}}
\multiput(395.55,198.17)(-1.547,-2.000){2}{\rule{0.350pt}{0.400pt}}
\put(392.17,197){\rule{0.400pt}{1.300pt}}
\multiput(393.17,197.00)(-2.000,3.302){2}{\rule{0.400pt}{0.650pt}}
\multiput(389.92,203.61)(-0.462,0.447){3}{\rule{0.500pt}{0.108pt}}
\multiput(390.96,202.17)(-1.962,3.000){2}{\rule{0.250pt}{0.400pt}}
\multiput(387.95,202.82)(-0.447,-0.909){3}{\rule{0.108pt}{0.767pt}}
\multiput(388.17,204.41)(-3.000,-3.409){2}{\rule{0.400pt}{0.383pt}}
\put(384,199.17){\rule{0.482pt}{0.400pt}}
\multiput(385.00,200.17)(-1.000,-2.000){2}{\rule{0.241pt}{0.400pt}}
\put(381,199.17){\rule{0.700pt}{0.400pt}}
\multiput(382.55,198.17)(-1.547,2.000){2}{\rule{0.350pt}{0.400pt}}
\put(378,199.67){\rule{0.723pt}{0.400pt}}
\multiput(379.50,200.17)(-1.500,-1.000){2}{\rule{0.361pt}{0.400pt}}
\multiput(376.95,196.26)(-0.447,-1.132){3}{\rule{0.108pt}{0.900pt}}
\multiput(377.17,198.13)(-3.000,-4.132){2}{\rule{0.400pt}{0.450pt}}
\put(373.17,191){\rule{0.400pt}{0.700pt}}
\multiput(374.17,192.55)(-2.000,-1.547){2}{\rule{0.400pt}{0.350pt}}
\put(370,189.67){\rule{0.723pt}{0.400pt}}
\multiput(371.50,190.17)(-1.500,-1.000){2}{\rule{0.361pt}{0.400pt}}
\multiput(368.95,186.26)(-0.447,-1.132){3}{\rule{0.108pt}{0.900pt}}
\multiput(369.17,188.13)(-3.000,-4.132){2}{\rule{0.400pt}{0.450pt}}
\put(365.17,177){\rule{0.400pt}{1.500pt}}
\multiput(366.17,180.89)(-2.000,-3.887){2}{\rule{0.400pt}{0.750pt}}
\put(362,175.67){\rule{0.723pt}{0.400pt}}
\multiput(363.50,176.17)(-1.500,-1.000){2}{\rule{0.361pt}{0.400pt}}
\put(359,174.17){\rule{0.700pt}{0.400pt}}
\multiput(360.55,175.17)(-1.547,-2.000){2}{\rule{0.350pt}{0.400pt}}
\put(357.17,169){\rule{0.400pt}{1.100pt}}
\multiput(358.17,171.72)(-2.000,-2.717){2}{\rule{0.400pt}{0.550pt}}
\multiput(355.95,165.82)(-0.447,-0.909){3}{\rule{0.108pt}{0.767pt}}
\multiput(356.17,167.41)(-3.000,-3.409){2}{\rule{0.400pt}{0.383pt}}
\put(351,163.67){\rule{0.723pt}{0.400pt}}
\multiput(352.50,163.17)(-1.500,1.000){2}{\rule{0.361pt}{0.400pt}}
\put(349.17,160){\rule{0.400pt}{1.100pt}}
\multiput(350.17,162.72)(-2.000,-2.717){2}{\rule{0.400pt}{0.550pt}}
\multiput(347.95,157.37)(-0.447,-0.685){3}{\rule{0.108pt}{0.633pt}}
\multiput(348.17,158.69)(-3.000,-2.685){2}{\rule{0.400pt}{0.317pt}}
\put(343,155.67){\rule{0.723pt}{0.400pt}}
\multiput(344.50,155.17)(-1.500,1.000){2}{\rule{0.361pt}{0.400pt}}
\put(402.0,202.0){\rule[-0.200pt]{0.723pt}{0.400pt}}
\multiput(339.95,153.82)(-0.447,-0.909){3}{\rule{0.108pt}{0.767pt}}
\multiput(340.17,155.41)(-3.000,-3.409){2}{\rule{0.400pt}{0.383pt}}
\put(335,150.17){\rule{0.700pt}{0.400pt}}
\multiput(336.55,151.17)(-1.547,-2.000){2}{\rule{0.350pt}{0.400pt}}
\put(333.17,150){\rule{0.400pt}{0.700pt}}
\multiput(334.17,150.00)(-2.000,1.547){2}{\rule{0.400pt}{0.350pt}}
\multiput(330.92,151.95)(-0.462,-0.447){3}{\rule{0.500pt}{0.108pt}}
\multiput(331.96,152.17)(-1.962,-3.000){2}{\rule{0.250pt}{0.400pt}}
\multiput(327.92,148.95)(-0.462,-0.447){3}{\rule{0.500pt}{0.108pt}}
\multiput(328.96,149.17)(-1.962,-3.000){2}{\rule{0.250pt}{0.400pt}}
\put(324,147.17){\rule{0.700pt}{0.400pt}}
\multiput(325.55,146.17)(-1.547,2.000){2}{\rule{0.350pt}{0.400pt}}
\put(322,147.17){\rule{0.482pt}{0.400pt}}
\multiput(323.00,148.17)(-1.000,-2.000){2}{\rule{0.241pt}{0.400pt}}
\multiput(319.92,145.95)(-0.462,-0.447){3}{\rule{0.500pt}{0.108pt}}
\multiput(320.96,146.17)(-1.962,-3.000){2}{\rule{0.250pt}{0.400pt}}
\put(316,142.67){\rule{0.723pt}{0.400pt}}
\multiput(317.50,143.17)(-1.500,-1.000){2}{\rule{0.361pt}{0.400pt}}
\put(341.0,157.0){\rule[-0.200pt]{0.482pt}{0.400pt}}
\put(311,141.67){\rule{0.723pt}{0.400pt}}
\multiput(312.50,142.17)(-1.500,-1.000){2}{\rule{0.361pt}{0.400pt}}
\put(314.0,143.0){\rule[-0.200pt]{0.482pt}{0.400pt}}
\put(306,140.17){\rule{0.482pt}{0.400pt}}
\multiput(307.00,141.17)(-1.000,-2.000){2}{\rule{0.241pt}{0.400pt}}
\put(308.0,142.0){\rule[-0.200pt]{0.723pt}{0.400pt}}
\put(300,139.67){\rule{0.723pt}{0.400pt}}
\multiput(301.50,139.17)(-1.500,1.000){2}{\rule{0.361pt}{0.400pt}}
\put(298.17,137){\rule{0.400pt}{0.900pt}}
\multiput(299.17,139.13)(-2.000,-2.132){2}{\rule{0.400pt}{0.450pt}}
\put(295,137.17){\rule{0.700pt}{0.400pt}}
\multiput(296.55,136.17)(-1.547,2.000){2}{\rule{0.350pt}{0.400pt}}
\put(303.0,140.0){\rule[-0.200pt]{0.723pt}{0.400pt}}
\put(290.17,135){\rule{0.400pt}{0.900pt}}
\multiput(291.17,137.13)(-2.000,-2.132){2}{\rule{0.400pt}{0.450pt}}
\multiput(288.95,135.00)(-0.447,0.909){3}{\rule{0.108pt}{0.767pt}}
\multiput(289.17,135.00)(-3.000,3.409){2}{\rule{0.400pt}{0.383pt}}
\multiput(285.95,135.71)(-0.447,-1.355){3}{\rule{0.108pt}{1.033pt}}
\multiput(286.17,137.86)(-3.000,-4.855){2}{\rule{0.400pt}{0.517pt}}
\put(282.17,133){\rule{0.400pt}{0.900pt}}
\multiput(283.17,133.00)(-2.000,2.132){2}{\rule{0.400pt}{0.450pt}}
\multiput(280.95,134.37)(-0.447,-0.685){3}{\rule{0.108pt}{0.633pt}}
\multiput(281.17,135.69)(-3.000,-2.685){2}{\rule{0.400pt}{0.317pt}}
\multiput(276.92,133.61)(-0.462,0.447){3}{\rule{0.500pt}{0.108pt}}
\multiput(277.96,132.17)(-1.962,3.000){2}{\rule{0.250pt}{0.400pt}}
\multiput(273.92,134.95)(-0.462,-0.447){3}{\rule{0.500pt}{0.108pt}}
\multiput(274.96,135.17)(-1.962,-3.000){2}{\rule{0.250pt}{0.400pt}}
\put(271,132.67){\rule{0.482pt}{0.400pt}}
\multiput(272.00,132.17)(-1.000,1.000){2}{\rule{0.241pt}{0.400pt}}
\put(268,132.17){\rule{0.700pt}{0.400pt}}
\multiput(269.55,133.17)(-1.547,-2.000){2}{\rule{0.350pt}{0.400pt}}
\put(265,131.67){\rule{0.723pt}{0.400pt}}
\multiput(266.50,131.17)(-1.500,1.000){2}{\rule{0.361pt}{0.400pt}}
\put(292.0,139.0){\rule[-0.200pt]{0.723pt}{0.400pt}}
\put(257,132.67){\rule{0.723pt}{0.400pt}}
\multiput(258.50,132.17)(-1.500,1.000){2}{\rule{0.361pt}{0.400pt}}
\put(255,133.67){\rule{0.482pt}{0.400pt}}
\multiput(256.00,133.17)(-1.000,1.000){2}{\rule{0.241pt}{0.400pt}}
\put(260.0,133.0){\rule[-0.200pt]{1.204pt}{0.400pt}}
\put(249,134.67){\rule{0.723pt}{0.400pt}}
\multiput(250.50,134.17)(-1.500,1.000){2}{\rule{0.361pt}{0.400pt}}
\put(252.0,135.0){\rule[-0.200pt]{0.723pt}{0.400pt}}
\put(244,134.67){\rule{0.723pt}{0.400pt}}
\multiput(245.50,135.17)(-1.500,-1.000){2}{\rule{0.361pt}{0.400pt}}
\put(247.0,136.0){\rule[-0.200pt]{0.482pt}{0.400pt}}
\put(239,133.67){\rule{0.482pt}{0.400pt}}
\multiput(240.00,134.17)(-1.000,-1.000){2}{\rule{0.241pt}{0.400pt}}
\put(236,132.67){\rule{0.723pt}{0.400pt}}
\multiput(237.50,133.17)(-1.500,-1.000){2}{\rule{0.361pt}{0.400pt}}
\put(233,131.67){\rule{0.723pt}{0.400pt}}
\multiput(234.50,132.17)(-1.500,-1.000){2}{\rule{0.361pt}{0.400pt}}
\put(241.0,135.0){\rule[-0.200pt]{0.723pt}{0.400pt}}
\put(225,132.17){\rule{0.700pt}{0.400pt}}
\multiput(226.55,131.17)(-1.547,2.000){2}{\rule{0.350pt}{0.400pt}}
\multiput(222.92,134.61)(-0.462,0.447){3}{\rule{0.500pt}{0.108pt}}
\multiput(223.96,133.17)(-1.962,3.000){2}{\rule{0.250pt}{0.400pt}}
\put(220,137.17){\rule{0.482pt}{0.400pt}}
\multiput(221.00,136.17)(-1.000,2.000){2}{\rule{0.241pt}{0.400pt}}
\put(228.0,132.0){\rule[-0.200pt]{1.204pt}{0.400pt}}
\put(217.0,139.0){\rule[-0.200pt]{0.723pt}{0.400pt}}
\put(879,672){\makebox(0,0)[r]{ { $P_z^c(m_l)$ } }}
\multiput(899,672)(20.756,0.000){5}{\usebox{\plotpoint}}
\put(999,672){\usebox{\plotpoint}}
\put(1044,190){\usebox{\plotpoint}}
\multiput(1044,190)(-2.935,20.547){2}{\usebox{\plotpoint}}
\multiput(1039,195)(-2.473,20.608){2}{\usebox{\plotpoint}}
\multiput(1036,220)(-1.036,20.730){2}{\usebox{\plotpoint}}
\put(1032.71,281.92){\usebox{\plotpoint}}
\put(1029.52,302.42){\usebox{\plotpoint}}
\put(1023.58,297.74){\usebox{\plotpoint}}
\put(1017.71,277.85){\usebox{\plotpoint}}
\put(1011.66,258.08){\usebox{\plotpoint}}
\put(1004.98,238.45){\usebox{\plotpoint}}
\put(993.01,222.02){\usebox{\plotpoint}}
\put(980.13,207.00){\usebox{\plotpoint}}
\put(965.75,197.83){\usebox{\plotpoint}}
\put(951.12,190.24){\usebox{\plotpoint}}
\put(934.71,186.24){\usebox{\plotpoint}}
\put(918.89,180.45){\usebox{\plotpoint}}
\put(900.35,175.90){\usebox{\plotpoint}}
\put(882.48,172.83){\usebox{\plotpoint}}
\put(865.55,173.48){\usebox{\plotpoint}}
\put(849.13,173.24){\usebox{\plotpoint}}
\put(829.77,173.00){\usebox{\plotpoint}}
\put(811.83,172.83){\usebox{\plotpoint}}
\put(795.64,173.90){\usebox{\plotpoint}}
\put(779.15,175.47){\usebox{\plotpoint}}
\put(760.87,177.51){\usebox{\plotpoint}}
\put(742.52,179.97){\usebox{\plotpoint}}
\put(725.72,185.00){\usebox{\plotpoint}}
\put(708.99,192.34){\usebox{\plotpoint}}
\put(693.91,203.73){\usebox{\plotpoint}}
\put(681.36,207.43){\usebox{\plotpoint}}
\put(664.34,204.66){\usebox{\plotpoint}}
\put(646.82,212.88){\usebox{\plotpoint}}
\put(629.20,216.00){\usebox{\plotpoint}}
\put(611.36,224.06){\usebox{\plotpoint}}
\put(598.39,228.00){\usebox{\plotpoint}}
\put(584.71,240.58){\usebox{\plotpoint}}
\put(572.20,253.80){\usebox{\plotpoint}}
\put(560.81,268.70){\usebox{\plotpoint}}
\put(554.02,287.40){\usebox{\plotpoint}}
\put(546.93,306.63){\usebox{\plotpoint}}
\put(544.40,327.23){\usebox{\plotpoint}}
\put(541.07,347.70){\usebox{\plotpoint}}
\put(537.57,368.12){\usebox{\plotpoint}}
\put(534.65,388.66){\usebox{\plotpoint}}
\put(531.57,409.19){\usebox{\plotpoint}}
\put(528.81,429.76){\usebox{\plotpoint}}
\put(526.80,450.41){\usebox{\plotpoint}}
\put(524.56,471.04){\usebox{\plotpoint}}
\multiput(523,483)(-1.592,20.694){2}{\usebox{\plotpoint}}
\put(518.84,533.03){\usebox{\plotpoint}}
\put(516.11,553.60){\usebox{\plotpoint}}
\put(512.02,573.93){\usebox{\plotpoint}}
\put(508.67,594.21){\usebox{\plotpoint}}
\put(503.92,614.23){\usebox{\plotpoint}}
\put(497.29,633.84){\usebox{\plotpoint}}
\put(493.42,654.16){\usebox{\plotpoint}}
\put(486.30,663.36){\usebox{\plotpoint}}
\put(479.30,665.14){\usebox{\plotpoint}}
\put(473.39,645.74){\usebox{\plotpoint}}
\put(469.13,625.47){\usebox{\plotpoint}}
\put(464.15,606.96){\usebox{\plotpoint}}
\multiput(461,588)(-1.724,-20.684){2}{\usebox{\plotpoint}}
\put(455.50,545.36){\usebox{\plotpoint}}
\put(452.43,524.86){\usebox{\plotpoint}}
\put(450.25,504.26){\usebox{\plotpoint}}
\put(446.19,483.94){\usebox{\plotpoint}}
\put(443.79,463.33){\usebox{\plotpoint}}
\put(441.07,442.77){\usebox{\plotpoint}}
\put(436.43,422.58){\usebox{\plotpoint}}
\put(433.21,402.08){\usebox{\plotpoint}}
\put(429.53,381.66){\usebox{\plotpoint}}
\put(425.64,361.28){\usebox{\plotpoint}}
\put(421.59,340.98){\usebox{\plotpoint}}
\put(417.12,320.73){\usebox{\plotpoint}}
\put(412.88,300.42){\usebox{\plotpoint}}
\put(408.09,280.26){\usebox{\plotpoint}}
\put(402.80,260.19){\usebox{\plotpoint}}
\put(398.48,239.90){\usebox{\plotpoint}}
\put(391.55,220.65){\usebox{\plotpoint}}
\put(384.77,201.07){\usebox{\plotpoint}}
\put(374.69,183.69){\usebox{\plotpoint}}
\put(364.16,167.32){\usebox{\plotpoint}}
\put(350.52,152.52){\usebox{\plotpoint}}
\put(334.72,142.72){\usebox{\plotpoint}}
\put(317.97,137.66){\usebox{\plotpoint}}
\put(298.50,134.25){\usebox{\plotpoint}}
\put(278.53,134.00){\usebox{\plotpoint}}
\put(260.13,134.04){\usebox{\plotpoint}}
\put(240.12,135.44){\usebox{\plotpoint}}
\put(220.45,134.22){\usebox{\plotpoint}}
\put(217,134){\usebox{\plotpoint}}
\end{picture}

%% file: probdist15.tex
\setlength{\unitlength}{0.240900pt}
\ifx\plotpoint\undefined\newsavebox{\plotpoint}\fi
\begin{picture}(1081,900)(0,0)
\font\gnuplot=cmr10 at 10pt
\gnuplot
\sbox{\plotpoint}{\rule[-0.200pt]{0.400pt}{0.400pt}}%
\put(201.0,123.0){\rule[-0.200pt]{4.818pt}{0.400pt}}
\put(181,123){\makebox(0,0)[r]{0}}
\put(1040.0,123.0){\rule[-0.200pt]{4.818pt}{0.400pt}}
\put(201.0,307.0){\rule[-0.200pt]{4.818pt}{0.400pt}}
\put(181,307){\makebox(0,0)[r]{0.005}}
\put(1040.0,307.0){\rule[-0.200pt]{4.818pt}{0.400pt}}
\put(201.0,492.0){\rule[-0.200pt]{4.818pt}{0.400pt}}
\put(181,492){\makebox(0,0)[r]{0.01}}
\put(1040.0,492.0){\rule[-0.200pt]{4.818pt}{0.400pt}}
\put(201.0,676.0){\rule[-0.200pt]{4.818pt}{0.400pt}}
\put(181,676){\makebox(0,0)[r]{0.015}}
\put(1040.0,676.0){\rule[-0.200pt]{4.818pt}{0.400pt}}
\put(201.0,860.0){\rule[-0.200pt]{4.818pt}{0.400pt}}
\put(181,860){\makebox(0,0)[r]{0.02}}
\put(1040.0,860.0){\rule[-0.200pt]{4.818pt}{0.400pt}}
\put(228.0,123.0){\rule[-0.200pt]{0.400pt}{4.818pt}}
\put(228,82){\makebox(0,0){-150}}
\put(228.0,840.0){\rule[-0.200pt]{0.400pt}{4.818pt}}
\put(362.0,123.0){\rule[-0.200pt]{0.400pt}{4.818pt}}
\put(362,82){\makebox(0,0){-100}}
\put(362.0,840.0){\rule[-0.200pt]{0.400pt}{4.818pt}}
\put(496.0,123.0){\rule[-0.200pt]{0.400pt}{4.818pt}}
\put(496,82){\makebox(0,0){-50}}
\put(496.0,840.0){\rule[-0.200pt]{0.400pt}{4.818pt}}
\put(631.0,123.0){\rule[-0.200pt]{0.400pt}{4.818pt}}
\put(631,82){\makebox(0,0){0}}
\put(631.0,840.0){\rule[-0.200pt]{0.400pt}{4.818pt}}
\put(765.0,123.0){\rule[-0.200pt]{0.400pt}{4.818pt}}
\put(765,82){\makebox(0,0){50}}
\put(765.0,840.0){\rule[-0.200pt]{0.400pt}{4.818pt}}
\put(899.0,123.0){\rule[-0.200pt]{0.400pt}{4.818pt}}
\put(899,82){\makebox(0,0){100}}
\put(899.0,840.0){\rule[-0.200pt]{0.400pt}{4.818pt}}
\put(1033.0,123.0){\rule[-0.200pt]{0.400pt}{4.818pt}}
\put(1033,82){\makebox(0,0){150}}
\put(1033.0,840.0){\rule[-0.200pt]{0.400pt}{4.818pt}}
\put(201.0,123.0){\rule[-0.200pt]{206.933pt}{0.400pt}}
\put(1060.0,123.0){\rule[-0.200pt]{0.400pt}{177.543pt}}
\put(201.0,860.0){\rule[-0.200pt]{206.933pt}{0.400pt}}
\put(630,21){\makebox(0,0){$m_l$}}
\put(201.0,123.0){\rule[-0.200pt]{0.400pt}{177.543pt}}
\put(879,713){\makebox(0,0)[r]{ { $P_z(m_l)$ } }}
\put(899.0,713.0){\rule[-0.200pt]{24.090pt}{0.400pt}}
\put(1044,319){\usebox{\plotpoint}}
\multiput(1042.95,302.53)(-0.447,-6.267){3}{\rule{0.108pt}{3.967pt}}
\multiput(1043.17,310.77)(-3.000,-20.767){2}{\rule{0.400pt}{1.983pt}}
\put(1039.17,284){\rule{0.400pt}{1.300pt}}
\multiput(1040.17,287.30)(-2.000,-3.302){2}{\rule{0.400pt}{0.650pt}}
\multiput(1037.95,278.05)(-0.447,-2.025){3}{\rule{0.108pt}{1.433pt}}
\multiput(1038.17,281.03)(-3.000,-7.025){2}{\rule{0.400pt}{0.717pt}}
\multiput(1034.95,271.37)(-0.447,-0.685){3}{\rule{0.108pt}{0.633pt}}
\multiput(1035.17,272.69)(-3.000,-2.685){2}{\rule{0.400pt}{0.317pt}}
\multiput(1031.95,263.50)(-0.447,-2.248){3}{\rule{0.108pt}{1.567pt}}
\multiput(1032.17,266.75)(-3.000,-7.748){2}{\rule{0.400pt}{0.783pt}}
\put(1028.17,259){\rule{0.400pt}{2.300pt}}
\multiput(1029.17,259.00)(-2.000,6.226){2}{\rule{0.400pt}{1.150pt}}
\put(1025,270.17){\rule{0.700pt}{0.400pt}}
\multiput(1026.55,269.17)(-1.547,2.000){2}{\rule{0.350pt}{0.400pt}}
\multiput(1023.95,268.82)(-0.447,-0.909){3}{\rule{0.108pt}{0.767pt}}
\multiput(1024.17,270.41)(-3.000,-3.409){2}{\rule{0.400pt}{0.383pt}}
\multiput(1018.95,261.60)(-0.447,-1.802){3}{\rule{0.108pt}{1.300pt}}
\multiput(1019.17,264.30)(-3.000,-6.302){2}{\rule{0.400pt}{0.650pt}}
\multiput(1015.95,258.00)(-0.447,2.695){3}{\rule{0.108pt}{1.833pt}}
\multiput(1016.17,258.00)(-3.000,9.195){2}{\rule{0.400pt}{0.917pt}}
\put(1012.17,260){\rule{0.400pt}{2.300pt}}
\multiput(1013.17,266.23)(-2.000,-6.226){2}{\rule{0.400pt}{1.150pt}}
\multiput(1010.95,260.00)(-0.447,1.132){3}{\rule{0.108pt}{0.900pt}}
\multiput(1011.17,260.00)(-3.000,4.132){2}{\rule{0.400pt}{0.450pt}}
\multiput(1007.95,255.07)(-0.447,-4.034){3}{\rule{0.108pt}{2.633pt}}
\multiput(1008.17,260.53)(-3.000,-13.534){2}{\rule{0.400pt}{1.317pt}}
\put(1004.17,247){\rule{0.400pt}{1.700pt}}
\multiput(1005.17,247.00)(-2.000,4.472){2}{\rule{0.400pt}{0.850pt}}
\multiput(1002.95,255.00)(-0.447,1.579){3}{\rule{0.108pt}{1.167pt}}
\multiput(1003.17,255.00)(-3.000,5.579){2}{\rule{0.400pt}{0.583pt}}
\multiput(999.95,254.84)(-0.447,-2.918){3}{\rule{0.108pt}{1.967pt}}
\multiput(1000.17,258.92)(-3.000,-9.918){2}{\rule{0.400pt}{0.983pt}}
\put(996.17,245){\rule{0.400pt}{0.900pt}}
\multiput(997.17,247.13)(-2.000,-2.132){2}{\rule{0.400pt}{0.450pt}}
\put(993,244.67){\rule{0.723pt}{0.400pt}}
\multiput(994.50,244.17)(-1.500,1.000){2}{\rule{0.361pt}{0.400pt}}
\multiput(991.95,246.00)(-0.447,3.141){3}{\rule{0.108pt}{2.100pt}}
\multiput(992.17,246.00)(-3.000,10.641){2}{\rule{0.400pt}{1.050pt}}
\put(988.17,257){\rule{0.400pt}{0.900pt}}
\multiput(989.17,259.13)(-2.000,-2.132){2}{\rule{0.400pt}{0.450pt}}
\multiput(986.95,257.00)(-0.447,2.025){3}{\rule{0.108pt}{1.433pt}}
\multiput(987.17,257.00)(-3.000,7.025){2}{\rule{0.400pt}{0.717pt}}
\multiput(983.95,267.00)(-0.447,0.685){3}{\rule{0.108pt}{0.633pt}}
\multiput(984.17,267.00)(-3.000,2.685){2}{\rule{0.400pt}{0.317pt}}
\multiput(980.95,261.18)(-0.447,-3.588){3}{\rule{0.108pt}{2.367pt}}
\multiput(981.17,266.09)(-3.000,-12.088){2}{\rule{0.400pt}{1.183pt}}
\put(977.17,245){\rule{0.400pt}{1.900pt}}
\multiput(978.17,250.06)(-2.000,-5.056){2}{\rule{0.400pt}{0.950pt}}
\multiput(975.95,245.00)(-0.447,4.704){3}{\rule{0.108pt}{3.033pt}}
\multiput(976.17,245.00)(-3.000,15.704){2}{\rule{0.400pt}{1.517pt}}
\multiput(972.95,259.39)(-0.447,-2.695){3}{\rule{0.108pt}{1.833pt}}
\multiput(973.17,263.19)(-3.000,-9.195){2}{\rule{0.400pt}{0.917pt}}
\put(969.17,254){\rule{0.400pt}{0.700pt}}
\multiput(970.17,254.00)(-2.000,1.547){2}{\rule{0.400pt}{0.350pt}}
\multiput(967.95,257.00)(-0.447,3.588){3}{\rule{0.108pt}{2.367pt}}
\multiput(968.17,257.00)(-3.000,12.088){2}{\rule{0.400pt}{1.183pt}}
\multiput(964.95,274.00)(-0.447,1.579){3}{\rule{0.108pt}{1.167pt}}
\multiput(965.17,274.00)(-3.000,5.579){2}{\rule{0.400pt}{0.583pt}}
\put(961.17,265){\rule{0.400pt}{3.500pt}}
\multiput(962.17,274.74)(-2.000,-9.736){2}{\rule{0.400pt}{1.750pt}}
\multiput(959.95,265.00)(-0.447,1.579){3}{\rule{0.108pt}{1.167pt}}
\multiput(960.17,265.00)(-3.000,5.579){2}{\rule{0.400pt}{0.583pt}}
\multiput(956.95,266.50)(-0.447,-2.248){3}{\rule{0.108pt}{1.567pt}}
\multiput(957.17,269.75)(-3.000,-7.748){2}{\rule{0.400pt}{0.783pt}}
\put(953,260.67){\rule{0.482pt}{0.400pt}}
\multiput(954.00,261.17)(-1.000,-1.000){2}{\rule{0.241pt}{0.400pt}}
\multiput(951.95,261.00)(-0.447,3.141){3}{\rule{0.108pt}{2.100pt}}
\multiput(952.17,261.00)(-3.000,10.641){2}{\rule{0.400pt}{1.050pt}}
\multiput(948.95,268.39)(-0.447,-2.695){3}{\rule{0.108pt}{1.833pt}}
\multiput(949.17,272.19)(-3.000,-9.195){2}{\rule{0.400pt}{0.917pt}}
\put(945,263.17){\rule{0.482pt}{0.400pt}}
\multiput(946.00,262.17)(-1.000,2.000){2}{\rule{0.241pt}{0.400pt}}
\put(1020.0,267.0){\rule[-0.200pt]{0.482pt}{0.400pt}}
\put(939,263.17){\rule{0.700pt}{0.400pt}}
\multiput(940.55,264.17)(-1.547,-2.000){2}{\rule{0.350pt}{0.400pt}}
\put(937.17,256){\rule{0.400pt}{1.500pt}}
\multiput(938.17,259.89)(-2.000,-3.887){2}{\rule{0.400pt}{0.750pt}}
\multiput(935.95,252.26)(-0.447,-1.132){3}{\rule{0.108pt}{0.900pt}}
\multiput(936.17,254.13)(-3.000,-4.132){2}{\rule{0.400pt}{0.450pt}}
\multiput(931.92,250.61)(-0.462,0.447){3}{\rule{0.500pt}{0.108pt}}
\multiput(932.96,249.17)(-1.962,3.000){2}{\rule{0.250pt}{0.400pt}}
\put(928,253.17){\rule{0.700pt}{0.400pt}}
\multiput(929.55,252.17)(-1.547,2.000){2}{\rule{0.350pt}{0.400pt}}
\put(926.17,249){\rule{0.400pt}{1.300pt}}
\multiput(927.17,252.30)(-2.000,-3.302){2}{\rule{0.400pt}{0.650pt}}
\put(923,248.67){\rule{0.723pt}{0.400pt}}
\multiput(924.50,248.17)(-1.500,1.000){2}{\rule{0.361pt}{0.400pt}}
\put(920,249.67){\rule{0.723pt}{0.400pt}}
\multiput(921.50,249.17)(-1.500,1.000){2}{\rule{0.361pt}{0.400pt}}
\put(918,251.17){\rule{0.482pt}{0.400pt}}
\multiput(919.00,250.17)(-1.000,2.000){2}{\rule{0.241pt}{0.400pt}}
\multiput(916.95,253.00)(-0.447,2.918){3}{\rule{0.108pt}{1.967pt}}
\multiput(917.17,253.00)(-3.000,9.918){2}{\rule{0.400pt}{0.983pt}}
\multiput(913.95,259.39)(-0.447,-2.695){3}{\rule{0.108pt}{1.833pt}}
\multiput(914.17,263.19)(-3.000,-9.195){2}{\rule{0.400pt}{0.917pt}}
\put(910.17,254){\rule{0.400pt}{0.700pt}}
\multiput(911.17,254.00)(-2.000,1.547){2}{\rule{0.400pt}{0.350pt}}
\multiput(908.95,257.00)(-0.447,2.695){3}{\rule{0.108pt}{1.833pt}}
\multiput(909.17,257.00)(-3.000,9.195){2}{\rule{0.400pt}{0.917pt}}
\multiput(905.95,265.16)(-0.447,-1.579){3}{\rule{0.108pt}{1.167pt}}
\multiput(906.17,267.58)(-3.000,-5.579){2}{\rule{0.400pt}{0.583pt}}
\put(902.17,253){\rule{0.400pt}{1.900pt}}
\multiput(903.17,258.06)(-2.000,-5.056){2}{\rule{0.400pt}{0.950pt}}
\multiput(899.92,253.61)(-0.462,0.447){3}{\rule{0.500pt}{0.108pt}}
\multiput(900.96,252.17)(-1.962,3.000){2}{\rule{0.250pt}{0.400pt}}
\multiput(897.95,252.82)(-0.447,-0.909){3}{\rule{0.108pt}{0.767pt}}
\multiput(898.17,254.41)(-3.000,-3.409){2}{\rule{0.400pt}{0.383pt}}
\put(894.17,248){\rule{0.400pt}{0.700pt}}
\multiput(895.17,249.55)(-2.000,-1.547){2}{\rule{0.400pt}{0.350pt}}
\put(942.0,265.0){\rule[-0.200pt]{0.723pt}{0.400pt}}
\put(886.17,238){\rule{0.400pt}{2.100pt}}
\multiput(887.17,243.64)(-2.000,-5.641){2}{\rule{0.400pt}{1.050pt}}
\multiput(884.95,238.00)(-0.447,2.248){3}{\rule{0.108pt}{1.567pt}}
\multiput(885.17,238.00)(-3.000,7.748){2}{\rule{0.400pt}{0.783pt}}
\multiput(881.95,249.00)(-0.447,1.132){3}{\rule{0.108pt}{0.900pt}}
\multiput(882.17,249.00)(-3.000,4.132){2}{\rule{0.400pt}{0.450pt}}
\multiput(878.95,246.84)(-0.447,-2.918){3}{\rule{0.108pt}{1.967pt}}
\multiput(879.17,250.92)(-3.000,-9.918){2}{\rule{0.400pt}{0.983pt}}
\put(888.0,248.0){\rule[-0.200pt]{1.445pt}{0.400pt}}
\multiput(873.95,241.00)(-0.447,1.132){3}{\rule{0.108pt}{0.900pt}}
\multiput(874.17,241.00)(-3.000,4.132){2}{\rule{0.400pt}{0.450pt}}
\multiput(870.95,243.82)(-0.447,-0.909){3}{\rule{0.108pt}{0.767pt}}
\multiput(871.17,245.41)(-3.000,-3.409){2}{\rule{0.400pt}{0.383pt}}
\put(867,242.17){\rule{0.482pt}{0.400pt}}
\multiput(868.00,241.17)(-1.000,2.000){2}{\rule{0.241pt}{0.400pt}}
\multiput(865.95,236.94)(-0.447,-2.472){3}{\rule{0.108pt}{1.700pt}}
\multiput(866.17,240.47)(-3.000,-8.472){2}{\rule{0.400pt}{0.850pt}}
\multiput(862.95,232.00)(-0.447,1.355){3}{\rule{0.108pt}{1.033pt}}
\multiput(863.17,232.00)(-3.000,4.855){2}{\rule{0.400pt}{0.517pt}}
\put(859.17,239){\rule{0.400pt}{2.500pt}}
\multiput(860.17,239.00)(-2.000,6.811){2}{\rule{0.400pt}{1.250pt}}
\multiput(857.95,244.50)(-0.447,-2.248){3}{\rule{0.108pt}{1.567pt}}
\multiput(858.17,247.75)(-3.000,-7.748){2}{\rule{0.400pt}{0.783pt}}
\multiput(854.95,236.82)(-0.447,-0.909){3}{\rule{0.108pt}{0.767pt}}
\multiput(855.17,238.41)(-3.000,-3.409){2}{\rule{0.400pt}{0.383pt}}
\put(851.17,235){\rule{0.400pt}{1.700pt}}
\multiput(852.17,235.00)(-2.000,4.472){2}{\rule{0.400pt}{0.850pt}}
\multiput(849.95,236.50)(-0.447,-2.248){3}{\rule{0.108pt}{1.567pt}}
\multiput(850.17,239.75)(-3.000,-7.748){2}{\rule{0.400pt}{0.783pt}}
\multiput(846.95,232.00)(-0.447,2.248){3}{\rule{0.108pt}{1.567pt}}
\multiput(847.17,232.00)(-3.000,7.748){2}{\rule{0.400pt}{0.783pt}}
\put(843.17,243){\rule{0.400pt}{2.300pt}}
\multiput(844.17,243.00)(-2.000,6.226){2}{\rule{0.400pt}{1.150pt}}
\multiput(841.95,248.05)(-0.447,-2.025){3}{\rule{0.108pt}{1.433pt}}
\multiput(842.17,251.03)(-3.000,-7.025){2}{\rule{0.400pt}{0.717pt}}
\multiput(838.95,244.00)(-0.447,0.685){3}{\rule{0.108pt}{0.633pt}}
\multiput(839.17,244.00)(-3.000,2.685){2}{\rule{0.400pt}{0.317pt}}
\put(835.17,248){\rule{0.400pt}{0.700pt}}
\multiput(836.17,248.00)(-2.000,1.547){2}{\rule{0.400pt}{0.350pt}}
\multiput(833.95,247.82)(-0.447,-0.909){3}{\rule{0.108pt}{0.767pt}}
\multiput(834.17,249.41)(-3.000,-3.409){2}{\rule{0.400pt}{0.383pt}}
\multiput(830.95,238.39)(-0.447,-2.695){3}{\rule{0.108pt}{1.833pt}}
\multiput(831.17,242.19)(-3.000,-9.195){2}{\rule{0.400pt}{0.917pt}}
\multiput(827.95,228.16)(-0.447,-1.579){3}{\rule{0.108pt}{1.167pt}}
\multiput(828.17,230.58)(-3.000,-5.579){2}{\rule{0.400pt}{0.583pt}}
\put(824.17,225){\rule{0.400pt}{3.900pt}}
\multiput(825.17,225.00)(-2.000,10.905){2}{\rule{0.400pt}{1.950pt}}
\multiput(822.95,238.05)(-0.447,-2.025){3}{\rule{0.108pt}{1.433pt}}
\multiput(823.17,241.03)(-3.000,-7.025){2}{\rule{0.400pt}{0.717pt}}
\multiput(819.95,234.00)(-0.447,2.472){3}{\rule{0.108pt}{1.700pt}}
\multiput(820.17,234.00)(-3.000,8.472){2}{\rule{0.400pt}{0.850pt}}
\put(816.17,241){\rule{0.400pt}{1.100pt}}
\multiput(817.17,243.72)(-2.000,-2.717){2}{\rule{0.400pt}{0.550pt}}
\multiput(814.95,237.82)(-0.447,-0.909){3}{\rule{0.108pt}{0.767pt}}
\multiput(815.17,239.41)(-3.000,-3.409){2}{\rule{0.400pt}{0.383pt}}
\multiput(811.95,236.00)(-0.447,0.909){3}{\rule{0.108pt}{0.767pt}}
\multiput(812.17,236.00)(-3.000,3.409){2}{\rule{0.400pt}{0.383pt}}
\put(808,240.67){\rule{0.482pt}{0.400pt}}
\multiput(809.00,240.17)(-1.000,1.000){2}{\rule{0.241pt}{0.400pt}}
\multiput(806.95,233.28)(-0.447,-3.141){3}{\rule{0.108pt}{2.100pt}}
\multiput(807.17,237.64)(-3.000,-10.641){2}{\rule{0.400pt}{1.050pt}}
\multiput(803.95,227.00)(-0.447,1.355){3}{\rule{0.108pt}{1.033pt}}
\multiput(804.17,227.00)(-3.000,4.855){2}{\rule{0.400pt}{0.517pt}}
\put(800.17,234){\rule{0.400pt}{1.100pt}}
\multiput(801.17,234.00)(-2.000,2.717){2}{\rule{0.400pt}{0.550pt}}
\multiput(798.95,228.07)(-0.447,-4.034){3}{\rule{0.108pt}{2.633pt}}
\multiput(799.17,233.53)(-3.000,-13.534){2}{\rule{0.400pt}{1.317pt}}
\multiput(795.95,220.00)(-0.447,1.579){3}{\rule{0.108pt}{1.167pt}}
\multiput(796.17,220.00)(-3.000,5.579){2}{\rule{0.400pt}{0.583pt}}
\put(792.17,228){\rule{0.400pt}{0.900pt}}
\multiput(793.17,228.00)(-2.000,2.132){2}{\rule{0.400pt}{0.450pt}}
\multiput(790.95,228.82)(-0.447,-0.909){3}{\rule{0.108pt}{0.767pt}}
\multiput(791.17,230.41)(-3.000,-3.409){2}{\rule{0.400pt}{0.383pt}}
\put(786,225.17){\rule{0.700pt}{0.400pt}}
\multiput(787.55,226.17)(-1.547,-2.000){2}{\rule{0.350pt}{0.400pt}}
\put(784,224.67){\rule{0.482pt}{0.400pt}}
\multiput(785.00,224.17)(-1.000,1.000){2}{\rule{0.241pt}{0.400pt}}
\multiput(782.95,220.60)(-0.447,-1.802){3}{\rule{0.108pt}{1.300pt}}
\multiput(783.17,223.30)(-3.000,-6.302){2}{\rule{0.400pt}{0.650pt}}
\put(778,216.67){\rule{0.723pt}{0.400pt}}
\multiput(779.50,216.17)(-1.500,1.000){2}{\rule{0.361pt}{0.400pt}}
\multiput(776.95,218.00)(-0.447,1.355){3}{\rule{0.108pt}{1.033pt}}
\multiput(777.17,218.00)(-3.000,4.855){2}{\rule{0.400pt}{0.517pt}}
\put(773,225.17){\rule{0.482pt}{0.400pt}}
\multiput(774.00,224.17)(-1.000,2.000){2}{\rule{0.241pt}{0.400pt}}
\multiput(771.95,227.00)(-0.447,2.248){3}{\rule{0.108pt}{1.567pt}}
\multiput(772.17,227.00)(-3.000,7.748){2}{\rule{0.400pt}{0.783pt}}
\multiput(768.95,233.71)(-0.447,-1.355){3}{\rule{0.108pt}{1.033pt}}
\multiput(769.17,235.86)(-3.000,-4.855){2}{\rule{0.400pt}{0.517pt}}
\put(765.17,231){\rule{0.400pt}{2.700pt}}
\multiput(766.17,231.00)(-2.000,7.396){2}{\rule{0.400pt}{1.350pt}}
\multiput(763.95,241.37)(-0.447,-0.685){3}{\rule{0.108pt}{0.633pt}}
\multiput(764.17,242.69)(-3.000,-2.685){2}{\rule{0.400pt}{0.317pt}}
\multiput(759.92,238.95)(-0.462,-0.447){3}{\rule{0.500pt}{0.108pt}}
\multiput(760.96,239.17)(-1.962,-3.000){2}{\rule{0.250pt}{0.400pt}}
\put(757,236.67){\rule{0.482pt}{0.400pt}}
\multiput(758.00,236.17)(-1.000,1.000){2}{\rule{0.241pt}{0.400pt}}
\put(754,236.67){\rule{0.723pt}{0.400pt}}
\multiput(755.50,237.17)(-1.500,-1.000){2}{\rule{0.361pt}{0.400pt}}
\multiput(752.95,229.39)(-0.447,-2.695){3}{\rule{0.108pt}{1.833pt}}
\multiput(753.17,233.19)(-3.000,-9.195){2}{\rule{0.400pt}{0.917pt}}
\put(749.17,224){\rule{0.400pt}{0.900pt}}
\multiput(750.17,224.00)(-2.000,2.132){2}{\rule{0.400pt}{0.450pt}}
\put(746,227.67){\rule{0.723pt}{0.400pt}}
\multiput(747.50,227.17)(-1.500,1.000){2}{\rule{0.361pt}{0.400pt}}
\multiput(744.95,226.37)(-0.447,-0.685){3}{\rule{0.108pt}{0.633pt}}
\multiput(745.17,227.69)(-3.000,-2.685){2}{\rule{0.400pt}{0.317pt}}
\put(741.17,225){\rule{0.400pt}{2.300pt}}
\multiput(742.17,225.00)(-2.000,6.226){2}{\rule{0.400pt}{1.150pt}}
\multiput(739.95,230.05)(-0.447,-2.025){3}{\rule{0.108pt}{1.433pt}}
\multiput(740.17,233.03)(-3.000,-7.025){2}{\rule{0.400pt}{0.717pt}}
\put(875.0,241.0){\rule[-0.200pt]{0.482pt}{0.400pt}}
\put(733.17,220){\rule{0.400pt}{1.300pt}}
\multiput(734.17,223.30)(-2.000,-3.302){2}{\rule{0.400pt}{0.650pt}}
\put(730,219.67){\rule{0.723pt}{0.400pt}}
\multiput(731.50,219.17)(-1.500,1.000){2}{\rule{0.361pt}{0.400pt}}
\put(727,219.17){\rule{0.700pt}{0.400pt}}
\multiput(728.55,220.17)(-1.547,-2.000){2}{\rule{0.350pt}{0.400pt}}
\multiput(725.95,219.00)(-0.447,1.132){3}{\rule{0.108pt}{0.900pt}}
\multiput(726.17,219.00)(-3.000,4.132){2}{\rule{0.400pt}{0.450pt}}
\put(722,225.17){\rule{0.482pt}{0.400pt}}
\multiput(723.00,224.17)(-1.000,2.000){2}{\rule{0.241pt}{0.400pt}}
\multiput(720.95,227.00)(-0.447,1.355){3}{\rule{0.108pt}{1.033pt}}
\multiput(721.17,227.00)(-3.000,4.855){2}{\rule{0.400pt}{0.517pt}}
\put(716,232.17){\rule{0.700pt}{0.400pt}}
\multiput(717.55,233.17)(-1.547,-2.000){2}{\rule{0.350pt}{0.400pt}}
\put(714.17,222){\rule{0.400pt}{2.100pt}}
\multiput(715.17,227.64)(-2.000,-5.641){2}{\rule{0.400pt}{1.050pt}}
\multiput(712.95,222.00)(-0.447,2.025){3}{\rule{0.108pt}{1.433pt}}
\multiput(713.17,222.00)(-3.000,7.025){2}{\rule{0.400pt}{0.717pt}}
\put(708,230.67){\rule{0.723pt}{0.400pt}}
\multiput(709.50,231.17)(-1.500,-1.000){2}{\rule{0.361pt}{0.400pt}}
\put(706.17,220){\rule{0.400pt}{2.300pt}}
\multiput(707.17,226.23)(-2.000,-6.226){2}{\rule{0.400pt}{1.150pt}}
\multiput(704.95,220.00)(-0.447,4.481){3}{\rule{0.108pt}{2.900pt}}
\multiput(705.17,220.00)(-3.000,14.981){2}{\rule{0.400pt}{1.450pt}}
\multiput(701.95,237.82)(-0.447,-0.909){3}{\rule{0.108pt}{0.767pt}}
\multiput(702.17,239.41)(-3.000,-3.409){2}{\rule{0.400pt}{0.383pt}}
\put(698.17,217){\rule{0.400pt}{3.900pt}}
\multiput(699.17,227.91)(-2.000,-10.905){2}{\rule{0.400pt}{1.950pt}}
\put(695,216.67){\rule{0.723pt}{0.400pt}}
\multiput(696.50,216.17)(-1.500,1.000){2}{\rule{0.361pt}{0.400pt}}
\multiput(693.95,218.00)(-0.447,3.141){3}{\rule{0.108pt}{2.100pt}}
\multiput(694.17,218.00)(-3.000,10.641){2}{\rule{0.400pt}{1.050pt}}
\put(690.17,228){\rule{0.400pt}{1.100pt}}
\multiput(691.17,230.72)(-2.000,-2.717){2}{\rule{0.400pt}{0.550pt}}
\multiput(688.95,224.26)(-0.447,-1.132){3}{\rule{0.108pt}{0.900pt}}
\multiput(689.17,226.13)(-3.000,-4.132){2}{\rule{0.400pt}{0.450pt}}
\multiput(685.95,222.00)(-0.447,1.579){3}{\rule{0.108pt}{1.167pt}}
\multiput(686.17,222.00)(-3.000,5.579){2}{\rule{0.400pt}{0.583pt}}
\put(682.17,216){\rule{0.400pt}{2.900pt}}
\multiput(683.17,223.98)(-2.000,-7.981){2}{\rule{0.400pt}{1.450pt}}
\multiput(680.95,216.00)(-0.447,1.132){3}{\rule{0.108pt}{0.900pt}}
\multiput(681.17,216.00)(-3.000,4.132){2}{\rule{0.400pt}{0.450pt}}
\multiput(677.95,222.00)(-0.447,1.802){3}{\rule{0.108pt}{1.300pt}}
\multiput(678.17,222.00)(-3.000,6.302){2}{\rule{0.400pt}{0.650pt}}
\multiput(674.95,226.16)(-0.447,-1.579){3}{\rule{0.108pt}{1.167pt}}
\multiput(675.17,228.58)(-3.000,-5.579){2}{\rule{0.400pt}{0.583pt}}
\put(671.17,223){\rule{0.400pt}{1.500pt}}
\multiput(672.17,223.00)(-2.000,3.887){2}{\rule{0.400pt}{0.750pt}}
\multiput(669.95,223.50)(-0.447,-2.248){3}{\rule{0.108pt}{1.567pt}}
\multiput(670.17,226.75)(-3.000,-7.748){2}{\rule{0.400pt}{0.783pt}}
\multiput(666.95,219.00)(-0.447,1.355){3}{\rule{0.108pt}{1.033pt}}
\multiput(667.17,219.00)(-3.000,4.855){2}{\rule{0.400pt}{0.517pt}}
\put(663.17,221){\rule{0.400pt}{1.100pt}}
\multiput(664.17,223.72)(-2.000,-2.717){2}{\rule{0.400pt}{0.550pt}}
\multiput(661.95,221.00)(-0.447,2.918){3}{\rule{0.108pt}{1.967pt}}
\multiput(662.17,221.00)(-3.000,9.918){2}{\rule{0.400pt}{0.983pt}}
\multiput(658.95,226.84)(-0.447,-2.918){3}{\rule{0.108pt}{1.967pt}}
\multiput(659.17,230.92)(-3.000,-9.918){2}{\rule{0.400pt}{0.983pt}}
\put(655.17,221){\rule{0.400pt}{1.300pt}}
\multiput(656.17,221.00)(-2.000,3.302){2}{\rule{0.400pt}{0.650pt}}
\multiput(653.95,227.00)(-0.447,1.132){3}{\rule{0.108pt}{0.900pt}}
\multiput(654.17,227.00)(-3.000,4.132){2}{\rule{0.400pt}{0.450pt}}
\multiput(650.95,230.37)(-0.447,-0.685){3}{\rule{0.108pt}{0.633pt}}
\multiput(651.17,231.69)(-3.000,-2.685){2}{\rule{0.400pt}{0.317pt}}
\put(647.17,219){\rule{0.400pt}{2.100pt}}
\multiput(648.17,224.64)(-2.000,-5.641){2}{\rule{0.400pt}{1.050pt}}
\multiput(645.95,219.00)(-0.447,1.579){3}{\rule{0.108pt}{1.167pt}}
\multiput(646.17,219.00)(-3.000,5.579){2}{\rule{0.400pt}{0.583pt}}
\multiput(642.95,227.00)(-0.447,1.132){3}{\rule{0.108pt}{0.900pt}}
\multiput(643.17,227.00)(-3.000,4.132){2}{\rule{0.400pt}{0.450pt}}
\put(639,231.17){\rule{0.482pt}{0.400pt}}
\multiput(640.00,232.17)(-1.000,-2.000){2}{\rule{0.241pt}{0.400pt}}
\multiput(636.92,229.95)(-0.462,-0.447){3}{\rule{0.500pt}{0.108pt}}
\multiput(637.96,230.17)(-1.962,-3.000){2}{\rule{0.250pt}{0.400pt}}
\multiput(634.95,228.00)(-0.447,3.141){3}{\rule{0.108pt}{2.100pt}}
\multiput(635.17,228.00)(-3.000,10.641){2}{\rule{0.400pt}{1.050pt}}
\put(631.17,229){\rule{0.400pt}{2.900pt}}
\multiput(632.17,236.98)(-2.000,-7.981){2}{\rule{0.400pt}{1.450pt}}
\multiput(629.95,223.05)(-0.447,-2.025){3}{\rule{0.108pt}{1.433pt}}
\multiput(630.17,226.03)(-3.000,-7.025){2}{\rule{0.400pt}{0.717pt}}
\multiput(626.95,213.60)(-0.447,-1.802){3}{\rule{0.108pt}{1.300pt}}
\multiput(627.17,216.30)(-3.000,-6.302){2}{\rule{0.400pt}{0.650pt}}
\multiput(623.95,210.00)(-0.447,2.472){3}{\rule{0.108pt}{1.700pt}}
\multiput(624.17,210.00)(-3.000,8.472){2}{\rule{0.400pt}{0.850pt}}
\put(620.17,222){\rule{0.400pt}{1.100pt}}
\multiput(621.17,222.00)(-2.000,2.717){2}{\rule{0.400pt}{0.550pt}}
\multiput(618.95,223.26)(-0.447,-1.132){3}{\rule{0.108pt}{0.900pt}}
\multiput(619.17,225.13)(-3.000,-4.132){2}{\rule{0.400pt}{0.450pt}}
\multiput(615.95,221.00)(-0.447,2.248){3}{\rule{0.108pt}{1.567pt}}
\multiput(616.17,221.00)(-3.000,7.748){2}{\rule{0.400pt}{0.783pt}}
\put(612.17,222){\rule{0.400pt}{2.100pt}}
\multiput(613.17,227.64)(-2.000,-5.641){2}{\rule{0.400pt}{1.050pt}}
\multiput(610.95,222.00)(-0.447,2.025){3}{\rule{0.108pt}{1.433pt}}
\multiput(611.17,222.00)(-3.000,7.025){2}{\rule{0.400pt}{0.717pt}}
\multiput(607.95,228.82)(-0.447,-0.909){3}{\rule{0.108pt}{0.767pt}}
\multiput(608.17,230.41)(-3.000,-3.409){2}{\rule{0.400pt}{0.383pt}}
\put(604,227.17){\rule{0.482pt}{0.400pt}}
\multiput(605.00,226.17)(-1.000,2.000){2}{\rule{0.241pt}{0.400pt}}
\multiput(602.95,224.16)(-0.447,-1.579){3}{\rule{0.108pt}{1.167pt}}
\multiput(603.17,226.58)(-3.000,-5.579){2}{\rule{0.400pt}{0.583pt}}
\multiput(599.95,221.00)(-0.447,2.025){3}{\rule{0.108pt}{1.433pt}}
\multiput(600.17,221.00)(-3.000,7.025){2}{\rule{0.400pt}{0.717pt}}
\put(596.17,225){\rule{0.400pt}{1.300pt}}
\multiput(597.17,228.30)(-2.000,-3.302){2}{\rule{0.400pt}{0.650pt}}
\multiput(594.95,225.00)(-0.447,2.248){3}{\rule{0.108pt}{1.567pt}}
\multiput(595.17,225.00)(-3.000,7.748){2}{\rule{0.400pt}{0.783pt}}
\multiput(591.95,229.50)(-0.447,-2.248){3}{\rule{0.108pt}{1.567pt}}
\multiput(592.17,232.75)(-3.000,-7.748){2}{\rule{0.400pt}{0.783pt}}
\put(588.17,225){\rule{0.400pt}{3.100pt}}
\multiput(589.17,225.00)(-2.000,8.566){2}{\rule{0.400pt}{1.550pt}}
\multiput(586.95,232.39)(-0.447,-2.695){3}{\rule{0.108pt}{1.833pt}}
\multiput(587.17,236.19)(-3.000,-9.195){2}{\rule{0.400pt}{0.917pt}}
\multiput(583.95,227.00)(-0.447,0.909){3}{\rule{0.108pt}{0.767pt}}
\multiput(584.17,227.00)(-3.000,3.409){2}{\rule{0.400pt}{0.383pt}}
\multiput(580.95,229.37)(-0.447,-0.685){3}{\rule{0.108pt}{0.633pt}}
\multiput(581.17,230.69)(-3.000,-2.685){2}{\rule{0.400pt}{0.317pt}}
\put(577.17,223){\rule{0.400pt}{1.100pt}}
\multiput(578.17,225.72)(-2.000,-2.717){2}{\rule{0.400pt}{0.550pt}}
\multiput(575.95,223.00)(-0.447,1.802){3}{\rule{0.108pt}{1.300pt}}
\multiput(576.17,223.00)(-3.000,6.302){2}{\rule{0.400pt}{0.650pt}}
\put(571,230.17){\rule{0.700pt}{0.400pt}}
\multiput(572.55,231.17)(-1.547,-2.000){2}{\rule{0.350pt}{0.400pt}}
\put(569.17,226){\rule{0.400pt}{0.900pt}}
\multiput(570.17,228.13)(-2.000,-2.132){2}{\rule{0.400pt}{0.450pt}}
\multiput(567.95,226.00)(-0.447,2.248){3}{\rule{0.108pt}{1.567pt}}
\multiput(568.17,226.00)(-3.000,7.748){2}{\rule{0.400pt}{0.783pt}}
\put(563,236.67){\rule{0.723pt}{0.400pt}}
\multiput(564.50,236.17)(-1.500,1.000){2}{\rule{0.361pt}{0.400pt}}
\put(561.17,226){\rule{0.400pt}{2.500pt}}
\multiput(562.17,232.81)(-2.000,-6.811){2}{\rule{0.400pt}{1.250pt}}
\multiput(559.95,226.00)(-0.447,1.355){3}{\rule{0.108pt}{1.033pt}}
\multiput(560.17,226.00)(-3.000,4.855){2}{\rule{0.400pt}{0.517pt}}
\multiput(556.95,225.94)(-0.447,-2.472){3}{\rule{0.108pt}{1.700pt}}
\multiput(557.17,229.47)(-3.000,-8.472){2}{\rule{0.400pt}{0.850pt}}
\put(553,221.17){\rule{0.482pt}{0.400pt}}
\multiput(554.00,220.17)(-1.000,2.000){2}{\rule{0.241pt}{0.400pt}}
\multiput(551.95,223.00)(-0.447,2.472){3}{\rule{0.108pt}{1.700pt}}
\multiput(552.17,223.00)(-3.000,8.472){2}{\rule{0.400pt}{0.850pt}}
\put(547,233.67){\rule{0.723pt}{0.400pt}}
\multiput(548.50,234.17)(-1.500,-1.000){2}{\rule{0.361pt}{0.400pt}}
\put(545,234.17){\rule{0.482pt}{0.400pt}}
\multiput(546.00,233.17)(-1.000,2.000){2}{\rule{0.241pt}{0.400pt}}
\multiput(543.95,231.71)(-0.447,-1.355){3}{\rule{0.108pt}{1.033pt}}
\multiput(544.17,233.86)(-3.000,-4.855){2}{\rule{0.400pt}{0.517pt}}
\multiput(540.95,229.00)(-0.447,0.685){3}{\rule{0.108pt}{0.633pt}}
\multiput(541.17,229.00)(-3.000,2.685){2}{\rule{0.400pt}{0.317pt}}
\put(537,233.17){\rule{0.482pt}{0.400pt}}
\multiput(538.00,232.17)(-1.000,2.000){2}{\rule{0.241pt}{0.400pt}}
\put(534,234.67){\rule{0.723pt}{0.400pt}}
\multiput(535.50,234.17)(-1.500,1.000){2}{\rule{0.361pt}{0.400pt}}
\multiput(532.95,228.39)(-0.447,-2.695){3}{\rule{0.108pt}{1.833pt}}
\multiput(533.17,232.19)(-3.000,-9.195){2}{\rule{0.400pt}{0.917pt}}
\multiput(529.95,223.00)(-0.447,1.579){3}{\rule{0.108pt}{1.167pt}}
\multiput(530.17,223.00)(-3.000,5.579){2}{\rule{0.400pt}{0.583pt}}
\put(526.17,225){\rule{0.400pt}{1.300pt}}
\multiput(527.17,228.30)(-2.000,-3.302){2}{\rule{0.400pt}{0.650pt}}
\multiput(524.95,225.00)(-0.447,2.025){3}{\rule{0.108pt}{1.433pt}}
\multiput(525.17,225.00)(-3.000,7.025){2}{\rule{0.400pt}{0.717pt}}
\multiput(521.95,229.60)(-0.447,-1.802){3}{\rule{0.108pt}{1.300pt}}
\multiput(522.17,232.30)(-3.000,-6.302){2}{\rule{0.400pt}{0.650pt}}
\put(518.17,226){\rule{0.400pt}{1.700pt}}
\multiput(519.17,226.00)(-2.000,4.472){2}{\rule{0.400pt}{0.850pt}}
\multiput(516.95,234.00)(-0.447,3.141){3}{\rule{0.108pt}{2.100pt}}
\multiput(517.17,234.00)(-3.000,10.641){2}{\rule{0.400pt}{1.050pt}}
\multiput(513.95,234.19)(-0.447,-5.597){3}{\rule{0.108pt}{3.567pt}}
\multiput(514.17,241.60)(-3.000,-18.597){2}{\rule{0.400pt}{1.783pt}}
\put(510.17,223){\rule{0.400pt}{1.300pt}}
\multiput(511.17,223.00)(-2.000,3.302){2}{\rule{0.400pt}{0.650pt}}
\multiput(508.95,229.00)(-0.447,3.365){3}{\rule{0.108pt}{2.233pt}}
\multiput(509.17,229.00)(-3.000,11.365){2}{\rule{0.400pt}{1.117pt}}
\multiput(505.95,237.39)(-0.447,-2.695){3}{\rule{0.108pt}{1.833pt}}
\multiput(506.17,241.19)(-3.000,-9.195){2}{\rule{0.400pt}{0.917pt}}
\put(502.17,225){\rule{0.400pt}{1.500pt}}
\multiput(503.17,228.89)(-2.000,-3.887){2}{\rule{0.400pt}{0.750pt}}
\multiput(499.92,225.61)(-0.462,0.447){3}{\rule{0.500pt}{0.108pt}}
\multiput(500.96,224.17)(-1.962,3.000){2}{\rule{0.250pt}{0.400pt}}
\multiput(497.95,228.00)(-0.447,2.025){3}{\rule{0.108pt}{1.433pt}}
\multiput(498.17,228.00)(-3.000,7.025){2}{\rule{0.400pt}{0.717pt}}
\put(494.17,229){\rule{0.400pt}{1.900pt}}
\multiput(495.17,234.06)(-2.000,-5.056){2}{\rule{0.400pt}{0.950pt}}
\multiput(492.95,229.00)(-0.447,0.909){3}{\rule{0.108pt}{0.767pt}}
\multiput(493.17,229.00)(-3.000,3.409){2}{\rule{0.400pt}{0.383pt}}
\multiput(489.95,234.00)(-0.447,2.248){3}{\rule{0.108pt}{1.567pt}}
\multiput(490.17,234.00)(-3.000,7.748){2}{\rule{0.400pt}{0.783pt}}
\put(486.17,237){\rule{0.400pt}{1.700pt}}
\multiput(487.17,241.47)(-2.000,-4.472){2}{\rule{0.400pt}{0.850pt}}
\multiput(484.95,231.60)(-0.447,-1.802){3}{\rule{0.108pt}{1.300pt}}
\multiput(485.17,234.30)(-3.000,-6.302){2}{\rule{0.400pt}{0.650pt}}
\multiput(481.95,228.00)(-0.447,2.918){3}{\rule{0.108pt}{1.967pt}}
\multiput(482.17,228.00)(-3.000,9.918){2}{\rule{0.400pt}{0.983pt}}
\multiput(478.95,242.00)(-0.447,0.685){3}{\rule{0.108pt}{0.633pt}}
\multiput(479.17,242.00)(-3.000,2.685){2}{\rule{0.400pt}{0.317pt}}
\put(475.17,234){\rule{0.400pt}{2.500pt}}
\multiput(476.17,240.81)(-2.000,-6.811){2}{\rule{0.400pt}{1.250pt}}
\multiput(473.95,234.00)(-0.447,2.248){3}{\rule{0.108pt}{1.567pt}}
\multiput(474.17,234.00)(-3.000,7.748){2}{\rule{0.400pt}{0.783pt}}
\multiput(470.95,242.37)(-0.447,-0.685){3}{\rule{0.108pt}{0.633pt}}
\multiput(471.17,243.69)(-3.000,-2.685){2}{\rule{0.400pt}{0.317pt}}
\put(735.0,226.0){\rule[-0.200pt]{0.723pt}{0.400pt}}
\multiput(465.95,233.39)(-0.447,-2.695){3}{\rule{0.108pt}{1.833pt}}
\multiput(466.17,237.19)(-3.000,-9.195){2}{\rule{0.400pt}{0.917pt}}
\multiput(462.95,228.00)(-0.447,2.248){3}{\rule{0.108pt}{1.567pt}}
\multiput(463.17,228.00)(-3.000,7.748){2}{\rule{0.400pt}{0.783pt}}
\put(459.17,232){\rule{0.400pt}{1.500pt}}
\multiput(460.17,235.89)(-2.000,-3.887){2}{\rule{0.400pt}{0.750pt}}
\multiput(457.95,232.00)(-0.447,1.579){3}{\rule{0.108pt}{1.167pt}}
\multiput(458.17,232.00)(-3.000,5.579){2}{\rule{0.400pt}{0.583pt}}
\multiput(454.95,234.60)(-0.447,-1.802){3}{\rule{0.108pt}{1.300pt}}
\multiput(455.17,237.30)(-3.000,-6.302){2}{\rule{0.400pt}{0.650pt}}
\put(451.17,231){\rule{0.400pt}{4.300pt}}
\multiput(452.17,231.00)(-2.000,12.075){2}{\rule{0.400pt}{2.150pt}}
\multiput(449.95,247.16)(-0.447,-1.579){3}{\rule{0.108pt}{1.167pt}}
\multiput(450.17,249.58)(-3.000,-5.579){2}{\rule{0.400pt}{0.583pt}}
\multiput(446.95,239.71)(-0.447,-1.355){3}{\rule{0.108pt}{1.033pt}}
\multiput(447.17,241.86)(-3.000,-4.855){2}{\rule{0.400pt}{0.517pt}}
\put(443.17,237){\rule{0.400pt}{1.100pt}}
\multiput(444.17,237.00)(-2.000,2.717){2}{\rule{0.400pt}{0.550pt}}
\put(440,240.17){\rule{0.700pt}{0.400pt}}
\multiput(441.55,241.17)(-1.547,-2.000){2}{\rule{0.350pt}{0.400pt}}
\multiput(438.95,236.26)(-0.447,-1.132){3}{\rule{0.108pt}{0.900pt}}
\multiput(439.17,238.13)(-3.000,-4.132){2}{\rule{0.400pt}{0.450pt}}
\put(435.17,234){\rule{0.400pt}{1.700pt}}
\multiput(436.17,234.00)(-2.000,4.472){2}{\rule{0.400pt}{0.850pt}}
\multiput(433.95,236.05)(-0.447,-2.025){3}{\rule{0.108pt}{1.433pt}}
\multiput(434.17,239.03)(-3.000,-7.025){2}{\rule{0.400pt}{0.717pt}}
\multiput(430.95,232.00)(-0.447,1.802){3}{\rule{0.108pt}{1.300pt}}
\multiput(431.17,232.00)(-3.000,6.302){2}{\rule{0.400pt}{0.650pt}}
\multiput(427.95,241.00)(-0.447,2.248){3}{\rule{0.108pt}{1.567pt}}
\multiput(428.17,241.00)(-3.000,7.748){2}{\rule{0.400pt}{0.783pt}}
\put(424.17,230){\rule{0.400pt}{4.500pt}}
\multiput(425.17,242.66)(-2.000,-12.660){2}{\rule{0.400pt}{2.250pt}}
\multiput(422.95,230.00)(-0.447,4.704){3}{\rule{0.108pt}{3.033pt}}
\multiput(423.17,230.00)(-3.000,15.704){2}{\rule{0.400pt}{1.517pt}}
\multiput(419.95,246.60)(-0.447,-1.802){3}{\rule{0.108pt}{1.300pt}}
\multiput(420.17,249.30)(-3.000,-6.302){2}{\rule{0.400pt}{0.650pt}}
\put(416.17,239){\rule{0.400pt}{0.900pt}}
\multiput(417.17,241.13)(-2.000,-2.132){2}{\rule{0.400pt}{0.450pt}}
\put(413,237.17){\rule{0.700pt}{0.400pt}}
\multiput(414.55,238.17)(-1.547,-2.000){2}{\rule{0.350pt}{0.400pt}}
\multiput(410.92,235.95)(-0.462,-0.447){3}{\rule{0.500pt}{0.108pt}}
\multiput(411.96,236.17)(-1.962,-3.000){2}{\rule{0.250pt}{0.400pt}}
\put(467.0,241.0){\rule[-0.200pt]{0.482pt}{0.400pt}}
\multiput(406.95,230.82)(-0.447,-0.909){3}{\rule{0.108pt}{0.767pt}}
\multiput(407.17,232.41)(-3.000,-3.409){2}{\rule{0.400pt}{0.383pt}}
\multiput(402.92,229.61)(-0.462,0.447){3}{\rule{0.500pt}{0.108pt}}
\multiput(403.96,228.17)(-1.962,3.000){2}{\rule{0.250pt}{0.400pt}}
\put(400.17,232){\rule{0.400pt}{3.100pt}}
\multiput(401.17,232.00)(-2.000,8.566){2}{\rule{0.400pt}{1.550pt}}
\multiput(398.95,236.07)(-0.447,-4.034){3}{\rule{0.108pt}{2.633pt}}
\multiput(399.17,241.53)(-3.000,-13.534){2}{\rule{0.400pt}{1.317pt}}
\multiput(395.95,228.00)(-0.447,2.472){3}{\rule{0.108pt}{1.700pt}}
\multiput(396.17,228.00)(-3.000,8.472){2}{\rule{0.400pt}{0.850pt}}
\put(392,239.67){\rule{0.482pt}{0.400pt}}
\multiput(393.00,239.17)(-1.000,1.000){2}{\rule{0.241pt}{0.400pt}}
\put(389,241.17){\rule{0.700pt}{0.400pt}}
\multiput(390.55,240.17)(-1.547,2.000){2}{\rule{0.350pt}{0.400pt}}
\multiput(386.92,243.61)(-0.462,0.447){3}{\rule{0.500pt}{0.108pt}}
\multiput(387.96,242.17)(-1.962,3.000){2}{\rule{0.250pt}{0.400pt}}
\put(384.17,233){\rule{0.400pt}{2.700pt}}
\multiput(385.17,240.40)(-2.000,-7.396){2}{\rule{0.400pt}{1.350pt}}
\multiput(382.95,233.00)(-0.447,1.802){3}{\rule{0.108pt}{1.300pt}}
\multiput(383.17,233.00)(-3.000,6.302){2}{\rule{0.400pt}{0.650pt}}
\multiput(379.95,242.00)(-0.447,1.802){3}{\rule{0.108pt}{1.300pt}}
\multiput(380.17,242.00)(-3.000,6.302){2}{\rule{0.400pt}{0.650pt}}
\multiput(376.95,239.52)(-0.447,-4.258){3}{\rule{0.108pt}{2.767pt}}
\multiput(377.17,245.26)(-3.000,-14.258){2}{\rule{0.400pt}{1.383pt}}
\put(373.17,231){\rule{0.400pt}{6.500pt}}
\multiput(374.17,231.00)(-2.000,18.509){2}{\rule{0.400pt}{3.250pt}}
\put(370,261.17){\rule{0.700pt}{0.400pt}}
\multiput(371.55,262.17)(-1.547,-2.000){2}{\rule{0.350pt}{0.400pt}}
\multiput(368.95,255.60)(-0.447,-1.802){3}{\rule{0.108pt}{1.300pt}}
\multiput(369.17,258.30)(-3.000,-6.302){2}{\rule{0.400pt}{0.650pt}}
\put(365.17,248){\rule{0.400pt}{0.900pt}}
\multiput(366.17,250.13)(-2.000,-2.132){2}{\rule{0.400pt}{0.450pt}}
\multiput(363.95,243.71)(-0.447,-1.355){3}{\rule{0.108pt}{1.033pt}}
\multiput(364.17,245.86)(-3.000,-4.855){2}{\rule{0.400pt}{0.517pt}}
\put(408.0,234.0){\rule[-0.200pt]{0.482pt}{0.400pt}}
\put(357,240.67){\rule{0.482pt}{0.400pt}}
\multiput(358.00,240.17)(-1.000,1.000){2}{\rule{0.241pt}{0.400pt}}
\multiput(355.95,242.00)(-0.447,3.811){3}{\rule{0.108pt}{2.500pt}}
\multiput(356.17,242.00)(-3.000,12.811){2}{\rule{0.400pt}{1.250pt}}
\multiput(352.95,254.60)(-0.447,-1.802){3}{\rule{0.108pt}{1.300pt}}
\multiput(353.17,257.30)(-3.000,-6.302){2}{\rule{0.400pt}{0.650pt}}
\put(349.17,247){\rule{0.400pt}{0.900pt}}
\multiput(350.17,249.13)(-2.000,-2.132){2}{\rule{0.400pt}{0.450pt}}
\multiput(347.95,247.00)(-0.447,1.132){3}{\rule{0.108pt}{0.900pt}}
\multiput(348.17,247.00)(-3.000,4.132){2}{\rule{0.400pt}{0.450pt}}
\multiput(343.92,251.95)(-0.462,-0.447){3}{\rule{0.500pt}{0.108pt}}
\multiput(344.96,252.17)(-1.962,-3.000){2}{\rule{0.250pt}{0.400pt}}
\put(341.17,243){\rule{0.400pt}{1.500pt}}
\multiput(342.17,246.89)(-2.000,-3.887){2}{\rule{0.400pt}{0.750pt}}
\multiput(339.95,243.00)(-0.447,2.248){3}{\rule{0.108pt}{1.567pt}}
\multiput(340.17,243.00)(-3.000,7.748){2}{\rule{0.400pt}{0.783pt}}
\multiput(336.95,245.84)(-0.447,-2.918){3}{\rule{0.108pt}{1.967pt}}
\multiput(337.17,249.92)(-3.000,-9.918){2}{\rule{0.400pt}{0.983pt}}
\put(333.17,240){\rule{0.400pt}{3.100pt}}
\multiput(334.17,240.00)(-2.000,8.566){2}{\rule{0.400pt}{1.550pt}}
\put(359.0,241.0){\rule[-0.200pt]{0.723pt}{0.400pt}}
\put(327,253.67){\rule{0.723pt}{0.400pt}}
\multiput(328.50,254.17)(-1.500,-1.000){2}{\rule{0.361pt}{0.400pt}}
\put(324,253.67){\rule{0.723pt}{0.400pt}}
\multiput(325.50,253.17)(-1.500,1.000){2}{\rule{0.361pt}{0.400pt}}
\put(322.17,248){\rule{0.400pt}{1.500pt}}
\multiput(323.17,251.89)(-2.000,-3.887){2}{\rule{0.400pt}{0.750pt}}
\multiput(320.95,248.00)(-0.447,2.025){3}{\rule{0.108pt}{1.433pt}}
\multiput(321.17,248.00)(-3.000,7.025){2}{\rule{0.400pt}{0.717pt}}
\multiput(317.95,258.00)(-0.447,2.248){3}{\rule{0.108pt}{1.567pt}}
\multiput(318.17,258.00)(-3.000,7.748){2}{\rule{0.400pt}{0.783pt}}
\put(314.17,245){\rule{0.400pt}{4.900pt}}
\multiput(315.17,258.83)(-2.000,-13.830){2}{\rule{0.400pt}{2.450pt}}
\multiput(312.95,245.00)(-0.447,2.918){3}{\rule{0.108pt}{1.967pt}}
\multiput(313.17,245.00)(-3.000,9.918){2}{\rule{0.400pt}{0.983pt}}
\put(308,259.17){\rule{0.700pt}{0.400pt}}
\multiput(309.55,258.17)(-1.547,2.000){2}{\rule{0.350pt}{0.400pt}}
\put(306.17,261){\rule{0.400pt}{1.100pt}}
\multiput(307.17,261.00)(-2.000,2.717){2}{\rule{0.400pt}{0.550pt}}
\multiput(304.95,266.00)(-0.447,0.685){3}{\rule{0.108pt}{0.633pt}}
\multiput(305.17,266.00)(-3.000,2.685){2}{\rule{0.400pt}{0.317pt}}
\multiput(301.95,261.28)(-0.447,-3.141){3}{\rule{0.108pt}{2.100pt}}
\multiput(302.17,265.64)(-3.000,-10.641){2}{\rule{0.400pt}{1.050pt}}
\put(298.17,244){\rule{0.400pt}{2.300pt}}
\multiput(299.17,250.23)(-2.000,-6.226){2}{\rule{0.400pt}{1.150pt}}
\put(330.0,255.0){\rule[-0.200pt]{0.723pt}{0.400pt}}
\put(292,243.67){\rule{0.723pt}{0.400pt}}
\multiput(293.50,243.17)(-1.500,1.000){2}{\rule{0.361pt}{0.400pt}}
\put(290.17,231){\rule{0.400pt}{2.900pt}}
\multiput(291.17,238.98)(-2.000,-7.981){2}{\rule{0.400pt}{1.450pt}}
\multiput(288.95,231.00)(-0.447,6.490){3}{\rule{0.108pt}{4.100pt}}
\multiput(289.17,231.00)(-3.000,21.490){2}{\rule{0.400pt}{2.050pt}}
\multiput(285.95,251.18)(-0.447,-3.588){3}{\rule{0.108pt}{2.367pt}}
\multiput(286.17,256.09)(-3.000,-12.088){2}{\rule{0.400pt}{1.183pt}}
\put(282,242.67){\rule{0.482pt}{0.400pt}}
\multiput(283.00,243.17)(-1.000,-1.000){2}{\rule{0.241pt}{0.400pt}}
\put(295.0,244.0){\rule[-0.200pt]{0.723pt}{0.400pt}}
\multiput(277.95,243.00)(-0.447,3.811){3}{\rule{0.108pt}{2.500pt}}
\multiput(278.17,243.00)(-3.000,12.811){2}{\rule{0.400pt}{1.250pt}}
\multiput(274.95,257.82)(-0.447,-0.909){3}{\rule{0.108pt}{0.767pt}}
\multiput(275.17,259.41)(-3.000,-3.409){2}{\rule{0.400pt}{0.383pt}}
\put(271.17,256){\rule{0.400pt}{1.700pt}}
\multiput(272.17,256.00)(-2.000,4.472){2}{\rule{0.400pt}{0.850pt}}
\multiput(269.95,250.85)(-0.447,-4.927){3}{\rule{0.108pt}{3.167pt}}
\multiput(270.17,257.43)(-3.000,-16.427){2}{\rule{0.400pt}{1.583pt}}
\put(265,239.67){\rule{0.723pt}{0.400pt}}
\multiput(266.50,240.17)(-1.500,-1.000){2}{\rule{0.361pt}{0.400pt}}
\put(263.17,240){\rule{0.400pt}{4.100pt}}
\multiput(264.17,240.00)(-2.000,11.490){2}{\rule{0.400pt}{2.050pt}}
\put(260,258.67){\rule{0.723pt}{0.400pt}}
\multiput(261.50,259.17)(-1.500,-1.000){2}{\rule{0.361pt}{0.400pt}}
\multiput(258.95,251.39)(-0.447,-2.695){3}{\rule{0.108pt}{1.833pt}}
\multiput(259.17,255.19)(-3.000,-9.195){2}{\rule{0.400pt}{0.917pt}}
\put(255.17,246){\rule{0.400pt}{4.100pt}}
\multiput(256.17,246.00)(-2.000,11.490){2}{\rule{0.400pt}{2.050pt}}
\multiput(253.95,263.37)(-0.447,-0.685){3}{\rule{0.108pt}{0.633pt}}
\multiput(254.17,264.69)(-3.000,-2.685){2}{\rule{0.400pt}{0.317pt}}
\multiput(250.95,256.05)(-0.447,-2.025){3}{\rule{0.108pt}{1.433pt}}
\multiput(251.17,259.03)(-3.000,-7.025){2}{\rule{0.400pt}{0.717pt}}
\put(247,252.17){\rule{0.482pt}{0.400pt}}
\multiput(248.00,251.17)(-1.000,2.000){2}{\rule{0.241pt}{0.400pt}}
\multiput(244.92,252.95)(-0.462,-0.447){3}{\rule{0.500pt}{0.108pt}}
\multiput(245.96,253.17)(-1.962,-3.000){2}{\rule{0.250pt}{0.400pt}}
\multiput(242.95,251.00)(-0.447,2.695){3}{\rule{0.108pt}{1.833pt}}
\multiput(243.17,251.00)(-3.000,9.195){2}{\rule{0.400pt}{0.917pt}}
\put(239.17,245){\rule{0.400pt}{3.900pt}}
\multiput(240.17,255.91)(-2.000,-10.905){2}{\rule{0.400pt}{1.950pt}}
\put(236,244.67){\rule{0.723pt}{0.400pt}}
\multiput(237.50,244.17)(-1.500,1.000){2}{\rule{0.361pt}{0.400pt}}
\multiput(233.92,246.61)(-0.462,0.447){3}{\rule{0.500pt}{0.108pt}}
\multiput(234.96,245.17)(-1.962,3.000){2}{\rule{0.250pt}{0.400pt}}
\put(231,248.67){\rule{0.482pt}{0.400pt}}
\multiput(232.00,248.17)(-1.000,1.000){2}{\rule{0.241pt}{0.400pt}}
\multiput(229.95,243.50)(-0.447,-2.248){3}{\rule{0.108pt}{1.567pt}}
\multiput(230.17,246.75)(-3.000,-7.748){2}{\rule{0.400pt}{0.783pt}}
\multiput(226.95,239.00)(-0.447,3.588){3}{\rule{0.108pt}{2.367pt}}
\multiput(227.17,239.00)(-3.000,12.088){2}{\rule{0.400pt}{1.183pt}}
\multiput(223.95,250.60)(-0.447,-1.802){3}{\rule{0.108pt}{1.300pt}}
\multiput(224.17,253.30)(-3.000,-6.302){2}{\rule{0.400pt}{0.650pt}}
\put(220,246.67){\rule{0.482pt}{0.400pt}}
\multiput(221.00,246.17)(-1.000,1.000){2}{\rule{0.241pt}{0.400pt}}
\multiput(218.95,248.00)(-0.447,6.937){3}{\rule{0.108pt}{4.367pt}}
\multiput(219.17,248.00)(-3.000,22.937){2}{\rule{0.400pt}{2.183pt}}
\put(279.0,243.0){\rule[-0.200pt]{0.723pt}{0.400pt}}
\put(879,672){\makebox(0,0)[r]{ { $P_z^c(m_l)$ } }}
\multiput(899,672)(20.756,0.000){5}{\usebox{\plotpoint}}
\put(999,672){\usebox{\plotpoint}}
\put(1044,317){\usebox{\plotpoint}}
\multiput(1044,317)(-1.634,-20.691){2}{\usebox{\plotpoint}}
\put(1040.18,275.71){\usebox{\plotpoint}}
\put(1031.71,270.98){\usebox{\plotpoint}}
\put(1024.04,261.84){\usebox{\plotpoint}}
\put(1012.71,257.00){\usebox{\plotpoint}}
\put(996.45,245.68){\usebox{\plotpoint}}
\put(982.49,246.82){\usebox{\plotpoint}}
\put(970.53,254.36){\usebox{\plotpoint}}
\put(961.89,251.57){\usebox{\plotpoint}}
\put(951.45,261.52){\usebox{\plotpoint}}
\put(942.58,275.61){\usebox{\plotpoint}}
\put(932.67,272.46){\usebox{\plotpoint}}
\put(924.58,256.68){\usebox{\plotpoint}}
\put(912.59,249.98){\usebox{\plotpoint}}
\put(899.69,248.92){\usebox{\plotpoint}}
\put(884.41,242.06){\usebox{\plotpoint}}
\put(869.18,239.37){\usebox{\plotpoint}}
\put(856.66,243.44){\usebox{\plotpoint}}
\put(842.89,240.96){\usebox{\plotpoint}}
\put(825.28,237.45){\usebox{\plotpoint}}
\put(810.20,234.26){\usebox{\plotpoint}}
\put(798.81,241.60){\usebox{\plotpoint}}
\put(784.13,238.67){\usebox{\plotpoint}}
\put(766.83,235.74){\usebox{\plotpoint}}
\put(752.41,232.41){\usebox{\plotpoint}}
\put(737.01,232.02){\usebox{\plotpoint}}
\put(726.34,229.88){\usebox{\plotpoint}}
\put(712.32,227.76){\usebox{\plotpoint}}
\put(694.52,226.68){\usebox{\plotpoint}}
\put(678.88,228.04){\usebox{\plotpoint}}
\put(660.44,226.00){\usebox{\plotpoint}}
\put(643.99,225.01){\usebox{\plotpoint}}
\put(626.57,226.57){\usebox{\plotpoint}}
\put(612.24,224.61){\usebox{\plotpoint}}
\put(602.06,228.88){\usebox{\plotpoint}}
\put(587.23,226.77){\usebox{\plotpoint}}
\put(570.80,228.39){\usebox{\plotpoint}}
\put(557.14,230.57){\usebox{\plotpoint}}
\put(545.89,230.34){\usebox{\plotpoint}}
\put(533.27,234.05){\usebox{\plotpoint}}
\put(521.25,233.75){\usebox{\plotpoint}}
\put(503.27,237.64){\usebox{\plotpoint}}
\put(485.32,235.68){\usebox{\plotpoint}}
\put(470.01,236.33){\usebox{\plotpoint}}
\put(454.75,234.58){\usebox{\plotpoint}}
\put(439.58,237.00){\usebox{\plotpoint}}
\put(424.50,241.25){\usebox{\plotpoint}}
\put(411.04,242.22){\usebox{\plotpoint}}
\put(396.72,242.91){\usebox{\plotpoint}}
\put(379.92,241.92){\usebox{\plotpoint}}
\put(366.75,243.00){\usebox{\plotpoint}}
\put(351.49,248.35){\usebox{\plotpoint}}
\put(334.62,247.62){\usebox{\plotpoint}}
\put(320.80,252.20){\usebox{\plotpoint}}
\put(306.66,252.33){\usebox{\plotpoint}}
\put(295.23,251.38){\usebox{\plotpoint}}
\put(283.49,255.94){\usebox{\plotpoint}}
\put(274.22,253.76){\usebox{\plotpoint}}
\put(261.27,252.58){\usebox{\plotpoint}}
\put(248.36,257.28){\usebox{\plotpoint}}
\put(230.53,258.16){\usebox{\plotpoint}}
\put(219.31,261.68){\usebox{\plotpoint}}
\put(217,274){\usebox{\plotpoint}}
\end{picture}

%% file: delta.1.215.140.n200.tex
\setlength{\unitlength}{0.240900pt}
\ifx\plotpoint\undefined\newsavebox{\plotpoint}\fi
\sbox{\plotpoint}{\rule[-0.200pt]{0.400pt}{0.400pt}}%
\begin{picture}(1500,900)(0,0)
\font\gnuplot=cmr10 at 10pt
\gnuplot
\sbox{\plotpoint}{\rule[-0.200pt]{0.400pt}{0.400pt}}%
\put(201.0,206.0){\rule[-0.200pt]{4.818pt}{0.400pt}}
\put(181,206){\makebox(0,0)[r]{0.001}}
\put(1419.0,206.0){\rule[-0.200pt]{4.818pt}{0.400pt}}
\put(201.0,326.0){\rule[-0.200pt]{4.818pt}{0.400pt}}
\put(181,326){\makebox(0,0)[r]{0.01}}
\put(1419.0,326.0){\rule[-0.200pt]{4.818pt}{0.400pt}}
\put(201.0,445.0){\rule[-0.200pt]{4.818pt}{0.400pt}}
\put(181,445){\makebox(0,0)[r]{0.1}}
\put(1419.0,445.0){\rule[-0.200pt]{4.818pt}{0.400pt}}
\put(201.0,564.0){\rule[-0.200pt]{4.818pt}{0.400pt}}
\put(181,564){\makebox(0,0)[r]{1}}
\put(1419.0,564.0){\rule[-0.200pt]{4.818pt}{0.400pt}}
\put(201.0,684.0){\rule[-0.200pt]{4.818pt}{0.400pt}}
\put(181,684){\makebox(0,0)[r]{10}}
\put(1419.0,684.0){\rule[-0.200pt]{4.818pt}{0.400pt}}
\put(201.0,803.0){\rule[-0.200pt]{4.818pt}{0.400pt}}
\put(181,803){\makebox(0,0)[r]{100}}
\put(1419.0,803.0){\rule[-0.200pt]{4.818pt}{0.400pt}}
\put(204.0,123.0){\rule[-0.200pt]{0.400pt}{4.818pt}}
\put(204,82){\makebox(0,0){0}}
\put(204.0,840.0){\rule[-0.200pt]{0.400pt}{4.818pt}}
\put(513.0,123.0){\rule[-0.200pt]{0.400pt}{4.818pt}}
\put(513,82){\makebox(0,0){50}}
\put(513.0,840.0){\rule[-0.200pt]{0.400pt}{4.818pt}}
\put(822.0,123.0){\rule[-0.200pt]{0.400pt}{4.818pt}}
\put(822,82){\makebox(0,0){100}}
\put(822.0,840.0){\rule[-0.200pt]{0.400pt}{4.818pt}}
\put(1130.0,123.0){\rule[-0.200pt]{0.400pt}{4.818pt}}
\put(1130,82){\makebox(0,0){150}}
\put(1130.0,840.0){\rule[-0.200pt]{0.400pt}{4.818pt}}
\put(1439.0,123.0){\rule[-0.200pt]{0.400pt}{4.818pt}}
\put(1439,82){\makebox(0,0){200}}
\put(1439.0,840.0){\rule[-0.200pt]{0.400pt}{4.818pt}}
\put(201.0,123.0){\rule[-0.200pt]{298.234pt}{0.400pt}}
\put(1439.0,123.0){\rule[-0.200pt]{0.400pt}{177.543pt}}
\put(201.0,860.0){\rule[-0.200pt]{298.234pt}{0.400pt}}
\put(40,651){\makebox(0,0){ $ \delta L_z $ }}
\put(820,21){\makebox(0,0){ Kick Number }}
\put(201.0,123.0){\rule[-0.200pt]{0.400pt}{177.543pt}}
\put(201,184){\usebox{\plotpoint}}
\multiput(201.58,184.00)(0.493,1.805){23}{\rule{0.119pt}{1.515pt}}
\multiput(200.17,184.00)(13.000,42.855){2}{\rule{0.400pt}{0.758pt}}
\multiput(214.58,230.00)(0.492,1.918){21}{\rule{0.119pt}{1.600pt}}
\multiput(213.17,230.00)(12.000,41.679){2}{\rule{0.400pt}{0.800pt}}
\multiput(226.58,275.00)(0.493,1.805){23}{\rule{0.119pt}{1.515pt}}
\multiput(225.17,275.00)(13.000,42.855){2}{\rule{0.400pt}{0.758pt}}
\multiput(239.58,321.00)(0.492,1.961){21}{\rule{0.119pt}{1.633pt}}
\multiput(238.17,321.00)(12.000,42.610){2}{\rule{0.400pt}{0.817pt}}
\multiput(251.58,367.00)(0.493,1.765){23}{\rule{0.119pt}{1.485pt}}
\multiput(250.17,367.00)(13.000,41.919){2}{\rule{0.400pt}{0.742pt}}
\multiput(264.58,412.00)(0.492,1.961){21}{\rule{0.119pt}{1.633pt}}
\multiput(263.17,412.00)(12.000,42.610){2}{\rule{0.400pt}{0.817pt}}
\multiput(276.58,458.00)(0.493,1.765){23}{\rule{0.119pt}{1.485pt}}
\multiput(275.17,458.00)(13.000,41.919){2}{\rule{0.400pt}{0.742pt}}
\multiput(289.58,503.00)(0.492,1.961){21}{\rule{0.119pt}{1.633pt}}
\multiput(288.17,503.00)(12.000,42.610){2}{\rule{0.400pt}{0.817pt}}
\multiput(301.58,549.00)(0.493,1.765){23}{\rule{0.119pt}{1.485pt}}
\multiput(300.17,549.00)(13.000,41.919){2}{\rule{0.400pt}{0.742pt}}
\multiput(314.58,594.00)(0.492,1.961){21}{\rule{0.119pt}{1.633pt}}
\multiput(313.17,594.00)(12.000,42.610){2}{\rule{0.400pt}{0.817pt}}
\multiput(326.58,640.00)(0.493,1.765){23}{\rule{0.119pt}{1.485pt}}
\multiput(325.17,640.00)(13.000,41.919){2}{\rule{0.400pt}{0.742pt}}
\multiput(339.58,685.00)(0.492,1.961){21}{\rule{0.119pt}{1.633pt}}
\multiput(338.17,685.00)(12.000,42.610){2}{\rule{0.400pt}{0.817pt}}
\multiput(351.58,731.00)(0.493,1.805){23}{\rule{0.119pt}{1.515pt}}
\multiput(350.17,731.00)(13.000,42.855){2}{\rule{0.400pt}{0.758pt}}
\multiput(364.58,777.00)(0.492,1.918){21}{\rule{0.119pt}{1.600pt}}
\multiput(363.17,777.00)(12.000,41.679){2}{\rule{0.400pt}{0.800pt}}
\multiput(376.58,822.00)(0.491,1.955){17}{\rule{0.118pt}{1.620pt}}
\multiput(375.17,822.00)(10.000,34.638){2}{\rule{0.400pt}{0.810pt}}
\sbox{\plotpoint}{\rule[-0.500pt]{1.000pt}{1.000pt}}%
\put(201,445){\usebox{\plotpoint}}
\put(201.00,445.00){\usebox{\plotpoint}}
\put(221.76,445.00){\usebox{\plotpoint}}
\put(242.51,445.00){\usebox{\plotpoint}}
\put(263.27,445.00){\usebox{\plotpoint}}
\put(284.02,445.00){\usebox{\plotpoint}}
\put(304.78,445.00){\usebox{\plotpoint}}
\put(325.53,445.00){\usebox{\plotpoint}}
\put(346.29,445.00){\usebox{\plotpoint}}
\put(367.04,445.00){\usebox{\plotpoint}}
\put(387.80,445.00){\usebox{\plotpoint}}
\put(408.55,445.00){\usebox{\plotpoint}}
\put(429.31,445.00){\usebox{\plotpoint}}
\put(450.07,445.00){\usebox{\plotpoint}}
\put(470.82,445.00){\usebox{\plotpoint}}
\put(491.58,445.00){\usebox{\plotpoint}}
\put(512.33,445.00){\usebox{\plotpoint}}
\put(533.09,445.00){\usebox{\plotpoint}}
\put(553.84,445.00){\usebox{\plotpoint}}
\put(574.60,445.00){\usebox{\plotpoint}}
\put(595.35,445.00){\usebox{\plotpoint}}
\put(616.11,445.00){\usebox{\plotpoint}}
\put(636.87,445.00){\usebox{\plotpoint}}
\put(657.62,445.00){\usebox{\plotpoint}}
\put(678.38,445.00){\usebox{\plotpoint}}
\put(699.13,445.00){\usebox{\plotpoint}}
\put(719.89,445.00){\usebox{\plotpoint}}
\put(740.64,445.00){\usebox{\plotpoint}}
\put(761.40,445.00){\usebox{\plotpoint}}
\put(782.15,445.00){\usebox{\plotpoint}}
\put(802.91,445.00){\usebox{\plotpoint}}
\put(823.66,445.00){\usebox{\plotpoint}}
\put(844.42,445.00){\usebox{\plotpoint}}
\put(865.18,445.00){\usebox{\plotpoint}}
\put(885.93,445.00){\usebox{\plotpoint}}
\put(906.69,445.00){\usebox{\plotpoint}}
\put(927.44,445.00){\usebox{\plotpoint}}
\put(948.20,445.00){\usebox{\plotpoint}}
\put(968.95,445.00){\usebox{\plotpoint}}
\put(989.71,445.00){\usebox{\plotpoint}}
\put(1010.46,445.00){\usebox{\plotpoint}}
\put(1031.22,445.00){\usebox{\plotpoint}}
\put(1051.98,445.00){\usebox{\plotpoint}}
\put(1072.73,445.00){\usebox{\plotpoint}}
\put(1093.49,445.00){\usebox{\plotpoint}}
\put(1114.24,445.00){\usebox{\plotpoint}}
\put(1135.00,445.00){\usebox{\plotpoint}}
\put(1155.75,445.00){\usebox{\plotpoint}}
\put(1176.51,445.00){\usebox{\plotpoint}}
\put(1197.26,445.00){\usebox{\plotpoint}}
\put(1218.02,445.00){\usebox{\plotpoint}}
\put(1238.77,445.00){\usebox{\plotpoint}}
\put(1259.53,445.00){\usebox{\plotpoint}}
\put(1280.29,445.00){\usebox{\plotpoint}}
\put(1301.04,445.00){\usebox{\plotpoint}}
\put(1321.80,445.00){\usebox{\plotpoint}}
\put(1342.55,445.00){\usebox{\plotpoint}}
\put(1363.31,445.00){\usebox{\plotpoint}}
\put(1384.06,445.00){\usebox{\plotpoint}}
\put(1404.82,445.00){\usebox{\plotpoint}}
\put(1425.57,445.00){\usebox{\plotpoint}}
\put(1439,445){\usebox{\plotpoint}}
\put(201,705){\usebox{\plotpoint}}
\put(201.00,705.00){\usebox{\plotpoint}}
\put(242.51,705.00){\usebox{\plotpoint}}
\put(284.02,705.00){\usebox{\plotpoint}}
\put(325.53,705.00){\usebox{\plotpoint}}
\put(367.04,705.00){\usebox{\plotpoint}}
\put(408.56,705.00){\usebox{\plotpoint}}
\put(450.07,705.00){\usebox{\plotpoint}}
\put(491.58,705.00){\usebox{\plotpoint}}
\put(533.09,705.00){\usebox{\plotpoint}}
\put(574.60,705.00){\usebox{\plotpoint}}
\put(616.11,705.00){\usebox{\plotpoint}}
\put(657.62,705.00){\usebox{\plotpoint}}
\put(699.13,705.00){\usebox{\plotpoint}}
\put(740.64,705.00){\usebox{\plotpoint}}
\put(782.15,705.00){\usebox{\plotpoint}}
\put(823.66,705.00){\usebox{\plotpoint}}
\put(865.18,705.00){\usebox{\plotpoint}}
\put(906.69,705.00){\usebox{\plotpoint}}
\put(948.20,705.00){\usebox{\plotpoint}}
\put(989.71,705.00){\usebox{\plotpoint}}
\put(1031.22,705.00){\usebox{\plotpoint}}
\put(1072.73,705.00){\usebox{\plotpoint}}
\put(1114.24,705.00){\usebox{\plotpoint}}
\put(1155.75,705.00){\usebox{\plotpoint}}
\put(1197.26,705.00){\usebox{\plotpoint}}
\put(1238.77,705.00){\usebox{\plotpoint}}
\put(1280.29,705.00){\usebox{\plotpoint}}
\put(1321.80,705.00){\usebox{\plotpoint}}
\put(1363.31,705.00){\usebox{\plotpoint}}
\put(1404.82,705.00){\usebox{\plotpoint}}
\put(1439,705){\usebox{\plotpoint}}
\put(204,136){\circle*{12}}
\put(216,193){\circle*{12}}
\put(223,243){\circle*{12}}
\put(229,337){\circle*{12}}
\put(235,230){\circle*{12}}
\put(241,348){\circle*{12}}
\put(247,221){\circle*{12}}
\put(253,404){\circle*{12}}
\put(260,274){\circle*{12}}
\put(266,440){\circle*{12}}
\put(272,451){\circle*{12}}
\put(278,470){\circle*{12}}
\put(284,512){\circle*{12}}
\put(291,540){\circle*{12}}
\put(297,573){\circle*{12}}
\put(303,565){\circle*{12}}
\put(309,565){\circle*{12}}
\put(315,456){\circle*{12}}
\put(321,421){\circle*{12}}
\put(328,586){\circle*{12}}
\put(334,521){\circle*{12}}
\put(340,572){\circle*{12}}
\put(346,561){\circle*{12}}
\put(352,533){\circle*{12}}
\put(358,531){\circle*{12}}
\put(365,584){\circle*{12}}
\put(371,599){\circle*{12}}
\put(377,622){\circle*{12}}
\put(383,582){\circle*{12}}
\put(389,492){\circle*{12}}
\put(395,516){\circle*{12}}
\put(402,333){\circle*{12}}
\put(408,613){\circle*{12}}
\put(414,559){\circle*{12}}
\put(420,517){\circle*{12}}
\put(426,521){\circle*{12}}
\put(433,565){\circle*{12}}
\put(439,586){\circle*{12}}
\put(445,570){\circle*{12}}
\put(451,407){\circle*{12}}
\put(457,444){\circle*{12}}
\put(463,385){\circle*{12}}
\put(470,554){\circle*{12}}
\put(476,565){\circle*{12}}
\put(482,613){\circle*{12}}
\put(488,330){\circle*{12}}
\put(494,588){\circle*{12}}
\put(500,541){\circle*{12}}
\put(507,560){\circle*{12}}
\put(513,536){\circle*{12}}
\put(519,576){\circle*{12}}
\put(525,614){\circle*{12}}
\put(531,617){\circle*{12}}
\put(538,592){\circle*{12}}
\put(544,596){\circle*{12}}
\put(550,567){\circle*{12}}
\put(556,644){\circle*{12}}
\put(562,643){\circle*{12}}
\put(568,654){\circle*{12}}
\put(575,630){\circle*{12}}
\put(581,619){\circle*{12}}
\put(587,601){\circle*{12}}
\put(593,557){\circle*{12}}
\put(599,642){\circle*{12}}
\put(605,637){\circle*{12}}
\put(612,644){\circle*{12}}
\put(618,644){\circle*{12}}
\put(624,590){\circle*{12}}
\put(630,436){\circle*{12}}
\put(636,611){\circle*{12}}
\put(642,642){\circle*{12}}
\put(649,659){\circle*{12}}
\put(655,660){\circle*{12}}
\put(661,655){\circle*{12}}
\put(667,643){\circle*{12}}
\put(673,519){\circle*{12}}
\put(680,627){\circle*{12}}
\put(686,635){\circle*{12}}
\put(692,634){\circle*{12}}
\put(698,623){\circle*{12}}
\put(704,606){\circle*{12}}
\put(710,573){\circle*{12}}
\put(717,585){\circle*{12}}
\put(723,623){\circle*{12}}
\put(729,629){\circle*{12}}
\put(735,637){\circle*{12}}
\put(741,642){\circle*{12}}
\put(747,635){\circle*{12}}
\put(754,625){\circle*{12}}
\put(760,570){\circle*{12}}
\put(766,592){\circle*{12}}
\put(772,639){\circle*{12}}
\put(778,648){\circle*{12}}
\put(784,641){\circle*{12}}
\put(791,632){\circle*{12}}
\put(797,583){\circle*{12}}
\put(803,515){\circle*{12}}
\put(809,600){\circle*{12}}
\put(815,628){\circle*{12}}
\put(822,650){\circle*{12}}
\put(828,646){\circle*{12}}
\put(834,641){\circle*{12}}
\put(840,640){\circle*{12}}
\put(846,590){\circle*{12}}
\put(852,552){\circle*{12}}
\put(859,593){\circle*{12}}
\put(865,641){\circle*{12}}
\put(871,648){\circle*{12}}
\put(877,648){\circle*{12}}
\put(883,646){\circle*{12}}
\put(889,596){\circle*{12}}
\put(896,494){\circle*{12}}
\put(902,617){\circle*{12}}
\put(908,648){\circle*{12}}
\put(914,668){\circle*{12}}
\put(920,667){\circle*{12}}
\put(927,655){\circle*{12}}
\put(933,638){\circle*{12}}
\put(939,516){\circle*{12}}
\put(945,612){\circle*{12}}
\put(951,663){\circle*{12}}
\put(957,672){\circle*{12}}
\put(964,672){\circle*{12}}
\put(970,657){\circle*{12}}
\put(976,589){\circle*{12}}
\put(982,602){\circle*{12}}
\put(988,658){\circle*{12}}
\put(994,682){\circle*{12}}
\put(1001,684){\circle*{12}}
\put(1007,675){\circle*{12}}
\put(1013,656){\circle*{12}}
\put(1019,422){\circle*{12}}
\put(1025,616){\circle*{12}}
\put(1031,667){\circle*{12}}
\put(1038,674){\circle*{12}}
\put(1044,674){\circle*{12}}
\put(1050,658){\circle*{12}}
\put(1056,607){\circle*{12}}
\put(1062,584){\circle*{12}}
\put(1069,653){\circle*{12}}
\put(1075,670){\circle*{12}}
\put(1081,677){\circle*{12}}
\put(1087,673){\circle*{12}}
\put(1093,656){\circle*{12}}
\put(1099,594){\circle*{12}}
\put(1106,595){\circle*{12}}
\put(1112,658){\circle*{12}}
\put(1118,670){\circle*{12}}
\put(1124,669){\circle*{12}}
\put(1130,651){\circle*{12}}
\put(1136,620){\circle*{12}}
\put(1143,603){\circle*{12}}
\put(1149,648){\circle*{12}}
\put(1155,668){\circle*{12}}
\put(1161,669){\circle*{12}}
\put(1167,664){\circle*{12}}
\put(1173,642){\circle*{12}}
\put(1180,574){\circle*{12}}
\put(1186,595){\circle*{12}}
\put(1192,650){\circle*{12}}
\put(1198,661){\circle*{12}}
\put(1204,665){\circle*{12}}
\put(1211,651){\circle*{12}}
\put(1217,625){\circle*{12}}
\put(1223,528){\circle*{12}}
\put(1229,633){\circle*{12}}
\put(1235,655){\circle*{12}}
\put(1241,663){\circle*{12}}
\put(1248,658){\circle*{12}}
\put(1254,647){\circle*{12}}
\put(1260,584){\circle*{12}}
\put(1266,546){\circle*{12}}
\put(1272,628){\circle*{12}}
\put(1278,644){\circle*{12}}
\put(1285,657){\circle*{12}}
\put(1291,636){\circle*{12}}
\put(1297,590){\circle*{12}}
\put(1303,581){\circle*{12}}
\put(1309,620){\circle*{12}}
\put(1316,647){\circle*{12}}
\put(1322,646){\circle*{12}}
\put(1328,650){\circle*{12}}
\put(1334,624){\circle*{12}}
\put(1340,606){\circle*{12}}
\put(1346,437){\circle*{12}}
\put(1353,603){\circle*{12}}
\put(1359,620){\circle*{12}}
\put(1365,638){\circle*{12}}
\put(1371,626){\circle*{12}}
\put(1377,633){\circle*{12}}
\put(1383,600){\circle*{12}}
\put(1390,589){\circle*{12}}
\put(1396,604){\circle*{12}}
\put(1402,625){\circle*{12}}
\put(1408,647){\circle*{12}}
\put(1414,648){\circle*{12}}
\put(1420,654){\circle*{12}}
\put(1427,628){\circle*{12}}
\put(1433,604){\circle*{12}}
\put(1439,571){\circle*{12}}
\sbox{\plotpoint}{\rule[-0.200pt]{0.400pt}{0.400pt}}%
\put(204,205){\circle{12}}
\put(210,210){\circle{12}}
\put(216,158){\circle{12}}
\put(223,269){\circle{12}}
\put(229,237){\circle{12}}
\put(241,236){\circle{12}}
\put(247,246){\circle{12}}
\put(253,249){\circle{12}}
\put(260,241){\circle{12}}
\put(266,271){\circle{12}}
\put(272,226){\circle{12}}
\put(278,310){\circle{12}}
\put(284,308){\circle{12}}
\put(291,207){\circle{12}}
\put(297,290){\circle{12}}
\put(303,312){\circle{12}}
\put(309,142){\circle{12}}
\put(315,295){\circle{12}}
\put(321,326){\circle{12}}
\put(328,334){\circle{12}}
\put(334,250){\circle{12}}
\put(340,330){\circle{12}}
\put(346,344){\circle{12}}
\put(352,258){\circle{12}}
\put(358,363){\circle{12}}
\put(365,358){\circle{12}}
\put(371,325){\circle{12}}
\put(377,341){\circle{12}}
\put(383,376){\circle{12}}
\put(389,306){\circle{12}}
\put(395,306){\circle{12}}
\put(402,381){\circle{12}}
\put(408,373){\circle{12}}
\put(414,286){\circle{12}}
\put(420,363){\circle{12}}
\put(426,392){\circle{12}}
\put(433,283){\circle{12}}
\put(439,382){\circle{12}}
\put(445,390){\circle{12}}
\put(451,376){\circle{12}}
\put(457,375){\circle{12}}
\put(463,397){\circle{12}}
\put(470,370){\circle{12}}
\put(476,286){\circle{12}}
\put(482,411){\circle{12}}
\put(488,389){\circle{12}}
\put(494,309){\circle{12}}
\put(500,389){\circle{12}}
\put(507,417){\circle{12}}
\put(513,300){\circle{12}}
\put(519,372){\circle{12}}
\put(525,418){\circle{12}}
\put(531,399){\circle{12}}
\put(538,383){\circle{12}}
\put(544,402){\circle{12}}
\put(550,414){\circle{12}}
\put(556,294){\circle{12}}
\put(562,420){\circle{12}}
\put(568,405){\circle{12}}
\put(575,390){\circle{12}}
\put(581,411){\circle{12}}
\put(587,426){\circle{12}}
\put(593,369){\circle{12}}
\put(599,332){\circle{12}}
\put(605,439){\circle{12}}
\put(612,406){\circle{12}}
\put(618,331){\circle{12}}
\put(624,410){\circle{12}}
\put(630,440){\circle{12}}
\put(636,297){\circle{12}}
\put(642,407){\circle{12}}
\put(649,430){\circle{12}}
\put(655,420){\circle{12}}
\put(661,419){\circle{12}}
\put(667,424){\circle{12}}
\put(673,422){\circle{12}}
\put(680,289){\circle{12}}
\put(686,447){\circle{12}}
\put(692,415){\circle{12}}
\put(698,385){\circle{12}}
\put(704,426){\circle{12}}
\put(710,448){\circle{12}}
\put(717,340){\circle{12}}
\put(723,360){\circle{12}}
\put(729,453){\circle{12}}
\put(735,429){\circle{12}}
\put(741,395){\circle{12}}
\put(747,425){\circle{12}}
\put(754,453){\circle{12}}
\put(760,296){\circle{12}}
\put(766,438){\circle{12}}
\put(772,438){\circle{12}}
\put(778,430){\circle{12}}
\put(784,437){\circle{12}}
\put(791,443){\circle{12}}
\put(797,424){\circle{12}}
\put(809,464){\circle{12}}
\put(815,433){\circle{12}}
\put(822,363){\circle{12}}
\put(828,438){\circle{12}}
\put(834,466){\circle{12}}
\put(840,323){\circle{12}}
\put(846,404){\circle{12}}
\put(852,464){\circle{12}}
\put(859,443){\circle{12}}
\put(865,427){\circle{12}}
\put(871,437){\circle{12}}
\put(877,463){\circle{12}}
\put(883,262){\circle{12}}
\put(889,461){\circle{12}}
\put(896,448){\circle{12}}
\put(902,440){\circle{12}}
\put(908,451){\circle{12}}
\put(914,468){\circle{12}}
\put(920,422){\circle{12}}
\put(927,330){\circle{12}}
\put(933,480){\circle{12}}
\put(939,447){\circle{12}}
\put(945,324){\circle{12}}
\put(951,444){\circle{12}}
\put(957,482){\circle{12}}
\put(964,328){\circle{12}}
\put(970,441){\circle{12}}
\put(976,471){\circle{12}}
\put(982,463){\circle{12}}
\put(988,452){\circle{12}}
\put(994,465){\circle{12}}
\put(1001,467){\circle{12}}
\put(1007,169){\circle{12}}
\put(1013,484){\circle{12}}
\put(1019,459){\circle{12}}
\put(1025,430){\circle{12}}
\put(1031,461){\circle{12}}
\put(1038,488){\circle{12}}
\put(1044,415){\circle{12}}
\put(1050,401){\circle{12}}
\put(1056,491){\circle{12}}
\put(1062,471){\circle{12}}
\put(1069,423){\circle{12}}
\put(1075,469){\circle{12}}
\put(1081,492){\circle{12}}
\put(1087,175){\circle{12}}
\put(1093,477){\circle{12}}
\put(1099,480){\circle{12}}
\put(1106,464){\circle{12}}
\put(1112,472){\circle{12}}
\put(1118,487){\circle{12}}
\put(1124,466){\circle{12}}
\put(1130,368){\circle{12}}
\put(1136,502){\circle{12}}
\put(1143,477){\circle{12}}
\put(1149,366){\circle{12}}
\put(1155,482){\circle{12}}
\put(1161,504){\circle{12}}
\put(1167,366){\circle{12}}
\put(1173,464){\circle{12}}
\put(1180,501){\circle{12}}
\put(1186,475){\circle{12}}
\put(1192,470){\circle{12}}
\put(1198,489){\circle{12}}
\put(1204,495){\circle{12}}
\put(1211,390){\circle{12}}
\put(1217,504){\circle{12}}
\put(1223,489){\circle{12}}
\put(1229,448){\circle{12}}
\put(1235,497){\circle{12}}
\put(1241,508){\circle{12}}
\put(1248,437){\circle{12}}
\put(1254,457){\circle{12}}
\put(1260,516){\circle{12}}
\put(1266,477){\circle{12}}
\put(1272,454){\circle{12}}
\put(1278,498){\circle{12}}
\put(1285,509){\circle{12}}
\put(1291,414){\circle{12}}
\put(1297,501){\circle{12}}
\put(1303,505){\circle{12}}
\put(1309,464){\circle{12}}
\put(1316,505){\circle{12}}
\put(1322,509){\circle{12}}
\put(1328,474){\circle{12}}
\put(1334,468){\circle{12}}
\put(1340,523){\circle{12}}
\put(1346,480){\circle{12}}
\put(1353,423){\circle{12}}
\put(1359,511){\circle{12}}
\put(1365,513){\circle{12}}
\put(1371,414){\circle{12}}
\put(1377,498){\circle{12}}
\put(1383,518){\circle{12}}
\put(1390,456){\circle{12}}
\put(1396,505){\circle{12}}
\put(1402,512){\circle{12}}
\put(1408,491){\circle{12}}
\put(1414,485){\circle{12}}
\put(1420,522){\circle{12}}
\put(1427,487){\circle{12}}
\put(1433,402){\circle{12}}
\put(1439,521){\circle{12}}
\put(210,483){\makebox(0,0){$+$}}
\put(216,480){\makebox(0,0){$+$}}
\put(223,593){\makebox(0,0){$+$}}
\put(229,599){\makebox(0,0){$+$}}
\put(235,543){\makebox(0,0){$+$}}
\put(241,467){\makebox(0,0){$+$}}
\put(247,650){\makebox(0,0){$+$}}
\put(253,626){\makebox(0,0){$+$}}
\put(260,701){\makebox(0,0){$+$}}
\put(266,674){\makebox(0,0){$+$}}
\put(272,631){\makebox(0,0){$+$}}
\put(278,611){\makebox(0,0){$+$}}
\put(284,565){\makebox(0,0){$+$}}
\put(291,687){\makebox(0,0){$+$}}
\put(297,646){\makebox(0,0){$+$}}
\put(303,780){\makebox(0,0){$+$}}
\put(309,770){\makebox(0,0){$+$}}
\put(315,737){\makebox(0,0){$+$}}
\put(321,734){\makebox(0,0){$+$}}
\put(328,749){\makebox(0,0){$+$}}
\put(334,658){\makebox(0,0){$+$}}
\put(340,794){\makebox(0,0){$+$}}
\put(346,776){\makebox(0,0){$+$}}
\put(352,749){\makebox(0,0){$+$}}
\put(358,669){\makebox(0,0){$+$}}
\put(365,804){\makebox(0,0){$+$}}
\put(371,797){\makebox(0,0){$+$}}
\put(377,827){\makebox(0,0){$+$}}
\put(383,812){\makebox(0,0){$+$}}
\put(389,794){\makebox(0,0){$+$}}
\put(395,771){\makebox(0,0){$+$}}
\put(402,795){\makebox(0,0){$+$}}
\put(408,803){\makebox(0,0){$+$}}
\put(414,818){\makebox(0,0){$+$}}
\put(420,813){\makebox(0,0){$+$}}
\put(426,794){\makebox(0,0){$+$}}
\put(433,769){\makebox(0,0){$+$}}
\put(439,795){\makebox(0,0){$+$}}
\put(445,795){\makebox(0,0){$+$}}
\put(451,825){\makebox(0,0){$+$}}
\put(457,823){\makebox(0,0){$+$}}
\put(463,808){\makebox(0,0){$+$}}
\put(470,798){\makebox(0,0){$+$}}
\put(476,610){\makebox(0,0){$+$}}
\put(482,805){\makebox(0,0){$+$}}
\put(488,808){\makebox(0,0){$+$}}
\put(494,821){\makebox(0,0){$+$}}
\put(500,818){\makebox(0,0){$+$}}
\put(507,816){\makebox(0,0){$+$}}
\put(513,816){\makebox(0,0){$+$}}
\put(519,805){\makebox(0,0){$+$}}
\put(525,792){\makebox(0,0){$+$}}
\put(531,729){\makebox(0,0){$+$}}
\put(538,808){\makebox(0,0){$+$}}
\put(544,807){\makebox(0,0){$+$}}
\put(550,822){\makebox(0,0){$+$}}
\put(556,822){\makebox(0,0){$+$}}
\put(562,787){\makebox(0,0){$+$}}
\put(568,785){\makebox(0,0){$+$}}
\put(575,783){\makebox(0,0){$+$}}
\put(581,807){\makebox(0,0){$+$}}
\put(587,813){\makebox(0,0){$+$}}
\put(593,816){\makebox(0,0){$+$}}
\put(599,787){\makebox(0,0){$+$}}
\put(605,786){\makebox(0,0){$+$}}
\put(612,798){\makebox(0,0){$+$}}
\put(618,810){\makebox(0,0){$+$}}
\put(624,818){\makebox(0,0){$+$}}
\put(630,824){\makebox(0,0){$+$}}
\put(636,777){\makebox(0,0){$+$}}
\put(642,752){\makebox(0,0){$+$}}
\put(649,767){\makebox(0,0){$+$}}
\put(655,802){\makebox(0,0){$+$}}
\put(661,793){\makebox(0,0){$+$}}
\put(667,794){\makebox(0,0){$+$}}
\put(673,781){\makebox(0,0){$+$}}
\put(680,766){\makebox(0,0){$+$}}
\put(686,811){\makebox(0,0){$+$}}
\put(692,809){\makebox(0,0){$+$}}
\put(698,763){\makebox(0,0){$+$}}
\put(704,773){\makebox(0,0){$+$}}
\put(710,792){\makebox(0,0){$+$}}
\put(717,780){\makebox(0,0){$+$}}
\put(723,793){\makebox(0,0){$+$}}
\put(729,317){\makebox(0,0){$+$}}
\put(735,773){\makebox(0,0){$+$}}
\put(741,799){\makebox(0,0){$+$}}
\put(747,822){\makebox(0,0){$+$}}
\put(754,802){\makebox(0,0){$+$}}
\put(760,799){\makebox(0,0){$+$}}
\put(766,693){\makebox(0,0){$+$}}
\put(772,806){\makebox(0,0){$+$}}
\put(778,809){\makebox(0,0){$+$}}
\put(784,815){\makebox(0,0){$+$}}
\put(791,794){\makebox(0,0){$+$}}
\put(797,791){\makebox(0,0){$+$}}
\put(803,781){\makebox(0,0){$+$}}
\put(809,810){\makebox(0,0){$+$}}
\put(815,817){\makebox(0,0){$+$}}
\put(822,825){\makebox(0,0){$+$}}
\put(828,781){\makebox(0,0){$+$}}
\put(834,758){\makebox(0,0){$+$}}
\put(840,778){\makebox(0,0){$+$}}
\put(846,810){\makebox(0,0){$+$}}
\put(852,802){\makebox(0,0){$+$}}
\put(859,797){\makebox(0,0){$+$}}
\put(865,782){\makebox(0,0){$+$}}
\put(871,778){\makebox(0,0){$+$}}
\put(877,821){\makebox(0,0){$+$}}
\put(883,819){\makebox(0,0){$+$}}
\put(889,781){\makebox(0,0){$+$}}
\put(896,781){\makebox(0,0){$+$}}
\put(902,796){\makebox(0,0){$+$}}
\put(908,794){\makebox(0,0){$+$}}
\put(914,807){\makebox(0,0){$+$}}
\put(920,751){\makebox(0,0){$+$}}
\put(927,746){\makebox(0,0){$+$}}
\put(933,788){\makebox(0,0){$+$}}
\put(939,823){\makebox(0,0){$+$}}
\put(945,810){\makebox(0,0){$+$}}
\put(951,807){\makebox(0,0){$+$}}
\put(957,744){\makebox(0,0){$+$}}
\put(964,797){\makebox(0,0){$+$}}
\put(970,800){\makebox(0,0){$+$}}
\put(976,815){\makebox(0,0){$+$}}
\put(982,806){\makebox(0,0){$+$}}
\put(988,804){\makebox(0,0){$+$}}
\put(994,539){\makebox(0,0){$+$}}
\put(1001,797){\makebox(0,0){$+$}}
\put(1007,806){\makebox(0,0){$+$}}
\put(1013,824){\makebox(0,0){$+$}}
\put(1019,809){\makebox(0,0){$+$}}
\put(1025,787){\makebox(0,0){$+$}}
\put(1031,725){\makebox(0,0){$+$}}
\put(1038,807){\makebox(0,0){$+$}}
\put(1044,805){\makebox(0,0){$+$}}
\put(1050,816){\makebox(0,0){$+$}}
\put(1056,800){\makebox(0,0){$+$}}
\put(1062,792){\makebox(0,0){$+$}}
\put(1069,732){\makebox(0,0){$+$}}
\put(1075,813){\makebox(0,0){$+$}}
\put(1081,812){\makebox(0,0){$+$}}
\put(1087,825){\makebox(0,0){$+$}}
\put(1093,794){\makebox(0,0){$+$}}
\put(1099,752){\makebox(0,0){$+$}}
\put(1106,739){\makebox(0,0){$+$}}
\put(1112,807){\makebox(0,0){$+$}}
\put(1118,795){\makebox(0,0){$+$}}
\put(1124,798){\makebox(0,0){$+$}}
\put(1130,779){\makebox(0,0){$+$}}
\put(1136,776){\makebox(0,0){$+$}}
\put(1143,818){\makebox(0,0){$+$}}
\put(1149,819){\makebox(0,0){$+$}}
\put(1155,782){\makebox(0,0){$+$}}
\put(1161,786){\makebox(0,0){$+$}}
\put(1167,793){\makebox(0,0){$+$}}
\put(1173,793){\makebox(0,0){$+$}}
\put(1180,806){\makebox(0,0){$+$}}
\put(1186,751){\makebox(0,0){$+$}}
\put(1192,739){\makebox(0,0){$+$}}
\put(1198,787){\makebox(0,0){$+$}}
\put(1204,822){\makebox(0,0){$+$}}
\put(1211,812){\makebox(0,0){$+$}}
\put(1217,812){\makebox(0,0){$+$}}
\put(1223,757){\makebox(0,0){$+$}}
\put(1229,788){\makebox(0,0){$+$}}
\put(1235,794){\makebox(0,0){$+$}}
\put(1241,814){\makebox(0,0){$+$}}
\put(1248,806){\makebox(0,0){$+$}}
\put(1254,807){\makebox(0,0){$+$}}
\put(1260,505){\makebox(0,0){$+$}}
\put(1266,778){\makebox(0,0){$+$}}
\put(1272,803){\makebox(0,0){$+$}}
\put(1278,822){\makebox(0,0){$+$}}
\put(1285,814){\makebox(0,0){$+$}}
\put(1291,802){\makebox(0,0){$+$}}
\put(1297,778){\makebox(0,0){$+$}}
\put(1303,798){\makebox(0,0){$+$}}
\put(1309,798){\makebox(0,0){$+$}}
\put(1316,816){\makebox(0,0){$+$}}
\put(1322,817){\makebox(0,0){$+$}}
\put(1328,816){\makebox(0,0){$+$}}
\put(1334,815){\makebox(0,0){$+$}}
\put(1340,773){\makebox(0,0){$+$}}
\put(1346,775){\makebox(0,0){$+$}}
\put(1353,802){\makebox(0,0){$+$}}
\put(1359,812){\makebox(0,0){$+$}}
\put(1365,817){\makebox(0,0){$+$}}
\put(1371,819){\makebox(0,0){$+$}}
\put(1377,806){\makebox(0,0){$+$}}
\put(1383,689){\makebox(0,0){$+$}}
\put(1390,709){\makebox(0,0){$+$}}
\put(1396,810){\makebox(0,0){$+$}}
\put(1402,808){\makebox(0,0){$+$}}
\put(1408,811){\makebox(0,0){$+$}}
\put(1414,784){\makebox(0,0){$+$}}
\put(1420,605){\makebox(0,0){$+$}}
\put(1427,761){\makebox(0,0){$+$}}
\put(1433,820){\makebox(0,0){$+$}}
\put(1439,819){\makebox(0,0){$+$}}
\end{picture}

%% file: delta.1.215.dvsl.tex
\setlength{\unitlength}{0.240900pt}
\ifx\plotpoint\undefined\newsavebox{\plotpoint}\fi
\sbox{\plotpoint}{\rule[-0.200pt]{0.400pt}{0.400pt}}%
\begin{picture}(1500,900)(0,0)
\font\gnuplot=cmr10 at 10pt
\gnuplot
\sbox{\plotpoint}{\rule[-0.200pt]{0.400pt}{0.400pt}}%
\put(201.0,223.0){\rule[-0.200pt]{4.818pt}{0.400pt}}
\put(181,223){\makebox(0,0)[r]{0.001}}
\put(1419.0,223.0){\rule[-0.200pt]{4.818pt}{0.400pt}}
\put(201.0,365.0){\rule[-0.200pt]{4.818pt}{0.400pt}}
\put(181,365){\makebox(0,0)[r]{0.01}}
\put(1419.0,365.0){\rule[-0.200pt]{4.818pt}{0.400pt}}
\put(201.0,507.0){\rule[-0.200pt]{4.818pt}{0.400pt}}
\put(181,507){\makebox(0,0)[r]{0.1}}
\put(1419.0,507.0){\rule[-0.200pt]{4.818pt}{0.400pt}}
\put(201.0,650.0){\rule[-0.200pt]{4.818pt}{0.400pt}}
\put(181,650){\makebox(0,0)[r]{1}}
\put(1419.0,650.0){\rule[-0.200pt]{4.818pt}{0.400pt}}
\put(201.0,792.0){\rule[-0.200pt]{4.818pt}{0.400pt}}
\put(181,792){\makebox(0,0)[r]{10}}
\put(1419.0,792.0){\rule[-0.200pt]{4.818pt}{0.400pt}}
\put(218.0,123.0){\rule[-0.200pt]{0.400pt}{4.818pt}}
\put(218,82){\makebox(0,0){0}}
\put(218.0,840.0){\rule[-0.200pt]{0.400pt}{4.818pt}}
\put(393.0,123.0){\rule[-0.200pt]{0.400pt}{4.818pt}}
\put(393,82){\makebox(0,0){5}}
\put(393.0,840.0){\rule[-0.200pt]{0.400pt}{4.818pt}}
\put(567.0,123.0){\rule[-0.200pt]{0.400pt}{4.818pt}}
\put(567,82){\makebox(0,0){10}}
\put(567.0,840.0){\rule[-0.200pt]{0.400pt}{4.818pt}}
\put(742.0,123.0){\rule[-0.200pt]{0.400pt}{4.818pt}}
\put(742,82){\makebox(0,0){15}}
\put(742.0,840.0){\rule[-0.200pt]{0.400pt}{4.818pt}}
\put(916.0,123.0){\rule[-0.200pt]{0.400pt}{4.818pt}}
\put(916,82){\makebox(0,0){20}}
\put(916.0,840.0){\rule[-0.200pt]{0.400pt}{4.818pt}}
\put(1090.0,123.0){\rule[-0.200pt]{0.400pt}{4.818pt}}
\put(1090,82){\makebox(0,0){25}}
\put(1090.0,840.0){\rule[-0.200pt]{0.400pt}{4.818pt}}
\put(1265.0,123.0){\rule[-0.200pt]{0.400pt}{4.818pt}}
\put(1265,82){\makebox(0,0){30}}
\put(1265.0,840.0){\rule[-0.200pt]{0.400pt}{4.818pt}}
\put(1439.0,123.0){\rule[-0.200pt]{0.400pt}{4.818pt}}
\put(1439,82){\makebox(0,0){35}}
\put(1439.0,840.0){\rule[-0.200pt]{0.400pt}{4.818pt}}
\put(201.0,123.0){\rule[-0.200pt]{298.234pt}{0.400pt}}
\put(1439.0,123.0){\rule[-0.200pt]{0.400pt}{177.543pt}}
\put(201.0,860.0){\rule[-0.200pt]{298.234pt}{0.400pt}}
\put(40,651){\makebox(0,0)[tl] { $ \delta L_z $ }}
\put(820,21){\makebox(0,0){ { Kick Number }}}
\put(201.0,123.0){\rule[-0.200pt]{0.400pt}{177.543pt}}
\put(201,317){\usebox{\plotpoint}}
\multiput(201.00,317.59)(0.728,0.489){15}{\rule{0.678pt}{0.118pt}}
\multiput(201.00,316.17)(11.593,9.000){2}{\rule{0.339pt}{0.400pt}}
\multiput(214.00,326.58)(0.600,0.491){17}{\rule{0.580pt}{0.118pt}}
\multiput(214.00,325.17)(10.796,10.000){2}{\rule{0.290pt}{0.400pt}}
\multiput(226.00,336.59)(0.728,0.489){15}{\rule{0.678pt}{0.118pt}}
\multiput(226.00,335.17)(11.593,9.000){2}{\rule{0.339pt}{0.400pt}}
\multiput(239.00,345.58)(0.600,0.491){17}{\rule{0.580pt}{0.118pt}}
\multiput(239.00,344.17)(10.796,10.000){2}{\rule{0.290pt}{0.400pt}}
\multiput(251.00,355.58)(0.652,0.491){17}{\rule{0.620pt}{0.118pt}}
\multiput(251.00,354.17)(11.713,10.000){2}{\rule{0.310pt}{0.400pt}}
\multiput(264.00,365.59)(0.669,0.489){15}{\rule{0.633pt}{0.118pt}}
\multiput(264.00,364.17)(10.685,9.000){2}{\rule{0.317pt}{0.400pt}}
\multiput(276.00,374.58)(0.652,0.491){17}{\rule{0.620pt}{0.118pt}}
\multiput(276.00,373.17)(11.713,10.000){2}{\rule{0.310pt}{0.400pt}}
\multiput(289.00,384.58)(0.600,0.491){17}{\rule{0.580pt}{0.118pt}}
\multiput(289.00,383.17)(10.796,10.000){2}{\rule{0.290pt}{0.400pt}}
\multiput(301.00,394.59)(0.728,0.489){15}{\rule{0.678pt}{0.118pt}}
\multiput(301.00,393.17)(11.593,9.000){2}{\rule{0.339pt}{0.400pt}}
\multiput(314.00,403.58)(0.600,0.491){17}{\rule{0.580pt}{0.118pt}}
\multiput(314.00,402.17)(10.796,10.000){2}{\rule{0.290pt}{0.400pt}}
\multiput(326.00,413.59)(0.728,0.489){15}{\rule{0.678pt}{0.118pt}}
\multiput(326.00,412.17)(11.593,9.000){2}{\rule{0.339pt}{0.400pt}}
\multiput(339.00,422.58)(0.600,0.491){17}{\rule{0.580pt}{0.118pt}}
\multiput(339.00,421.17)(10.796,10.000){2}{\rule{0.290pt}{0.400pt}}
\multiput(351.00,432.58)(0.652,0.491){17}{\rule{0.620pt}{0.118pt}}
\multiput(351.00,431.17)(11.713,10.000){2}{\rule{0.310pt}{0.400pt}}
\multiput(364.00,442.59)(0.669,0.489){15}{\rule{0.633pt}{0.118pt}}
\multiput(364.00,441.17)(10.685,9.000){2}{\rule{0.317pt}{0.400pt}}
\multiput(376.00,451.58)(0.652,0.491){17}{\rule{0.620pt}{0.118pt}}
\multiput(376.00,450.17)(11.713,10.000){2}{\rule{0.310pt}{0.400pt}}
\multiput(389.00,461.58)(0.600,0.491){17}{\rule{0.580pt}{0.118pt}}
\multiput(389.00,460.17)(10.796,10.000){2}{\rule{0.290pt}{0.400pt}}
\multiput(401.00,471.59)(0.728,0.489){15}{\rule{0.678pt}{0.118pt}}
\multiput(401.00,470.17)(11.593,9.000){2}{\rule{0.339pt}{0.400pt}}
\multiput(414.00,480.58)(0.600,0.491){17}{\rule{0.580pt}{0.118pt}}
\multiput(414.00,479.17)(10.796,10.000){2}{\rule{0.290pt}{0.400pt}}
\multiput(426.00,490.59)(0.728,0.489){15}{\rule{0.678pt}{0.118pt}}
\multiput(426.00,489.17)(11.593,9.000){2}{\rule{0.339pt}{0.400pt}}
\multiput(439.00,499.58)(0.600,0.491){17}{\rule{0.580pt}{0.118pt}}
\multiput(439.00,498.17)(10.796,10.000){2}{\rule{0.290pt}{0.400pt}}
\multiput(451.00,509.58)(0.652,0.491){17}{\rule{0.620pt}{0.118pt}}
\multiput(451.00,508.17)(11.713,10.000){2}{\rule{0.310pt}{0.400pt}}
\multiput(464.00,519.59)(0.669,0.489){15}{\rule{0.633pt}{0.118pt}}
\multiput(464.00,518.17)(10.685,9.000){2}{\rule{0.317pt}{0.400pt}}
\multiput(476.00,528.58)(0.652,0.491){17}{\rule{0.620pt}{0.118pt}}
\multiput(476.00,527.17)(11.713,10.000){2}{\rule{0.310pt}{0.400pt}}
\multiput(489.00,538.59)(0.669,0.489){15}{\rule{0.633pt}{0.118pt}}
\multiput(489.00,537.17)(10.685,9.000){2}{\rule{0.317pt}{0.400pt}}
\multiput(501.00,547.58)(0.652,0.491){17}{\rule{0.620pt}{0.118pt}}
\multiput(501.00,546.17)(11.713,10.000){2}{\rule{0.310pt}{0.400pt}}
\multiput(514.00,557.58)(0.600,0.491){17}{\rule{0.580pt}{0.118pt}}
\multiput(514.00,556.17)(10.796,10.000){2}{\rule{0.290pt}{0.400pt}}
\multiput(526.00,567.59)(0.728,0.489){15}{\rule{0.678pt}{0.118pt}}
\multiput(526.00,566.17)(11.593,9.000){2}{\rule{0.339pt}{0.400pt}}
\multiput(539.00,576.58)(0.600,0.491){17}{\rule{0.580pt}{0.118pt}}
\multiput(539.00,575.17)(10.796,10.000){2}{\rule{0.290pt}{0.400pt}}
\multiput(551.00,586.58)(0.652,0.491){17}{\rule{0.620pt}{0.118pt}}
\multiput(551.00,585.17)(11.713,10.000){2}{\rule{0.310pt}{0.400pt}}
\multiput(564.00,596.59)(0.669,0.489){15}{\rule{0.633pt}{0.118pt}}
\multiput(564.00,595.17)(10.685,9.000){2}{\rule{0.317pt}{0.400pt}}
\multiput(576.00,605.58)(0.652,0.491){17}{\rule{0.620pt}{0.118pt}}
\multiput(576.00,604.17)(11.713,10.000){2}{\rule{0.310pt}{0.400pt}}
\multiput(589.00,615.59)(0.669,0.489){15}{\rule{0.633pt}{0.118pt}}
\multiput(589.00,614.17)(10.685,9.000){2}{\rule{0.317pt}{0.400pt}}
\multiput(601.00,624.58)(0.652,0.491){17}{\rule{0.620pt}{0.118pt}}
\multiput(601.00,623.17)(11.713,10.000){2}{\rule{0.310pt}{0.400pt}}
\multiput(614.00,634.58)(0.600,0.491){17}{\rule{0.580pt}{0.118pt}}
\multiput(614.00,633.17)(10.796,10.000){2}{\rule{0.290pt}{0.400pt}}
\multiput(626.00,644.59)(0.728,0.489){15}{\rule{0.678pt}{0.118pt}}
\multiput(626.00,643.17)(11.593,9.000){2}{\rule{0.339pt}{0.400pt}}
\multiput(639.00,653.58)(0.600,0.491){17}{\rule{0.580pt}{0.118pt}}
\multiput(639.00,652.17)(10.796,10.000){2}{\rule{0.290pt}{0.400pt}}
\multiput(651.00,663.58)(0.652,0.491){17}{\rule{0.620pt}{0.118pt}}
\multiput(651.00,662.17)(11.713,10.000){2}{\rule{0.310pt}{0.400pt}}
\multiput(664.00,673.59)(0.669,0.489){15}{\rule{0.633pt}{0.118pt}}
\multiput(664.00,672.17)(10.685,9.000){2}{\rule{0.317pt}{0.400pt}}
\multiput(676.00,682.58)(0.652,0.491){17}{\rule{0.620pt}{0.118pt}}
\multiput(676.00,681.17)(11.713,10.000){2}{\rule{0.310pt}{0.400pt}}
\multiput(689.00,692.59)(0.669,0.489){15}{\rule{0.633pt}{0.118pt}}
\multiput(689.00,691.17)(10.685,9.000){2}{\rule{0.317pt}{0.400pt}}
\multiput(701.00,701.58)(0.652,0.491){17}{\rule{0.620pt}{0.118pt}}
\multiput(701.00,700.17)(11.713,10.000){2}{\rule{0.310pt}{0.400pt}}
\multiput(714.00,711.58)(0.600,0.491){17}{\rule{0.580pt}{0.118pt}}
\multiput(714.00,710.17)(10.796,10.000){2}{\rule{0.290pt}{0.400pt}}
\multiput(726.00,721.59)(0.728,0.489){15}{\rule{0.678pt}{0.118pt}}
\multiput(726.00,720.17)(11.593,9.000){2}{\rule{0.339pt}{0.400pt}}
\multiput(739.00,730.58)(0.600,0.491){17}{\rule{0.580pt}{0.118pt}}
\multiput(739.00,729.17)(10.796,10.000){2}{\rule{0.290pt}{0.400pt}}
\multiput(751.00,740.58)(0.652,0.491){17}{\rule{0.620pt}{0.118pt}}
\multiput(751.00,739.17)(11.713,10.000){2}{\rule{0.310pt}{0.400pt}}
\multiput(764.00,750.59)(0.669,0.489){15}{\rule{0.633pt}{0.118pt}}
\multiput(764.00,749.17)(10.685,9.000){2}{\rule{0.317pt}{0.400pt}}
\multiput(776.00,759.58)(0.652,0.491){17}{\rule{0.620pt}{0.118pt}}
\multiput(776.00,758.17)(11.713,10.000){2}{\rule{0.310pt}{0.400pt}}
\multiput(789.00,769.59)(0.669,0.489){15}{\rule{0.633pt}{0.118pt}}
\multiput(789.00,768.17)(10.685,9.000){2}{\rule{0.317pt}{0.400pt}}
\multiput(801.00,778.58)(0.652,0.491){17}{\rule{0.620pt}{0.118pt}}
\multiput(801.00,777.17)(11.713,10.000){2}{\rule{0.310pt}{0.400pt}}
\multiput(814.00,788.58)(0.600,0.491){17}{\rule{0.580pt}{0.118pt}}
\multiput(814.00,787.17)(10.796,10.000){2}{\rule{0.290pt}{0.400pt}}
\multiput(826.00,798.59)(0.728,0.489){15}{\rule{0.678pt}{0.118pt}}
\multiput(826.00,797.17)(11.593,9.000){2}{\rule{0.339pt}{0.400pt}}
\multiput(839.00,807.58)(0.600,0.491){17}{\rule{0.580pt}{0.118pt}}
\multiput(839.00,806.17)(10.796,10.000){2}{\rule{0.290pt}{0.400pt}}
\multiput(851.00,817.58)(0.652,0.491){17}{\rule{0.620pt}{0.118pt}}
\multiput(851.00,816.17)(11.713,10.000){2}{\rule{0.310pt}{0.400pt}}
\multiput(864.00,827.59)(0.669,0.489){15}{\rule{0.633pt}{0.118pt}}
\multiput(864.00,826.17)(10.685,9.000){2}{\rule{0.317pt}{0.400pt}}
\multiput(876.00,836.58)(0.652,0.491){17}{\rule{0.620pt}{0.118pt}}
\multiput(876.00,835.17)(11.713,10.000){2}{\rule{0.310pt}{0.400pt}}
\multiput(889.00,846.59)(0.669,0.489){15}{\rule{0.633pt}{0.118pt}}
\multiput(889.00,845.17)(10.685,9.000){2}{\rule{0.317pt}{0.400pt}}
\multiput(901.00,855.59)(0.599,0.477){7}{\rule{0.580pt}{0.115pt}}
\multiput(901.00,854.17)(4.796,5.000){2}{\rule{0.290pt}{0.400pt}}
\sbox{\plotpoint}{\rule[-0.500pt]{1.000pt}{1.000pt}}%
\put(201,174){\usebox{\plotpoint}}
\put(201.00,174.00){\usebox{\plotpoint}}
\put(217.48,186.61){\usebox{\plotpoint}}
\put(234.01,199.16){\usebox{\plotpoint}}
\put(250.11,212.26){\usebox{\plotpoint}}
\put(266.91,224.42){\usebox{\plotpoint}}
\put(283.07,237.44){\usebox{\plotpoint}}
\put(299.62,249.97){\usebox{\plotpoint}}
\put(316.10,262.58){\usebox{\plotpoint}}
\put(332.65,275.11){\usebox{\plotpoint}}
\put(348.79,288.16){\usebox{\plotpoint}}
\put(365.59,300.32){\usebox{\plotpoint}}
\put(381.92,313.10){\usebox{\plotpoint}}
\put(398.33,325.77){\usebox{\plotpoint}}
\put(414.70,338.53){\usebox{\plotpoint}}
\put(431.26,351.04){\usebox{\plotpoint}}
\put(447.44,364.03){\usebox{\plotpoint}}
\put(464.24,376.20){\usebox{\plotpoint}}
\put(480.48,389.10){\usebox{\plotpoint}}
\put(496.98,401.65){\usebox{\plotpoint}}
\put(513.31,414.47){\usebox{\plotpoint}}
\put(529.87,426.97){\usebox{\plotpoint}}
\put(546.09,439.91){\usebox{\plotpoint}}
\put(562.81,452.18){\usebox{\plotpoint}}
\put(579.04,465.10){\usebox{\plotpoint}}
\put(595.63,477.53){\usebox{\plotpoint}}
\put(611.92,490.40){\usebox{\plotpoint}}
\put(628.48,502.91){\usebox{\plotpoint}}
\put(644.75,515.79){\usebox{\plotpoint}}
\put(661.37,528.18){\usebox{\plotpoint}}
\put(677.59,541.10){\usebox{\plotpoint}}
\put(694.29,553.41){\usebox{\plotpoint}}
\put(710.53,566.33){\usebox{\plotpoint}}
\put(727.09,578.84){\usebox{\plotpoint}}
\put(743.40,591.67){\usebox{\plotpoint}}
\put(759.93,604.18){\usebox{\plotpoint}}
\put(776.15,617.11){\usebox{\plotpoint}}
\put(792.94,629.28){\usebox{\plotpoint}}
\put(809.14,642.26){\usebox{\plotpoint}}
\put(825.69,654.77){\usebox{\plotpoint}}
\put(842.05,667.54){\usebox{\plotpoint}}
\put(858.49,680.18){\usebox{\plotpoint}}
\put(874.79,693.00){\usebox{\plotpoint}}
\put(891.59,705.16){\usebox{\plotpoint}}
\put(907.75,718.19){\usebox{\plotpoint}}
\put(924.29,730.72){\usebox{\plotpoint}}
\put(940.70,743.42){\usebox{\plotpoint}}
\put(957.05,756.19){\usebox{\plotpoint}}
\put(973.45,768.87){\usebox{\plotpoint}}
\put(990.25,781.04){\usebox{\plotpoint}}
\put(1006.36,794.12){\usebox{\plotpoint}}
\put(1022.89,806.67){\usebox{\plotpoint}}
\put(1039.36,819.30){\usebox{\plotpoint}}
\put(1055.60,832.19){\usebox{\plotpoint}}
\put(1072.10,844.75){\usebox{\plotpoint}}
\put(1088.89,856.92){\usebox{\plotpoint}}
\put(1092,860){\usebox{\plotpoint}}
\put(201,186){\usebox{\plotpoint}}
\put(201.00,186.00){\usebox{\plotpoint}}
\put(242.38,189.28){\usebox{\plotpoint}}
\put(283.76,192.60){\usebox{\plotpoint}}
\put(325.18,194.93){\usebox{\plotpoint}}
\put(366.56,198.21){\usebox{\plotpoint}}
\put(407.96,201.00){\usebox{\plotpoint}}
\put(449.35,203.86){\usebox{\plotpoint}}
\put(490.73,207.14){\usebox{\plotpoint}}
\put(532.15,209.47){\usebox{\plotpoint}}
\put(573.53,212.79){\usebox{\plotpoint}}
\put(614.91,216.00){\usebox{\plotpoint}}
\put(656.33,218.41){\usebox{\plotpoint}}
\put(697.71,221.73){\usebox{\plotpoint}}
\put(739.13,224.01){\usebox{\plotpoint}}
\put(780.50,227.35){\usebox{\plotpoint}}
\put(821.88,230.66){\usebox{\plotpoint}}
\put(863.30,232.95){\usebox{\plotpoint}}
\put(904.68,236.28){\usebox{\plotpoint}}
\put(946.06,239.59){\usebox{\plotpoint}}
\put(987.48,241.88){\usebox{\plotpoint}}
\put(1028.85,245.22){\usebox{\plotpoint}}
\put(1070.25,248.00){\usebox{\plotpoint}}
\put(1111.65,250.82){\usebox{\plotpoint}}
\put(1153.03,254.16){\usebox{\plotpoint}}
\put(1194.45,256.45){\usebox{\plotpoint}}
\put(1235.83,259.76){\usebox{\plotpoint}}
\put(1277.21,263.00){\usebox{\plotpoint}}
\put(1318.62,265.39){\usebox{\plotpoint}}
\put(1360.00,268.69){\usebox{\plotpoint}}
\put(1401.42,271.03){\usebox{\plotpoint}}
\put(1439,274){\usebox{\plotpoint}}
\sbox{\plotpoint}{\rule[-0.600pt]{1.200pt}{1.200pt}}%
\put(218,335){\circle*{18}}
\put(253,327){\circle*{18}}
\put(288,375){\circle*{18}}
\put(323,410){\circle*{18}}
\put(358,457){\circle*{18}}
\put(393,391){\circle*{18}}
\put(428,497){\circle*{18}}
\put(463,528){\circle*{18}}
\put(497,416){\circle*{18}}
\put(532,576){\circle*{18}}
\put(567,656){\circle*{18}}
\put(602,475){\circle*{18}}
\put(637,614){\circle*{18}}
\put(672,582){\circle*{18}}
\put(707,562){\circle*{18}}
\put(742,616){\circle*{18}}
\put(776,649){\circle*{18}}
\put(811,689){\circle*{18}}
\put(846,638){\circle*{18}}
\put(881,618){\circle*{18}}
\put(916,535){\circle*{18}}
\put(951,594){\circle*{18}}
\put(986,658){\circle*{18}}
\put(1021,685){\circle*{18}}
\put(1055,692){\circle*{18}}
\put(1090,674){\circle*{18}}
\put(1125,671){\circle*{18}}
\put(1160,636){\circle*{18}}
\put(1195,616){\circle*{18}}
\put(1230,571){\circle*{18}}
\put(1265,562){\circle*{18}}
\put(1300,655){\circle*{18}}
\put(1334,679){\circle*{18}}
\put(1369,656){\circle*{18}}
\put(1404,684){\circle*{18}}
\put(1439,660){\circle*{18}}
\sbox{\plotpoint}{\rule[-0.200pt]{0.400pt}{0.400pt}}%
\put(218,194){\circle{18}}
\put(253,208){\circle{18}}
\put(288,246){\circle{18}}
\put(323,298){\circle{18}}
\put(358,349){\circle{18}}
\put(393,285){\circle{18}}
\put(428,368){\circle{18}}
\put(463,335){\circle{18}}
\put(497,447){\circle{18}}
\put(532,427){\circle{18}}
\put(567,373){\circle{18}}
\put(602,503){\circle{18}}
\put(637,526){\circle{18}}
\put(672,576){\circle{18}}
\put(707,631){\circle{18}}
\put(742,655){\circle{18}}
\put(776,636){\circle{18}}
\put(811,647){\circle{18}}
\put(846,625){\circle{18}}
\put(881,595){\circle{18}}
\put(916,687){\circle{18}}
\put(951,649){\circle{18}}
\put(986,680){\circle{18}}
\put(1021,636){\circle{18}}
\put(1055,482){\circle{18}}
\put(1090,638){\circle{18}}
\put(1125,692){\circle{18}}
\put(1160,707){\circle{18}}
\put(1195,731){\circle{18}}
\put(1230,711){\circle{18}}
\put(1265,672){\circle{18}}
\put(1300,679){\circle{18}}
\put(1334,616){\circle{18}}
\put(1369,669){\circle{18}}
\put(1404,562){\circle{18}}
\put(1439,697){\circle{18}}
\end{picture}

%% file: delta.2.835.140.tex
\setlength{\unitlength}{0.240900pt}
\ifx\plotpoint\undefined\newsavebox{\plotpoint}\fi
\begin{picture}(1500,900)(0,0)
\font\gnuplot=cmr10 at 10pt
\gnuplot
\sbox{\plotpoint}{\rule[-0.200pt]{0.400pt}{0.400pt}}%
\put(201.0,223.0){\rule[-0.200pt]{4.818pt}{0.400pt}}
\put(181,223){\makebox(0,0)[r]{0.001}}
\put(1419.0,223.0){\rule[-0.200pt]{4.818pt}{0.400pt}}
\put(201.0,365.0){\rule[-0.200pt]{4.818pt}{0.400pt}}
\put(181,365){\makebox(0,0)[r]{0.01}}
\put(1419.0,365.0){\rule[-0.200pt]{4.818pt}{0.400pt}}
\put(201.0,507.0){\rule[-0.200pt]{4.818pt}{0.400pt}}
\put(181,507){\makebox(0,0)[r]{0.1}}
\put(1419.0,507.0){\rule[-0.200pt]{4.818pt}{0.400pt}}
\put(201.0,650.0){\rule[-0.200pt]{4.818pt}{0.400pt}}
\put(181,650){\makebox(0,0)[r]{1}}
\put(1419.0,650.0){\rule[-0.200pt]{4.818pt}{0.400pt}}
\put(201.0,792.0){\rule[-0.200pt]{4.818pt}{0.400pt}}
\put(181,792){\makebox(0,0)[r]{10}}
\put(1419.0,792.0){\rule[-0.200pt]{4.818pt}{0.400pt}}
\put(218.0,123.0){\rule[-0.200pt]{0.400pt}{4.818pt}}
\put(218,82){\makebox(0,0){0}}
\put(218.0,840.0){\rule[-0.200pt]{0.400pt}{4.818pt}}
\put(393.0,123.0){\rule[-0.200pt]{0.400pt}{4.818pt}}
\put(393,82){\makebox(0,0){5}}
\put(393.0,840.0){\rule[-0.200pt]{0.400pt}{4.818pt}}
\put(567.0,123.0){\rule[-0.200pt]{0.400pt}{4.818pt}}
\put(567,82){\makebox(0,0){10}}
\put(567.0,840.0){\rule[-0.200pt]{0.400pt}{4.818pt}}
\put(742.0,123.0){\rule[-0.200pt]{0.400pt}{4.818pt}}
\put(742,82){\makebox(0,0){15}}
\put(742.0,840.0){\rule[-0.200pt]{0.400pt}{4.818pt}}
\put(916.0,123.0){\rule[-0.200pt]{0.400pt}{4.818pt}}
\put(916,82){\makebox(0,0){20}}
\put(916.0,840.0){\rule[-0.200pt]{0.400pt}{4.818pt}}
\put(1090.0,123.0){\rule[-0.200pt]{0.400pt}{4.818pt}}
\put(1090,82){\makebox(0,0){25}}
\put(1090.0,840.0){\rule[-0.200pt]{0.400pt}{4.818pt}}
\put(1265.0,123.0){\rule[-0.200pt]{0.400pt}{4.818pt}}
\put(1265,82){\makebox(0,0){30}}
\put(1265.0,840.0){\rule[-0.200pt]{0.400pt}{4.818pt}}
\put(1439.0,123.0){\rule[-0.200pt]{0.400pt}{4.818pt}}
\put(1439,82){\makebox(0,0){35}}
\put(1439.0,840.0){\rule[-0.200pt]{0.400pt}{4.818pt}}
\put(201.0,123.0){\rule[-0.200pt]{298.234pt}{0.400pt}}
\put(1439.0,123.0){\rule[-0.200pt]{0.400pt}{177.543pt}}
\put(201.0,860.0){\rule[-0.200pt]{298.234pt}{0.400pt}}
\put(40,651){\makebox(0,0)[tl] { $ \delta L_z $ }}
\put(820,21){\makebox(0,0){ { Kick Number } }}
\put(201.0,123.0){\rule[-0.200pt]{0.400pt}{177.543pt}}
\put(201,175){\usebox{\plotpoint}}
\multiput(201.58,175.00)(0.493,0.972){23}{\rule{0.119pt}{0.869pt}}
\multiput(200.17,175.00)(13.000,23.196){2}{\rule{0.400pt}{0.435pt}}
\multiput(214.58,200.00)(0.492,1.013){21}{\rule{0.119pt}{0.900pt}}
\multiput(213.17,200.00)(12.000,22.132){2}{\rule{0.400pt}{0.450pt}}
\multiput(226.58,224.00)(0.493,0.933){23}{\rule{0.119pt}{0.838pt}}
\multiput(225.17,224.00)(13.000,22.260){2}{\rule{0.400pt}{0.419pt}}
\multiput(239.58,248.00)(0.492,1.056){21}{\rule{0.119pt}{0.933pt}}
\multiput(238.17,248.00)(12.000,23.063){2}{\rule{0.400pt}{0.467pt}}
\multiput(251.58,273.00)(0.493,0.933){23}{\rule{0.119pt}{0.838pt}}
\multiput(250.17,273.00)(13.000,22.260){2}{\rule{0.400pt}{0.419pt}}
\multiput(264.58,297.00)(0.492,1.056){21}{\rule{0.119pt}{0.933pt}}
\multiput(263.17,297.00)(12.000,23.063){2}{\rule{0.400pt}{0.467pt}}
\multiput(276.58,322.00)(0.493,0.933){23}{\rule{0.119pt}{0.838pt}}
\multiput(275.17,322.00)(13.000,22.260){2}{\rule{0.400pt}{0.419pt}}
\multiput(289.58,346.00)(0.492,1.013){21}{\rule{0.119pt}{0.900pt}}
\multiput(288.17,346.00)(12.000,22.132){2}{\rule{0.400pt}{0.450pt}}
\multiput(301.58,370.00)(0.493,0.972){23}{\rule{0.119pt}{0.869pt}}
\multiput(300.17,370.00)(13.000,23.196){2}{\rule{0.400pt}{0.435pt}}
\multiput(314.58,395.00)(0.492,1.013){21}{\rule{0.119pt}{0.900pt}}
\multiput(313.17,395.00)(12.000,22.132){2}{\rule{0.400pt}{0.450pt}}
\multiput(326.58,419.00)(0.493,0.972){23}{\rule{0.119pt}{0.869pt}}
\multiput(325.17,419.00)(13.000,23.196){2}{\rule{0.400pt}{0.435pt}}
\multiput(339.58,444.00)(0.492,1.013){21}{\rule{0.119pt}{0.900pt}}
\multiput(338.17,444.00)(12.000,22.132){2}{\rule{0.400pt}{0.450pt}}
\multiput(351.58,468.00)(0.493,0.933){23}{\rule{0.119pt}{0.838pt}}
\multiput(350.17,468.00)(13.000,22.260){2}{\rule{0.400pt}{0.419pt}}
\multiput(364.58,492.00)(0.492,1.056){21}{\rule{0.119pt}{0.933pt}}
\multiput(363.17,492.00)(12.000,23.063){2}{\rule{0.400pt}{0.467pt}}
\multiput(376.58,517.00)(0.493,0.933){23}{\rule{0.119pt}{0.838pt}}
\multiput(375.17,517.00)(13.000,22.260){2}{\rule{0.400pt}{0.419pt}}
\multiput(389.58,541.00)(0.492,1.013){21}{\rule{0.119pt}{0.900pt}}
\multiput(388.17,541.00)(12.000,22.132){2}{\rule{0.400pt}{0.450pt}}
\multiput(401.58,565.00)(0.493,0.972){23}{\rule{0.119pt}{0.869pt}}
\multiput(400.17,565.00)(13.000,23.196){2}{\rule{0.400pt}{0.435pt}}
\multiput(414.58,590.00)(0.492,1.013){21}{\rule{0.119pt}{0.900pt}}
\multiput(413.17,590.00)(12.000,22.132){2}{\rule{0.400pt}{0.450pt}}
\multiput(426.58,614.00)(0.493,0.972){23}{\rule{0.119pt}{0.869pt}}
\multiput(425.17,614.00)(13.000,23.196){2}{\rule{0.400pt}{0.435pt}}
\multiput(439.58,639.00)(0.492,1.013){21}{\rule{0.119pt}{0.900pt}}
\multiput(438.17,639.00)(12.000,22.132){2}{\rule{0.400pt}{0.450pt}}
\multiput(451.58,663.00)(0.493,0.933){23}{\rule{0.119pt}{0.838pt}}
\multiput(450.17,663.00)(13.000,22.260){2}{\rule{0.400pt}{0.419pt}}
\multiput(464.58,687.00)(0.492,1.056){21}{\rule{0.119pt}{0.933pt}}
\multiput(463.17,687.00)(12.000,23.063){2}{\rule{0.400pt}{0.467pt}}
\multiput(476.58,712.00)(0.493,0.933){23}{\rule{0.119pt}{0.838pt}}
\multiput(475.17,712.00)(13.000,22.260){2}{\rule{0.400pt}{0.419pt}}
\multiput(489.58,736.00)(0.492,1.056){21}{\rule{0.119pt}{0.933pt}}
\multiput(488.17,736.00)(12.000,23.063){2}{\rule{0.400pt}{0.467pt}}
\multiput(501.58,761.00)(0.493,0.933){23}{\rule{0.119pt}{0.838pt}}
\multiput(500.17,761.00)(13.000,22.260){2}{\rule{0.400pt}{0.419pt}}
\multiput(514.58,785.00)(0.492,1.013){21}{\rule{0.119pt}{0.900pt}}
\multiput(513.17,785.00)(12.000,22.132){2}{\rule{0.400pt}{0.450pt}}
\multiput(526.58,809.00)(0.493,0.972){23}{\rule{0.119pt}{0.869pt}}
\multiput(525.17,809.00)(13.000,23.196){2}{\rule{0.400pt}{0.435pt}}
\multiput(539.58,834.00)(0.492,1.013){21}{\rule{0.119pt}{0.900pt}}
\multiput(538.17,834.00)(12.000,22.132){2}{\rule{0.400pt}{0.450pt}}
\put(550.67,858){\rule{0.400pt}{0.482pt}}
\multiput(550.17,858.00)(1.000,1.000){2}{\rule{0.400pt}{0.241pt}}
\sbox{\plotpoint}{\rule[-0.500pt]{1.000pt}{1.000pt}}%
\put(201,195){\usebox{\plotpoint}}
\put(201.00,195.00){\usebox{\plotpoint}}
\put(217.35,207.79){\usebox{\plotpoint}}
\put(233.52,220.79){\usebox{\plotpoint}}
\put(249.63,233.86){\usebox{\plotpoint}}
\put(265.98,246.65){\usebox{\plotpoint}}
\put(282.11,259.70){\usebox{\plotpoint}}
\put(298.27,272.72){\usebox{\plotpoint}}
\put(314.61,285.51){\usebox{\plotpoint}}
\put(330.70,298.62){\usebox{\plotpoint}}
\put(346.90,311.59){\usebox{\plotpoint}}
\put(363.23,324.40){\usebox{\plotpoint}}
\put(379.30,337.54){\usebox{\plotpoint}}
\put(395.54,350.45){\usebox{\plotpoint}}
\put(411.82,363.32){\usebox{\plotpoint}}
\put(427.89,376.45){\usebox{\plotpoint}}
\put(444.17,389.31){\usebox{\plotpoint}}
\put(460.41,402.24){\usebox{\plotpoint}}
\put(476.48,415.37){\usebox{\plotpoint}}
\put(492.81,428.17){\usebox{\plotpoint}}
\put(509.00,441.15){\usebox{\plotpoint}}
\put(525.10,454.25){\usebox{\plotpoint}}
\put(541.44,467.04){\usebox{\plotpoint}}
\put(557.59,480.07){\usebox{\plotpoint}}
\put(573.73,493.11){\usebox{\plotpoint}}
\put(590.08,505.90){\usebox{\plotpoint}}
\put(606.18,518.99){\usebox{\plotpoint}}
\put(622.37,531.97){\usebox{\plotpoint}}
\put(638.71,544.77){\usebox{\plotpoint}}
\put(654.78,557.90){\usebox{\plotpoint}}
\put(671.29,570.47){\usebox{\plotpoint}}
\put(687.79,583.07){\usebox{\plotpoint}}
\put(703.86,596.20){\usebox{\plotpoint}}
\put(720.12,609.10){\usebox{\plotpoint}}
\put(736.38,621.98){\usebox{\plotpoint}}
\put(752.45,635.12){\usebox{\plotpoint}}
\put(768.75,647.96){\usebox{\plotpoint}}
\put(784.97,660.90){\usebox{\plotpoint}}
\put(801.04,674.03){\usebox{\plotpoint}}
\put(817.39,686.82){\usebox{\plotpoint}}
\put(833.56,699.82){\usebox{\plotpoint}}
\put(849.68,712.90){\usebox{\plotpoint}}
\put(866.02,725.68){\usebox{\plotpoint}}
\put(882.15,738.73){\usebox{\plotpoint}}
\put(898.31,751.76){\usebox{\plotpoint}}
\put(914.66,764.55){\usebox{\plotpoint}}
\put(930.75,777.65){\usebox{\plotpoint}}
\put(946.95,790.62){\usebox{\plotpoint}}
\put(963.27,803.44){\usebox{\plotpoint}}
\put(979.34,816.57){\usebox{\plotpoint}}
\put(995.58,829.48){\usebox{\plotpoint}}
\put(1011.86,842.35){\usebox{\plotpoint}}
\put(1027.95,855.46){\usebox{\plotpoint}}
\put(1034,860){\usebox{\plotpoint}}
\sbox{\plotpoint}{\rule[-0.600pt]{1.200pt}{1.200pt}}%
\put(218,222){\circle*{18}}
\put(253,285){\circle*{18}}
\put(288,386){\circle*{18}}
\put(323,372){\circle*{18}}
\put(358,516){\circle*{18}}
\put(393,557){\circle*{18}}
\put(428,459){\circle*{18}}
\put(463,500){\circle*{18}}
\put(497,569){\circle*{18}}
\put(532,494){\circle*{18}}
\put(567,621){\circle*{18}}
\put(602,677){\circle*{18}}
\put(637,616){\circle*{18}}
\put(672,565){\circle*{18}}
\put(707,522){\circle*{18}}
\put(742,611){\circle*{18}}
\put(776,560){\circle*{18}}
\put(811,414){\circle*{18}}
\put(846,615){\circle*{18}}
\put(881,465){\circle*{18}}
\put(916,430){\circle*{18}}
\put(951,458){\circle*{18}}
\put(986,557){\circle*{18}}
\put(1021,478){\circle*{18}}
\put(1055,500){\circle*{18}}
\put(1090,586){\circle*{18}}
\put(1125,551){\circle*{18}}
\put(1160,530){\circle*{18}}
\put(1195,583){\circle*{18}}
\put(1230,618){\circle*{18}}
\put(1265,571){\circle*{18}}
\put(1300,512){\circle*{18}}
\put(1334,529){\circle*{18}}
\put(1369,542){\circle*{18}}
\put(1404,621){\circle*{18}}
\put(1439,611){\circle*{18}}
\sbox{\plotpoint}{\rule[-0.200pt]{0.400pt}{0.400pt}}%
\put(218,209){\circle{18}}
\put(253,221){\circle{18}}
\put(288,198){\circle{18}}
\put(323,419){\circle{18}}
\put(358,536){\circle{18}}
\put(393,489){\circle{18}}
\put(428,689){\circle{18}}
\put(463,647){\circle{18}}
\put(497,821){\circle{18}}
\put(532,628){\circle{18}}
\put(567,764){\circle{18}}
\put(602,715){\circle{18}}
\put(637,480){\circle{18}}
\put(672,676){\circle{18}}
\put(707,687){\circle{18}}
\put(742,657){\circle{18}}
\put(776,653){\circle{18}}
\put(811,629){\circle{18}}
\put(846,558){\circle{18}}
\put(881,598){\circle{18}}
\put(916,611){\circle{18}}
\put(951,616){\circle{18}}
\put(986,646){\circle{18}}
\put(1021,580){\circle{18}}
\put(1055,572){\circle{18}}
\put(1090,478){\circle{18}}
\put(1125,597){\circle{18}}
\put(1160,517){\circle{18}}
\put(1195,551){\circle{18}}
\put(1230,574){\circle{18}}
\put(1265,623){\circle{18}}
\put(1300,603){\circle{18}}
\put(1334,618){\circle{18}}
\put(1369,532){\circle{18}}
\put(1404,563){\circle{18}}
\put(1439,613){\circle{18}}
\end{picture}

%% file: break_time_p0.1.tex
\setlength{\unitlength}{0.240900pt}
\ifx\plotpoint\undefined\newsavebox{\plotpoint}\fi
\sbox{\plotpoint}{\rule[-0.200pt]{0.400pt}{0.400pt}}%
\begin{picture}(1500,900)(0,0)
\font\gnuplot=cmr10 at 10pt
\gnuplot
\sbox{\plotpoint}{\rule[-0.200pt]{0.400pt}{0.400pt}}%
\put(141.0,184.0){\rule[-0.200pt]{4.818pt}{0.400pt}}
\put(121,184){\makebox(0,0)[r]{2}}
\put(1419.0,184.0){\rule[-0.200pt]{4.818pt}{0.400pt}}
\put(141.0,307.0){\rule[-0.200pt]{4.818pt}{0.400pt}}
\put(121,307){\makebox(0,0)[r]{4}}
\put(1419.0,307.0){\rule[-0.200pt]{4.818pt}{0.400pt}}
\put(141.0,430.0){\rule[-0.200pt]{4.818pt}{0.400pt}}
\put(121,430){\makebox(0,0)[r]{6}}
\put(1419.0,430.0){\rule[-0.200pt]{4.818pt}{0.400pt}}
\put(141.0,553.0){\rule[-0.200pt]{4.818pt}{0.400pt}}
\put(121,553){\makebox(0,0)[r]{8}}
\put(1419.0,553.0){\rule[-0.200pt]{4.818pt}{0.400pt}}
\put(141.0,676.0){\rule[-0.200pt]{4.818pt}{0.400pt}}
\put(121,676){\makebox(0,0)[r]{10}}
\put(1419.0,676.0){\rule[-0.200pt]{4.818pt}{0.400pt}}
\put(141.0,799.0){\rule[-0.200pt]{4.818pt}{0.400pt}}
\put(121,799){\makebox(0,0)[r]{12}}
\put(1419.0,799.0){\rule[-0.200pt]{4.818pt}{0.400pt}}
\put(141.0,123.0){\rule[-0.200pt]{0.400pt}{4.818pt}}
\put(141,82){\makebox(0,0){0}}
\put(141.0,840.0){\rule[-0.200pt]{0.400pt}{4.818pt}}
\put(401.0,123.0){\rule[-0.200pt]{0.400pt}{4.818pt}}
\put(401,82){\makebox(0,0){50}}
\put(401.0,840.0){\rule[-0.200pt]{0.400pt}{4.818pt}}
\put(660.0,123.0){\rule[-0.200pt]{0.400pt}{4.818pt}}
\put(660,82){\makebox(0,0){100}}
\put(660.0,840.0){\rule[-0.200pt]{0.400pt}{4.818pt}}
\put(920.0,123.0){\rule[-0.200pt]{0.400pt}{4.818pt}}
\put(920,82){\makebox(0,0){150}}
\put(920.0,840.0){\rule[-0.200pt]{0.400pt}{4.818pt}}
\put(1179.0,123.0){\rule[-0.200pt]{0.400pt}{4.818pt}}
\put(1179,82){\makebox(0,0){200}}
\put(1179.0,840.0){\rule[-0.200pt]{0.400pt}{4.818pt}}
\put(1439.0,123.0){\rule[-0.200pt]{0.400pt}{4.818pt}}
\put(1439,82){\makebox(0,0){250}}
\put(1439.0,840.0){\rule[-0.200pt]{0.400pt}{4.818pt}}
\put(141.0,123.0){\rule[-0.200pt]{312.688pt}{0.400pt}}
\put(1439.0,123.0){\rule[-0.200pt]{0.400pt}{177.543pt}}
\put(141.0,860.0){\rule[-0.200pt]{312.688pt}{0.400pt}}
\put(40,491){\makebox(0,0){ { $t_b$ } }}
\put(790,21){\makebox(0,0){ { $l$ } }}
\put(141.0,123.0){\rule[-0.200pt]{0.400pt}{177.543pt}}
\put(198,246){\circle{18}}
\put(255,307){\circle{18}}
\put(369,307){\circle{18}}
\put(484,307){\circle{18}}
\put(598,307){\circle{18}}
\put(712,307){\circle{18}}
\put(826,307){\circle{18}}
\put(941,307){\circle{18}}
\put(1055,307){\circle{18}}
\put(1169,307){\circle{18}}
\put(1283,307){\circle{18}}
\multiput(162.59,123.00)(0.477,1.712){7}{\rule{0.115pt}{1.380pt}}
\multiput(161.17,123.00)(5.000,13.136){2}{\rule{0.400pt}{0.690pt}}
\multiput(167.58,139.00)(0.493,0.853){23}{\rule{0.119pt}{0.777pt}}
\multiput(166.17,139.00)(13.000,20.387){2}{\rule{0.400pt}{0.388pt}}
\multiput(180.58,161.00)(0.493,0.616){23}{\rule{0.119pt}{0.592pt}}
\multiput(179.17,161.00)(13.000,14.771){2}{\rule{0.400pt}{0.296pt}}
\multiput(193.00,177.58)(0.536,0.493){23}{\rule{0.531pt}{0.119pt}}
\multiput(193.00,176.17)(12.898,13.000){2}{\rule{0.265pt}{0.400pt}}
\multiput(207.00,190.58)(0.652,0.491){17}{\rule{0.620pt}{0.118pt}}
\multiput(207.00,189.17)(11.713,10.000){2}{\rule{0.310pt}{0.400pt}}
\multiput(220.00,200.59)(0.824,0.488){13}{\rule{0.750pt}{0.117pt}}
\multiput(220.00,199.17)(11.443,8.000){2}{\rule{0.375pt}{0.400pt}}
\multiput(233.00,208.59)(0.824,0.488){13}{\rule{0.750pt}{0.117pt}}
\multiput(233.00,207.17)(11.443,8.000){2}{\rule{0.375pt}{0.400pt}}
\multiput(246.00,216.59)(1.123,0.482){9}{\rule{0.967pt}{0.116pt}}
\multiput(246.00,215.17)(10.994,6.000){2}{\rule{0.483pt}{0.400pt}}
\multiput(259.00,222.59)(1.123,0.482){9}{\rule{0.967pt}{0.116pt}}
\multiput(259.00,221.17)(10.994,6.000){2}{\rule{0.483pt}{0.400pt}}
\multiput(272.00,228.59)(1.378,0.477){7}{\rule{1.140pt}{0.115pt}}
\multiput(272.00,227.17)(10.634,5.000){2}{\rule{0.570pt}{0.400pt}}
\multiput(285.00,233.59)(1.378,0.477){7}{\rule{1.140pt}{0.115pt}}
\multiput(285.00,232.17)(10.634,5.000){2}{\rule{0.570pt}{0.400pt}}
\multiput(298.00,238.60)(1.797,0.468){5}{\rule{1.400pt}{0.113pt}}
\multiput(298.00,237.17)(10.094,4.000){2}{\rule{0.700pt}{0.400pt}}
\multiput(311.00,242.59)(1.489,0.477){7}{\rule{1.220pt}{0.115pt}}
\multiput(311.00,241.17)(11.468,5.000){2}{\rule{0.610pt}{0.400pt}}
\multiput(325.00,247.61)(2.695,0.447){3}{\rule{1.833pt}{0.108pt}}
\multiput(325.00,246.17)(9.195,3.000){2}{\rule{0.917pt}{0.400pt}}
\multiput(338.00,250.60)(1.797,0.468){5}{\rule{1.400pt}{0.113pt}}
\multiput(338.00,249.17)(10.094,4.000){2}{\rule{0.700pt}{0.400pt}}
\multiput(351.00,254.61)(2.695,0.447){3}{\rule{1.833pt}{0.108pt}}
\multiput(351.00,253.17)(9.195,3.000){2}{\rule{0.917pt}{0.400pt}}
\multiput(364.00,257.61)(2.695,0.447){3}{\rule{1.833pt}{0.108pt}}
\multiput(364.00,256.17)(9.195,3.000){2}{\rule{0.917pt}{0.400pt}}
\multiput(377.00,260.61)(2.695,0.447){3}{\rule{1.833pt}{0.108pt}}
\multiput(377.00,259.17)(9.195,3.000){2}{\rule{0.917pt}{0.400pt}}
\multiput(390.00,263.61)(2.695,0.447){3}{\rule{1.833pt}{0.108pt}}
\multiput(390.00,262.17)(9.195,3.000){2}{\rule{0.917pt}{0.400pt}}
\multiput(403.00,266.61)(2.695,0.447){3}{\rule{1.833pt}{0.108pt}}
\multiput(403.00,265.17)(9.195,3.000){2}{\rule{0.917pt}{0.400pt}}
\multiput(416.00,269.61)(2.695,0.447){3}{\rule{1.833pt}{0.108pt}}
\multiput(416.00,268.17)(9.195,3.000){2}{\rule{0.917pt}{0.400pt}}
\put(429,272.17){\rule{2.900pt}{0.400pt}}
\multiput(429.00,271.17)(7.981,2.000){2}{\rule{1.450pt}{0.400pt}}
\put(443,274.17){\rule{2.700pt}{0.400pt}}
\multiput(443.00,273.17)(7.396,2.000){2}{\rule{1.350pt}{0.400pt}}
\multiput(456.00,276.61)(2.695,0.447){3}{\rule{1.833pt}{0.108pt}}
\multiput(456.00,275.17)(9.195,3.000){2}{\rule{0.917pt}{0.400pt}}
\put(469,279.17){\rule{2.700pt}{0.400pt}}
\multiput(469.00,278.17)(7.396,2.000){2}{\rule{1.350pt}{0.400pt}}
\put(482,281.17){\rule{2.700pt}{0.400pt}}
\multiput(482.00,280.17)(7.396,2.000){2}{\rule{1.350pt}{0.400pt}}
\put(495,283.17){\rule{2.700pt}{0.400pt}}
\multiput(495.00,282.17)(7.396,2.000){2}{\rule{1.350pt}{0.400pt}}
\put(508,285.17){\rule{2.700pt}{0.400pt}}
\multiput(508.00,284.17)(7.396,2.000){2}{\rule{1.350pt}{0.400pt}}
\put(521,287.17){\rule{2.700pt}{0.400pt}}
\multiput(521.00,286.17)(7.396,2.000){2}{\rule{1.350pt}{0.400pt}}
\put(534,288.67){\rule{3.132pt}{0.400pt}}
\multiput(534.00,288.17)(6.500,1.000){2}{\rule{1.566pt}{0.400pt}}
\put(547,290.17){\rule{2.900pt}{0.400pt}}
\multiput(547.00,289.17)(7.981,2.000){2}{\rule{1.450pt}{0.400pt}}
\put(561,292.17){\rule{2.700pt}{0.400pt}}
\multiput(561.00,291.17)(7.396,2.000){2}{\rule{1.350pt}{0.400pt}}
\put(574,294.17){\rule{2.700pt}{0.400pt}}
\multiput(574.00,293.17)(7.396,2.000){2}{\rule{1.350pt}{0.400pt}}
\put(587,295.67){\rule{3.132pt}{0.400pt}}
\multiput(587.00,295.17)(6.500,1.000){2}{\rule{1.566pt}{0.400pt}}
\put(600,297.17){\rule{2.700pt}{0.400pt}}
\multiput(600.00,296.17)(7.396,2.000){2}{\rule{1.350pt}{0.400pt}}
\put(613,298.67){\rule{3.132pt}{0.400pt}}
\multiput(613.00,298.17)(6.500,1.000){2}{\rule{1.566pt}{0.400pt}}
\put(626,300.17){\rule{2.700pt}{0.400pt}}
\multiput(626.00,299.17)(7.396,2.000){2}{\rule{1.350pt}{0.400pt}}
\put(639,301.67){\rule{3.132pt}{0.400pt}}
\multiput(639.00,301.17)(6.500,1.000){2}{\rule{1.566pt}{0.400pt}}
\put(652,303.17){\rule{2.700pt}{0.400pt}}
\multiput(652.00,302.17)(7.396,2.000){2}{\rule{1.350pt}{0.400pt}}
\put(665,304.67){\rule{3.373pt}{0.400pt}}
\multiput(665.00,304.17)(7.000,1.000){2}{\rule{1.686pt}{0.400pt}}
\put(679,305.67){\rule{3.132pt}{0.400pt}}
\multiput(679.00,305.17)(6.500,1.000){2}{\rule{1.566pt}{0.400pt}}
\put(692,307.17){\rule{2.700pt}{0.400pt}}
\multiput(692.00,306.17)(7.396,2.000){2}{\rule{1.350pt}{0.400pt}}
\put(705,308.67){\rule{3.132pt}{0.400pt}}
\multiput(705.00,308.17)(6.500,1.000){2}{\rule{1.566pt}{0.400pt}}
\put(718,309.67){\rule{3.132pt}{0.400pt}}
\multiput(718.00,309.17)(6.500,1.000){2}{\rule{1.566pt}{0.400pt}}
\put(731,310.67){\rule{3.132pt}{0.400pt}}
\multiput(731.00,310.17)(6.500,1.000){2}{\rule{1.566pt}{0.400pt}}
\put(744,312.17){\rule{2.700pt}{0.400pt}}
\multiput(744.00,311.17)(7.396,2.000){2}{\rule{1.350pt}{0.400pt}}
\put(757,313.67){\rule{3.132pt}{0.400pt}}
\multiput(757.00,313.17)(6.500,1.000){2}{\rule{1.566pt}{0.400pt}}
\put(770,314.67){\rule{3.132pt}{0.400pt}}
\multiput(770.00,314.17)(6.500,1.000){2}{\rule{1.566pt}{0.400pt}}
\put(783,315.67){\rule{3.373pt}{0.400pt}}
\multiput(783.00,315.17)(7.000,1.000){2}{\rule{1.686pt}{0.400pt}}
\put(797,316.67){\rule{3.132pt}{0.400pt}}
\multiput(797.00,316.17)(6.500,1.000){2}{\rule{1.566pt}{0.400pt}}
\put(810,317.67){\rule{3.132pt}{0.400pt}}
\multiput(810.00,317.17)(6.500,1.000){2}{\rule{1.566pt}{0.400pt}}
\put(823,318.67){\rule{3.132pt}{0.400pt}}
\multiput(823.00,318.17)(6.500,1.000){2}{\rule{1.566pt}{0.400pt}}
\put(836,319.67){\rule{3.132pt}{0.400pt}}
\multiput(836.00,319.17)(6.500,1.000){2}{\rule{1.566pt}{0.400pt}}
\put(849,320.67){\rule{3.132pt}{0.400pt}}
\multiput(849.00,320.17)(6.500,1.000){2}{\rule{1.566pt}{0.400pt}}
\put(862,321.67){\rule{3.132pt}{0.400pt}}
\multiput(862.00,321.17)(6.500,1.000){2}{\rule{1.566pt}{0.400pt}}
\put(875,322.67){\rule{3.132pt}{0.400pt}}
\multiput(875.00,322.17)(6.500,1.000){2}{\rule{1.566pt}{0.400pt}}
\put(888,323.67){\rule{3.132pt}{0.400pt}}
\multiput(888.00,323.17)(6.500,1.000){2}{\rule{1.566pt}{0.400pt}}
\put(901,324.67){\rule{3.373pt}{0.400pt}}
\multiput(901.00,324.17)(7.000,1.000){2}{\rule{1.686pt}{0.400pt}}
\put(915,325.67){\rule{3.132pt}{0.400pt}}
\multiput(915.00,325.17)(6.500,1.000){2}{\rule{1.566pt}{0.400pt}}
\put(928,326.67){\rule{3.132pt}{0.400pt}}
\multiput(928.00,326.17)(6.500,1.000){2}{\rule{1.566pt}{0.400pt}}
\put(941,327.67){\rule{3.132pt}{0.400pt}}
\multiput(941.00,327.17)(6.500,1.000){2}{\rule{1.566pt}{0.400pt}}
\put(954,328.67){\rule{3.132pt}{0.400pt}}
\multiput(954.00,328.17)(6.500,1.000){2}{\rule{1.566pt}{0.400pt}}
\put(967,329.67){\rule{3.132pt}{0.400pt}}
\multiput(967.00,329.17)(6.500,1.000){2}{\rule{1.566pt}{0.400pt}}
\put(993,330.67){\rule{3.132pt}{0.400pt}}
\multiput(993.00,330.17)(6.500,1.000){2}{\rule{1.566pt}{0.400pt}}
\put(1006,331.67){\rule{3.132pt}{0.400pt}}
\multiput(1006.00,331.17)(6.500,1.000){2}{\rule{1.566pt}{0.400pt}}
\put(1019,332.67){\rule{3.373pt}{0.400pt}}
\multiput(1019.00,332.17)(7.000,1.000){2}{\rule{1.686pt}{0.400pt}}
\put(1033,333.67){\rule{3.132pt}{0.400pt}}
\multiput(1033.00,333.17)(6.500,1.000){2}{\rule{1.566pt}{0.400pt}}
\put(1046,334.67){\rule{3.132pt}{0.400pt}}
\multiput(1046.00,334.17)(6.500,1.000){2}{\rule{1.566pt}{0.400pt}}
\put(980.0,331.0){\rule[-0.200pt]{3.132pt}{0.400pt}}
\put(1072,335.67){\rule{3.132pt}{0.400pt}}
\multiput(1072.00,335.17)(6.500,1.000){2}{\rule{1.566pt}{0.400pt}}
\put(1085,336.67){\rule{3.132pt}{0.400pt}}
\multiput(1085.00,336.17)(6.500,1.000){2}{\rule{1.566pt}{0.400pt}}
\put(1098,337.67){\rule{3.132pt}{0.400pt}}
\multiput(1098.00,337.17)(6.500,1.000){2}{\rule{1.566pt}{0.400pt}}
\put(1059.0,336.0){\rule[-0.200pt]{3.132pt}{0.400pt}}
\put(1124,338.67){\rule{3.132pt}{0.400pt}}
\multiput(1124.00,338.17)(6.500,1.000){2}{\rule{1.566pt}{0.400pt}}
\put(1137,339.67){\rule{3.373pt}{0.400pt}}
\multiput(1137.00,339.17)(7.000,1.000){2}{\rule{1.686pt}{0.400pt}}
\put(1151,340.67){\rule{3.132pt}{0.400pt}}
\multiput(1151.00,340.17)(6.500,1.000){2}{\rule{1.566pt}{0.400pt}}
\put(1111.0,339.0){\rule[-0.200pt]{3.132pt}{0.400pt}}
\put(1177,341.67){\rule{3.132pt}{0.400pt}}
\multiput(1177.00,341.17)(6.500,1.000){2}{\rule{1.566pt}{0.400pt}}
\put(1190,342.67){\rule{3.132pt}{0.400pt}}
\multiput(1190.00,342.17)(6.500,1.000){2}{\rule{1.566pt}{0.400pt}}
\put(1164.0,342.0){\rule[-0.200pt]{3.132pt}{0.400pt}}
\put(1216,343.67){\rule{3.132pt}{0.400pt}}
\multiput(1216.00,343.17)(6.500,1.000){2}{\rule{1.566pt}{0.400pt}}
\put(1229,344.67){\rule{3.132pt}{0.400pt}}
\multiput(1229.00,344.17)(6.500,1.000){2}{\rule{1.566pt}{0.400pt}}
\put(1203.0,344.0){\rule[-0.200pt]{3.132pt}{0.400pt}}
\put(1255,345.67){\rule{3.373pt}{0.400pt}}
\multiput(1255.00,345.17)(7.000,1.000){2}{\rule{1.686pt}{0.400pt}}
\put(1269,346.67){\rule{3.132pt}{0.400pt}}
\multiput(1269.00,346.17)(6.500,1.000){2}{\rule{1.566pt}{0.400pt}}
\put(1242.0,346.0){\rule[-0.200pt]{3.132pt}{0.400pt}}
\put(1295,347.67){\rule{3.132pt}{0.400pt}}
\multiput(1295.00,347.17)(6.500,1.000){2}{\rule{1.566pt}{0.400pt}}
\put(1282.0,348.0){\rule[-0.200pt]{3.132pt}{0.400pt}}
\put(1321,348.67){\rule{3.132pt}{0.400pt}}
\multiput(1321.00,348.17)(6.500,1.000){2}{\rule{1.566pt}{0.400pt}}
\put(1334,349.67){\rule{3.132pt}{0.400pt}}
\multiput(1334.00,349.17)(6.500,1.000){2}{\rule{1.566pt}{0.400pt}}
\put(1308.0,349.0){\rule[-0.200pt]{3.132pt}{0.400pt}}
\put(1360,350.67){\rule{3.132pt}{0.400pt}}
\multiput(1360.00,350.17)(6.500,1.000){2}{\rule{1.566pt}{0.400pt}}
\put(1347.0,351.0){\rule[-0.200pt]{3.132pt}{0.400pt}}
\put(1387,351.67){\rule{3.132pt}{0.400pt}}
\multiput(1387.00,351.17)(6.500,1.000){2}{\rule{1.566pt}{0.400pt}}
\put(1400,352.67){\rule{3.132pt}{0.400pt}}
\multiput(1400.00,352.17)(6.500,1.000){2}{\rule{1.566pt}{0.400pt}}
\put(1373.0,352.0){\rule[-0.200pt]{3.373pt}{0.400pt}}
\put(1426,353.67){\rule{3.132pt}{0.400pt}}
\multiput(1426.00,353.17)(6.500,1.000){2}{\rule{1.566pt}{0.400pt}}
\put(1413.0,354.0){\rule[-0.200pt]{3.132pt}{0.400pt}}
\sbox{\plotpoint}{\rule[-0.600pt]{1.200pt}{1.200pt}}%
\put(198,430){\circle*{18}}
\put(255,491){\circle*{18}}
\put(312,614){\circle*{18}}
\put(369,614){\circle*{18}}
\put(427,614){\circle*{18}}
\put(484,614){\circle*{18}}
\put(541,676){\circle*{18}}
\put(598,676){\circle*{18}}
\put(655,676){\circle*{18}}
\put(712,676){\circle*{18}}
\put(769,676){\circle*{18}}
\put(826,676){\circle*{18}}
\put(883,676){\circle*{18}}
\put(941,737){\circle*{18}}
\put(998,737){\circle*{18}}
\put(1055,737){\circle*{18}}
\put(1112,737){\circle*{18}}
\put(1169,799){\circle*{18}}
\put(1226,799){\circle*{18}}
\put(1283,799){\circle*{18}}
\sbox{\plotpoint}{\rule[-0.200pt]{0.400pt}{0.400pt}}%
\put(154,161){\usebox{\plotpoint}}
\multiput(154.58,161.00)(0.493,3.867){23}{\rule{0.119pt}{3.115pt}}
\multiput(153.17,161.00)(13.000,91.534){2}{\rule{0.400pt}{1.558pt}}
\multiput(167.58,259.00)(0.493,2.281){23}{\rule{0.119pt}{1.885pt}}
\multiput(166.17,259.00)(13.000,54.088){2}{\rule{0.400pt}{0.942pt}}
\multiput(180.58,317.00)(0.493,1.567){23}{\rule{0.119pt}{1.331pt}}
\multiput(179.17,317.00)(13.000,37.238){2}{\rule{0.400pt}{0.665pt}}
\multiput(193.58,357.00)(0.494,1.158){25}{\rule{0.119pt}{1.014pt}}
\multiput(192.17,357.00)(14.000,29.895){2}{\rule{0.400pt}{0.507pt}}
\multiput(207.58,389.00)(0.493,1.012){23}{\rule{0.119pt}{0.900pt}}
\multiput(206.17,389.00)(13.000,24.132){2}{\rule{0.400pt}{0.450pt}}
\multiput(220.58,415.00)(0.493,0.814){23}{\rule{0.119pt}{0.746pt}}
\multiput(219.17,415.00)(13.000,19.451){2}{\rule{0.400pt}{0.373pt}}
\multiput(233.58,436.00)(0.493,0.734){23}{\rule{0.119pt}{0.685pt}}
\multiput(232.17,436.00)(13.000,17.579){2}{\rule{0.400pt}{0.342pt}}
\multiput(246.58,455.00)(0.493,0.655){23}{\rule{0.119pt}{0.623pt}}
\multiput(245.17,455.00)(13.000,15.707){2}{\rule{0.400pt}{0.312pt}}
\multiput(259.58,472.00)(0.493,0.576){23}{\rule{0.119pt}{0.562pt}}
\multiput(258.17,472.00)(13.000,13.834){2}{\rule{0.400pt}{0.281pt}}
\multiput(272.00,487.58)(0.497,0.493){23}{\rule{0.500pt}{0.119pt}}
\multiput(272.00,486.17)(11.962,13.000){2}{\rule{0.250pt}{0.400pt}}
\multiput(285.00,500.58)(0.497,0.493){23}{\rule{0.500pt}{0.119pt}}
\multiput(285.00,499.17)(11.962,13.000){2}{\rule{0.250pt}{0.400pt}}
\multiput(298.00,513.58)(0.590,0.492){19}{\rule{0.573pt}{0.118pt}}
\multiput(298.00,512.17)(11.811,11.000){2}{\rule{0.286pt}{0.400pt}}
\multiput(311.00,524.58)(0.637,0.492){19}{\rule{0.609pt}{0.118pt}}
\multiput(311.00,523.17)(12.736,11.000){2}{\rule{0.305pt}{0.400pt}}
\multiput(325.00,535.59)(0.728,0.489){15}{\rule{0.678pt}{0.118pt}}
\multiput(325.00,534.17)(11.593,9.000){2}{\rule{0.339pt}{0.400pt}}
\multiput(338.00,544.59)(0.728,0.489){15}{\rule{0.678pt}{0.118pt}}
\multiput(338.00,543.17)(11.593,9.000){2}{\rule{0.339pt}{0.400pt}}
\multiput(351.00,553.59)(0.728,0.489){15}{\rule{0.678pt}{0.118pt}}
\multiput(351.00,552.17)(11.593,9.000){2}{\rule{0.339pt}{0.400pt}}
\multiput(364.00,562.59)(0.824,0.488){13}{\rule{0.750pt}{0.117pt}}
\multiput(364.00,561.17)(11.443,8.000){2}{\rule{0.375pt}{0.400pt}}
\multiput(377.00,570.59)(0.824,0.488){13}{\rule{0.750pt}{0.117pt}}
\multiput(377.00,569.17)(11.443,8.000){2}{\rule{0.375pt}{0.400pt}}
\multiput(390.00,578.59)(0.950,0.485){11}{\rule{0.843pt}{0.117pt}}
\multiput(390.00,577.17)(11.251,7.000){2}{\rule{0.421pt}{0.400pt}}
\multiput(403.00,585.59)(0.950,0.485){11}{\rule{0.843pt}{0.117pt}}
\multiput(403.00,584.17)(11.251,7.000){2}{\rule{0.421pt}{0.400pt}}
\multiput(416.00,592.59)(0.950,0.485){11}{\rule{0.843pt}{0.117pt}}
\multiput(416.00,591.17)(11.251,7.000){2}{\rule{0.421pt}{0.400pt}}
\multiput(429.00,599.59)(1.214,0.482){9}{\rule{1.033pt}{0.116pt}}
\multiput(429.00,598.17)(11.855,6.000){2}{\rule{0.517pt}{0.400pt}}
\multiput(443.00,605.59)(1.123,0.482){9}{\rule{0.967pt}{0.116pt}}
\multiput(443.00,604.17)(10.994,6.000){2}{\rule{0.483pt}{0.400pt}}
\multiput(456.00,611.59)(1.123,0.482){9}{\rule{0.967pt}{0.116pt}}
\multiput(456.00,610.17)(10.994,6.000){2}{\rule{0.483pt}{0.400pt}}
\multiput(469.00,617.59)(1.378,0.477){7}{\rule{1.140pt}{0.115pt}}
\multiput(469.00,616.17)(10.634,5.000){2}{\rule{0.570pt}{0.400pt}}
\multiput(482.00,622.59)(1.378,0.477){7}{\rule{1.140pt}{0.115pt}}
\multiput(482.00,621.17)(10.634,5.000){2}{\rule{0.570pt}{0.400pt}}
\multiput(495.00,627.59)(1.123,0.482){9}{\rule{0.967pt}{0.116pt}}
\multiput(495.00,626.17)(10.994,6.000){2}{\rule{0.483pt}{0.400pt}}
\multiput(508.00,633.59)(1.378,0.477){7}{\rule{1.140pt}{0.115pt}}
\multiput(508.00,632.17)(10.634,5.000){2}{\rule{0.570pt}{0.400pt}}
\multiput(521.00,638.60)(1.797,0.468){5}{\rule{1.400pt}{0.113pt}}
\multiput(521.00,637.17)(10.094,4.000){2}{\rule{0.700pt}{0.400pt}}
\multiput(534.00,642.59)(1.378,0.477){7}{\rule{1.140pt}{0.115pt}}
\multiput(534.00,641.17)(10.634,5.000){2}{\rule{0.570pt}{0.400pt}}
\multiput(547.00,647.59)(1.489,0.477){7}{\rule{1.220pt}{0.115pt}}
\multiput(547.00,646.17)(11.468,5.000){2}{\rule{0.610pt}{0.400pt}}
\multiput(561.00,652.60)(1.797,0.468){5}{\rule{1.400pt}{0.113pt}}
\multiput(561.00,651.17)(10.094,4.000){2}{\rule{0.700pt}{0.400pt}}
\multiput(574.00,656.60)(1.797,0.468){5}{\rule{1.400pt}{0.113pt}}
\multiput(574.00,655.17)(10.094,4.000){2}{\rule{0.700pt}{0.400pt}}
\multiput(587.00,660.60)(1.797,0.468){5}{\rule{1.400pt}{0.113pt}}
\multiput(587.00,659.17)(10.094,4.000){2}{\rule{0.700pt}{0.400pt}}
\multiput(600.00,664.60)(1.797,0.468){5}{\rule{1.400pt}{0.113pt}}
\multiput(600.00,663.17)(10.094,4.000){2}{\rule{0.700pt}{0.400pt}}
\multiput(613.00,668.60)(1.797,0.468){5}{\rule{1.400pt}{0.113pt}}
\multiput(613.00,667.17)(10.094,4.000){2}{\rule{0.700pt}{0.400pt}}
\multiput(626.00,672.60)(1.797,0.468){5}{\rule{1.400pt}{0.113pt}}
\multiput(626.00,671.17)(10.094,4.000){2}{\rule{0.700pt}{0.400pt}}
\multiput(639.00,676.60)(1.797,0.468){5}{\rule{1.400pt}{0.113pt}}
\multiput(639.00,675.17)(10.094,4.000){2}{\rule{0.700pt}{0.400pt}}
\multiput(652.00,680.61)(2.695,0.447){3}{\rule{1.833pt}{0.108pt}}
\multiput(652.00,679.17)(9.195,3.000){2}{\rule{0.917pt}{0.400pt}}
\multiput(665.00,683.60)(1.943,0.468){5}{\rule{1.500pt}{0.113pt}}
\multiput(665.00,682.17)(10.887,4.000){2}{\rule{0.750pt}{0.400pt}}
\multiput(679.00,687.61)(2.695,0.447){3}{\rule{1.833pt}{0.108pt}}
\multiput(679.00,686.17)(9.195,3.000){2}{\rule{0.917pt}{0.400pt}}
\multiput(692.00,690.61)(2.695,0.447){3}{\rule{1.833pt}{0.108pt}}
\multiput(692.00,689.17)(9.195,3.000){2}{\rule{0.917pt}{0.400pt}}
\multiput(705.00,693.60)(1.797,0.468){5}{\rule{1.400pt}{0.113pt}}
\multiput(705.00,692.17)(10.094,4.000){2}{\rule{0.700pt}{0.400pt}}
\multiput(718.00,697.61)(2.695,0.447){3}{\rule{1.833pt}{0.108pt}}
\multiput(718.00,696.17)(9.195,3.000){2}{\rule{0.917pt}{0.400pt}}
\multiput(731.00,700.61)(2.695,0.447){3}{\rule{1.833pt}{0.108pt}}
\multiput(731.00,699.17)(9.195,3.000){2}{\rule{0.917pt}{0.400pt}}
\multiput(744.00,703.61)(2.695,0.447){3}{\rule{1.833pt}{0.108pt}}
\multiput(744.00,702.17)(9.195,3.000){2}{\rule{0.917pt}{0.400pt}}
\multiput(757.00,706.61)(2.695,0.447){3}{\rule{1.833pt}{0.108pt}}
\multiput(757.00,705.17)(9.195,3.000){2}{\rule{0.917pt}{0.400pt}}
\multiput(770.00,709.61)(2.695,0.447){3}{\rule{1.833pt}{0.108pt}}
\multiput(770.00,708.17)(9.195,3.000){2}{\rule{0.917pt}{0.400pt}}
\multiput(783.00,712.61)(2.918,0.447){3}{\rule{1.967pt}{0.108pt}}
\multiput(783.00,711.17)(9.918,3.000){2}{\rule{0.983pt}{0.400pt}}
\put(797,715.17){\rule{2.700pt}{0.400pt}}
\multiput(797.00,714.17)(7.396,2.000){2}{\rule{1.350pt}{0.400pt}}
\multiput(810.00,717.61)(2.695,0.447){3}{\rule{1.833pt}{0.108pt}}
\multiput(810.00,716.17)(9.195,3.000){2}{\rule{0.917pt}{0.400pt}}
\multiput(823.00,720.61)(2.695,0.447){3}{\rule{1.833pt}{0.108pt}}
\multiput(823.00,719.17)(9.195,3.000){2}{\rule{0.917pt}{0.400pt}}
\multiput(836.00,723.61)(2.695,0.447){3}{\rule{1.833pt}{0.108pt}}
\multiput(836.00,722.17)(9.195,3.000){2}{\rule{0.917pt}{0.400pt}}
\put(849,726.17){\rule{2.700pt}{0.400pt}}
\multiput(849.00,725.17)(7.396,2.000){2}{\rule{1.350pt}{0.400pt}}
\multiput(862.00,728.61)(2.695,0.447){3}{\rule{1.833pt}{0.108pt}}
\multiput(862.00,727.17)(9.195,3.000){2}{\rule{0.917pt}{0.400pt}}
\put(875,731.17){\rule{2.700pt}{0.400pt}}
\multiput(875.00,730.17)(7.396,2.000){2}{\rule{1.350pt}{0.400pt}}
\multiput(888.00,733.61)(2.695,0.447){3}{\rule{1.833pt}{0.108pt}}
\multiput(888.00,732.17)(9.195,3.000){2}{\rule{0.917pt}{0.400pt}}
\put(901,736.17){\rule{2.900pt}{0.400pt}}
\multiput(901.00,735.17)(7.981,2.000){2}{\rule{1.450pt}{0.400pt}}
\put(915,738.17){\rule{2.700pt}{0.400pt}}
\multiput(915.00,737.17)(7.396,2.000){2}{\rule{1.350pt}{0.400pt}}
\multiput(928.00,740.61)(2.695,0.447){3}{\rule{1.833pt}{0.108pt}}
\multiput(928.00,739.17)(9.195,3.000){2}{\rule{0.917pt}{0.400pt}}
\put(941,743.17){\rule{2.700pt}{0.400pt}}
\multiput(941.00,742.17)(7.396,2.000){2}{\rule{1.350pt}{0.400pt}}
\put(954,745.17){\rule{2.700pt}{0.400pt}}
\multiput(954.00,744.17)(7.396,2.000){2}{\rule{1.350pt}{0.400pt}}
\multiput(967.00,747.61)(2.695,0.447){3}{\rule{1.833pt}{0.108pt}}
\multiput(967.00,746.17)(9.195,3.000){2}{\rule{0.917pt}{0.400pt}}
\put(980,750.17){\rule{2.700pt}{0.400pt}}
\multiput(980.00,749.17)(7.396,2.000){2}{\rule{1.350pt}{0.400pt}}
\put(993,752.17){\rule{2.700pt}{0.400pt}}
\multiput(993.00,751.17)(7.396,2.000){2}{\rule{1.350pt}{0.400pt}}
\put(1006,754.17){\rule{2.700pt}{0.400pt}}
\multiput(1006.00,753.17)(7.396,2.000){2}{\rule{1.350pt}{0.400pt}}
\put(1019,756.17){\rule{2.900pt}{0.400pt}}
\multiput(1019.00,755.17)(7.981,2.000){2}{\rule{1.450pt}{0.400pt}}
\put(1033,758.17){\rule{2.700pt}{0.400pt}}
\multiput(1033.00,757.17)(7.396,2.000){2}{\rule{1.350pt}{0.400pt}}
\put(1046,760.17){\rule{2.700pt}{0.400pt}}
\multiput(1046.00,759.17)(7.396,2.000){2}{\rule{1.350pt}{0.400pt}}
\put(1059,762.17){\rule{2.700pt}{0.400pt}}
\multiput(1059.00,761.17)(7.396,2.000){2}{\rule{1.350pt}{0.400pt}}
\put(1072,764.17){\rule{2.700pt}{0.400pt}}
\multiput(1072.00,763.17)(7.396,2.000){2}{\rule{1.350pt}{0.400pt}}
\put(1085,766.17){\rule{2.700pt}{0.400pt}}
\multiput(1085.00,765.17)(7.396,2.000){2}{\rule{1.350pt}{0.400pt}}
\put(1098,768.17){\rule{2.700pt}{0.400pt}}
\multiput(1098.00,767.17)(7.396,2.000){2}{\rule{1.350pt}{0.400pt}}
\put(1111,770.17){\rule{2.700pt}{0.400pt}}
\multiput(1111.00,769.17)(7.396,2.000){2}{\rule{1.350pt}{0.400pt}}
\put(1124,772.17){\rule{2.700pt}{0.400pt}}
\multiput(1124.00,771.17)(7.396,2.000){2}{\rule{1.350pt}{0.400pt}}
\put(1137,774.17){\rule{2.900pt}{0.400pt}}
\multiput(1137.00,773.17)(7.981,2.000){2}{\rule{1.450pt}{0.400pt}}
\put(1151,776.17){\rule{2.700pt}{0.400pt}}
\multiput(1151.00,775.17)(7.396,2.000){2}{\rule{1.350pt}{0.400pt}}
\put(1164,777.67){\rule{3.132pt}{0.400pt}}
\multiput(1164.00,777.17)(6.500,1.000){2}{\rule{1.566pt}{0.400pt}}
\put(1177,779.17){\rule{2.700pt}{0.400pt}}
\multiput(1177.00,778.17)(7.396,2.000){2}{\rule{1.350pt}{0.400pt}}
\put(1190,781.17){\rule{2.700pt}{0.400pt}}
\multiput(1190.00,780.17)(7.396,2.000){2}{\rule{1.350pt}{0.400pt}}
\put(1203,783.17){\rule{2.700pt}{0.400pt}}
\multiput(1203.00,782.17)(7.396,2.000){2}{\rule{1.350pt}{0.400pt}}
\put(1216,784.67){\rule{3.132pt}{0.400pt}}
\multiput(1216.00,784.17)(6.500,1.000){2}{\rule{1.566pt}{0.400pt}}
\put(1229,786.17){\rule{2.700pt}{0.400pt}}
\multiput(1229.00,785.17)(7.396,2.000){2}{\rule{1.350pt}{0.400pt}}
\put(1242,788.17){\rule{2.700pt}{0.400pt}}
\multiput(1242.00,787.17)(7.396,2.000){2}{\rule{1.350pt}{0.400pt}}
\put(1255,789.67){\rule{3.373pt}{0.400pt}}
\multiput(1255.00,789.17)(7.000,1.000){2}{\rule{1.686pt}{0.400pt}}
\put(1269,791.17){\rule{2.700pt}{0.400pt}}
\multiput(1269.00,790.17)(7.396,2.000){2}{\rule{1.350pt}{0.400pt}}
\put(1282,793.17){\rule{2.700pt}{0.400pt}}
\multiput(1282.00,792.17)(7.396,2.000){2}{\rule{1.350pt}{0.400pt}}
\put(1295,794.67){\rule{3.132pt}{0.400pt}}
\multiput(1295.00,794.17)(6.500,1.000){2}{\rule{1.566pt}{0.400pt}}
\put(1308,796.17){\rule{2.700pt}{0.400pt}}
\multiput(1308.00,795.17)(7.396,2.000){2}{\rule{1.350pt}{0.400pt}}
\put(1321,797.67){\rule{3.132pt}{0.400pt}}
\multiput(1321.00,797.17)(6.500,1.000){2}{\rule{1.566pt}{0.400pt}}
\put(1334,799.17){\rule{2.700pt}{0.400pt}}
\multiput(1334.00,798.17)(7.396,2.000){2}{\rule{1.350pt}{0.400pt}}
\put(1347,801.17){\rule{2.700pt}{0.400pt}}
\multiput(1347.00,800.17)(7.396,2.000){2}{\rule{1.350pt}{0.400pt}}
\put(1360,802.67){\rule{3.132pt}{0.400pt}}
\multiput(1360.00,802.17)(6.500,1.000){2}{\rule{1.566pt}{0.400pt}}
\put(1373,804.17){\rule{2.900pt}{0.400pt}}
\multiput(1373.00,803.17)(7.981,2.000){2}{\rule{1.450pt}{0.400pt}}
\put(1387,805.67){\rule{3.132pt}{0.400pt}}
\multiput(1387.00,805.17)(6.500,1.000){2}{\rule{1.566pt}{0.400pt}}
\put(1400,806.67){\rule{3.132pt}{0.400pt}}
\multiput(1400.00,806.17)(6.500,1.000){2}{\rule{1.566pt}{0.400pt}}
\put(1413,808.17){\rule{2.700pt}{0.400pt}}
\multiput(1413.00,807.17)(7.396,2.000){2}{\rule{1.350pt}{0.400pt}}
\put(1426,809.67){\rule{3.132pt}{0.400pt}}
\multiput(1426.00,809.17)(6.500,1.000){2}{\rule{1.566pt}{0.400pt}}
\end{picture}

%% file: deltamax.mod.l.tex
\setlength{\unitlength}{0.240900pt}
\ifx\plotpoint\undefined\newsavebox{\plotpoint}\fi
\sbox{\plotpoint}{\rule[-0.200pt]{0.400pt}{0.400pt}}%
\begin{picture}(1500,900)(0,0)
\font\gnuplot=cmr10 at 10pt
\gnuplot
\sbox{\plotpoint}{\rule[-0.200pt]{0.400pt}{0.400pt}}%
\put(161.0,123.0){\rule[-0.200pt]{4.818pt}{0.400pt}}
\put(141,123){\makebox(0,0)[r]{0.1}}
\put(1419.0,123.0){\rule[-0.200pt]{4.818pt}{0.400pt}}
\put(161.0,197.0){\rule[-0.200pt]{2.409pt}{0.400pt}}
\put(1429.0,197.0){\rule[-0.200pt]{2.409pt}{0.400pt}}
\put(161.0,240.0){\rule[-0.200pt]{2.409pt}{0.400pt}}
\put(1429.0,240.0){\rule[-0.200pt]{2.409pt}{0.400pt}}
\put(161.0,271.0){\rule[-0.200pt]{2.409pt}{0.400pt}}
\put(1429.0,271.0){\rule[-0.200pt]{2.409pt}{0.400pt}}
\put(161.0,295.0){\rule[-0.200pt]{2.409pt}{0.400pt}}
\put(1429.0,295.0){\rule[-0.200pt]{2.409pt}{0.400pt}}
\put(161.0,314.0){\rule[-0.200pt]{2.409pt}{0.400pt}}
\put(1429.0,314.0){\rule[-0.200pt]{2.409pt}{0.400pt}}
\put(161.0,331.0){\rule[-0.200pt]{2.409pt}{0.400pt}}
\put(1429.0,331.0){\rule[-0.200pt]{2.409pt}{0.400pt}}
\put(161.0,345.0){\rule[-0.200pt]{2.409pt}{0.400pt}}
\put(1429.0,345.0){\rule[-0.200pt]{2.409pt}{0.400pt}}
\put(161.0,357.0){\rule[-0.200pt]{2.409pt}{0.400pt}}
\put(1429.0,357.0){\rule[-0.200pt]{2.409pt}{0.400pt}}
\put(161.0,369.0){\rule[-0.200pt]{4.818pt}{0.400pt}}
\put(141,369){\makebox(0,0)[r]{1}}
\put(1419.0,369.0){\rule[-0.200pt]{4.818pt}{0.400pt}}
\put(161.0,443.0){\rule[-0.200pt]{2.409pt}{0.400pt}}
\put(1429.0,443.0){\rule[-0.200pt]{2.409pt}{0.400pt}}
\put(161.0,486.0){\rule[-0.200pt]{2.409pt}{0.400pt}}
\put(1429.0,486.0){\rule[-0.200pt]{2.409pt}{0.400pt}}
\put(161.0,517.0){\rule[-0.200pt]{2.409pt}{0.400pt}}
\put(1429.0,517.0){\rule[-0.200pt]{2.409pt}{0.400pt}}
\put(161.0,540.0){\rule[-0.200pt]{2.409pt}{0.400pt}}
\put(1429.0,540.0){\rule[-0.200pt]{2.409pt}{0.400pt}}
\put(161.0,560.0){\rule[-0.200pt]{2.409pt}{0.400pt}}
\put(1429.0,560.0){\rule[-0.200pt]{2.409pt}{0.400pt}}
\put(161.0,576.0){\rule[-0.200pt]{2.409pt}{0.400pt}}
\put(1429.0,576.0){\rule[-0.200pt]{2.409pt}{0.400pt}}
\put(161.0,591.0){\rule[-0.200pt]{2.409pt}{0.400pt}}
\put(1429.0,591.0){\rule[-0.200pt]{2.409pt}{0.400pt}}
\put(161.0,603.0){\rule[-0.200pt]{2.409pt}{0.400pt}}
\put(1429.0,603.0){\rule[-0.200pt]{2.409pt}{0.400pt}}
\put(161.0,614.0){\rule[-0.200pt]{4.818pt}{0.400pt}}
\put(141,614){\makebox(0,0)[r]{10}}
\put(1419.0,614.0){\rule[-0.200pt]{4.818pt}{0.400pt}}
\put(161.0,688.0){\rule[-0.200pt]{2.409pt}{0.400pt}}
\put(1429.0,688.0){\rule[-0.200pt]{2.409pt}{0.400pt}}
\put(161.0,732.0){\rule[-0.200pt]{2.409pt}{0.400pt}}
\put(1429.0,732.0){\rule[-0.200pt]{2.409pt}{0.400pt}}
\put(161.0,762.0){\rule[-0.200pt]{2.409pt}{0.400pt}}
\put(1429.0,762.0){\rule[-0.200pt]{2.409pt}{0.400pt}}
\put(161.0,786.0){\rule[-0.200pt]{2.409pt}{0.400pt}}
\put(1429.0,786.0){\rule[-0.200pt]{2.409pt}{0.400pt}}
\put(161.0,805.0){\rule[-0.200pt]{2.409pt}{0.400pt}}
\put(1429.0,805.0){\rule[-0.200pt]{2.409pt}{0.400pt}}
\put(161.0,822.0){\rule[-0.200pt]{2.409pt}{0.400pt}}
\put(1429.0,822.0){\rule[-0.200pt]{2.409pt}{0.400pt}}
\put(161.0,836.0){\rule[-0.200pt]{2.409pt}{0.400pt}}
\put(1429.0,836.0){\rule[-0.200pt]{2.409pt}{0.400pt}}
\put(161.0,849.0){\rule[-0.200pt]{2.409pt}{0.400pt}}
\put(1429.0,849.0){\rule[-0.200pt]{2.409pt}{0.400pt}}
\put(161.0,860.0){\rule[-0.200pt]{4.818pt}{0.400pt}}
\put(141,860){\makebox(0,0)[r]{100}}
\put(1419.0,860.0){\rule[-0.200pt]{4.818pt}{0.400pt}}
\put(161.0,123.0){\rule[-0.200pt]{0.400pt}{4.818pt}}
\put(161,82){\makebox(0,0){0}}
\put(161.0,840.0){\rule[-0.200pt]{0.400pt}{4.818pt}}
\put(417.0,123.0){\rule[-0.200pt]{0.400pt}{4.818pt}}
\put(417,82){\makebox(0,0){50}}
\put(417.0,840.0){\rule[-0.200pt]{0.400pt}{4.818pt}}
\put(672.0,123.0){\rule[-0.200pt]{0.400pt}{4.818pt}}
\put(672,82){\makebox(0,0){100}}
\put(672.0,840.0){\rule[-0.200pt]{0.400pt}{4.818pt}}
\put(928.0,123.0){\rule[-0.200pt]{0.400pt}{4.818pt}}
\put(928,82){\makebox(0,0){150}}
\put(928.0,840.0){\rule[-0.200pt]{0.400pt}{4.818pt}}
\put(1183.0,123.0){\rule[-0.200pt]{0.400pt}{4.818pt}}
\put(1183,82){\makebox(0,0){200}}
\put(1183.0,840.0){\rule[-0.200pt]{0.400pt}{4.818pt}}
\put(1439.0,123.0){\rule[-0.200pt]{0.400pt}{4.818pt}}
\put(1439,82){\makebox(0,0){250}}
\put(1439.0,840.0){\rule[-0.200pt]{0.400pt}{4.818pt}}
\put(161.0,123.0){\rule[-0.200pt]{307.870pt}{0.400pt}}
\put(1439.0,123.0){\rule[-0.200pt]{0.400pt}{177.543pt}}
\put(161.0,860.0){\rule[-0.200pt]{307.870pt}{0.400pt}}
\put(40,491){\makebox(0,0){ $\delta L_z^{max}$ } }
\put(800,21){\makebox(0,0){ $l$ }}
\put(161.0,123.0){\rule[-0.200pt]{0.400pt}{177.543pt}}
\put(192,357){\circle*{18}}
\put(220,357){\circle*{18}}
\put(248,362){\circle*{18}}
\put(276,349){\circle*{18}}
\put(304,332){\circle*{18}}
\put(332,365){\circle*{18}}
\put(360,358){\circle*{18}}
\put(388,356){\circle*{18}}
\put(417,348){\circle*{18}}
\put(445,383){\circle*{18}}
\put(501,392){\circle*{18}}
\put(557,383){\circle*{18}}
\put(613,392){\circle*{18}}
\put(670,368){\circle*{18}}
\put(723,370){\circle*{18}}
\put(780,354){\circle*{18}}
\put(836,356){\circle*{18}}
\put(892,422){\circle*{18}}
\put(948,407){\circle*{18}}
\put(1004,432){\circle*{18}}
\put(1061,442){\circle*{18}}
\put(1117,411){\circle*{18}}
\put(1173,441){\circle*{18}}
\put(1229,433){\circle*{18}}
\put(1286,427){\circle*{18}}
\put(192,317){\circle{18}}
\put(220,373){\circle{18}}
\put(248,403){\circle{18}}
\put(276,423){\circle{18}}
\put(304,451){\circle{18}}
\put(332,490){\circle{18}}
\put(360,489){\circle{18}}
\put(388,487){\circle{18}}
\put(417,434){\circle{18}}
\put(445,418){\circle{18}}
\put(501,512){\circle{18}}
\put(557,528){\circle{18}}
\put(613,571){\circle{18}}
\put(670,594){\circle{18}}
\put(723,614){\circle{18}}
\put(780,634){\circle{18}}
\put(836,647){\circle{18}}
\put(892,656){\circle{18}}
\put(948,665){\circle{18}}
\put(1004,675){\circle{18}}
\put(1061,684){\circle{18}}
\put(1117,681){\circle{18}}
\put(1173,676){\circle{18}}
\put(1229,678){\circle{18}}
\put(1286,670){\circle{18}}
\end{picture}